\documentclass[3p, 12pt, sort&compress]{elsarticle}

\pdfoutput = 1

\usepackage{bm, dsfont, 
  amssymb, amsmath,
  graphicx,
  hyperref,
  color,
  braket,
  array,
  slashed,
  mathrsfs,
  subfig,
  upgreek,
  slashbox}

\DeclareMathAlphabet{\mathpzc}{OT1}{pzc}{m}{it}
\numberwithin{equation}{section}

\journal{Nuclear Physics B}

\makeatletter

\def\ps@pprintTitle{
 \let\@oddhead\@empty
 \let\@evenhead\@empty
 \def\@oddfoot{}
 \let\@evenfoot\@oddfoot}

\long\def\MaketitleBox{
  \resetTitleCounters
  \def\baselinestretch{1}
  \begin{center}
   \def\baselinestretch{1}
    \Large\@title\par\vskip18pt
    \normalsize\elsauthors\par\vskip10pt
    \footnotesize\itshape\elsaddress\par\vskip36pt
    \rule{\textwidth}{1.5pt}\vskip12pt
    \ifvoid\absbox\else\unvbox\absbox\par\vskip10pt\fi
    \ifvoid\keybox\else\unvbox\keybox\par\vskip10pt\fi
    \rule{\textwidth}{1.5pt}\vskip12pt
    \end{center}
  }

\renewcommand\subsection{\@startsection{subsection}{2}{\z@}
           {18\p@ \@plus 6\p@ \@minus 3\p@}
           {9\p@ \@plus 6\p@ \@minus 3\p@}
           {\normalfont\normalsize\itshape\bfseries}}

\gdef\emailauthor#1#2{\stepcounter{ead}
     \g@addto@macro\@elseads{\raggedright
      \let\corref\@gobble
      \eadsep\newline\texttt{#1} (#2)\def\eadsep{\unskip,\space}}
}

\def\appendixname{Appendix}
\renewcommand\@makefntext[1]{#1}

\makeatother

\DeclareMathOperator{\Tr}{Tr}
\newcommand{\mat}[1]{\bm{#1}}
\newcommand{\maf}[1]{\mathbf{#1}}

\newcommand{\ve}[1]{\mathbf{#1}}

\newcommand{\D}[2]{\mathrm{d}^{#1}{#2}}
\newcommand{\smrm}[1]{{\scriptscriptstyle{\mathrm{#1}}}}

\newcommand{\UP}{{\cal U}_P}
\newcommand{\UPd}{{\cal U}^\dag_P}
\newcommand{\T}{{\cal A}_T}
\newcommand{\Td}{{\cal A}^\dag_T}

\newcommand{\lsim}{\lesssim}
\newcommand{\gsim}{\gtrsim}
\newcommand{\CP}{C\!P}
\newcommand{\gCP}{\widetilde{C}\!P}
\newcommand{\C}{{\cal U}_C}
\newcommand{\Cd}{{\cal U}^\dag_C}

\newcommand{\Edu}[4]{[E_{#1}(\ve #2)]_{#3}^{\phantom{#3} #4}}
\newcommand{\Esud}[4]
{\left[(2 E_{#1}(\ve #2))^{-{1}/{2}}\right]^{#3}_{\phantom{#3}#4}}
\newcommand{\Esdu}[4]
{\left[(2 E_{#1}(\ve #2))^{-{1}/{2}}\right]_{#3}^{\phantom{#3}#4}}

\newcommand{\su}[4]{[u(\ve #2, #1)]_{#3}^{\phantom #3 #4}}
\newcommand{\sub}[4]{[\bar{u}(\ve #2, #1)]^{#3}_{\phantom #3 #4}}
\newcommand{\sv}[4]{[v(\ve #2,#1)]_{#3}^{\phantom #3 #4}}
\newcommand{\svb}[4]{[\bar{v}(\ve #2,#1)]^{#3}_{\phantom #3 #4}}

\newcommand{\N}[3]{[n^{#1}]_{#2}^{\phantom{#2} #3}}
\newcommand{\Nb}[3]{[\bar{n}^{#1}]_{#2}^{\phantom{#2} #3}}
\newcommand{\n}[6]{[n_{#2}^{#1}(\ve #3,#6)]_{#4}^{\phantom{#4}#5}}
\newcommand{\nb}[6]
{[\bar{n}_{#2}^{#1}(\ve #3,#6)]_{#4}^{\phantom{#4}#5}}
\newcommand{\dn}[6]{[\delta n_{#2}^{#1}(\ve #3,#6)]_{#4}
  ^{\phantom{#4}#5}}
\newcommand{\nn}[6]{[\underline{n}_{#2}^{#1}(\ve #3,#6)]
  _{#4}^{\phantom{#4}#5}}

\newcommand{\DiT}{(2 \pi)^3 \delta^{(3)}}
\newcommand{\DiFud}[5]
{{(2 \pi)^4 [\delta^{(4)}(#1)]^{#2 \phantom{#3 #4} #5}
    _{\phantom #2 #3 #4}}}
\newcommand{\DiFdu}[5]
{{(2 \pi)^4 [\delta^{(4)}(#1)]_{#2 \phantom{#3 #4} #5}
    ^{\phantom #2 #3 #4}}}

\newcommand{\eud}[4]{\big[e^{#1 i #2 \cdot x}\big]^{#3}
  _{\phantom{#3}#4}}
\newcommand{\edu}[4]{\big[e^{#1 i #2 \cdot x}\big]_{#3}
  ^{\phantom{#3} #4}}

\newcommand{\h}[2]{h_{#1}^{\phantom{#1}#2}}
\newcommand{\hs}[2]{h_{\phantom{#1}#2}^{#1}}
\newcommand{\hr}[2]{\mathbf{h}_{#1}^{\phantom{#1}#2}}
\newcommand{\hrs}[2]{\mathbf{h}_{\phantom{#1}#2}^{#1}}
\newcommand{\hrc}[2]{[\mathbf{h}^{\tilde{c}}]^{#1}
  _{\phantom{#1}#2}}
\newcommand{\hrcs}[2]{[\mathbf{h}^{\tilde{c}}]
  ^{\phantom{#1}#2}_{#1}}
\newcommand{\chr}[2]{\mathbf{\widehat h}_{#1}
  ^{\phantom{#1}#2}}
\newcommand{\chrs}[2]{\mathbf{\widehat h}
  _{\phantom{#1}#2}^{#1}}

\newcommand{\chrct}[2]{[\mathbf{\widehat h}^{\tilde{c}}]^{#1}
  _{\phantom{#1}#2}}
\newcommand{\chrcst}[2]{[\mathbf{\widehat h}^{\tilde{c}}]
  ^{\phantom{#1}#2}_{#1}}

\newcommand{\Tud}[5]
{{#1}^{#2 \phantom{#3} #4 \phantom{#5}}
  _{\phantom{#2} #3 \phantom{#4}  #5}}
\newcommand{\Tdu}[5]
{{#1}_{#2 \phantom{#3} #4 \phantom{#5}}
  ^{\phantom{#2} #3 \phantom{#4}  #5}}

\begin{document}

\begin{frontmatter}
\begin{flushright}
\begin{footnotesize}
  MAN/HEP/2014/01,  
  IPPP/14/20, DCPT/14/40\\
  April 2014
\end{footnotesize} 
\end{flushright}

\vspace{-0.5cm}

\title{{\bf {\LARGE Flavour Covariant Transport Equations: \\[1mm]
    an Application to Resonant Leptogenesis}}\medskip}

\author[a]{\large P.~S.~Bhupal Dev}

\author[a,b]{\large Peter Millington}

\author[a]{\large Apostolos Pilaftsis}

\author[a]{\large Daniele Teresi}

\address[a]{\smallskip ~Consortium for Fundamental Physics,
  School of Physics and Astronomy, \\ 
  University of Manchester, Manchester M13 9PL, United Kingdom.}

\address[b]{~Institute for Particle Physics Phenomenology, 
  Durham University, Durham DH1 3LE, United Kingdom.}

\begin{abstract}

\noindent  We present  a {\em  fully} flavour-covariant  formalism for
transport  phenomena,  by  deriving  Markovian master  equations  that
describe  the  time-evolution  of   particle  number  densities  in  a
statistical   ensemble  with   arbitrary  flavour   content.    As  an
application of this  general formalism, we study flavour  effects in a
scenario    of    resonant    leptogenesis~(RL)   and    obtain    the
flavour-covariant  evolution  equations  for  heavy-neutrino  and
 lepton number densities.  This provides a complete and unified
description of  RL, capturing  three {\em distinct} physical  phenomena: (i)
the resonant  mixing between the heavy-neutrino  states, (ii) coherent
oscillations  between  different heavy-neutrino  flavours, and  (iii)
quantum  decoherence   effects  in  the   charged-lepton  sector.   To
illustrate the  importance of this formalism, we  numerically solve the
flavour-covariant rate equations for a  minimal RL model and show that
the total lepton  asymmetry can be enhanced by up to one order of
magnitude,  as  compared to  that  obtained  from flavour-diagonal  or
partially  flavour  off-diagonal rate  equations.   Thus, the  viable 
RL model  parameter  space is  enlarged,  thereby  enhancing further  the
prospects of  probing a common origin of neutrino masses and the baryon asymmetry 
in the Universe at  the LHC, as well  as in low-energy
experiments searching for lepton flavour and number violation. The key 
new ingredients
in our flavour-covariant formalism are rank-4 rate  tensors, 
which  are required for  the consistency  of our
flavour-mixing treatment,  as shown by an explicit  calculation of the
relevant  transition amplitudes by  generalizing the  optical theorem.
We  also  provide  a  geometric  and physical  interpretation  of  the
heavy-neutrino degeneracy limits in the minimal RL scenario.  Finally,
we comment on the consistency of various suggested
forms for  the heavy-neutrino self-energy  regulator 
in the  lepton-number conserving limit.

\end{abstract}

\medskip

\begin{keyword}
\begin{footnotesize}
Flavour Covariance, Discrete Symmetries, Transport Equations,
  Resonant~Leptogenesis
\end{footnotesize}
\end{keyword}

\end{frontmatter}

\makeatletter
\def\appendixname{Appendix}
\renewcommand\@makefntext[1]
{\leftskip=0em\hskip1em\@makefnmark\space #1}
\makeatother


\begin{footnotesize}
\tableofcontents
\end{footnotesize}


\section{Introduction}
\label{sec:1}

The  observed  matter-antimatter asymmetry  in  the  Universe and  the
observation  of non-zero  neutrino masses  and mixing  (for  a review,
see~\cite{pdg}) provide  two of  the strongest pieces  of experimental
evidence    for   physics    beyond   the    Standard    Model   (SM).
Leptogenesis~\cite{Fukugita:1986hr}  is   an  elegant  framework  that
satisfies   the    basic   Sakharov   conditions~\cite{Sakharov:1967},
dynamically  generating   the  observed  matter-antimatter  asymmetry.
According to  the standard paradigm of leptogenesis  (for reviews, see
e.g.~\cite{Pilaftsis:1998pd,    Buchmuller:2005eh,   Davidson:2008bu,
Blanchet:2012bk}),  there exist  heavy Majorana  neutrinos  in minimal
extensions of  the SM, whose out-of-equilibrium decays  in an expanding
Universe  create  a net  excess  of  lepton  number ($L$),  which  is
reprocessed  into  the  observed   baryon  number  ($B$)  through  the
equilibrated    $(B+L)$-violating    electroweak    sphaleron
interactions~\cite{Kuzmin:1985mm}.    In  addition,  these   heavy  
SM-singlet Majorana  neutrinos $N_\alpha$ (with $\alpha  = 1,...,\mathcal{N}_N$) 
could explain
the  observed smallness  of the light neutrino  masses by  the seesaw
mechanism~\cite{seesaw1,    seesaw2,   seesaw4,    seesaw5, seesaw6}.    Hence,
leptogenesis  can be  regarded as  a cosmological  consequence  of the
seesaw  mechanism,  thus  providing  an attractive  link  between  two
seemingly disparate pieces of evidence for new physics at or above the
electroweak scale.

In      the      original   scenario
of   thermal      leptogenesis~\cite{Fukugita:1986hr}, 
the heavy  Majorana neutrino masses are typically
close  to the  Grand Unified  Theory  (GUT) scale,  $M_{\rm GUT}  \sim
10^{16}$  GeV, as  suggested by  natural GUT  embedding of  the seesaw
mechanism~\cite{seesaw2,   seesaw4,    seesaw5}.      
In a   `vanilla'     leptogenesis 
scenario~\cite{Buchmuller:2004nz}, where the  heavy
neutrino masses are hierarchical ($m_{N_1} \ll  m_{N_{2}} < m_{N_{3}}$), 
the  solar  and atmospheric  neutrino  oscillation  data impose  a  
{\it lower} limit    on    $m_{N_1}    \gsim   10^9$    GeV~\cite{Davidson:2002qv,
Buchmuller:2002rq,  Hambye:2003rt,   Branco:2006ce}. As a consequence, 
such leptogenesis models are difficult  to  test  in  foreseeable  laboratory
experiments. Moreover,  these high-scale thermal  leptogenesis scenarios,
when   embedded  within  supergravity   models  of   inflation,  could
potentially lead to  a conflict with the upper  bound on the reheating
temperature of the Universe, $T_R  \lsim 10^6$--$10^9$ GeV, required
to avoid overproduction of  gravitinos whose late decays may otherwise
spoil     the      success     of      Big     Bang
Nucleosynthesis~\cite{Khlopov:1984pf,    Ellis:1984eq,   Ellis:1984er,
Kawasaki:1994af,    Cyburt:2002uv,    Kawasaki:2004qu,   
Kawasaki:2008qe}.  In  general, it is  difficult to build  a 
{\it    testable}  low-scale model    of   leptogenesis, with a hierarchical 
heavy neutrino mass spectrum~\cite{Pilaftsis:1998pd,
Hambye:2001eu}.

A potentially interesting solution  to the aforementioned problems may
be   obtained   within   the   framework  of   resonant   leptogenesis
(RL)~\cite{Pilaftsis:1997dr,  Pilaftsis:1997jf, Pilaftsis:2003gt}. The
key   aspect  of   RL   is  that   the   heavy  Majorana neutrino   self-energy
effects~\cite{Liu:1993tg}  on  the  leptonic $  \CP$-asymmetry  become
dominant~\cite{Flanz:1994yx, Covi:1996wh} and get resonantly enhanced,
even up to order one~\cite{Pilaftsis:1997dr, Pilaftsis:1997jf}, when 
at least two of the heavy  neutrinos have a  small mass difference  
comparable to
their  decay  widths.  As  a  consequence  of  thermal RL,  the  heavy
Majorana  neutrino  mass  scale  can  be as  low  as  the  electroweak
scale~\cite{Pilaftsis:2005rv},  while  maintaining complete  agreement
with the neutrino oscillation data~\cite{pdg}.

A crucial model-building aspect of RL is the {\it quasi-degeneracy} of
the heavy neutrino mass spectrum, which could be obtained as a natural
consequence  of  the approximate  breaking  of  some  symmetry in  the
leptonic  sector.  In  minimal  extensions  of the  SM,  there  is  no
theoretically  or phenomenologically  compelling reason  that prevents
the singlet  neutrino sector from  possessing such a symmetry  and, in
fact,  in realistic  ultraviolet-complete  extensions of  the SM,
such a symmetry can often  be realized naturally.  For instance, the
RL  model discussed  in~\cite{Pilaftsis:1997dr,  Pilaftsis:1997jf} was
based  on a  $U(1)_L$ lepton  symmetry in  the heavy  neutrino sector,
motivated       by       superstring-inspired      $E_6$  GUTs~\cite{Mohapatra:1986aw,  Nandi:1985uh, Mohapatra:1986bd}. The
small  mass splitting  between the  heavy neutrinos  was  generated by
approximate  breaking   of  this  lepton  symmetry   via  GUT-  and/or
Planck-scale-suppressed higher-dimensional  operators.  The RL  model
discussed in~\cite{Pilaftsis:2003gt} was based on the Froggatt-Nielsen
(FN)  mechanism~\cite{Froggatt:1978nt}  in  which  two  of  the  heavy
Majorana  neutrinos, having  opposite charges  under  $U(1)_{\rm FN}$,
naturally   had  a   mass   difference  comparable   to  their   decay
widths. There is a vast literature on other viable constructions of RL
models, e.g.~within  minimal extensions of  the SM~\cite{Xing:2006ms,
Hambye:2006zn,      Blanchet:2009bu,     Iso:2010mv,     Okada:2012fs,
Haba:2013pca},  with   approximate    flavour
symmetries~\cite{Ellis:2002eh,     Araki:2005ec,    Cirigliano:2006nu,
Chun:2007vh,  Babu:2007zm,  Branco:2009by}, with  variations  of  the
minimal      type-I      seesaw~\cite{Ma:1998dx,      
Albright:2003xb,     Hambye:2003rt,    Hambye:2000ui,    Asaka:2008bj,
Blanchet:2009kk}, 
within $SO(10)$ GUTs~\cite{Akhmedov:2003dg,
Albright:2004ws,  Majee:2007uv,  Blanchet:2010kw},        within    the   context    of
supersymmetric      theories~\cite{Dar:2003cr,      Allahverdi:2004ix,
Hambye:2004jf,  West:2004me,  West:2006fs},  and in  extra-dimensional
theories~\cite{Pilaftsis:1999jk,   Gherghetta:2007au,   Eisele:2007ws,
Gu:2010ye, Bechinger:2009qk}.  There also  exist other variants of the
RL    scenario,     such    as    radiative    RL~\cite{Felipe:2003fi,
Turzynski:2004xy,     Branco:2005ye,    Branco:2006hz}     and    soft
RL~\cite{Grossman:2003jv, D'Ambrosio:2003wy}.

In another  important  variant  of  RL,  a {\it  single}
lepton-flavour asymmetry is  resonantly produced by out-of-equilibrium
decays of heavy Majorana  neutrinos of a particular family type~\cite{
Pilaftsis:2004xx, Deppisch:2010fr}.  This mechanism uses the fact that
the sphaleron processes preserve, in addition to $B-L$, the individual
quantum numbers $X_i =  {B}/3 - {L}_i$~\cite{Khlebnikov:1988sr,   Harvey:1990qw,  Dreiner:1992vm,
Cline:1993vv,Laine:1999wv}, where $i = 1,  2, 3$ is the SM
family index and $L_i$ is the lepton asymmetry in the $i$th family.  
Therefore, it  is important to  estimate the  net baryon
number $B$ created by sphalerons just before they freeze out. In particular,  
a  generated baryon 
asymmetry can be protected from potentially large washout effects due
to  sphalerons  if  an individual lepton flavour  $\ell$  is  out  of
equilibrium.  We  refer to such scenarios of RL as {\it  resonant  $\ell$-genesis}
(RL$_\ell$). In  this case, the  heavy Majorana neutrinos could  be as
light as the  electroweak scale~\cite{Pilaftsis:2005rv} and still have
sizable couplings to other  charged-lepton flavours $\ell' \neq \ell$.
This     enables     the     modelling     of     minimal     RL$_\ell$
scenarios~\cite{Deppisch:2010fr} with electroweak-scale heavy Majorana
neutrinos  that  could be  {\it tested}  at the  LHC~\cite{Dev:2013wba}, 
while being  consistent with
the indirect  constraints from  various low-energy experiments  at the
intensity frontier~\cite{deGouvea:2013zba}.

Flavour effects play an important role in determining the final
lepton asymmetry in RL models. There are two kinds of flavour effects,
which  are  usually  ignored  in vanilla  leptogenesis  scenarios,
namely: (i)  heavy neutrino flavour  effects, assuming that  the final
asymmetry is  produced dominantly  by the out-of-equilibrium  decay of
only  one  (usually  the  lightest) heavy  neutrino,  with  negligible
contributions  from  heavier  species;  and (ii)  charged-lepton
flavour effects,  assuming that the flavour composition  of the lepton
quantum states produced  by (or producing) the heavy  neutrinos can be
neglected and all  leptons can be treated as  having the same flavour.
Neglecting (i) can be justified in 
`vanilla' scenarios, because the $\CP$ asymmetries due to the heavier 
Majorana neutrinos are  usually suppressed  in the hierarchical  limit $m_{N_1}
\ll m_{N_{2,3}}$. Moreover, even if a sizable asymmetry is produced by
these  effects,  it is  washed  out  by  the processes  involving  the
lightest  heavy  neutrino~\cite{Buchmuller:2004nz}.\footnote{There is  an
exception  to this  case depending  on  the flavour  structure of  the
neutrino   Yukawa   couplings,   when   the  contribution   from   the
next-to-lightest       heavy      neutrino      decay       could      be
dominant~\cite{DiBari:2005st,     Vives:2005ra}.}      However,    for
quasi-degenerate heavy neutrinos,  as in the RL case,  the flavour effects
due  to the  neutrino  Yukawa couplings  do  play an  important
role~\cite{Pilaftsis:2004xx, Endoh:2003mz}. In  fact, a sizable lepton
asymmetry   can  be  generated   through  $\CP$-violating
oscillations  of sterile neutrinos~\cite{Akhmedov:1998qx,
Asaka:2005, Shaposhnikov:2008pf, Canetti:2012kh, Drewes:2012ma},  which  is then  communicated  to  the  SM lepton  sector
through their Yukawa couplings.
 
On  the other  hand, the  lepton flavour  effects, as identified in (ii) above, 
are  related  to the
interactions       mediated       by       charged-lepton       Yukawa
couplings~\cite{Barbieri:1999ma}.    Depending    on   whether   these
interactions are in or out  of thermal equilibrium at the leptogenesis
scale,  the  predicted  value  for  the  baryon  asymmetry  could  get
significantly  modified,   as  already  shown   by  various  partially
flavour-dependent     treatments~\cite{Abada:2006ea,     Abada:2006fw,
Nardi:2006fx,   Blanchet:2006be,  De  Simone:2006dd, Pascoli:2006ie}.\footnote{Similar
partial flavour  effects have also been considered  for other variants
of leptogenesis models, e.g.~with type-II seesaw~\cite{Abada:2008gs, Felipe:2013kk,
Sierra:2014tqa} and soft leptogenesis~\cite{Fong:2008mu}.}  The
lepton flavour effects can be neglected only when the heavy neutrino mass
scale  $m_{N_\alpha}  \gsim  10^{12}$   GeV,  in  which  case  all  the
charged-lepton  Yukawa  interactions  are out-of-equilibrium  and  the
quantum states of all charged-lepton flavours evolve {\it coherently},
i.e.~effectively as a single lepton flavour, between their production
from $N_\alpha  \to L_l \Phi$  and subsequent inverse decay  $L_l \Phi
\to N_\alpha$. Here,  $L_l = (\begin{array}{cc}\nu_{lL} & l_L\end{array})^{\sf T}$  
is the $SU(2)_L$ 
lepton doublet (with  flavour index $l =  e, \mu,
\tau$)  and  $\Phi$  is  the   SM  Higgs  doublet.   For  $10^9  \lsim
m_{N_\alpha} \lsim 10^{12}$ GeV, the $\tau$-lepton Yukawa interactions
are in thermal  equilibrium, and hence, the lepton  quantum states are
an incoherent mixture of $\tau$-lepton and a coherent superposition of
electron and  muon.  Finally, for  $m_{N_\alpha} \lsim 10^9$  GeV, since the
muon  and electron Yukawa  interactions are  also in  equilibrium, 
their impact  on the final lepton asymmetry  must be taken into
account in low-scale RL models. Note that flavour effects also play an
important  role in  the  collision  terms describing  $\Delta  L =  1$
scatterings  that involve Yukawa  and  gauge interactions,  as  well  as
$\Delta  L  =  0$ and  $\Delta  L  =  2$  scatterings mediated  by  heavy
neutrinos~\cite{Pilaftsis:2005rv}.

Therefore, a  {\it flavour-covariant} formalism is required,  
in  order  to  consistently capture  all  the flavour  effects, including  
flavour mixing, oscillations and (de)coherence. 
These intrinsically quantum effects  can be accounted for by extending
the classical  Boltzmann equations for number  densities of individual
flavour species to a  semi-classical evolution equation containing a
matrix  of  number densities,  analogous  to  the formalism  presented
in~\cite{Sigl:1993} for  light neutrinos.  Following  this approach, a
matrix  Boltzmann equation in  the lepton  flavour space  was obtained
in~\cite{Abada:2006fw,  De  Simone:2006dd}. Similar considerations 
were made in~\cite{Blanchet:2011xq}  to  include  heavy  neutrino
flavour  effects  in a  hierarchical  scenario.   However,  in RL
scenarios, the interplay between  heavy-neutrino and lepton
flavour effects are  important. With these observations, a {\it fully} 
flavour-covariant treatment of the quantum statistical evolution of all 
relevant number densities, including their off-diagonal coherences, 
is entirely necessary. This is the main objective of this long article.

To this  end, we derive  a set of general  flavour-covariant transport
equations  for the  number densities  of any population of lepton and 
heavy-neutrino flavours in a quantum-statistical
ensemble.  This set of transport equations are obtained from a set of master 
equations for number density matrices derived in the Markovian approximation, 
in which quantum `memory' effects are ignored (see e.g.~\cite{Bellac}). 
We demonstrate  the necessary  appearance of  rank-4 tensor
rates  in  flavour  space  that  properly  account  for  the  statistical
evolution of  off-diagonal flavour coherences.  This novel formalism enables
us to capture three important flavour effects pertinent to RL: (i) the
resonant  mixing of  heavy neutrinos,  (ii) the  coherent oscillations
between heavy neutrino flavours, and (iii) quantum (de)coherence effects
in the charged-lepton sector.   In addition, we describe the structure
of  generalized flavour-covariant  discrete  symmetry transformations 
$C$, $P$ and $T$, ensuring definite transformation properties of the transport equations
and the generated lepton asymmetries in arbitrary flavour bases. 
Subsequently,  we obtain a  simplified version of
the general  transport equations in the heavy-neutrino mass eigenbasis, 
but retaining {\it  all} the flavour
effects.  We  further check  that these rate  equations reduce  to the
well-known  Boltzmann  equations in  the  flavour-diagonal limit.  

To illustrate  the  importance of  the effects captured {\it only} in this flavour-covariant treatment,  we
consider a  minimal low-scale  RL
scenario in which  the baryon asymmetry is  generated from
and protected in a {\it single} lepton flavour~\cite{Pilaftsis:2004xx}. 
As a  concrete example, we  consider a
minimal         model        of         resonant        $\tau$-genesis
(RL$_\tau$)~\cite{Pilaftsis:2004xx},  involving  three quasi-degenerate  
heavy neutrinos,  at or above the  electroweak  scale, with  sizable  couplings to  the
electron  and  muon, while  satisfying  all  the current  experimental
constraints.  We show that the  final lepton asymmetry obtained in our
flavour-covariant formalism can  be significantly enhanced (by roughly
one order of magnitude), as  compared to the partially flavour-dependent limits. 

We  should emphasize  that our  flavour-covariant formalism  is rather
general, and its applicability is not limited only to the RL phenomenon. 
The flavour-covariant transport equations presented here provide a 
complete description of the leptogenesis mechanism in all relevant temperature regimes. 
In addition, this formalism can be used to study other physical 
phenomena, in which flavour effects may be important, such as the evolution of multiple  
jet flavours in a dense   QCD    medium   in the    quark-gluon   plasma   (see
e.g.~\cite{Blaizot:2013vha}), the evolution of  neutrino flavours in a supernova
core   collapse  (see   e.g.~\cite{Zhang:2013lka}), or the scenario of  
$\CP T$-violation induced by the propagation of neutrinos in 
gravitational backgrounds~\cite{Mavromatos:2012ii}.   We   have  also
developed a flavour-covariant generalization of the helicity amplitude
technique,  and a  generalized optical  theorem in  the presence  of a
non-homogeneous background ensemble, which may find applications in non-equilibrium Quantum Field Theory (QFT).

It is worth  mentioning here that there have been  a number of studies
(see     e.g.~\cite{Buchmuller:2000nd,    De     Simone:2007rw,    De
Simone:2007pa,    Cirigliano:2007hb,    Garny:2009qn,   Cirigliano:2009yt, 
Beneke:2010dz,
Anisimov:2010dk,   Garbrecht:2011aw,   Garny:2011hg,  Frossard:2012pc,
Iso:2013lba}), aspiring  to go  beyond the semi-classical  approach to
Boltzmann equations  in order to understand  the transport phenomena
from `first  principles' within the framework  of non-equilibrium QFT.
Such  approaches are  commonly based  on the  Schwinger-Keldysh Closed
Time  Path (CTP) formalism~\cite{Schwinger:1961,  Keldysh:1964}.  This
real-time  framework  allows  one  to derive  quantum  field-theoretic
analogues   of  the  Boltzmann   equations,  known   as  Kadanoff-Baym
equations~\cite{Kadanoff:1962}, obtained  from the CTP Schwinger-Dyson
equation  and  describing the  non-equilibrium  time-evolution of  the
two-point correlation functions. The Kadanoff-Baym
equations are  manifestly non-Markovian, accounting  for the so-called
`memory' effects that  depend on the history  of the system. 
These equations can, in principle, account consistently for all 
flavour and thermal effects. However, one  should note that in order
to  define  particle  number  densities and  solve  the  Kadanoff-Baym
equations  for  their out-of-equilibrium  evolution  (as e.g.~in  the
context  of leptogenesis), particular  approximations are  often made. 
These specifically include quasi-particle     approximation     and     gradient
expansion  in time  derivatives~\cite{Bornath:1996zz}.   Moreover, the
loopwise perturbative  expansion of non-equilibrium  propagators 
are      normally     spoiled      by     the      so-called     pinch
singularities~\cite{Altherr:1994fx},    which    are   
mathematical  pathologies arising from  ill-defined products  of delta
functions  with identical  arguments.  Recently,  a new  formalism was
developed   for   a   perturbative   non-equilibrium   thermal   field
theory~\cite{Millington:2012pf},   which  makes   use   of  physically
meaningful particle number densities  that are directly derivable from
the Noether  charge. This approach  allows the loopwise  truncation of
the  resulting transport  equations  without the  appearance of  pinch
singularities,  while  maintaining all  orders  in gradients,  thereby
capturing more  accurately the early-time non-Markovian  regime of the
non-equilibrium dynamics.   An application of this  approach to study
the impact  of thermal effects  on the flavour-covariant  RL formalism
presented here lies beyond the scope of this article.
    
The rest of the paper is organized as follows: in Section~\ref{sec:2},
we  review  the  main   features  of  the  flavour-diagonal  Boltzmann
equations.   In  Section~\ref{sec:3},  we  derive  a  set  of  general
flavour-covariant  transport equations  in the  Markovian  regime.  In
Section~\ref{sec:4},   we    apply   the   formalism    developed   in
Section~\ref{sec:3}  to  a generic  RL  scenario and  derive  the  relevant
flavour-covariant  evolution  equations  for  the  heavy-neutrino  and
lepton-doublet number densities.  In Section~\ref{sec:5}, we present 
a geometric understanding of the degeneracy limit in minimal RL 
scenarios and also discuss an explicit  model of RL$_\tau$.  
In  Section~\ref{sec:6},  we present numerical results for three 
benchmark points, which   illustrate  the  impact  of  flavour
off-diagonal effects on the final lepton asymmetry.
We summarize our conclusions in Section~\ref{sec:7}.  In~\ref{app:cp}, we
comment on different forms of  the self-energy regulator used in the literature to
calculate      the     leptonic $\CP$-asymmetry      in      RL models and check their consistency in the $L$-conserving limit.
In~\ref{app:propagator}, we develop a flavour-covariant generalization
of    the   helicity   amplitude    formalism   and    describe   the
flavour-covariant quantization of spinorial  fields in the presence of
time-dependent     and     spatially-inhomogeneous     backgrounds.
In~\ref{app:optical},  we justify the  tensorial flavour  structure of
the  transport equations  introduced in  Section~\ref{sec:3},  by means of 
a generalization of the optical theorem.  Finally, in \ref{app:loop}, we 
exhibit the form factors relevant for the lepton flavour violating decay
rates discussed in Section~\ref{sec:6}.

\section{Flavour Diagonal Boltzmann Equations}
\label{sec:2}

The time-evolution of the number density $n^a$ of any particle species
$a$  can be  modelled  by a  set  of coupled  Boltzmann equations  (see
e.g.~\cite{kolb}).      Adopting     the      formalism     described
in~\cite{Kolb:1980, Luty:1992un}, this may be written down
in the generic form
\begin{eqnarray}
  \frac{\D{}{n^a}}{\D{}{t}} \: + \: 3Hn^a \ = \
  -  \sum_{aX  \leftrightarrow  Y}
  \Bigg[
    \frac{n^a n^X}{n^a_{\rm eq}n^X_{\rm eq}} \: \gamma(aX  \to  Y) \:
    - \: \frac{n^Y}{n^Y_{\rm eq}} \ \gamma(Y  \to aX)
  \Bigg] \; ,
  \label{be1}
\end{eqnarray}
where  the drift  terms on  the left-hand  side (LHS)  arise  from the
covariant  hydrodynamic derivative  and  include the  dilution of  the
number density due  to the expansion of the  Universe, parametrized by
the  Hubble  expansion  rate   $H$.   The  right-hand  side  (RHS)  of
\eqref{be1}   comprises  the  collision   terms  accounting   for  the
interactions  that change  the number  density $n^a$.   Here,  we have
summed over  all possible reactions of the  form $aX \to Y$  or $Y \to
aX$,  in  which  the  species  $a$  can  be  annihilated  or  created,
respectively.  If the species $a$ is  unstable, it can occur as a real
intermediate state  (RIS) in resonant processes  of the form  $X \to a
\to Y$, which must be  properly taken into  account in order  to avoid
double-counting  of  this  contribution  from the  already  considered
decays and inverse decays in the Boltzmann equations~\cite{Kolb:1980}.
At this point,  it is important to note that  the formalism leading to
\eqref{be1}   neglects  both   the  coherent   time-oscillatory  terms,
describing  particle  oscillations   between  different  flavours,  and
off-diagonal   correlations  in   the  matrix   of   number  densities
$n^{a\bar{b}}$,  corresponding  to  the  annihilation  of  a  particle
species $b$  and the  correlated creation of  a particle  species $a$.
For  this  reason,   we  refer  to  \eqref{be1}  as   a  set  of  {\it
flavour-diagonal} Boltzmann equations.

It  is  useful to  summarize  the  notation  and definitions  used  in
\eqref{be1}.  Firstly, the Hubble expansion rate in the early Universe
is given as a function of the temperature $T$ by~\cite{kolb}
\begin{equation}
  H(T) \ = \ \left(\frac{4\pi^3}{45}\right)^{1/2}  g_*^{1/2}  
\frac{T^2}{M_{\rm Pl}} \ ,
  \label{hub}
\end{equation}
where $M_{\rm  Pl} = 1.2  \times 10^{19}$ GeV  is the Planck  mass and
$g_*(T)$  is  the  number   of  relativistic  degrees  of  freedom  at
temperature $T$.  Throughout our  discussions, all species are assumed
to be in  kinetic (but not necessarily chemical)  equilibrium. In this
case, the number  density of a particle species $a$
is given by
\begin{equation}
  n^a(T) \ = \ g_a \int \frac{\mathrm{d}^3\ve p}{(2\pi)^3} \ 
  \frac{1}{\exp{[(E_a  - \mu_a)/T]}
    \: \pm \: 1} 
  \ \equiv \ g_a\int_{\mathbf{p}}\ 
  \frac{1}{\exp{[(E_a  - \mu_a)/T]}
    \: \pm \: 1} \; ,
  \label{nat0}
\end{equation}
where  $\int_{\ve{p}}\equiv \int  \mathrm{d}^3  \ve{p}/(2\pi)^3$ is  a
short-hand  notation for the  three-momentum integral,  the $-  \ (+)$
sign in the denominator corresponds to particles obeying Bose-Einstein
(Fermi-Dirac) quantum statistics,  $E_a(\mathbf{p}) = (|{\mathbf p}|^2
+ m_a^2)^{1/2}$ is  the relativistic energy of the  species $a$, $m_a$
being its  rest mass, $g_a  = g_a^{\rm hel}  \, g_a^{\rm iso}$  is the
total degeneracy factor of  the internal degrees of freedom, $g_a^{\rm
hel}$ and $g_a^{\rm iso}$ being the degenerate helicity and degenerate
isospin degrees  of freedom respectively, and  $\mu_a \equiv \mu_a(T)$
is   the  temperature-dependent   chemical  potential,   encoding  the
deviation from  local thermodynamic equilibrium.   It will prove
convenient in our later discussions to define an in-equilibrium number
density   $n^a_{\mathrm{eq}}$  as   the   limit  $\mu_a   \to  0$   in
\eqref{nat0}. We  note however that the {\it  true} equilibrium number
density   will   depend   on   the  equilibrium   chemical   potential
$\mu_{a}^{\mathrm{eq}}$, which may not be zero in general.

There  are  two limits  of  \eqref{nat0}  of  interest here:  (i)  the
Maxwell-Boltzmann          classical         statistical         limit
$(E_a(\mathbf{p})-\mu_a)/T\gg 1$ in which we can drop the $\pm 1$ term
in the denominator of \eqref{nat0}, giving
\begin{equation}
  n^a(T)\ = \ g_a\int_{\mathbf{p}}e^{-\: [E_a(\mathbf p)  - \mu_a(T)]/T} 
\  = \ \frac{g_am_a^2T}{2\pi^2}\,K_2
  \Big(\frac{m_a}{T}\Big)\: e^{\mu_a(T)/T}\;,
  \label{nat}
\end{equation}
where  $K_n(x)$ is  the $n$th-order  modified Bessel  function  of the
second kind;  (ii) the relativistic limit ($T\gg  m_a,\mu_a$) in which
case
\begin{eqnarray}
  n^a(T) \ = \  \frac{\sigma_\chi \zeta(3)}{\pi^2} \:  g_a T^3 \; ,
  \label{nrel}
\end{eqnarray}
where $\sigma_\chi = 1 \  (3/4)$ for bosons (fermions), and $\zeta(x)$
is the Riemann zeta function, with $\zeta(3) \approx 1.20206$.

Following~\cite{Pilaftsis:2003gt},  we  define  the $  \CP$-conserving
collision  rate   for  a  generic  process   $X  \to  Y$   and  its  $
\CP$-conjugate $X^c \to Y^c$ as
\begin{equation}
  \gamma^X_Y \ \equiv \ \gamma(X  \to  Y)\:
  + \: \gamma(X^c  \to Y^c)\;,
  \label{cpc}
\end{equation}
where we have used the shorthand superscript $c$ to denote $\CP$ conjugation, and
\begin{align}
  \gamma(X  \to  Y) \ & = \ \int\! \D{}{\Pi_X} \, \D{}{\Pi_Y}
  (2\pi)^4\delta^{(4)}(p_X  - p_Y) \: e^{- p^0_X/T}
  \big|{\cal M}(X  \to  Y)\big|^2
  \nonumber\\ &
  \ \equiv \ \int_{XY}\big|{\cal M}(X  \to Y)\big|^2\;.
  \label{col}
\end{align}
Here,  the  squared  matrix   element  $\left|{\cal  M}(X    \to  
Y)\right|^2$ is summed, but not averaged, over the internal degrees of
freedom of the initial and  final multiparticle states $X$ and $Y$. We
have introduced an abbreviated notation $\int_{XY}$ in \eqref{col} for
the phase-space  integrals over $X$  and $Y$. The  phase-space measure
for the multiparticle state $X$, containing ${\cal N}_X$ particles, is
defined as
\begin{equation}
  \D{}{\Pi_X} \ = \ \frac{1}{{\cal N}_{\rm id}!} \prod_{i  = 1}^{{\cal N}_X}
  \frac{\D{4}{p_i}}{(2\pi)^4} \: 2\pi \delta(p_i^2  - m_i^2) \: 
  \theta(p_i^0)\;,
\end{equation} 
where  $\delta(x)$  and $\theta(x)$  are  the  usual  Dirac delta  and
Heaviside step functions, respectively,  and ${\cal N}_{\rm id}!$ is a
symmetry factor in the case  that the multiparticle state $X$ contains
${\cal  N}_{\rm id}$  identical  particles.  In  a $\CP  T$-conserving
theory, the $ \CP$-conserving collision rates must obey $\gamma_Y^X 
=    \gamma^Y_X$.   Analogous  to \eqref{cpc},  a  $  \CP$-violating
collision rate can be defined as~\cite{Pilaftsis:2003gt}
\begin{equation}
  \delta \gamma^X_Y \ = \
  \gamma(X  \to Y) \:
  - \: \gamma(X^c  \to Y^c)\;, 
  \label{cpv}
\end{equation}
which obeys $\delta \gamma^X_Y  = - \delta\gamma^Y_X$, following $ \CP
T$ invariance.

The  relevant Boltzmann  equations for describing leptogenesis are  those
involving the number densities $n^N_\alpha$ (with $\alpha = 1, \dots, {\cal
N}_N)$ of the heavy Majorana  neutrinos, $n^L_l$ (with $l = 1, \dots, {\cal
N}_L$)  of  the  lepton-doublets  and  $\bar{n}^L_l$  of  their  $\CP$
conjugates.    When  solving   the  coupled   system   of  first-order
differential  equations  \eqref{be1}  for  $n^N_\alpha$,  $n^L_l$  and
$\bar{n}^L_l$,  it is  convenient to  introduce a  new variable  $z =
m_{N_1}/T$.   In  the   radiation-dominated  epoch,  relevant  to  the
production of lepton asymmetry,  $z$ is related to the cosmic time
$t$ via the relation $t = z^2/2H_N$, where
\begin{align}
  H_N  \ \equiv \ H(z = 1)  \: \simeq  \: 17\:
  \frac{m_{N_1}^2}{M_{\rm  Pl}}
\end{align}
is  the Hubble  parameter \eqref{hub}  at $z  = 1$,  assuming  only SM
relativistic degrees of freedom.  We also normalize the number density
of species $a$ to the number density of photons, defining $\eta^a(z) =
n^a(z)  /  n^\gamma(z)$, with  $n^\gamma$  given  by \eqref{nrel}  for
$\sigma_\chi=1$ and $g_\gamma=2$, i.e.
\begin{equation}
  n^\gamma(z) \ =\ \frac{2T^3\zeta(3)}{\pi^2}\ 
  = \ \frac{2m_{N_1}^3\zeta(3)}{\pi^2 z^3}\;. 
  \label{ngamma}
\end{equation}
With these  definitions, we write down  the flavour-diagonal Boltzmann
equations \eqref{be1}  in terms of the normalized  number densities of
heavy neutrinos $\eta^N_\alpha$  and the normalized lepton asymmetries
$\delta   \eta^L_l   =   (n_l^L   -  \bar{n}_l^L)   /   n^\gamma$   as
follows~\cite{Pilaftsis:2005rv}:
\begin{align} 
 \frac{n^\gamma H_N}{z} \frac{\D{}{\eta^N_\alpha}}{\D{}{z}} \ & = 
  \bigg(1  -  \frac{\eta^N_\alpha}{\eta^N_{\rm eq}}\bigg)
  \sum_l\gamma^{N_\alpha}_{L_l\Phi}\;,
  \label{be2} \\
  \frac{n^\gamma H_N}{z}\frac{\D{}{\delta \eta^L_l}}{\D{}{z}} \ & =
  \sum_{\alpha}
    \bigg(\frac{\eta^N_\alpha}{\eta^N_{\rm eq}}  -  1\bigg)
   \delta\gamma^{N_\alpha}_{L_l\Phi}\:
    - \: \frac{2}{3}\delta\eta^L_l \sum_{k}
    \Big(\gamma^{L_l\Phi}_{L^c_k\Phi^c} 
    +  \gamma^{L_l\Phi}_{L_k\Phi}\Big)
  \nonumber \\
  & \qquad \qquad \quad
- \: \frac{2}{3}\sum_{k}\delta\eta^L_k
    \Big(\gamma^{L_k\Phi}_{L^c_l\Phi^c} 
    -  \gamma^{L_k\Phi}_{L_l\Phi}\Big)\;,
  \label{be3}
\end{align}
where $\eta^N_{\rm  eq} \approx  z^2K_2(z)/2$ is  the normalized
equilibrium  number density  of  the heavy  neutrinos, obtained  using
\eqref{ngamma}  and \eqref{nat} with  $g_N=2$.  The  various collision
rates  appearing  in  \eqref{be2}   and  \eqref{be3}  can  be  readily
understood   from   the  general   definitions   in  \eqref{cpc}   and
\eqref{cpv};  their  explicit  expressions  in  terms  of  the  Yukawa
couplings  will  be  given  in  Section~\ref{sec:2.2}.  Here  we  have
included only  the dominant contributions  arising from the $1  \to 2$
decays  and  $2  \to  1$   inverse  decays  of  the  heavy  neutrinos,
proportional to  the rate $\gamma^{N_\alpha}_{L_l\Phi}$,  and the {\it
resonant} part of the $2 \leftrightarrow 2$ $\Delta L = 0$ and $\Delta
L =  2$ scatterings, proportional  to $\gamma^{L_k\Phi}_{L_l\Phi}$ and
$\gamma^{L_k\Phi}_{L^c_l\Phi^c}$   respectively.     We   ignore   the
sub-dominant  chemical  potential contributions  from  the right-handed (RH) 
charged-lepton, quark  and the  Higgs fields, as  well as  the $\Delta L  = 1$
Yukawa and gauge scattering terms~\cite{Pilaftsis:2005rv}.

Note that for the collision rate pertinent to the heavy neutrino decay
in  \eqref{be2},  we  have   summed  over  the  lepton  flavours,  and
similarly, for  the charged-lepton rate equation  \eqref{be3}, we have
summed  over  the  heavy  neutrino  flavours; therefore,  these  are  still
designated  as  flavour-diagonal  Boltzmann equations.  The  $\CP$-odd
collision rate  $\delta\gamma^{N_\alpha}_{L_l\Phi}$ in \eqref{be3} can
be   expressed  in   terms   of  the   flavour-dependent  leptonic   $
\CP$-asymmetries $\varepsilon_{l\alpha}$  and the $\CP$-even collision
rate         $\gamma^{N_\alpha}_{L_l\Phi}$,         as        follows:
$\delta\gamma^{N_\alpha}_{L_l\Phi}                                    =
\varepsilon_{l\alpha}\sum_k\gamma^{N_\alpha}_{L_k\Phi}$,          where
$\varepsilon_{l\alpha}$  is  defined in  terms  of  the partial  decay
widths  $\Gamma_{l\alpha}  \equiv  \Gamma(N_\alpha \to  L_l\Phi)$  and
their $\CP$-conjugates  $\Gamma^c_{l\alpha} \equiv \Gamma(N_\alpha \to
L^c_l\Phi^c)$:
\begin{equation}
  \varepsilon_{l\alpha} \ = \ \frac{\Gamma_{l\alpha} 
    -  \Gamma^c_{l\alpha}}
  {\sum_{k}\big(\Gamma_{k\alpha} 
      +  \Gamma^{c}_{k\alpha}\big)} \
  \equiv \ \frac{\Delta \Gamma_{l\alpha}}{\Gamma_{N_\alpha}}\;,
  \label{eps}
\end{equation}
where  $\Gamma_{N_\alpha}$  is the  total  decay  width  of the  heavy
Majorana neutrino $N_\alpha$. Since we are interested in the
heavy  neutrino decay  for  temperatures above  the electroweak  phase
transition,  where  the  SM   Higgs  vacuum  expectation  value  (VEV)
vanishes,  only  the  would-be   Goldstone  and  Higgs  modes  of  the
$\Phi$-doublet  contribute predominantly to  the partial  decay widths
$\Gamma_{l\alpha}$ and the total decay width $\Gamma_{N_\alpha}$ in \eqref{eps}.

\subsection{Resummed Effective Yukawa Couplings}
\label{sec:2.1}

\begin{figure}[t!]
  \centering
  \includegraphics[width = 8cm]{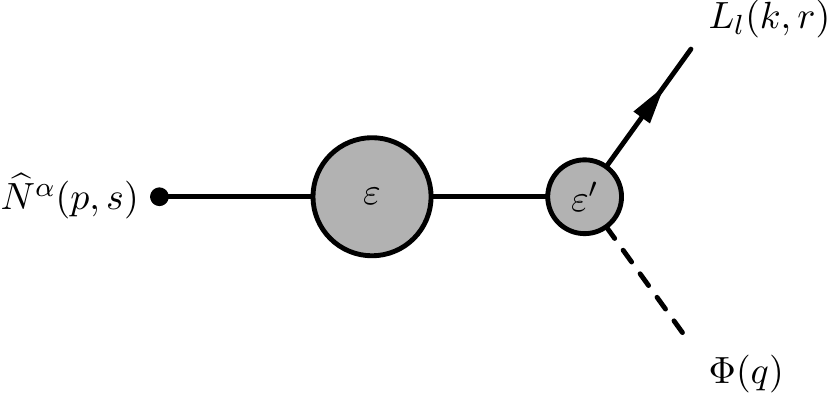}
  \caption{\it  The two  types of  $ \CP$-violation  due to  the heavy
Majorana neutrino decay  $N\to L\Phi$. The notation used  here will be
explained in Section~\ref{sec:3}.}
  \label{figcp}
\end{figure}

The  physical   $\CP$-violating  observable  defined   in  \eqref{eps}
receives   contributions   from    two   different   mechanisms   (see
Figure~\ref{figcp}):  (i) $\varepsilon$-type $  \CP$ violation  due to
the  interference between the  tree-level and  absorptive part  of the
self-energy   graphs   in   the   heavy-neutrino   decay,   and   (ii)
$\varepsilon'$-type $  \CP$ violation due to  the interference between
the  tree-level  graph  and   the  absorptive  part  of  the  one-loop
vertex. This  terminology is in analogy  with the two kinds  of $ \CP$
violation    in   the    $K^0\overline{K}^0$-system    (for   reviews,
see~\cite{pdg, kabir}), where  $\varepsilon$ represents the indirect $
\CP$  violation   through  $K^0$  --  $\overline{K}^0$  mixing,  while
$\varepsilon'$ represents the direct $\ \CP$ violation entirely due to
the decay amplitude.

The contribution  of the self-energy diagrams to  the $ \CP$-asymmetry
can  in  principle  be   calculated  using  an  effective  Hamiltonian
approach,      similar      to      that     applied      for      the
$K^0\overline{K}^0$-system~\cite{kabir}.      However,    the    heavy
neutrinos,  being  unstable  particles,  cannot be  described  by  the
asymptotic    (free)     in-    and    out-states     of  an  $S$-matrix
theory~\cite{Veltman:1963}.  Instead, their properties can be inferred
from   the  transition   matrix  elements   of   $2\leftrightarrow  2$
scatterings of stable particles,  and by identifying the resonant part
of  the   $2\leftrightarrow  2$   amplitude  that  contains   the  RIS
contributions  only.    This  allows  one  to   perform  an  effective
resummation of the heavy-neutrino self-energy diagrams contributing to
the     $\varepsilon$-type     $\CP$-asymmetry~\cite{Pilaftsis:1997jf,
Buchmuller:1997yu,   Pilaftsis:2003gt}.\footnote{For  other  effective
approaches  within   the  framework  of   perturbative  field  theory,
see~\cite{Flanz:1996fb, Flanz:1994yx, Liu:1993ds, Covi:1996fm, Sarkar:1998, Rangarajan:1999kt,
Anisimov:2005hr}. However, for a critical appraisal of the existing approaches, 
see~\ref{app:cp}. }

Neglecting   the  charged-lepton  and   light  neutrino   masses,  the
absorptive part of the heavy Majorana neutrino self-energy transitions
$N_\beta\to N_\alpha$ can be  written in a simple spinorial structure,
as follows:
\begin{equation}\label{self_LR}
  \Sigma^{\rm abs}_{\alpha \beta}(\slashed {p})\
  = \ A_{\alpha \beta}(p^2)\: \slashed {p} \: \mathrm {P_ L} \:
  + \: A^*_{\alpha \beta}(p^2) \: \slashed {p} \: \mathrm{P_R}\;, 
\end{equation} 
where $\mathrm{P_{L,R}} = (\mat{1}_4  \mp \gamma_5)/2$ are the left- and
right-chiral projection  operators respectively, and $A_{\alpha\beta}$
is the absorptive transition amplitude, summed over all charged-lepton
flavours running in the loop:
\begin{equation}
  A_{\alpha\beta}(\widehat{h}) \ 
  = \ \frac{(\widehat{h}^\dag \widehat{h})^*_{\alpha\beta}}{16\pi}
  = \ \frac{1}{16\pi}\sum_{l}\widehat{h}_{l \alpha}
  \widehat{h}^*_{l \beta} \
  \equiv \ \sum_l A^l_{\alpha \beta}(\widehat{h})\;.
  \label{Aab}
\end{equation}
Here  $h_{l\alpha}$  is the  Yukawa  coupling  of  the heavy  neutrino
$N_\alpha$   with   the    lepton-doublet   $L_l$,   and   the   caret
(\,$\widehat{~}$\,) denotes the fact that \eqref{Aab} was derived in a
basis   in  which  the   heavy  Majorana   neutrino  mass   matrix  is
diagonal. The  tree-level decay width  of the heavy  Majorana neutrino
$N_\alpha$   is   related  to   the   diagonal  transition   amplitude
$A_{\alpha\alpha}$ by
\begin{eqnarray}
  \Gamma^{(0)}_{N_\alpha} \
  = \ 2m_{N_\alpha}A_{\alpha\alpha}(\widehat{h}) \ = \ 
  \frac{m_{N_\alpha}}{8\pi}
  (\widehat{h}^\dag \widehat{h})_{\alpha\alpha} \;.
  \label{gammatree}
\end{eqnarray}

To  account for  unstable-particle  mixing effects  between the  heavy
Majorana neutrinos,  we define the one-loop  resummed effective Yukawa
couplings,  denoted  by  (bold-faced Latin)  $\maf{h}_{l\alpha}$,  and
their $  \CP$-conjugates $\maf{h}_{l\alpha}^c$, related  to the matrix
elements     $\mathcal{M}(N_\alpha      \to     L_l     \Phi)$     and
$\mathcal{M}(N_\alpha \to L^c_l \Phi^c)$ respectively.  This formalism
captures   all  dominant   effects  of   heavy  neutrino   mixing  and
$\CP$-violation,  and  has  been shown~\cite{Pilaftsis:2003gt}  to  be
equivalent      to      an      earlier      proposed      resummation
method~\cite{Pilaftsis:1997jf}          based          on          the
Lehmann-Symanzik-Zimmermann reduction formalism~\cite{LSZ}. Working in
the  heavy neutrino  mass  eigenbasis, the  resummed effective  Yukawa
couplings are given by~\cite{Pilaftsis:2003gt, Pilaftsis:2008qt}
\begin{align}
  &\widehat{\maf{h}}_{l \alpha} \ = \ \widehat{h}_{l \alpha} \:
  - \: i\sum_{\beta,\gamma}|\epsilon_{\alpha\beta\gamma}| \: 
  \widehat{h}_{l \beta}
  \nonumber \\
  & \ \times \frac{m_\alpha(m_\alpha A_{\alpha\beta}
    + m_\beta A_{\beta\alpha})
    \: - \: iR_{\alpha\gamma}[m_\alpha A_{\gamma\beta}
    (m_\alpha A_{\alpha\gamma}
    + m_\gamma A_{\gamma\alpha})\:
    + m_\beta A_{\beta\gamma}(m_\alpha A_{\gamma\alpha} \:
    + m_\gamma A_{\alpha\gamma})]}
  {m^2_\alpha \: - \: m^2_\beta \: + \: 2i m^2_\alpha A_{\beta\beta} \:
    + \: 2i {\rm Im}(R_{\alpha \gamma})
    [m^2_\alpha|A_{\beta\gamma}|^2 \:
    + \: m_\beta m_\gamma {\rm Re}(A^2_{\beta\gamma})]}\;, 
\label{resum1}
\end{align}
where   $\epsilon_{\alpha\beta\gamma}$   is   the  usual   Levi-Civita
anti-symmetric tensor,  $m^2_\alpha \equiv m^2_{N_\alpha}$ is a shorthand 
notation used here for brevity, 
and 
\begin{equation}
  R_{\alpha\beta} \ = \ \frac{m_\alpha^2}
  {m_\alpha^2 \: - \: m_\beta^2 \: + \: 2im_\alpha^2 A_{\beta\beta}}\;.
  \label{Rab}
\end{equation}
All   the  transition   amplitudes  $A_{\alpha\beta}      \equiv  
A_{\alpha\beta}(\widehat{h})$ in \eqref{resum1} are evaluated on-shell
with  $p^2   =    m^2_{N_\alpha}$.   The
respective  $  \CP$-conjugate   resummed  effective  Yukawa  couplings
$\widehat{\maf{h}}^c_{l\alpha}$ can be obtained from \eqref{resum1} by
replacing  the tree-level  Yukawa couplings  $h_{l\alpha}$  with their
complex conjugates  $h^*_{l\alpha}$.\footnote{Note that $\maf{h}^c\neq
\maf{h}^*$ in  general, whereas  for the tree-level  Yukawa couplings,
$h^c=h^*$ by  $\CP T$-invariance of the Lagrangian.}   We will neglect
the one-loop  corrections to the proper vertices  $L_l \Phi N_\alpha$,
whose  absorptive  parts are  numerically  insignificant  in RL.   The
partial  decay widths  $\Gamma_{\alpha l}$  and  $\Gamma_{\alpha l}^c$
appearing in \eqref{eps} can now be expressed in terms of the 
effective  Yukawa     couplings       $\widehat{\maf{h}}_{l\alpha}$      and
$\widehat{\maf{h}}^c_{l\alpha}$,  and the  flavour-dependent absorptive
transition  amplitudes   $A^l_{\alpha  \beta}(\widehat{\maf{h}})$,  as
follows:
\begin{equation}
  \Gamma_{l\alpha} \
  = \ m_{N_\alpha}A^l_{\alpha\alpha}(\widehat{\maf{h}})\;,
  \quad 
  \Gamma^c_{l\alpha} \ 
  = \ m_{N_\alpha}A^l_{\alpha\alpha}(\widehat{\maf{h}}^c)\;.
  \label{gamma}
\end{equation}
Note the  explicit dependence of the  absorptive transition amplitudes
on   the    effective   Yukawa couplings    $\widehat{\maf{h}}$   in
\eqref{gamma}. The  total decay  width of the  heavy neutrino  is thus
obtained by summing over all lepton flavours:
\begin{equation}
  \Gamma_{N_\alpha} \ = \ \sum_l \left(\Gamma_{l\alpha}
    +\Gamma^c_{l\alpha}\right) 
  \ = \ 
  \frac{m_{N_\alpha}}{16 \pi}\,\left[ (\mathbf{\widehat{h}}^{\dagger} \,
    \mathbf{\widehat{h}})_{\alpha \alpha}  \;
    + \; (\mathbf{\widehat{h}}^{c\dagger} \,
    \mathbf{\widehat{h}}^c)_{\alpha \alpha} \right]. 
  \label{resum_width}
\end{equation}
Replacing $\widehat{\maf{h}}$ by  the tree-level Yukawa coupling $\widehat{h}$
in \eqref{resum_width},  we can  reproduce the tree-level  decay width
given    by   \eqref{gammatree}.    Substituting    \eqref{gamma}   in
\eqref{eps}, the flavour-dependent leptonic $ \CP$-asymmetry in RL can
be written as
\begin{equation}
  \varepsilon_{l\alpha} \ = \ \frac{|\widehat{\maf{h}}_{l\alpha}|^2 \: 
    - \: |\widehat{\maf{h}}^c_{l\alpha}|^2}
  {\sum_{k}\left(|\widehat{\maf{h}}_{k\alpha}|^2 \: + \: 
      |\widehat{\maf{h}}^c_{k\alpha}|^2 \right)} \
  = \ \frac{|\widehat{\maf{h}}_{l\alpha}|^2 \: - \: 
    |\widehat{\maf{h}}^c_{l\alpha}|^2}
  {(\widehat{\maf{h}}^\dag\widehat{\maf{h}})_{\alpha\alpha} \:
    + \: (\widehat{\maf{h}}^{c^\dag}\widehat{\maf{h}}^c)_{\alpha\alpha}}\;.
  \label{cpasy}
\end{equation}
Note    that   \eqref{cpasy}    encodes   both    $\varepsilon$-   and
$\varepsilon'$-type $\CP$ asymmetries, although we simply denote it by
$\varepsilon$ for brevity.   The analytic results for both  types of $
\CP$-asymmetry and  their $L$-conserving  limits for a  simplified case
will be discussed in~\ref{app:cp}.

\subsection{Analytic Solutions}
\label{sec:2.2}

It  is instructive  to derive  approximate analytic  solutions  of the
Boltzmann  equations  \eqref{be2}  and  \eqref{be4}. To  do  this,  we
express  \eqref{be2}  in  terms  of the  non-equilibrium  deviation
parameter  $\upeta^N_\alpha  =  \big(\eta^N_\alpha  / \eta^N_{\rm
eq}  - 1\big)$, thus obtaining
\begin{equation}
  \frac{\D{}\upeta^N_{\alpha}}{\D{}z} \ = \
  \frac{K_1(z)}{K_2(z)}\left[1  + 
    \left(1  -  {\rm K}_\alpha z\right)
    \upeta^N_\alpha \: \right]\; .
  \label{be5}
\end{equation}
where    the   K-factors,    defined   by    $   {\rm    K}_\alpha   =
{\Gamma_{N_\alpha}}/\zeta(3)H_N$,  determine  the  depletion of  the
lepton asymmetry  due to inverse  decays. In deriving  \eqref{be5}, we
have  used  the  analytic  expression  for the  total  collision  rate
$\gamma^{N_\alpha}_{L\Phi}$ pertinent to the heavy neutrino decay
\begin{equation}
  \gamma^{N_\alpha}_{L\Phi} \: \equiv  \: \sum_l\gamma^{N_\alpha}_{L_l\Phi} 
  \ = \ \frac{m^3_{N_\alpha}}{\pi^2 z}K_1(z)\Gamma_{N_\alpha}\;.
\label{gam12}
\end{equation}
In   the   kinematic   regime    $z   >   z_1^\alpha   \approx   2\:{\rm
K}^{-1/3}_\alpha$, \eqref{be5} has an approximate attractor solution
\begin{eqnarray}
  \label{eq:N_anal_diag}
  \upeta^N_\alpha(z) \ \simeq \ \frac{1}{{\rm K}_\alpha z} \; , 
  \label{etaN_at}
\end{eqnarray}
independent of the initial conditions.

The  collision rates  for  the  $\Delta L  =  0$ and  $\Delta  L =  2$
scatterings are given by~\cite{Deppisch:2010fr}
\begin{align}
  \gamma^{L_k\Phi}_{L_l\Phi} & \ = \ \sum_{\alpha,\beta}
  \frac{\left(\gamma^{N_\alpha}_{L\Phi}+
      \gamma^{N_\beta}_{L\Phi}\right)}
  {\left(1-2i \: \frac{m_{N_\alpha}\: - \: m_{N_\beta}}
      {\Gamma_{N_\alpha} \: + \: \Gamma_{N_\beta}}\right)}
  \frac{2\left(\widehat{\maf{h}}^*_{l\alpha} \widehat{\maf{h}}^{c^*}_{k\alpha}
      \widehat{\maf{h}}_{l\beta}\widehat{\maf{h}}^c_{k\beta}
      \: + \: \widehat{\maf{h}}^{c^*}_{l\alpha}\widehat{\maf{h}}^*_{k\alpha}
      \widehat{\maf{h}}^c_{l\beta}\widehat{\maf{h}}_{k\beta}\right)}
  {\left[(\widehat{\maf{h}}^\dag\widehat{\maf{h}})_{\alpha\alpha}
      + (\widehat{\maf{h}}^{c^\dag}\widehat{\maf{h}}^c)_{\alpha\alpha}
      + (\widehat{\maf{h}}^\dag\widehat{\maf{h}})_{\beta\beta}
      + (\widehat{\maf{h}}^{c^\dag}\widehat{\maf{h}}^c)_{\beta\beta} \right]^2}\;,
  \label{scat1}\\
  \gamma^{L_k\Phi}_{L^c_l\Phi^c} & \ = \ \sum_{\alpha,\beta}
  \frac{\left(\gamma^{N_\alpha}_{L\Phi}+
      \gamma^{N_\beta}_{L\Phi}\right)}
  {\left(1-2i \: \frac{m_{N_\alpha}\: - \: m_{N_\beta}}
      {\Gamma_{N_\alpha}\: + \: \Gamma_{N_\beta}}\right)}
  \frac{2\left(\widehat{\maf{h}}^*_{l\alpha} \widehat{\maf{h}}^*_{k\alpha}
      \widehat{\maf{h}}_{l\beta}\widehat{\maf{h}}_{k\beta}
      \: + \: \widehat{\maf{h}}^{c^*}_{l\alpha}\widehat{\maf{h}}^{c^*}_{k\alpha}
      \widehat{\maf{h}}^c_{l\beta}\widehat{\maf{h}}^c_{k\beta}\right)}
  {\left[(\widehat{\maf{h}}^\dag\widehat{\maf{h}})_{\alpha\alpha}
      + (\widehat{\maf{h}}^{c^\dag}\widehat{\maf{h}}^c)_{\alpha\alpha}
      + (\widehat{\maf{h}}^\dag\widehat{\maf{h}})_{\beta\beta}
      + (\widehat{\maf{h}}^{c^\dag}\widehat{\maf{h}}^c)_{\beta\beta}
    \right]^2}\; ,
  \label{scat2}
\end{align}
where  we  have used  the  narrow-width  approximation  (NWA) for  the
resummed heavy neutrino propagators in the pole-dominance region, i.e.
\begin{eqnarray}
  \frac{1}{(s-m_{N_\alpha}^2)^2+m_{N_\alpha}^2\Gamma_{N_\alpha}^2} \
  \approx \ 
  \frac{\pi}{m_{N_\alpha}\Gamma_{N_\alpha}}\: \delta(s-m^2_{N_\alpha}) \: 
  \theta(\sqrt s)\; ,
  \label{nwa}
\end{eqnarray}
since  we  are   only  interested  in  the  resonant   part  of  these
$2\leftrightarrow  2$  scatterings  in  the RL  case.  Separating  the
diagonal  $\alpha =  \beta$  RIS contributions  from the  off-diagonal
$\alpha \neq  \beta$ terms  in the sum,  \eqref{be3} can  be rewritten
as~\cite{Pilaftsis:2005rv}
\begin{align}
 \frac{n^\gamma H_N}{z} \frac{\D{}{\delta\eta^L_l}}{\D{}{z}} \ & = \ 
  \sum_{\alpha}
    \bigg(\frac{\eta^N_\alpha}{\eta^N_{\rm eq}} \: - \: 1\bigg)
    \varepsilon_{l\alpha}\gamma^{N_\alpha}_{L\Phi} \:
    - \: \frac{2}{3}\delta\eta^L_l
    \bigg[\sum_{\alpha} B_{l\alpha}\gamma^{N_\alpha}_{L\Phi} \:
    + \sum_{k}\Big(\gamma'^{L_l\Phi}_{L^c_k\Phi^c} \:
        + \: \gamma'^{L_l\Phi}_{L_k\Phi}\Big)\bigg]
  \nonumber\\
  & \qquad \qquad \quad - \: \frac{2}{3}
\sum_{k}\delta\eta^L_k \left[\sum_{\alpha} \varepsilon_{l\alpha}
      \delta\gamma^{N_\alpha}_{L_k\Phi} + 
      \Big(\gamma'^{L_k\Phi}_{L^c_l\Phi^c} \:
        - \: \gamma'^{L_k\Phi}_{L_l\Phi}\Big) \right]\;,
  \label{be4}
\end{align}
where $B_{l\alpha} = \big(\Gamma_{l\alpha} + \Gamma^c_{l\alpha}\big) /
\Gamma_{N_\alpha}$ is the heavy  neutrino decay branching ratio, and
$\gamma'^X_Y  \equiv   \gamma^X_Y  -  \big(\gamma^X_Y\big)_{\rm  RIS}$
denote the RIS-subtracted collision  rates, which can be obtained from
\eqref{scat1} and  \eqref{scat2} taking $\alpha\neq  \beta$. Including
only  the  important RIS-subtracted  collision  rates proportional  to
$\delta\eta^L_l$,   and   neglecting   the   terms   proportional   to
$\delta\eta^L_k$ (for $k\neq l$)  which are numerically small for the
minimal  RL$_\ell$ scenarios~\cite{Deppisch:2010fr},  we  can simplify
\eqref{be4} to
\begin{eqnarray}
  \frac{\D{}\delta \eta^L_l}{\D{}z} \ = \ z^3K_1(z)
  \sum_\alpha {\rm K}_\alpha
  \left(\varepsilon_{l\alpha}\upeta^N_\alpha
    -\frac{2}{3}B_{l\alpha}\kappa_l\delta\eta^L_l\right) \; ,
  \label{be6}
\end{eqnarray}
where we have introduced a flavour-dependent parameter 
\begin{align}
  \kappa_l \ & \equiv \ \frac{\sum_k\left(\gamma^{L_l\Phi}_{L^c_k\Phi^c}
      +\gamma^{L_l\Phi}_{L_k\Phi}\right)+\gamma^{L_l\Phi}_{L^c_l\Phi^c}
    -\gamma^{L_l\Phi}_{L_l\Phi}}
  {\sum_\alpha \gamma^{N_\alpha}_{L\Phi}B_{l\alpha}} 
  \ = \ 2  \sum_{\alpha,\beta} \left(1-2i \: \frac{m_{N_\alpha}-m_{N_\beta}}
    {\Gamma_{N_\alpha}+\Gamma_{N_\beta}}\right)^{-1} \notag \\
  \ & \ \times \; \frac{\left(\widehat{\maf{h}}^*_{l\alpha}
      \widehat{\maf{h}}_{l\beta}+\widehat{\maf{h}}^{c^*}_{l\alpha}
      \widehat{\maf{h}}^c_{l\beta}\right)
    \left[(\widehat{\maf{h}}^\dag\widehat{\maf{h}})_{\alpha\beta}
      +(\widehat{\maf{h}}^{c^\dag}\widehat{\maf{h}}^c)_{\alpha\beta}\right]
    +\left(\widehat{\maf{h}}^*_{l\alpha}
      \widehat{\maf{h}}_{l\beta}-\widehat{\maf{h}}^{c^*}_{l\alpha}
      \widehat{\maf{h}}^c_{l\beta}\right)^2}
  {\left[(\widehat{\maf{h}}\widehat{\maf{h}}^\dag)_{ll}
      +(\widehat{\maf{h}}^c\widehat{\maf{h}}^{c^\dag})_{ll}\right]
    \left[(\widehat{\maf{h}}^\dag\widehat{\maf{h}})_{\alpha\alpha}
      +(\widehat{\maf{h}}^{c^\dag}\widehat{\maf{h}}^c)_{\alpha\alpha}
      +(\widehat{\maf{h}}^\dag\widehat{\maf{h}})_{\beta\beta}
      +(\widehat{\maf{h}}^{c^\dag}\widehat{\maf{h}}^c)_{\beta\beta}
    \right]} \; .
  \label{kappal}
\end{align}
Using the  attractor solution \eqref{etaN_at} in  the kinematic regime
$z > z_1^\alpha$, \eqref{be6} can be written as
\begin{eqnarray}
  \frac{\D{}\delta \eta^L_l}{\D{}z} \ = \ z^2K_1(z)
  \left(\varepsilon_l-\frac{2}{3}z{\rm K}^{\rm eff}_l
    \delta \eta^L_l\right), 
  \label{be7}
\end{eqnarray} 
where $\varepsilon_l = \sum_\alpha \varepsilon_{l\alpha}$ is the total
leptonic $\CP$-asymmetry stored in a given lepton flavour $l$,
and   ${\rm  K}^{\rm   eff}_l=\kappa_l   \sum_\alpha  {\rm   K}_\alpha
B_{l\alpha}\equiv  \kappa_l  {\rm   K}_l$  is  the  effective  washout
parameter due  to $2\leftrightarrow  2$ scatterings mediated  by heavy
neutrinos. Note  that if we only consider  the diagonal $\alpha=\beta$
terms representing the RIS contributions in the sum in \eqref{kappal},
$\kappa_l$   reaches   its   maximum  value,   i.e.~$\kappa_l=1+{\cal
O}(\varepsilon_l^2)$.   On  the other  hand,  in the  $L_l$-conserving
limit,  $\kappa_l$  vanishes at  a  rate at  least  equal  to that  of
$\varepsilon_l$  (see~\ref{app:cp}).   In  the regime  $z>z_2^l\approx
2({\rm  K}_l^{\rm eff})^{-1/3}$,  the total lepton asymmetry, dominated by 
$\varepsilon$-type mixing effects, can be approximated by the analytic 
solution  to \eqref{be7}: 
\begin{eqnarray}
  \label{eq:anal_diag}
  \delta \eta^L \  \simeq \ \delta \eta^L_{\rm mix} \ = \ 
  \frac{3}{2z} \sum_l \frac{\varepsilon_l}{{\rm K}_l^{\rm eff}}
\end{eqnarray}
up to a point  $z=z_3^l\approx 1.25\ln(25\: {\rm K}_l^{\rm eff})$, beyond
which  the lepton asymmetry  freezes out  and approaches  a constant
value    $\delta    \eta^L_{\rm mix}    =   (3/2)\sum_l\varepsilon_l/({\rm    K}_l^{\rm
eff}z_3^l)$~\cite{Deppisch:2010fr}.

\subsection{Observed Lepton Asymmetry}
\label{sec:2.3}

Having obtained  the net lepton asymmetry  $\sum_l \delta \eta^L_l$,
the  next  step  is to  convert  it  to  the  asymmetry in  the  total
baryon-to-photon  ratio  $\delta\eta^B\equiv (n^B-\bar{n}^B)/n^\gamma$
via   $(B+L)$-violating  sphaleron   interactions.   In   a  sphaleron
transition, an  $SU(3)_{\rm c}$ and $SU(2)_L$-singlet neutral  object from each
generation of the SM is created out of the vacuum~\cite{Kuzmin:1985mm,
'tHooft:1976up,  Klinkhamer:1984di, Dimopoulos:1978kv}.   The operator
responsible for sphaleron transitions can be written as
\begin{eqnarray}
  O_{B+L} \ = \ \prod_{i=1}^3 \epsilon_{kl}\epsilon_{mn}\epsilon_{def}
  \left[Q_{k}^d Q^e_{l}Q^f_{m}L_n \right]_i \; ,
  \label{spha}
\end{eqnarray}
where  $i$ is  the  family  index; $d,e,f$  are  the $SU(3)_{\rm c}$  colour
indices;  $k,l,m,n$  are  the  $SU(2)_L$  isospin indices;  and  $Q  =
(\begin{array}{cc} u & d\end{array})^{\sf  T}_L$  is  the   $SU(2)_L$  quark  doublet.  The operator $O_{B+L}$ is invariant under both gauge transformations and $U(3)$ flavour rotations. For the case of our interest, the latter freedom can be used to make the charged lepton Yukawa matrix positive and diagonal. Above the
electroweak  phase transition,  all  the SM  processes, including  the
sphaleron  interactions in  \eqref{spha}, are  assumed to  be  in full
thermal  equilibrium, which  leads  to the  following relations  among
their chemical potentials~\cite{Harvey:1990qw}:
\begin{align}
 & \mu_V \  = \ 0 \; , 
  \quad & \mu_\Phi \  = \ \frac{4}{21}\sum_l \mu_{L_l} \; , 
  \quad & \mu_{e_{R,l}} \  = \ \mu_{L_l}-\frac{4}{21} \sum_l \mu_{L_l} \; , 
  \nonumber\\
  & \mu_{Q_L} \  = \ -\frac{1}{9}\sum_l \mu_{L_l} \; , 
  \quad & \mu_{u_R} \  = \ \frac{5}{63}\sum_l\mu_{L_l} \; , 
  \quad & \mu_{d_R} \  = \ -\frac{19}{63}\sum_l \mu_{L_l} \; ,
\end{align}
where $V$  stands for all vector  bosons, $u_R$ and $d_R$  for up and
down-type  $SU(2)_L$ quark  singlets, and  $e_R$ for  $SU(2)_L$ lepton
singlets in the SM.  The total chemical potentials of the baryonic and
leptonic doublet fields are then given by
\begin{eqnarray}
  \mu_B \ = \ 3(2\mu_{Q_L}+\mu_{u_R}+\mu_{d_R}) \ 
  = \ -\frac{4}{3}\sum_l\mu_{L_l} \; , 
  \qquad
  \mu_L \ = \ 2\sum_l \mu_{L_l} \; . 
  \label{chem}
\end{eqnarray}

Using  the  relations \eqref{chem}  into  \eqref{nat},  in the  linear
approximation  of $\mu_a/T$,  we obtain  the conversion  of  the total
lepton asymmetry  stored in  the SM lepton-doublet to  the baryon 
asymmetry\footnote{Note  that since  we are  converting  the asymmetry
stored in the {\it lepton-doublet}, the conversion coefficient derived
here    is    different     from    28/51    used    elsewhere    (see
e.g.~\cite{Deppisch:2010fr}),  which corresponds  to the  {\it total}
lepton asymmetry, including the RH leptons.}
\begin{eqnarray}
  \delta\eta^B \ = \ -\frac{2}{3}\sum_l \delta\eta^L_l \; , 
  \label{etaB} 
\end{eqnarray}
assuming  a  rapid  sphaleron  transition  rate  $\Gamma_{\rm  sph}\gg
H(z=1)$. This is  valid at temperatures $T > T_c$,  where $T_c$ is the
critical temperature  for the  electroweak phase transition,  given at
one loop by~\cite{Cline:1993bd}
\begin{eqnarray}
  T_c^2 \  = \ \frac{1}{4D_c}\left[M_H^2-\frac{3}{8\pi^2v^2}
    \left(2M_W^4+M_Z^4-4M_t^4\right)
    -\frac{1}{8\pi^2v^4 D_c}\left(2M_W^3+M_Z^3\right)^2 \right].
  \label{Tcrit}
\end{eqnarray}
Here,     $D_c\equiv    (2M_W^2+M_Z^2+2M_t^2+M_H^2)/8v^2$, where    $v    =
2^{-1/4}G_F^{-1/2} =  246.2$ GeV is  the electroweak VEV  ($G_F$ being
the  Fermi coupling  constant),  $M_H$ is  the zero-temperature  Higgs
boson mass, and  $M_W,~M_Z,~M_t$ are the $W$, $Z$ boson masses and 
top-quark mass respectively,  defined at the  electroweak scale. Using
the latest experimental values of the SM mass parameters $M_W = 80.385
(15)$ GeV, $M_Z = 91.1876 (21)$ GeV~\cite{pdg}, $M_t = 173.34\pm 0.76$
GeV~\cite{ATLAS:2014wva},    and     $M_H    =    125.5_{-0.6}^{+0.5}$
GeV~\cite{ATLAS:2013mma}, we obtain
\begin{eqnarray}
T_c \ = \ 149.4_{\: - \: 0.8}^{\: + \: 0.7}~{\rm GeV} \; .
\label{Tcritnum}
\end{eqnarray} 

For  $T<T_c$,  the  so-generated  baryon asymmetry  \eqref{etaB}  gets
diluted by standard photon  interactions until the recombination epoch
at temperature  $T_0$. Assuming that  there is no source  or mechanism
for  significant entropy release  while the  Universe is  cooling down
from $T_c$ to $T_0$, the  baryon number in a comoving volume, $n^B/s$,
is  constant during  this  epoch. Here,  $s=(2\pi^2/45)g_sT^3$ is  the
entropy  density and $g_s$  is the  corresponding effective  number of
relativistic  degrees of freedom.   Thus, the  baryon-to-photon ratio,
$\eta^B  = \pi^4g_s/45\zeta(3)s$, at  the recombination  epoch is
different from its  value predicted by \eqref{etaB} due  to the change
in  $g_s$ with  temperature.  The  prediction for  the  current baryon
asymmetry is thus given by
\begin{eqnarray}
  \delta\eta^B_0 \ = \ \frac{g_s(T_0)}{g_s(T_c)}\delta\eta^B \ 
  \simeq \ \frac{\delta\eta^B}{27.3} \; ,
  \label{etaB0}
\end{eqnarray}
where we have used $g_s(T_c)=106.75$ and $g_s(T_0)=3.91$~\cite{kolb}. 

The theoretical  prediction \eqref{etaB0} has to be  compared with the
observed  value  today,  which  remains  almost unchanged  from  the  end  of
recombination epoch until the present. The latter
can be  expressed in terms of directly  measurable quantities, namely,
the  baryon density  $\Omega_B  h^2$ and  the  primordial $^4$He  mass
fraction $Y_{\rm P}$, as follows~\cite{Steigman:2006nf}:
\begin{eqnarray}
  \frac{\delta\eta^B_0}{\Omega_B h^2} \
  = \ \Big[273.9\pm 0.3 + 1.95 \ (Y_{\rm P}-0.25)\Big]\times 10^{-10} \; .
\label{etaB3}
\end{eqnarray}
Using  the recent results  for the  Planck temperature  power spectrum
data,  combined with  the WMAP  polarization data  at  low multipoles,
which give $\Omega_B h^2=0.02205\pm 0.00028$ and $Y_{\rm P}=0.24770\pm
0.00012$  at 68\% CL~\cite{Ade:2013zuv},  we infer  from \eqref{etaB3}
the observed value of the baryon-to-photon ratio at 68\% CL
\begin{eqnarray}
  \delta\eta^B_{\rm obs} \ = \ (6.04\pm 0.08)\times 10^{-10} \; , 
  \label{eta_obs0}
\end{eqnarray} 
from which we can estimate the necessary lepton asymmetry 
using \eqref{etaB} and \eqref{etaB0}, i.e.
\begin{eqnarray}
  \delta \eta^L_{\rm obs} \ = \ -(2.47\pm 0.03)\times 10^{-8} \; .
  \label{eta_obs}
\end{eqnarray}
Note  the sign  difference between  $\delta\eta^B$  and $\delta\eta^L$
[cf.~\eqref{etaB}].   The  numerical  value  of  the  total  lepton
asymmetry  in a given  leptogenesis model  should  be compatible  with the
observed  value \eqref{eta_obs},  thus  constraining  the relevant
model parameter space.

\section{Flavour Covariant Transport Equations}
\label{sec:3}

As  discussed  in  Section~\ref{sec:2}, the  semi-classical  Boltzmann
equations \eqref{be1} and,  in particular \eqref{be2} and \eqref{be3},
do not take into account  quantum flavour effects such as the coherent
oscillations  between different  flavours of  heavy neutrinos  and the
quantum-statistical decoherence of  flavour off-diagonal matrix number
densities.  In order to  capture these effects consistently,
we will derive
a set of {\it  fully}  flavour-covariant  transport  equations for  the  matrix
number densities describing  the statistical content of the system.
In the next section and as an application of our general formalism, we
will consider the specific  case of RL, subsequently demonstrating the
importance of the flavour effects captured here, but missed in earlier
treatments  of the  subject.  Keeping  this particular  application in
mind,  we will  consider a  specific system  of  $\mathcal{N}_L$ Dirac
lepton   isospin  doublets   $L_l$,  $\mathcal{N}_N$   heavy  Majorana
neutrinos    $N_\alpha$,    and    an    $SU(2)_L$    Higgs    doublet
$\Phi$. Nevertheless,  we emphasize that the analysis  of this section
can  be  easily generalized  to  other  physical situations  involving
flavour  effects,  such  as  the   evolution  of  jet  flavours  in  a
quark-gluon  plasma  or  of  neutrino  flavours in  a  
core-collapse supernova.

In a general flavour basis, the  relevant part  of the  Lagrangian, involving  the  heavy Majorana
neutrinos, is given by
\begin{equation}
  -\mathcal{L}_N  \ = \ \h{l}{\alpha} \, \overline L^{l} \,
  \widetilde{\Phi} \, N_{\rm R, \alpha} \;
  +\; \frac{1}{2}\,\overline{N}_{\rm R, \alpha}^C \, [M_N]^{\alpha \beta} \,
  N_{\rm R, \beta} \;+\; {\rm H.c.}\;,
  \label{eq:Lagr}
\end{equation}
where  $N_{\rm R, \alpha}  \equiv (\mathrm{P_R}  N)_\alpha$ are  the 
RH heavy  neutrino  fields  and $\widetilde{\Phi}\equiv  i\sigma_2\Phi^*$.
Unless otherwise stated, the  Einstein summation convention is implied
henceforth in the summations over  the lepton flavour indices (lower-case
Latin) $l,m,...$  and the  heavy-neutrino flavour  indices (lower-case
Greek) $\alpha,\beta,...$.   In order  to familiarize the  reader with
the covariant convention  used here and throughout this  text, we will
first  discuss  the  flavour-covariant  transformations of  the  field
operators   appearing  in  \eqref{eq:Lagr}.   In  addition,   we  will
illustrate  the  flavour-covariant  generalizations  of  the  discrete
symmetry  transformations  $C$, $P$  and  $T$.  Subsequently, we  will
derive  general  Markovian  master  equations for  the  matrix  number
densities,  and use them to  describe the  statistical evolution  of our
model system  due to the out-of-equilibrium decays  and inverse decays
of the heavy neutrinos.

\subsection{Flavour Transformations}
\label{sec:3.1}

Under  the $U(\mathcal{N}_L) \:  \otimes \:  U(\mathcal{N}_N)$ flavour
transformations,  the  lepton fields  transform  as  follows in  their
fundamental representation:
\begin{alignat}{5}
  L_l \ &\rightarrow \ L'_l \ = \ V_l^{\phantom l m} \; L_m \;, 
  &\qquad &&
  L^{l} \ & \equiv\ (L_l)^\dagger \ \rightarrow \ L'^l \ = \ 
  V^l_{\phantom{l} m} \; L^{m} \;, \label{flavtrans_L} \\
  N_{\rm R, \alpha} \ &\rightarrow \ N'_{\rm R, \alpha} \ = \ 
  U_\alpha^{\phantom \alpha \beta} \;
  N_{\rm R, \beta} \;, &\qquad &&
  N_{\rm R}^{\phantom{\rm R} \alpha} \ & \equiv \
  (N_{\rm R, \alpha})^\dagger \ \rightarrow \ 
  N_{\rm R}'^{\phantom{\rm R} \alpha} \ = \ 
  U^\alpha_{\phantom{\alpha} \beta} \;
  N_{\rm R}^{\phantom{\rm R} \beta} \;,
  \label{flavtrans_N}
\end{alignat}
with  the  unitary transformation  matrices  $V_l^{\phantom  l m}  \in
U(\mathcal{N}_L)$   and    $U_\alpha^{\phantom   \alpha   \beta}   \in
U(\mathcal{N}_N)$  for  which  the  operation of  complex  conjugation
exchanges  subscripts   and  superscripts,  i.e.~$V^l_{\phantom{l}  m}
\equiv (V_l^{\phantom l  m})^*$ and $U^\alpha_{\phantom{\alpha} \beta}
\equiv  (U_\alpha^{\phantom \alpha  \beta})^*$.  We  note that  the RH
part of  the Majorana neutrino fields  $N_{{\rm R},\alpha}$ transforms
covariantly,  as shown in  \eqref{flavtrans_N}.  The  left-handed (LH) part,  
on the
other  hand, transforms  contravariantly  and, as  such, the  Majorana
fields   $N_\alpha$  do   not  have   definite  flavour-transformation
properties.   The   Lagrangian  \eqref{eq:Lagr}  is   invariant  under
$U(\mathcal{N}_L)    \otimes     U(\mathcal{N}_N)$,    provided    the
heavy-neutrino  Yukawa  couplings and  the  Majorana masses  transform
appropriately, as indicated by the relative position of the indices in
\eqref{eq:Lagr}, i.e.
\begin{equation}
  \h{l}{\alpha} \ \rightarrow \ h_{l}'^{\ \alpha} \ = \ V_l^{\phantom l m}
  \;
  U^\alpha_{\phantom{\alpha} \beta} \; \h{m}{\beta} \; , 
\qquad\qquad
  [M_N]^{\alpha \beta} \ \rightarrow \ 
[M'_N]^{\alpha \beta} \ = \ U^\alpha_{\phantom{\alpha} \gamma} \;
  U^\beta_{\phantom{\beta} \delta} \; [M_N]^{\gamma \delta} \;.
\end{equation}

In the  physical mass eigenbasis, the  Dirac field can  be expanded in
a basis of plane waves:
\begin{equation}
  \widehat{L}_l(x;\tilde{t}_i) \ = \ \sum_s \int_{\ve p} \,
  (2 \widehat{E}_{L,l}(\ve p))^{-1/2} \,
  \Big(  e^{-i\widehat{p}_l\cdot x}\,
    \widehat{u}_l(\ve p, s) \,
    \widehat{b}_l(\ve p, s,0;\tilde{t}_i)\:+\:  
     e^{i \widehat{p}_l\cdot x}\,
    \widehat{v}_l(\ve p, s) \, 
    \widehat{d}^{\dagger}_l(\ve p, s,0;\tilde{t}_i)  \Big) \;,
    \label{Lhat1}
\end{equation}
where      $\widehat{b}_l(\ve      p,s,\tilde{t};\tilde{t}_i)$     and
$\widehat{d}_l^{\dag}(\ve p,s,\tilde{t};\tilde{t}_i)$ are respectively
the   interaction-picture  particle   annihilation   and  antiparticle
creation operators  evaluated at the time $\tilde{t}  = 0$. Hereafter,
for  notational convenience, we  will suppress  the dependence  of the
operators  on  the boundary  time  $\tilde{t}_i$  at  which the  three
pictures  of quantum  mechanics, viz.~Schr\"{o}dinger,  Heisenberg and
interaction (Dirac) pictures, are coincident~\cite{Millington:2012pf}.
The index $s = \pm$ denotes the two helicity states with the unit spin
vector  $\maf{n} =  s\maf{s}=s\maf{p}/|\maf{p}|$  aligned parallel  or
anti-parallel  to  the   three  momentum  $\mathbf{p}$,  respectively.
Herein, we  have suppressed the  isospin index of the  lepton doublet.
Notice also that we have  chosen the normalization of the creation and
annihilation  operators, such  that they  have mass  dimension $-3/2$.
This choice  of normalization is made  so that the  bilinears of these
creation and  annihilation operators have the dimension  of the number
density  operator, i.e.~mass  dimension  $-3$.  A  discussion of  the
flavour-covariant Bogoliubov  transformations relating this  choice of
normalization  to  the more  common  and manifestly  Lorentz-invariant
normalization, in  which the creation and  annihilation operators have
mass dimension $-1$, is included in \ref{app:propagator}.

It  is  now  important  to  note  that  $b_k(\ve  p,s,\tilde{t})$  and
$d^{\dagger}_k(\ve   p,s,\tilde{t})$    transform   under   the   same
representation  of  $U(\mathcal{N}_L)$   and  so  also  do  $b^{k}(\ve
p,s,\tilde{t})  \equiv  \big(b_k(\ve p,s,\tilde{t})\big)^\dagger$  and
$d^{\dagger, k}(\ve  p,s,\tilde{t}) \ \equiv  \ \big(d_k^{\dagger}(\ve
p,s,\tilde{t})\big)^\dagger$.\footnote{Hence,    $d^{\dagger,   k}(\ve
p,s,\tilde{t})$ is an annihilation operator, i.e.~$d^{\dagger, k}(\ve
p,s,\tilde{t})  |0\rangle =  b_k(\ve p,s,\tilde{t})  |0\rangle  = 0$.}
The  equal-time  anti-commutation  relations  for these  operators  are
obtained by  a flavour-transformation from the mass  eigenbasis of the
corresponding flavour-diagonal anti-commutators, i.e.
\begin{align}
  \label{eq:b_d_anticomm}
  \big\{ b_l(\ve p,s,\tilde{t}), \,
  b^{m}(\ve p',s',\tilde{t}) \big\} \
  & = \ \big\{d^{\dagger, m}(\ve p,s,\tilde{t}) , \,
  d_l^{\dagger}(\ve p',s',\tilde{t}) \big\} 
  \ = \ \DiT{(\ve p - \ve p')} \, \delta_{s s'}\,
  \delta_l^{\phantom l m} \; .
\end{align}
Note  that due  to the  choice of  normalization of  the  creation and
annihilation  operators, the  anti-commutation relations  obtained here
are  isotropic in  flavour space,  which  does not  occur for  the alternate
choice    of    normalization    discussed    in~\ref{app:propagator} 
[cf.~\eqref{eq:bdcom}].

By applying a flavour  transformation \eqref{flavtrans_L} to a general
basis,  we  obtain  the  covariant  expansion of  the  lepton  doublet
\eqref{Lhat1}:
\begin{align}
  L_l(x) \ &= \ \sum_s \int_{\ve p} \, \Esdu{L}{p}{l}{i}
  \Big( \edu{-}{p}{i}{j} \, \su{s}{p}{j}{k} \,
    b_k(\ve p,s,0)  \;
    \nonumber\\&\qquad \qquad +\: \edu{}{p}{i}{j} \, \sv{s}{p}{j}{k} \,
    d_k^{\dagger}(\ve p,s,0)
  \Big)\;,
  \label{eq:expansion_Dirac1}
  \\
  \overline{L}^l(x) \ &= \ \sum_s \int_{\ve p} \, \Esud{L}{p}{l}{i}
  \Big( \eud{}{p}{i}{j} \, \sub{s}{p}{j}{k} \,
    b^{k}(\ve p,s,0) \nonumber\\&\qquad \qquad
    +\: \eud{-}{p}{i}{j} \, \svb{s}{p}{j}{k} \,
    d^{\dagger, k}(\ve p,s,0) \Big) \;,
  \label{eq:expansion_Dirac2}
\end{align}
where   the   rank-2   tensors  in   \eqref{eq:expansion_Dirac1}   and
\eqref{eq:expansion_Dirac2}  are  defined  by  means  of  the  flavour
transformations \eqref{flavtrans_L} from the mass eigenbasis, i.e.
\begin{equation}
  [(E_L(\ve p))^2]_l^{\phantom l m} \ \equiv\ V_l^{\phantom l k} \;
  V^m_{\phantom m k} \; (\widehat{E}_{L,{k}}(\ve p) )^2 \
  = \ |\ve p|^2 \delta_l^{\phantom{l} m} \;+\;
  [M_{L}^\dag M_L]_l^{\phantom l m} \; ,
  \label{ELmatx}
\end{equation}
in   which    $M_L$   is   the   charged-lepton    mass   matrix   and
$\delta_l^{\phantom{l}  m}$  is  the  usual  Kronecker  delta.   Since
$\mat{E}_L(\ve        p)$        is       Hermitian,        $([E_L(\ve
p)]_{l}^{\phantom{l}m})^*=[E_L(\ve  p)]^{l}_{\phantom{l}m}$.   For the
Dirac spinors, our notation is such that
\begin{equation}
  \su{s}{p}{l}{m}\ =\ V_l^{\phantom{l} k}\; V^{m}_{\phantom{m} k}\; 
  \widehat{u}_k(\ve p, s)\;,\qquad
  \sub{s}{p}{l}{m}\ =\ V^l_{\phantom{l} k}\; V_{m}^{\phantom{m} k}\; 
  \widehat{\bar{u}}_k(s,\ve p)\;.
\end{equation}
Full  details  of  the  flavour-covariant  spinor  algebra  are  given
in~\ref{app:propagator}.

In order to write down the flavour covariant expansion of the Majorana
field,  we recall  that  in the  mass  eigenbasis the  expansion of  a
Majorana  fermion can be  obtained from  a Dirac  one by  imposing the
Majorana condition
\begin{equation}
  \widehat d^{\dagger,\alpha}(\ve k,-\,r,\tilde{t}) \
  = \ \widehat b_\alpha(\ve k,r,\tilde{t}) \
  \equiv \ \widehat a_\alpha(\ve k,r,\tilde{t}) \;.
\end{equation}
Since   $b_\alpha(\ve   k,r,\tilde{t})$  and   $d^{\dagger,\alpha}(\ve
k,-\,r,\tilde{t})$  transform  differently  under the  transformations
given in  \eqref{flavtrans_N}, this condition  cannot be imposed  in a
general flavour basis.  Instead, writing the mass-eigenbasis operators
in   terms   of   the   ones   in   a   general   basis,   we   obtain
$U^{\beta}_{\phantom \beta \alpha} \, b_\beta(\ve k,r,\tilde{t}) \ = \
U_{\gamma}^{\phantom   \gamma  \alpha}   \,   d^{\dagger  ,\gamma}(\ve
k,-\,r,\tilde{t})$,  where $U_\alpha^{\phantom  \alpha \beta}$  is the
flavour transformation that connects  the mass eigenbasis to the basis
under  consideration.  We thus  obtain the  flavour-covariant Majorana
condition
\begin{equation}
  d^{\dagger , \alpha}(\ve k,-\,r,\tilde{t})\ 
  = \ ( U^* U^\dagger )^{\alpha \beta} \, b_\beta(\ve k,r,\tilde{t}) \
  \equiv \ G^{\alpha \beta} \, b_\beta(\ve k,r,\tilde{t}) \;,
  \label{eq:def_G}
\end{equation}
where $G^{\alpha\beta}$ denote the elements of a unitary and symmetric
matrix.  It  can be  shown  that this  matrix  $\mat{G}$  is a  rank-2
contravariant  tensor,\footnote{Performing  a  flavour  transformation
$U'$ on $G^{\alpha\beta}$ defined in \eqref{eq:def_G}, we get
\[ G^{\alpha\beta} \ \rightarrow  \ G'^{\alpha\beta}\ = \ [(U' U)^*(U'
U)^\dagger]^{\alpha \beta} \ = \ [ U'^{*} \, G \, U'^\dagger ]^{\alpha
\beta}    \    =     \    {U'}^{\alpha}_{\phantom    \alpha    \gamma}
{U'}^{\beta}_{\phantom \beta \delta} G^{\gamma\delta} \;,
\] which is the transformation  law of a rank-2 contravariant tensor.}
which  is  equal  to  the   identity  matrix  $\mat{1}$  in  the  mass
eigenbasis.   Combining  the   constraint  \eqref{eq:def_G}  with  the
expansions               \eqref{eq:expansion_Dirac1}               and
\eqref{eq:expansion_Dirac2}, the flavour covariant expansion of the RH
part of the Majorana neutrino  field and its Dirac conjugate are given
by
\begin{align}
  \label{expand:Majo1}
  N_{\rm R, \, \alpha}(x) \ &= \ \sum_r \int_{ \ve k} \,
  \Esdu{N}{k}{\alpha}{\beta}\Big(
  \edu{-}{k}{\beta}{\gamma} \, \mathrm {P_R} \,
    \su{r}{k}{\gamma}{\delta} \, a_\delta(\ve k, r,0) 
    \nonumber\\&\qquad \qquad
    +\: \edu{}{k}{\beta}{\gamma} \, \mathrm {P_R}\,
    \sv{-\,r}{k}{\gamma}{\delta} \, G_{\delta \epsilon} \,
    a^{\epsilon}(\ve k,r,0) \Big) \;,\\
  \overline{N}_{\rm R}^{\alpha}(x) \ &= \ \sum_r \int_{ \ve k} \, 
  \Esud{N}{k}{\alpha}{\beta} \Big( \eud{}{k}{\beta}{\gamma} \,
    \sub{r}{k}{\gamma}{\delta} \, {\rm {P_L}} \,
    a^{\delta}(\ve k, r,0)
    \nonumber\\&\qquad \qquad
    +\: \eud{-}{k}{\beta}{\gamma} \,
    \svb{-\,r}{k}{\gamma}{\delta} \, {\rm {P_L}} \,
    G^{\delta \epsilon} a_\delta(\ve k,r,0) \Big) \; .
  \label{expand:Majo2}
\end{align}
Notice that the helicity of the $v$ spinors is different from those of
the   corresponding   creation   and   annihilation   operators   (see
e.g.~\cite{Pal}).  The  rank-2 tensors in  \eqref{expand:Majo1}  and
\eqref{expand:Majo2} can be  defined using the flavour transformations
\eqref{flavtrans_N} from the mass eigenbasis, e.g.  
\begin{equation}
  [(E_N(\ve k))^2]_\alpha^{\phantom \alpha \beta} \
  = \ |\ve k|^2 \delta_\alpha^{\phantom{\alpha}\beta} \;
  +\; [M_N^\dag M_N]_{\alpha}^{\ \beta} \;.
  \label{ENmatx}
\end{equation}
The  anti-commutation  relation  for  the heavy-neutrino  creation  and
annihilation operators are given by
\begin{equation}
  \big\{ a_\alpha(\ve k,r,\tilde{t}), \,
  a^{\beta}(\ve k',r',\tilde{t}) \big\} \
  = \ \DiT{(\ve k - \ve k')} \, \delta_{r r'} \;
  \delta_\alpha^{\phantom \alpha \beta} \; .
\end{equation}
From  \eqref{expand:Majo1} and \eqref{expand:Majo2},  we see  that the
elements  $G_{\alpha \beta}$  play  the role  of generalized  Majorana
creation  phases.\footnote{With  the  necessary  introduction  of  the
matrix $\mat{G}$, we may decompose  the Majorana field in terms of its
RH  and  LH  components  as   $N_{\alpha}  =  N_{\rm  R,\,  \alpha}  +
G_{\alpha\beta}N_{\rm  L}^{\beta}$,   having  definite  transformation
properties.  However,  we see that  there remains an ambiguity  in any
such  decomposition,   since  we  could  equally   well  have  written
$N^{\alpha} = G^{\alpha\beta}N_{\rm R,\,\beta} + N_{\rm L}^{\alpha}$.}

Finally, the  complex scalar field in \eqref{eq:Lagr}  can be expanded
as
\begin{equation}
  \widetilde\Phi(x) \ = \ \int_{\ve q} (2 E_{\Phi}(\ve q))^{-{1}/{2}}
  \Big( e^{-i q \cdot x}\,\bar c(\ve q,0) \:
    +\: e^{i q \cdot x}\,c^{\dagger}(\ve q,0) \Big)  \;,
\end{equation}
where the interaction-picture  creation and annihilation operators for
the scalar field satisfy the commutation relations
\begin{equation}
  \big[ c(\ve q,\tilde{t}), \,
  c^{\dagger}(\ve q',\tilde{t}) \big] \
  = \ \big[ \bar{c}(\ve q,\tilde{t}), \,
  \bar{c}^{\dagger}(\ve q',\tilde{t}) \big] \ 
  = \ \DiT{(\ve q - \ve q')} \;.
\end{equation}

In  a general  flavour  basis,  the free  Hamiltonians  of the  lepton
doublet and heavy neutrino fields are
\begin{align}
  H^0_L & \ = \ \sum_s\int_{\ve p} \,
  \Edu{L}{p}{m}{l}
  \Big(b^{m}(\ve p,s,\tilde{t}) \,
  b_l(\ve p,s,\tilde{t}) \;
    +\; d^{\dagger}_l(\ve p,s,\tilde{t}) \,
    d^{\dagger,m}(\ve p,s,\tilde{t}) \Big)\; ,
  \label{free_HamL}\\
  H^0_N & \ = \ \sum_r \int_{\ve k} \,
  \Edu{N}{k}{\beta}{\alpha} \,
  a^{\dagger,\beta}(\ve k,r,\tilde{t}) \,
  a_\alpha(\ve k,r,\tilde{t}) \; , 
  \label{free_HamN}
\end{align}
as can  readily be verified  by flavour transformations from  the mass
eigenbasis in which the Hamiltonians are flavour-diagonal.

The  flavour  occupancies  and  coherences  in the  evolution  of  our
multiparticle system  can be  described in terms  of flavour-covariant
matrix  number densities,  analogous to  the ones  for  light neutrino
flavours introduced in~\cite{Sigl:1993}.   For the lepton doublets, we
define
\begin{align}
  \n{L}{s_1 s_2}{p}{l}{m}{t} \ & \equiv \ \frac{1}{\mathcal V_3} \,
  \langle b^m(\ve p, s_2,\tilde{t}) \,
  b_l(\ve p, s_1,\tilde{t})\rangle_t \;,
  \label{eq:def_n_1}\\
  \nb{L}{s_1 s_2}{p}{l}{m}{t} \ & \equiv \ \frac{1}{\mathcal V_3} \,
  \langle d_l^{\dagger}(\ve p,s_1,\tilde{t}) \,
  d^{\dagger,m}(\ve p,s_2,\tilde{t})\rangle_t \; ,
  \label{eq:def_n_2}
\end{align}
where  $\mathcal V_3\:  =\: \DiT(\ve  0)$ is  the  infinite coordinate
three-volume    of    the   system    and    the   macroscopic    time
$t=\tilde{t}-\tilde{t}_i$  is  the   interval  of  microscopic  time
between  the specification of  the initial  conditions ($\tilde{t}_i$)
and  the  observation  of  the  system ($\tilde{t}$).   We  note  in
particular the  relative reversed order  of indices in the  lepton and
anti-lepton number densities, which  guarantees that the two quantities
transform in the same representation  and thus can be combined to form
a  flavour-covariant  lepton  asymmetry.   Similarly,  we  define  the
heavy-neutrino number densities
\begin{align}
  \n{N}{r_1 r_2}{k}{\alpha}{\beta}{t} \ & \equiv \
  \frac{1}{\mathcal V_3} \,
  \langle a^{\beta}(\ve k,r_2,\tilde{t}) \,
  a_\alpha(\ve k,r_1,\tilde{t})\rangle_t \;,
  \label{eq:def_n_3} \\
  \nb{N}{r_1 r_2}{k}{\alpha}{\beta}{t} \ & \equiv
  \frac{1}{\mathcal V_3} \,
  \langle G_{\alpha \gamma}\, a^\gamma(\ve k, r_1,\tilde{t}) \;
  G^{\beta \delta} \, a_\delta(\ve k, r_2,\tilde{t}) \rangle_t\;,
  \label{eq:def_n_4}
\end{align}
and the scalar number densities
\begin{align}
  {n}^{\Phi}(\ve q,t) \  \equiv \ \frac{1}{\mathcal V_3} \,
  \langle c^{\dagger}(\ve q,\tilde{t}) \,
  c(\ve q,\tilde{t})\rangle_t \;, \qquad 
  {\bar{n}}^{\Phi}(\ve q,t) \ \equiv \ \frac{1}{\mathcal V_3} \,
  \langle \bar c^{\dagger}(\ve q,\tilde{t}) \,
  \bar c(\ve q,\tilde{t})\rangle_t \; .
  \label{eq:def_n_5}
\end{align}

The  total   number  densities  $\mat   n^X$  (without  three-momentum
argument)         are          obtained         by         integrating
\eqref{eq:def_n_1}--\eqref{eq:def_n_5}    over    the    corresponding
three-momenta and tracing over helicity and isospin, i.e.
\begin{equation}
  \mat{n}^{N}(t) \ \equiv \ \sum_{r = -,+} \int_{\ve k} \,
  \mat{n}_{r r}^N(\ve k,t)\;, \quad
  \mat{n}^{L}(t) \ \equiv \ \underset{\rm iso}{\Tr} \sum_{s = -,+}
  \int_{\ve p} \, \mat{n}_{s s}^L(\ve p,t) \;, \quad
  \mat{n}^{\Phi}(t) \ \equiv \ \underset{\rm iso}{\Tr}
  \int_{\ve q} \, \mat{n}^\Phi(\ve q,t) \;,
  \label{def_n_tot}
\end{equation}
where  we have  identified explicitly  that  the traces  are taken  in
isospin space. Analogous definitions  are assumed for the antiparticle
number densities.   Note that all the matrix  number densities defined
above,   as   well  as   the   energy   matrices  \eqref{ELmatx}   and
\eqref{ENmatx}, are Hermitian in flavour space.

\subsection{Flavour Covariant Discrete Symmetries} 
\label{sec:3.2}

It is useful  to derive the transformation properties  of the discrete
symmetries $C$,  $P$, and $T$  in the flavour-covariant  formalism. We
assume that the action of these operators on the fermion fields in the
mass eigenbasis  is the  standard one (see  e.g.~\cite{Pokorski}), and
find its generalization to an  arbitrary flavour basis by means of the
appropriate       flavour      transformations       discussed      in
Section~\ref{sec:3.1}.   In the  mass  eigenbasis, the  action of  the
unitary charge-conjugation operator $\C$ on elements of the Fock space
is given by~\cite{Pokorski}
\begin{equation}
  \widehat{b}_l(\ve p,s,\tilde{t})^C \ \equiv \ {\C} \,
  \widehat{b}_l(\ve p,s,\tilde{t}) \, {\Cd} \
  = \ - i \,\widehat{d}^{\dagger,l}(\ve p,s,\tilde{t}) \;.
  \label{eq:b_C}
\end{equation}
Note  that the  phase convention  for the  operators used  here  is in
accordance with those used for the spinors in~\ref{app:propagator}. By
writing the mass-eigenbasis  operators in terms of those  in a general
basis, i.e.~$  \widehat{b}_l(\ve p,s,\tilde{t}) =  V^m_{\phantom m l}
\,      b_m(\ve      p,s,\tilde{t})$,     $\widehat{d}^{\dagger,l}(\ve
p,s,\tilde{t})  = V_n^{\phantom  n  l} \,  \widehat{d}^{\dagger,n}(\ve
p,s,\tilde{t})$, we find the flavour-covariant $C$-transformation
\begin{equation}\label{eq:C_trans}
  b_l(\ve p,s,\tilde{t})^C \ = \
  {\C} \, b_l(\ve p,s,\tilde{t}) \, {\Cd} \ 
  = \ - i \, (V V^{\mathsf T})_{lm} \,
  d^{\dagger,m}(\ve p,s,\tilde{t})
  \ \equiv \ - i \, \mathcal{G}_{lm} \,
  d^{\dagger,m}(\ve p,s,\tilde{t}) \;, 
\end{equation}
where we have been required to introduce the matrix $\bm{\mathcal{G}}$
for the  charged leptons,  analogous to $\mat{G}$  in \eqref{eq:def_G}
for  heavy  neutrinos.  Thus,  we  see  that  in  a  flavour-covariant
formulation  the action  of $C$  necessarily involves  the appropriate
flavour rotation.  We find it useful to  define the \emph{generalized}
$C$-transformation $\widetilde{C}$, i.e.
\begin{equation}
  b_l(\ve p,s,\tilde{t})^{\widetilde{C}} \ \equiv \ \mathcal{G}^{lm} \,
  b_m(\ve p,s,\tilde{t})^{C} \
  = \ -i \, d^{\dagger,l}(\ve p,s,\tilde{t}) \; ,
\end{equation}
which is  a combination of the $C$-transformation  and the appropriate
flavour rotation.\footnote{These are equivalent to the transformations
considered  in~\cite{Ecker:1981wv, Bernabeu:1986fc}.  However,  in our
case,  the  appropriate flavour  rotations  are  {\it  forced} by  the
flavour-covariance  of  the  formalism,  once the  canonical  discrete
transformations are defined in the  mass basis.}  Thus we see that, in
an arbitrary  flavour basis,  the particle and  antiparticle operators
are related by a  $\widetilde{C}$-transformation, which reduces to the
usual charge-conjugation operation in  the mass eigenbasis. The action
of $C$ on the fermion fields is obtained analogously, i.e.
\begin{equation}\label{eq:C_field}
  L_l(x)^C \ = \ {\C} \, L_l(x) \, {\Cd} \
  = \ \mathcal{G}_{lm} \, {C} \, \bar{L}^{m,\mathsf{T}}(x) \;,
\end{equation}
with  ${C} =  i  \gamma^0 \gamma^2$  (in  the helicity  basis),
whereas the action of $\widetilde{C}$ gives the more familiar result
\begin{equation}
  \label{eq:Ct_field}
  L_l(x)^{\widetilde{C}} \ = \ \mathcal{G}^{lm}\, L_m(x)^C \
  = \ {C} \, \bar{L}^{l,\mathsf{T}}(x) \;,
\end{equation}
using the fact that $\mathcal{G}^{lm}\mathcal{G}_{mn} = \delta^l_n$. 

Similarly, the  parity transformation,  given by the  unitary operator
$\UP$  in Fock space,  can be  generalized straightforwardly  from the
mass eigenbasis, i.e.
\begin{align}
  \label{eq:P_trans1}
  b_l(\ve p,s,\tilde{t})^P \ & \equiv \
  {\UP} \, b_l(\ve p,s,\tilde{t}) \, {\UPd} \ 
  = - s \, b_l(-\ve p,-s,\tilde{t})\;, \\
  L_l(x_0, \ve x)^P \ & \equiv \ {\UP} \, L_l(x_0, \ve x) \, {\UPd} \
  = \ {P} \, L_l(x_0, - \ve x) \;, 
  \label{eq:P_trans2}
\end{align}
with ${P} = \gamma^0$.  Under  $\CP$, the action of the heavy-neutrino
interaction Lagrangian \eqref{eq:Lagr} transforms as
\begin{equation} \label{lag3.33}
  \C \, \UP \left(\int_x \hs{l}{\alpha} \,
    \overline{N}_{\rm R}^{\alpha} \, \widetilde{\Phi}^\dagger \, L_l  \;
    + \; {\rm H.c.}\right) \UPd \,\Cd \
  = \ \int_x \mathcal{G}^{l m} \, \h{m}{\beta} \,
  G_{\beta \alpha} \, \bar L^{l} \, \widetilde{\Phi} \,
  N_{\rm R, \alpha} \; + \mathrm{H.c.} \;,
\end{equation}
where we  have introduced the short-hand notation  $\int_x \equiv \int
\mathrm{d}^4x$   for  the   integration   over  space-time.   Equation
\eqref{lag3.33}  defines  the   $\CP$-transformations  of  the  Yukawa
couplings:
\begin{equation}
  (\h{l}{\alpha})^{\CP} \ = \ \mathcal{G}_{lm} \, \hs{m}{\beta} \, 
  G^{\beta \alpha}\;, \qquad
  (\h{l}{\alpha})^{\widetilde{C}P} \ 
  \equiv \ \mathcal{G}^{lm} \, (\h{m}{\beta})^{\CP} \,
  G_{\beta \alpha} \ = \ \hs{l}{\alpha} \;.
  \label{CP_yuk}
\end{equation}
For  a general  matrix  element, the  relation  \eqref{CP_yuk} can  be
generalized to
\begin{equation}
  \label{amp_CP}
  \mathcal{M}(X \to Y)^{\widetilde{C}P} \
  = \ \mathcal{M}(X^{\tilde{c}} \to Y^{\tilde{c}}) \;,
\end{equation}
where $X^{\tilde{c}}  \equiv X^{\widetilde{C}\!P}$ is  the generalized
$\CP$-transformation  of the state  $X$, which  can, for  instance, be
obtained from \eqref{eq:Ct_field} and \eqref{eq:P_trans2}.

The action of the time-reversal  transformation $T$ is described by an
anti-unitary operator $\T$ in the  Fock space. Again, starting from the
mass-eigenbasis relation
\begin{equation}
  \label{eq:b_T}
  \widehat{b}_l(\ve p,s,\tilde{t})^T \ \equiv \ {\T} \,
  \widehat{b}_l(\ve p,s,\tilde{t}) \, {\Td} \
  = \ \widehat{b}_l(-\ve p,s,-\tilde{t}) \;,
\end{equation}
we find, because of the anti-linearity of $\T$,
\begin{align}
  b_l(\ve p,s,\tilde{t})^T \ & = \ {\T} \,
  b_l(\ve p,s,\tilde{t}) \, {\Td} \
  = \ \mathcal{G}^{lm} \,  b_m(-\ve p,s,-\tilde{t}) \;, \\
  L_l(x_0, \ve x)^T \ & = \ {\T} \, L_l(x_0, \ve x) \, {\Td} \
  = \ \mathcal{G}^{lm} \, {T} \,L_m(-x_0, \ve x) \;,
\end{align}
with  ${T}   =  i   \gamma^1  \gamma^3$.  Thus,   we may introduce
the generalized $T$-transformations $\widetilde{T}$ as follows:
\begin{align}
  b_l(\ve p,s,\tilde{t})^{\widetilde{T}} \ &\equiv \
  \mathcal{G}_{lm} \, b_m(\ve p,s,\tilde{t})^{T} \
  = \   b_l(-\ve p,s,-\tilde{t}) \;, \\
  L_l(x_0, \ve x)^{\widetilde{T}} \ &\equiv \
  \mathcal{G}_{lm} \, L_m(x_0, \ve x)^{T} \
  = \ {T} \,L_l(-x_0, \ve x)\;.
\label{eq:T_trans2}
\end{align}
Hence, in a general basis,  incoming and outgoing states are exchanged
by a $\widetilde{T}$ operation.  This  can be seen by transforming the
interaction Lagrangian \eqref{eq:Lagr} under $\T$ from which we obtain
\begin{equation}
  (\h{l}{\alpha})^T \ = \ \mathcal{G}_{lm} \, \hs{m}{\beta} \,
  G^{\beta \alpha}\;, \qquad
  (\h{l}{\alpha})^{\widetilde{T}} \
  \equiv \ \mathcal{G}^{lm} \, (\h{m}{\beta})^T \, G_{\beta \alpha} \
  = \ \hs{l}{\alpha} \; .
\end{equation}
Generalizing the above transformations to the matrix elements gives
\begin{equation}
  \label{amp_T}
  \mathcal{M}(X \to Y)^{\widetilde{T}} \ = \ \mathcal{M}(Y \to X) \;.
\end{equation}
From  \eqref{eq:Ct_field}   and  \eqref{eq:T_trans2},  we   obtain  an
important equivalence relation:
\begin{align}
  L_l(x)^{\widetilde{C}\!P\widetilde{T}} \ = \
  \mathcal{G}^{lm}L_m(x)^{\CP\widetilde{T}} \ = \
  \mathcal{G}^{lm}\mathcal{G}_{mn} L_n(x)^{\CP T} \ = \
  \delta^l_{\phantom{l}n} L_n(x)^{\CP T} \ = \ L_l(x)^{\CP T} \; .
\end{align}
As a consequence, we have the identity 
\begin{equation}
  \widetilde{C}\! P \widetilde{T} \ = \ \CP T \;.
\label{cpt}
\end{equation}
Combining the results \eqref{amp_CP}  and \eqref{amp_T}, and using the
$CPT$-invariance  of  the  Lagrangian  \eqref{eq:Lagr},  the  identity
\eqref{cpt} allows us to relate the matrix elements as
\begin{align}
 \mathcal{M}(X \to Y)^{\widetilde{C}P\widetilde{T}} \ &= \ 
 \mathcal{M}(Y^{\tilde{c}} \to X^{\tilde{c}})\ =\ \mathcal{M}^*(X\to Y)\;. 
\end{align}

The        number       density       matrices        defined       in
\eqref{eq:def_n_1}--\eqref{eq:def_n_4}   have   simple  transformation
properties    under    $\widetilde{C}$.    Since    $d^{\dagger,l}(\ve
p,s,\tilde{t}) = i \, b_l(\ve p,s,\tilde{t})^{\widetilde{C}}$, for the
lepton number densities  \eqref{eq:def_n_1} and \eqref{eq:def_n_2}, we
have
\begin{equation}
  \mat{n}^{L}(\ve p,t)^{\widetilde{C}} \
  \equiv \  \mat{\mathcal{G}} \, \langle \, {\C} \,
  \mat{\check{n}}^{L}(\ve p,\tilde{t}) \, {\Cd} \, \rangle_t \,
  \mat{\mathcal{G}}^\dagger  \ = \ \mat{\bar{n}}^{L}(\ve p,t)^{\mathsf T} \;,
  \label{gC_L}
\end{equation}
where  the transposition on  the far  RHS of  \eqref{gC_L} is  on both
flavour and  helicity indices. Similarly,  for the heavy  neutrinos we
have $a_\alpha(\ve k,r)^{C} \ = - i \, a_\alpha(\ve k, r)$, and hence,
the transformation relation
\begin{equation}
  \mat{n}^{N}(\ve k,t)^{\widetilde{C}} \ \equiv \
  \mat{G} \, \langle \, {\C} \, 
  \mat{\check{n}}^{N}(\ve k,\tilde{t}) \,
  {\Cd} \,\rangle_t \, \mat{G}^\dagger \
  = \ \mat{\bar{n}}^{N}(\ve k,t)^{\sf T} \;.
  \label{gC_N}
\end{equation}
Thus,     $\mat{\bar{n}}^N(\ve    k,t    )^{\mathsf{T}}$     is    the
$\widetilde{C}$-conjugate of $\mat{n}^N(\ve k, t)$. For Majorana heavy
neutrinos,  the  two   $\widetilde{C}$-conjugate  quantities  are  not
independent, and are related by
\begin{equation}
  \label{eq:Majorana_bar_G}
  \nb{N}{r_1 r_2}{k}{\alpha}{\beta}{t} \
  = \ G_{\alpha \mu} \, \n{N}{r_2 r_1}{k}{\lambda}{\mu}{t} \,
  G^{\lambda \beta} \;.
\end{equation}

Using  the $\widetilde{C}$-transformation  relations  \eqref{gC_L} and
\eqref{gC_N}, we can construct  the following quantities with definite
$\widetilde{C}$ transformation properties:
\begin{align}
  \dn{L}{s_1 s_2}{p}{l}{m}{t} \ &= \ \n{L}{s_1 s_2}{p}{l}{m}{t} \ 
  - \ \nb{L}{s_1 s_2}{p}{l}{m}{t} \;, 
  \label{eq:L_CP} \\
  \dn{N}{r_1 r_2}{k}{\alpha}{\beta}{t} \ &
  = \ \n{N}{r_1 r_2}{k}{\alpha}{\beta}{t} \ 
  - \ \nb{N}{r_1 r_2}{k}{\alpha}{\beta}{t}\;,
  \label{eq:dN_CP}\\
  \nn{N}{r_1 r_2}{k}{\alpha}{\beta}{t} \ &
  = \ \frac{1}{2} \, \left( \n{N}{r_1 r_2}{k}{\alpha}{\beta}{t} \
    + \ \nb{N}{r_1 r_2}{k}{\alpha}{\beta}{t} \right) \;,
  \label{eq:N_CP}
\end{align}
which transform under $\widetilde{C}$ as
\begin{align}
  \mat{\delta n}^{L}(\ve p, t)^{\widetilde{C}} \ 
  = \ - \; \mat{\delta n}^{L}(\ve p, t)^{\sf T} , \quad
  \mat{\delta n}^{N}(\ve k, t)^{\widetilde{C}} \
  = \ - \; \mat{\delta n}^{N}(\ve k, t)^{\sf T}, \quad
  \mat{\underline n}^{N}(\ve k, t)^{\widetilde{C}} \ 
  = \ + \; \mat{\underline n}^{N}(\ve k, t)^{\sf T} .
  \label{gctrans}
\end{align}
Thus, the  quantities $\mat{\delta n}^X$  are $\widetilde{C}$-``odd'',
and  $\mat{\underline n}^{N}$  is $\widetilde{C}$-``even'',  where the
quotation marks refer to the fact that  this is not meant to be in the
usual  sense   due  to  the  transposition   involved.   The  definite
$\widetilde{C}$-properties of the above  quantities can be extended to
$\gCP$,  once  the  total  unpolarized  number  densities  defined  by
\eqref{def_n_tot}  are  considered. Note  that  we  did  not define  a
$\widetilde{C}$-even quantity  for lepton number  densities (analogous
to  $\mat{\underline n}^N$),  since this  can be  approximated  by the
equilibrium number density $n^{L}_{\rm eq}$, i.e.
\begin{equation}
  \mat{n}^L(t)  + \bar{\mat{n}}^L(t) \
  =  \ 2 \, n_{\rm eq}^L \: \mat{1} + {\cal O}(\mu_L^2/T^2) \; .
  \label{nL_tot}
\end{equation}
However, this is  not always true for the heavy   neutrinos,   i.e.~$\mat{\underline   n}^{N}(t)\neq   n^N_{\rm
eq}\mat{1}$, since  we must have a departure  from thermal equilibrium
in      order      to      satisfy     the      basic      Sakharov
conditions~\cite{Sakharov:1967}  for  the  generation  of  a  non-zero
lepton asymmetry.

In  the  heavy-neutrino  mass  eigenbasis  the  transformation  matrix
$\widehat{\mat{G}}$ reduces  to the identity  matrix $\mat{1}$, and
hence, the  transformations $C$ and $\widetilde{C}$  are identical for
the heavy neutrinos. In this basis, the heavy Majorana neutrino number
densities \eqref{eq:dN_CP} and \eqref{eq:N_CP} reduce to
\begin{equation}
  \mat{\widehat{\underline n}}^N(\ve k,t) \
  = \ {\rm Re}\left[\mat{\widehat{n}}^N(\ve k,t)\right] \;,
  \qquad\qquad
  \mat{\delta \widehat{n}}^N(\ve k,t) \
  = \ 2\, i \, {\rm Im}\left[\mat{\widehat{n}}^{N}(\ve k,t)\right] \;.
\end{equation}
It  should be  noted  that both  $\mat{\widehat{\underline n}}^N(t)$  and
$\mat{\delta  \widehat{n}}^N(t)$ are  {\it even}  under the  usual charge
conjugation operation in the mass eigenbasis, as expected for Majorana
fermions:\footnote{This   is  consistent   with   the  $\widetilde{C}$ 
transformations in a general basis, as given by \eqref{gctrans}, 
due to the transposition involved. 
However, under a naive $T$-reversal transformation $N_\alpha\leftrightarrow N_\beta$, 
we have
\[
\mat{\widehat{\underline n}}^N(\ve k,t) \ \to \ [\mat{\widehat{\underline n}}^N(\ve k,t)]^{\sf T} \ =  \ \mat{\widehat{\underline n}}^N(\ve k,t)\; , \qquad 
\mat{\delta \widehat{n}}^N(\ve k,t) \ \to \ [\mat{\delta \widehat{n}}^N(\ve k,t)]^{\sf T} \ = \ - \: \mat{\delta \widehat{n}}^N(\ve k,t) \; ,
\]
due to the fact that, in the mass eigenbasis, $\mat{\widehat{\underline n}}^N$ is a symmetric matrix, while $\mat{\delta \widehat{n}}^N$ is anti-symmetric.}
\begin{align}
  \mat{\widehat{\underline n}}^N(\ve k,t)^{\widetilde{C}} \ &
  = \ \mat{\widehat{\underline n}}^N(\ve k,t)^{C} \
  = \ + \; \mat{\widehat{\underline n}}^N(\ve k,t) \;,\\
  \mat{\delta\widehat{n}}^N(\ve k,t)^{\widetilde{C}} \ &
  = \ \mat{\delta\widehat{n}}^N(\ve k,t)^{C} \
  = \ +\; \mat{\delta\widehat{n}}^N(\ve k,t) \; .
\end{align}
In  addition, we  note that  the  total lepton  asymmetry $\delta  n^L(t)
\equiv \Tr[\mat{\delta n}^L(t)]$ is $\CP$-odd in any basis:
\begin{equation}
  \delta n^L(t)^{ \CP} \ \equiv \ \Tr\:[\mat{\delta n}^L(t)]^{CP} \
  = \ \Tr \:[\mat{\delta n}^L(t)]^{\widetilde{C}P} \
  = \ - \, \Tr\:[\mat{\delta n}^L(t)]^{\sf T} \ = \ - \, \delta n^L(t) \;. 
\end{equation}
We will use these definitions to write down the flavour-covariant rate
equations for RL in Section~\ref{sec:4}.

\subsection{Markovian Master Equation}
\label{sec:3.3}

In  this section,  we  derive  a master  equation  governing the  time
evolution  of the matrix  number densities  $\mat{n}^X(\ve p,t)$  in a
Markovian  approximation. We  will  work in  the interaction  picture,
beginning  from  the (picture-independent)  definition  of the  number
density   in  terms   of  the   quantum-mechanical   density  operator
$\rho(\tilde{t};\tilde{t}_i)$:
\begin{equation}
  \label{eq:def_n_rho}
  \mat{n}^{X}(t) \ \equiv \ \langle
  \mat{\check{n}}^{X}(\tilde{t};\tilde{t}_i) \rangle_t \
  = \ \Tr\left\{ \rho(\tilde{t};\tilde{t}_i) \;
    \mat{\check{n}}^{X}(\tilde{t};\tilde{t}_i) \right\} \;,
\end{equation}
where the trace is over the Fock space and, for notational simplicity,
we leave the momentum dependence implicit. Here we have used the
accent  (~$\check{}$~) to distinguish  the quantum-mechanical
number         density         operator         $\mat{\check{n}}^X(\ve
p,\tilde{t};\tilde{t}_i)$  from its  expectation  value $\mat{n}^X(\ve
p,t)$, where recall that $t=\tilde{t}-\tilde{t}_i$ is the macroscopic time. 
In addition,  for the purposes of this section,  it is necessary
to reintroduce the explicit dependence of the operators 
on the microscopic boundary time $\tilde{t}_i$.

Differentiating \eqref{eq:def_n_rho} with respect to time, we have
\begin{equation}
  \label{eq:diff}
  \frac{ \D{}{\mat{n}^X(t)}}{\D{}{t}} \ = \
  \Tr\left\{ \rho(\tilde{t};\tilde{t}_i) \,
    \frac{ \D{}{\mat{\check{n}}^{X}(\tilde{t};\tilde{t}_i)}}
    {\D{}{\tilde{t}}} \right\} \;
  + \; \Tr \left\{ \frac{\D{}{\rho(\tilde{t};\tilde{t}_i)}}
    {\D{}{\tilde{t}}} \,
    \mat{\check{n}}^{X}(\tilde{t};\tilde{t}_i) \right\} \ \equiv \
  \mat{\mathcal{I}}_1 \,
  + \, \mat{\mathcal{I}}_2 \;,
\end{equation}
where   we    have   used   the   fact    that   $\D{}{}/\D{}{t}\   =\
\D{}{}/\D{}{\tilde{t}}$ for  fixed $\tilde{t}_i$.   As we work  in the
interaction  picture,  the time  evolution  of the  quantum-mechanical
operator $\mat{\check{n}}^X(\tilde{t};\tilde{t}_i)$ is governed by the
free    Hamiltonian   $H_0^X$    given    by   \eqref{free_HamL}    or
\eqref{free_HamN} depending  on whether $X=L$  or $N$.  Hence,  we use
the  Heisenberg  equation  of  motion  to  write  the  first  term  in
\eqref{eq:diff} as
\begin{equation}
  \mat{\mathcal{I}}_1  \ = \ i \,
  \Tr\Big\{\rho(\tilde{t};\tilde{t}_i) \,
    [H_0^X,\
    \mat{\check{n}}^{X}(\tilde{t};\tilde{t}_i)]\Big\} \
  \equiv \ i \, \langle \, [H_0^X,\ 
  \mat{\check{n}}^{X}(\tilde{t};\tilde{t}_i) ] \, \rangle_t \; .
\end{equation}
This term generates flavour oscillations in the case of a non-diagonal
energy  matrix.   The  second  term in  \eqref{eq:diff}  involves  the
interaction Hamiltonian, e.g.
\begin{eqnarray}
  H_{\rm int} \ = \ \int_x \h{l}{\alpha} \,
  \bar L^{l} \, \widetilde{\Phi} \, N_{\rm R, \alpha} \,
  + \, {\rm H.c.} \label{Hint}
\end{eqnarray}
As we shall see below, this term will generate the collision terms for
the  generalized  Boltzmann   equations  (in  addition  to  dispersive
corrections). The starting point is the Liouville-von~Neumann equation
(see e.g.~\cite{sakurai})
\begin{equation}
  \label{eq:tovolt}
  \frac{\D{}{\rho(\tilde{t};\tilde{t}_i)}}
  {\D{}{\tilde{t}}} \ = \
  - \, i \, [H_{\rm int}(\tilde{t};\tilde{t}_i),\
  \rho(\tilde{t};\tilde{t}_i)] \;.
\end{equation}
Rewriting  \eqref{eq:tovolt}  in  the  form  of  a  Volterra  integral
equation  of   the  second   kind,  iterating  once   and  subsequently
differentiating with respect to time, we obtain the integral form
\begin{equation}
  \frac{\D{}{\rho(\tilde{t};\tilde{t}_i)}}{\D{}{\tilde{t}}} \
    = \ - \, i \, [H_{\rm int}(\tilde{t};\tilde{t}_i),
    \rho(\tilde{t}_i;\tilde{t}_i)] \;
  - \;  \int_{\tilde{t}_i}^{\tilde{t}} \D{}{\tilde{t}'}\;
  [H_{\rm int}(\tilde{t};\tilde{t}_i),\
  [H_{\rm int}(\tilde{t}';\tilde{t}_i),\
  \rho(\tilde{t}';\tilde{t}_i)]] \;.
  \label{drho_dt}
\end{equation}
Inserting \eqref{drho_dt} into \eqref{eq:diff}, we obtain 
\begin{align}
  \label{eq:I2int1}
  \mat{\mathcal{I}}_2 \ &= \ -\:i\Tr\Big\{[
  H_{\rm int}(\tilde{t};\tilde{t}_i),\
  \rho(\tilde{t}_i;\tilde{t}_i)]
  \mat{\check{n}}^{X}(\tilde{t};\tilde{t}_i) \Big\} \nonumber\\&\qquad 
  - \int_{\tilde{t}_i}^{\tilde{t}} \D{}{\tilde{t}'}\; \Tr
  \Big\{ [H_{\rm int}(\tilde{t};\tilde{t}_i),\
  [H_{\rm int}(\tilde{t}';\tilde{t}_i),\ \rho(\tilde{t}';\tilde{t}_i)]] \, 
  \mat{\check{n}}^{X}(\tilde{t};\tilde{t}_i) \Big\}\;.
\end{align}
The  first term  on  the  RHS of  \eqref{eq:I2int1}  vanishes for  the
$H_{\rm  int}$  term given  by  \eqref{Hint},  since  it involves  the
product of an odd number of fields.  For the second term on the RHS of
\eqref{eq:I2int1}, we may use the cyclicity of the trace to obtain
\begin{align}
  \label{eq:colone}
  \mat{\mathcal{I}}_2 \ &= \ - \int_{\tilde{t}_i}^{\tilde{t}}
  \D{}{\tilde{t}'}\;
  \Tr\Big\{ \rho(\tilde{t}';\tilde{t}_i)
  [H_{\rm int}(\tilde{t}';\tilde{t}_i),\
  [H_{\rm int}(\tilde{t};\tilde{t}_i),\ 
  \mat{\check{n}}^{X}(\tilde{t};\tilde{t}_i)]] \Big\}
  \nonumber \\ &\qquad \equiv \ - \int_{\tilde{t}_i}^{\tilde{t}}
  \D{}{\tilde{t}'} 
  \; \langle \,
  [H_{\rm int}(\tilde{t}';\tilde{t}_i),\
  [H_{\rm int}(\tilde{t};\tilde{t}_i),
  \mat{\check{n}}^{X}(\tilde{t};\tilde{t}_i)]] \,
  \rangle_{\tilde{t}'-\tilde{t}_i} \;.
\end{align}
When used in \eqref{eq:diff},  this gives an exact and self-consistent
time-evolution equation,  which captures  the entire evolution  of the
system, including any non-Markovian memory effects.

We      now      perform      a      set      of      Wigner-Weisskopf
approximations~\cite{Weisskopf:1930au}  to  obtain  the  leading-order
Markovian    form   of    \eqref{eq:diff}.   Let    us    define   the
$\tilde{t}$-dependent function
\begin{equation}
  \mat{F}(\tilde{t};\tilde{t}_i) \ =\
  [H_{\mathrm{int}}(\tilde{t};\tilde{t}_i),\ 
  \mat{\check{n}}^{X}(\tilde{t};\tilde{t}_i)]\;.
\end{equation}
Inserting          the          Fourier         transforms          of
$H_{\mathrm{int}}(\tilde{t};\tilde{t}_i)$                           and
$\mat{F}(\tilde{t};\tilde{t}_i)$        with        respect       to
$\tilde{t}'-\tilde{t}_i$      and      $\tilde{t}-\tilde{t}_i$     in
\eqref{eq:colone}, we obtain
\begin{equation}
  \mat{\mathcal{I}}_2 \ =\ -\int_{\tilde{t}_i}^{\tilde{t}} 
  \D{}{\tilde{t}'} \int \frac{\D{} \omega}{2\pi}
  \int \frac{\D{} \omega'}{2\pi}\;
  e^{-i\omega' (\tilde{t}'-\tilde{t}_i)}\,e^{-i\omega (\tilde{t}-\tilde{t}_i)}
  \,\,
  \braket{[H_{\mathrm{int}}(\omega'),\
  \mat{F}(\omega)]}_{\tilde{t}'-\tilde{t}_i} \; .
\end{equation}
Making the change of variables $\omega = \omega'' - \omega'$, this may
be recast in the form
\begin{equation}
  \label{eq:coltwo}
  \mat{\mathcal{I}}_2\ =\ - \int_{\tilde{t}_i}^{\tilde{t}}
  \D{}{\tilde{t}'} \int
  \frac{\D{}\omega'}{2\pi} \int \frac{\D{} \omega''}{2\pi}\:
  e^{-i\omega'(\tilde{t}'-\tilde{t})}\,e^{-i\omega'' t}\,
  \braket{[H_{\mathrm{int}}(\omega'),\,
    \mat{F}(\omega''-\omega')]}_{\tilde{t}'-\tilde{t}_i}\;.
\end{equation}
As  long  as  $\mat{F}(\omega''-\omega')$  remains  dynamical  on
inverse   Fourier  transformation,  i.e.~$\omega''\neq   \omega'$,  the
dominant  contribution to  the integral  \eqref{eq:coltwo}  occurs for
$\tilde{t}'\sim     \tilde{t}$.     We    may     therefore    replace
$\rho(\tilde{t}';\tilde{t}_i)   \to  \rho(\tilde{t};\tilde{t}_i)$,  or
$\langle  \cdots  \rangle_{\tilde{t}'-\tilde{t}_i}\to  \langle  \cdots
\rangle_{t}$  in  \eqref{eq:coltwo}.   We   now  make  the  change  of
variables    $t'=\tilde{t}'-\tilde{t}$     and    take    the    limit
$\tilde{t}_i\:\to\:-\infty$.  Herein,  we assume that  the statistical
evolution is  slow compared to the  quantum-mechanical evolution, such
that   the  system   remains  out-of-equilibrium   in  spite   of  the
quantum-mechanical boundary time being in the infinitely distant past.
With this  approximation, we replace  the interaction-picture creation
and               annihilation               operators              in
$H_{\mathrm{int}}(\tilde{t};\tilde{t}_i)$                           and
$\mat{\check{n}}^X(\tilde{t};\tilde{t}_i)$  by  their asymptotic  `in'
counterparts via
\begin{equation}
  \label{eq:asymphase}
  c_{\mathrm{in}}^{(\dag)}(\mathbf{p})\ \equiv\
  Z^{-1/2}\lim_{\tilde{t}_i\:\to\:-\infty}e^{(-)iE(\mathbf{p})\tilde{t}}
  c^{(\dag)}(\mathbf{p},\tilde{t};\tilde{t}_i) 
\ = \ Z^{-1/2} \lim_{\tilde{t}_i\:\to\:-\infty}
  c^{(\dag)}(\mathbf{p},0;\tilde{t}_i) \;, 
\end{equation}
where $Z  = 1+\mathcal{O}(\hbar)$ is  the wavefunction renormalization
factor. Notice  that in the replacement  \eqref{eq:asymphase}, we must
account  for  the  free  phase evolution  of  the  interaction-picture
operators.

The contribution $\mat{\mathcal{I}}_2$ now takes the form
\begin{equation}
  \mat{\mathcal{I}}_2 \ \simeq\ - \int_{-\infty}^{0}\!\! \D{}{t'}
  \int \frac{\D{}{\omega'}}{2\pi} \int \frac{\D{}{\omega''}}{2\pi} \,
  e^{-i\omega' t'} \, e^{-i \omega'' t} \, 
  \braket{[H_{\mathrm{int}}(\omega'),\, \mat{F}(\omega''-\omega')]}_{t}\;.
\end{equation}
Performing the $t'$ integration, we find
\begin{equation}
  \int_{-\infty}^{0} \D{}{t'}\; e^{-i\omega' t'} \ =\
  \pi \, \delta(\omega') \; + \;  i \, \mathcal{P} \frac{1}{\omega'} \ 
  =\ \frac{1}{2}\int_{-\infty}^{+\infty} \D{}{t'} \,
  e^{-i\omega' t'} \;+\; i \, \mathcal{P}\frac{1}{\omega'}\;,
\end{equation}
where $\mathcal{P}$  denotes the Cauchy  principal value. We  are then
able to write the result
\begin{align}
  \label{eq:withstark}
  \mat{\mathcal{I}}_2 \ &\simeq \ - \,\frac{1}{2}
  \int_{-\infty}^{+\infty} \D{}{t'} \, \langle \,
  [H_{\rm int}(t'), \, [H_{\rm int}(t), \mat{\check{n}}^{X}(t)]] \,
  \rangle_{t}
   \nonumber\\&\qquad -\:  \mathcal{P}\int \frac{\D{}{\omega}}{2\pi} \,
  \frac{ie^{-i\omega t}}{\omega} \, \langle \, [H_{\rm int}(\omega), \,
   [H_{\rm int}(t), \mat{\check{n}}^{X}(t)]] \, \rangle_{t}\;,
\end{align}
where objects constructed from asymptotic operators depend only on the
time   $t$   and  the   $\mathcal{O}(\hbar)$   corrections  from   the
wavefunction renormalization  have been neglected.  The  first term in
\eqref{eq:withstark}   can  be  identified   as  the   collision  term
$\mat{C}^X$ and the second  term represents the dispersive self-energy
corrections arising from vacuum contributions (Lamb shift) and thermal
contributions  (Stark shift), which  we neglect  in the  following.  A
discussion  of  our  resummation  scheme in  relation  to  self-energy
corrections is included at the beginning of Section~\ref{sec:4}.

Restoring the implicit momentum dependence of the number densities, we
finally  obtain the  leading-order  master equation  in the  Markovian
approximation:\footnote{There  are  two   major  assumptions  in  this
approximation: (i)  separation of time scales, i.e.~the QFT processes
governed  by  $H_{\rm int}$  occur  at  much  smaller time  scales  as
compared  to  the  coarse-grained  statistical evolution  governed  by
$\rho(t)$; and  (ii) molecular chaos, i.e.~the velocity correlations
that may form between different  species in the QFT processes are lost
on  time-scales relevant for  the statistical  evolution, so  that the
background can be factorized at all times.}
\begin{eqnarray}
  \label{eq:master}
  \frac{\D{}{}}{\D{}{t}} \mat{n}^X(\ve k, t) \
  \simeq \ i  \langle \, [H_0^X,\  \mat{\check{n}}^{X}(\ve k, t) ] \,
  \rangle_t  -  \frac{1}{2} \int_{-\infty}^{+\infty} \D{}{t'} \;
  \langle \, [H_{\rm int}(t'),\
  [H_{\rm int}(t),\ \mat{\check{n}}^{X}(\ve k, t)]] \, \rangle_{t} \; .
\end{eqnarray}
This  is our  central  equation governing  the  time-evolution of  the
particle  number  densities  with  arbitrary  flavour  content.\footnote{This formalism 
is sometimes called the `density matrix formalism' in the literature. 
In our opinion, this terminology is misleading, since $\mat{n}^X(\ve k, t)$ is actually a 
{\it matrix of densities}~\cite{Sigl:1993}, which should be distinguished from the 
quantum-statistical {\it density matrix} $\rho(t)$. Such confusion could potentially 
lead to incorrect results, since there is a crucial sign difference in the time evolution equations for the 
two quantities, as can be seen by comparing~\eqref{eq:tovolt} with the first term on the RHS of~\eqref{eq:master}. Therefore, we avoid referring to our approach as the 
`density matrix formalism'.}   With
\eqref{eq:master}    no   longer    in    integro-differential   
form [cf.~\eqref{drho_dt}],  we   are  now  free  to   specify  the  initial
conditions at any  finite macroscopic time $t_0$. It  remains the case
however  that  the macroscopic  time  $t_0  =  0$ corresponds  to  the
microscopic time  $\tilde{t}=\tilde{t}_i \to -\infty$.

We note that although a similar form of~\eqref{eq:master} was also used in~\cite{Sigl:1993} 
to describe the time-evolution  of active neutrinos in a thermal bath, the  full  flavour  structure  contained  
in~\eqref{eq:master} to describe the simultaneous time-evolution of multiple species, e.g. 
heavy neutrinos and SM leptons as in the context of leptogenesis,  
was not discussed before  in the literature. In  order  to  elucidate these flavour 
effects, we  will explicitly derive  {\it fully} flavour-covariant  transport equations
for  the system  described by  the Lagrangian  \eqref{eq:Lagr}  in the
following subsection.

\subsection{Transport Equations}
\label{sec:3.4}

Using the expressions \eqref{free_HamL}  and \eqref{Hint} for the free
and interaction Hamiltonians respectively, we can explicitly calculate
the        oscillation       and       collision        terms       in
\eqref{eq:master}.  Specifically, we  obtain  the following  evolution
equations  for the  charged  lepton and  anti-lepton number  densities,
specifically
\begin{align}
  \frac{\D{}{} }{\D{}{t}} \, \n{L}{s_1 s_2}{p}{l}{m}{t} \ &
  = \ - \, i \,
  \Big[{E}_L(\ve p), \,{n}^{L}_{s_1 s_2}(\ve p,t)
  \Big]_{l}^{\phantom{l}m} \;
  + \; [{C}^L_{s_1 s_2}(\ve p,t)]_l^{\phantom l m} \;,
  \label{eq:evol_lept}\\
  \frac{\D{}{} }{\D{}{t}} \, \nb{L}{s_1 s_2}{p}{l}{m}{t} \ &
  = \ + \, i \,
  \Big[{E}_L(\ve p), \,{\bar{n}}^{L}_{s_1 s_2}(\ve p,t)
  \Big]_{l}^{\phantom{l}m} \; 
  + \; [\overline{{C}}^L_{s_1 s_2}(\ve p,t)]_l^{\phantom l m} \;,
  \label{eq:evol_lept2}
\end{align}
where  commutators  carrying  flavour  indices are  understood  to  be
commutators in flavour space. The collision terms are given by
\begin{align}
  \label{def_coll}
  [{C}^L_{s_1 s_2}(\ve p,t)]_l^{\phantom l m} \ &
  = \ - \frac{1}{2} \, \big[ \,{\mathcal{F}} \cdot {\Gamma} \,
  +\, {\Gamma^\dagger} \cdot
  {\mathcal{F}} \,\big]_{s_1 s_2, \,l}^{\phantom{s_1 s_2, \,l} m} \;,
  \\[6pt]
  [\overline{{C}}^L_{s_1 s_2}(\ve p,t)]_l^{\phantom l m} \ &
  = \ - \frac{1}{2} \, \big[\, {\overline{\mathcal{F}}}
  \cdot {\overline{\Gamma}} \,
  +\, {\overline{\Gamma}^\dagger} \cdot
  {\overline{\mathcal{F}}} \, \big]_{s_1 s_2, \,l}^{\phantom{s_1 s_2, \,l} m}
  \; , \label{C_l2}
\end{align}
where  we   have  suppressed  the  overall   momentum  dependence  and
introduced the compact notation
\begin{align}
  \label{compact1}
  \big[{\mathcal{F}}
  \cdot {\Gamma} \,\big]_{s_1 s_2, \,l}^{\phantom{s_1 s_2, \,l} m} \ &
  \equiv \ \sum_{s,r_1,r_2} \int_{\ve k, \, \ve q}
  \Tdu{\mathcal{F}_{s_1 s \,r_1 r_2}
    (\ve p, \ve q, \ve k,t)}{l}{n}{\alpha}{\beta}
  \;
  \Tdu{\Gamma_{s\, s_2 r_2 r_1}
    (\ve p, \ve q, \ve k)}{n}{m}{\beta}{\alpha}
  \;, \\
  \big[{\Gamma^\dag}
  \cdot {\mathcal{F}} \,\big]_{s_1 s_2, \,l}^{\phantom{s_1 s_2, \,l} m} \ &
  \equiv \ \sum_{s,r_1,r_2} \int_{\ve k, \, \ve q}  \;
  \Tdu{\Gamma^\dagger_{s_1 s \, r_2 r_1}
    (\ve p, \ve q, \ve k,t)}{l}{n}{\beta}{\alpha}
  \;
  \Tdu{\mathcal{F}_{s \, s_2 r_1 r_2}
    (\ve p, \ve q, \ve k)}{n}{m}{\alpha}{\beta}
  \; . \label{compact2}
\end{align} 
It is  important to  emphasize that our  flavour-covariant formulation
requires  {\it   new  rank-4  tensors}  in  flavour   space:  (i)  the
statistical number density tensors
\begin{align}
  \Tdu{\mat{\mathcal{F}}(\ve p, \ve q, \ve k,t)}{}{}{}{} \ &
  = \  n^{\Phi}(\ve q, t) \, \mat{{n}}^L(\ve p, t)  \otimes
  \left[{\mat 1} - \mat{{n}}^{N}(\ve k, t)\right] \nonumber \\
 \ & \qquad  
  - \left[1 + n^{\Phi}(\ve q, t)\right] \left[{\mat 1}
    - \mat{{n}}^L(\ve p, t)\right] \otimes  \mat{{n}}^{N}(\ve k, t)\, ,
  \label{Fstat} \\ 
  \Tdu{\overline{\mat{\mathcal{F}}}(\ve p, \ve q, \ve k,t)}{}{}{}{} \ &
  = \  \bar{n}^{\Phi}(\ve q, t) \, \bar{\mat{n}}^L(\ve p, t)  \otimes
  [{\mat 1} - \bar{\mat{n}}^{N}(\ve k, t)] \nonumber \\
\ & \qquad  
  - [1 + \bar{n}^{\Phi}(\ve q, t)] \,
  [{\mat 1} - \bar{\mat{n}}^L(\ve p, t)] \otimes 
  \bar{\mat{n}}^{N}(\ve k, t) 
  \; , \label{Fbarstat}
\end{align}
and (ii) the absorptive tensors
\begin{align}
  \Tdu{[\Gamma_{s_1 s_2 r_1 r_2}
    (\ve p, \ve q, \ve k)]}{l}{m}{\alpha}{\beta} \ & 
  = \ \hs{k}{\nu} \; \h{i}{\lambda} \
  \DiFud{k-p-q}{j}{p}{\mu}{\delta} \notag\\
  &\hspace{-8em}\times \
  \frac{1}{2 E_\Phi(\ve q)}\, \Esud{L}{p}{i}{j} \, \Esdu{L}{p}{k}{n} \,
  \Esdu{N}{k}{\lambda}{\mu} \, \Esud{N}{k}{\nu}{\gamma} \notag\\
  &\hspace{-8em} \times  \; \Tr\big\{\su{r_2}{k}{\delta}{\beta} \,
  \sub{r_1}{k}{\gamma}{\alpha} \, {\rm {P_L}} \,
  \su{s_2}{p}{n}{m} \, \sub{s_1}{p}{p}{l} \, {\rm {P_R}} \big\}
  \;,\label{eq:Gamma1}\\[9pt]
  \Tdu{[\overline{\Gamma}_{s_1 s_2 r_1 r_2}
    (\ve p, \ve q, \ve k)]}{l}{m}{\alpha}{\beta} \ &
  = \ \hs{k}{\nu} \; \h{i}{\lambda} \ \DiFud{k-p-q}{j}{p}{\mu}{\delta}
  \notag \\
  &\hspace{-8em}\times \ \frac{1}{2 E_\Phi(\ve q)} \, \Esud{L}{p}{i}{j} \, 
  \Esdu{L}{p}{k}{n} \, \Esdu{N}{k}{\lambda}{\mu} \, 
  \Esud{N}{k}{\nu}{\gamma} \notag\\
  &\hspace{-8em}\times \; \Tr\big\{\sv{r_2}{k}{\delta}{\beta} \,
  \svb{r_1}{k}{\gamma}{\alpha} \, {\rm {P_L}} \,
  \sv{s_2}{p}{n}{m} \, \svb{s_1}{p}{p}{l} \, {\rm {P_R}} \big\} \;,
  \label{eq:Gamma2}\\[9pt]
  \Tdu{[\Gamma_{s_1 s_2 r_1 r_2}^\dagger
    (\ve p, \ve q, \ve k)]}{l}{m}{\alpha}{\beta} \ &
  = \  \left(\Tdu{[\Gamma_{s_2 s_1 r_2 r_1}
      (\ve p, \ve q, \ve k)]}{m}{l}{\beta}{\alpha} \right)^* \;,\\[6pt]
  \Tdu{[\overline{\Gamma}_{s_1 s_2 r_1 r_2}^\dagger
    (\ve p, \ve q, \ve k)]}{l}{m}{\alpha}{\beta} \ &
  = \   \left(\Tdu{[\overline{\Gamma}_{s_2 s_1 r_2 r_1}
      (\ve p, \ve q, \ve k)]}{m}{l}{\beta}{\alpha} \right)^*\;.
\end{align}\\[-6pt]
In \eqref{eq:Gamma1} and  \eqref{eq:Gamma2}, the rank-4 delta function
of  on-shell four-momenta  originates from  the integration  of tensor
exponentials, such as
\begin{equation}
  \int_x \, \eud{-}{p}{l}{m} \, e^{- i q \cdot x} \,
  \edu{}{k}{\alpha}{\beta} \ 
  = \ \DiFud{k-p-q}{l}{m}{\alpha}{\beta} \; ,
\end{equation}
and is defined as a  linear combination of ordinary delta functions in
terms  of  the  appropriate   flavour  transformation  from  the  mass
eigenbasis, as defined in Section~\ref{sec:3.1}, i.e.
\begin{equation}
  [\delta^{(4)}(k-p-q)]^{l \phantom{m \alpha}\beta}_{\phantom l m \alpha} \ 
  \equiv \  \; 
  V^l_{\phantom l k} \, V_m^{\phantom m k} \, 
  U_\alpha^{\phantom \alpha \gamma} \, U^\beta_{\phantom \beta \gamma} \, 
  \delta(\widehat{E}_{N,{\gamma}}(\ve k) - \widehat{E}_{L,{k}}(\ve p) 
  - \widehat{E}_{\Phi}(\ve q))\,
  \delta^{(3)}(\ve{k} - \ve{p} - \ve{q}) \; .
\end{equation}
The   absorptive  tensors   \eqref{eq:Gamma1}   and  \eqref{eq:Gamma2}, 
obtained   from  the   Markovian  master   equation  \eqref{eq:master}, 
represent  the contributions  from decays  and inverse  decays  of the
heavy Majorana  neutrinos to the  statistical evolution of  the system
(see                     Section~\ref{sec:4.1}                     and
Figures~\ref{fig:2}(a)--\ref{fig:2}(b)).       As     shown
in~\ref{app:optical}, these rank-4 objects can be interpreted in terms
of  the   unitarity  cut   of  the  partial   one-loop  heavy-neutrino
self-energies,    using   a    generalized   optical    theorem   (see
Figure~\ref{fig:cuts1}). This justifies  the necessity of the tensorial
structure of these objects, and  also the form of the $2\leftrightarrow 2 $ 
scattering terms that will be  included later in the evolution  equations when directly
applied to the RL scenario in Section~\ref{sec:4}. This formalism 
can be generalized to include higher order processes involving multiple flavour 
degrees of freedom, e.g.~$LN\leftrightarrow Le_R$ and $LN\leftrightarrow LN$, 
by introducing the corresponding rate tensors of rank 6 and higher.   
	
\newpage
Proceeding analogously for the heavy neutrino number densities, we find
the evolution equations
\begin{align}
  \frac{\D{}{} }{\D{}{t}} \,\n{N}{r_1 r_2}{k}{\alpha}{\beta}{t} \ &
  = \ - \, i \, \Big[{E}_N(\ve k), \,
  {n}^{N}_{r_1 r_2}(\ve k,t)\Big]_{\alpha}^{\phantom{\alpha}\beta} \;
  + \; [{C}^{N}_{r_1 r_2}(\ve k,t)]_\alpha^{\phantom \alpha \beta} 
+ \; G_{\alpha \lambda} \,
  [\overline{{C}}^N_{r_2 r_1}(\ve k,t)]_{\mu}^{\phantom{\mu} \lambda} \,
  G^{\mu \beta}
 \;,
  \label{eq:evol_neu}\\
  \frac{\D{}{} }{\D{}{t}} \, \nb{N}{r_1 r_2}{k}{\alpha}{\beta}{t} \ &
  = \ + \, i \, \Big[{E}_N(\ve k), \,
  {\bar{n}}^{N}_{r_1 r_2}(\ve k,t)\Big]_{\alpha}^{\phantom{\alpha}\beta} \; 
  + \; [\overline{{C}}^{N}_{r_1 r_2}
  (\ve k,t)]_\alpha^{\phantom \alpha \beta} 
+ \; G_{\alpha \lambda} \,
  [{C}^N_{r_2 r_1}(\ve k,t)]_{\mu}^{\phantom{\mu} \lambda} \,
  G^{\mu \beta}
\;,
  \label{eq:evol_neu2}
\end{align} 
with the heavy-neutrino collision terms given by
\begin{align}
  [{C}^{N}_{r_1 r_2}(\ve k,t)]_\alpha^{\phantom \alpha \beta} \ &
  = \ + \, \frac{1}{2} \, \big[ \, {\mathcal{F}}
  \cdot {\Gamma^\dagger} \,
  +\, {\Gamma} \cdot {\mathcal{F}} 
  \, \big]_{r_1 r_2 , \, \alpha}^{\phantom{r_1 r_2,\, \alpha}  \beta} \;,
  \label{coll_N1} \\[6pt]
  [\overline{{C}}^{N}_{r_1 r_2}
  (\ve k,t)]_\alpha^{\phantom \alpha \beta} \ &
  = \ + \, \frac{1}{2} \, \big[ \, {\overline{\mathcal{F}}}
  \cdot {\overline{\Gamma}^\dagger} \,
  +\, {\overline{\Gamma}} \cdot {\overline{\mathcal{F}}} 
  \, \big]_{r_1 r_2 , \, \alpha}^{\phantom{r_1 r_2,\, \alpha}  \beta} \;,
  \label{coll_N2}
\end{align}
where  the tensor  contraction  is analogous  to \eqref{compact1}  and
\eqref{compact2}, with  the role of  charged-lepton and heavy-neutrino
indices exchanged.  Note the appearance of the matrix $\mat{G}$ in the
transport equations  \eqref{eq:evol_neu} and \eqref{eq:evol_neu2}, and
the  transposition  of  both  flavour  and  helicity  indices  in  the
corresponding  collision  terms.   One  should however  remember  that
\eqref{eq:evol_neu}  and  \eqref{eq:evol_neu2}  are  not  independent,
because  of the relation  \eqref{gC_N}. Note  also that  the transport
equations have an internal structure  in isospin space, which has been
suppressed here  for brevity.  In Section~\ref{sec:4},  when we derive
the rate equations for the  total number densities, we will explicitly
trace over these degenerate isospin degrees of freedom.

As a final  remark, we point out that  the flavour-covariant formalism
developed  so far  can  also  be applied  more  generally to  describe
quantum coherences between species  with different SM quantum numbers,
e.g.~$L_l$ and  $N_\alpha$.  In this case,  we may need  to study the
evolution of  the number densities  $[n^{LN}]_l^{\phantom{l} \alpha}$,
which   transforms  as   a  rank-2   tensor  in   the   flavour  space
$U(\mathcal{N}_L)  \otimes U^*(\mathcal{N}_N)$,  corresponding  to the
correlated creation of $L_l$  and the annihilation of $N_\alpha$.  The
evolution    equation   would    still   have    the    generic   form
\eqref{eq:evol_lept},    with   collision    terms    of   the    form
\eqref{def_coll}.   However,  we will  neglect  these  effects in  our
discussions, since they are not expected to play an important role for
the RL scenarios under study.

\section{Application to Resonant Leptogenesis}
\label{sec:4}

As already discussed in  Section~\ref{sec:2.1}, there are two types of
$\CP$-violation possible in the  out-of-equilibrium decay of the heavy
Majorana  neutrino. In  the limit  when two  (or more)  heavy Majorana
neutrinos become  degenerate, the $\varepsilon$-type  $ \CP$-violation
can be resonantly enhanced by several orders of magnitude, as compared
to   the   $\varepsilon'$-type  $CP$-violation~\cite{Pilaftsis:1997dr,
Pilaftsis:1997jf} (see also~\ref{app:cp}).  In this case, finite-order
perturbation  theory breaks  down and  it  is necessary  to perform  a
consistent field-theoretic  resummation of the  self-energy and vertex
corrections,  as shown  schematically in  Figure~\ref{figcp}. However,
such resummation can only be performed in a closed algebraic form when
working in  the mass eigenbasis and at  full thermodynamic equilibrium
in a  Markovian approximation. This fact is  illustrated explicitly in
\ref{sec:props},  where we derive  the most  general flavour-covariant
propagators    in   a    non-homogeneous    background   within    the
Schwinger-Keldysh   CTP  formalism.  Therein,   we  show   that,  when
out-of-equilibrium  (where   flavour-coherences  must  be  permitted),
canonical quantization  necessarily leads to  off-diagonal propagators
in flavour space.   As a result, the inversion  of the Schwinger-Dyson
equation for the resummed  propagators contains an infinite nesting of
convolution integrals, which does  not collapse to the usual algebraic
equation of resummation.

As identified in  \cite{Millington:2012pf}, the loopwise truncation of
quantum transport equations is two-fold, i.e.~the transport equations
may  be truncated  both  spectrally and  statistically.  The  spectral
truncation corresponds to choosing  those objects that will be counted
as  particulate degrees of  freedom; in  our case,  we will  choose to
count the number densities  of spectrally-free on-shell particles. The
statistical  truncation,  on  the   other  hand,  corresponds  to  the
restriction  of  the  set  of  processes that  drive  the  macroscopic
evolution  of  the  system.  It  is  the latter  that  we  will  treat
non-perturbatively   in  order   to  consistently   account   for  the
$\varepsilon$-     and     $\varepsilon'$-type    $     \CP$-violation
effects. Recalling the argument above, we ensure that this resummation
can be performed algebraically  by first flavour-rotating to the heavy
neutrino mass eigenbasis, maintaining the Markovian approximation used
in   Section~\ref{sec:3.3}   and   resumming   only   zero-temperature
contributions, thereby  neglecting thermal  loop effects~\cite{Giudice:2003jh}. 
As  we shall
see later in this section (see Section \ref{sec:4.4}), the omission of
these  thermal effects  is  appropriate in  the classical  statistical
limit,  as   long  as  one   accounts  systematically  for   both  the
zero-temperature and thermal RIS contributions. Lastly, we will assume
that  the  heavy-neutrino  helicity  states are  fully  decohered  and
equally populated \cite{Pilaftsis:2003gt}.

Hence,  we  proceed  by   replacing  the  tree-level  neutrino  Yukawa
couplings by  their resummed  counterparts in the  transport equations
given  in Section~\ref{sec:3.4}.  Specifically,  for the  processes $N
\to  L  \Phi$ and  $L^{\tilde{c}}  \Phi^{\tilde{c}}  \to  N$, we  have
$\h{l}{\alpha}  \to  \hr{l}{\alpha}$  and,  for $N  \to  L^{\tilde{c}}
\Phi^{\tilde{c}}$  and $L  \Phi \to  N$, we  have  $\hs{l}{\alpha} \to
\hrc{l}{\alpha}$,  corresponding  to  the effective  one-loop-resummed
matrix elements  $[\mathcal{M}(N \to L \Phi)]_{k}^{\phantom{l}\alpha}$
and            $[\mathcal{M}(N            \to            L^{\tilde{c}}
\Phi^{\tilde{c}})]^{k}_{\phantom{l}\alpha}$  respectively,  as defined
in~\ref{app:optical}.  Working in  the heavy-neutrino mass eigenbasis,
the resummed  neutrino Yukawa  couplings are given  by \eqref{resum1},
from which the covariant resummed  Yukawa couplings may be obtained by
the   appropriate  flavour  transformation,   i.e.~$\hr{l}{\alpha}  =
V_{l}^{\ m}U^{\alpha}_{\  \beta}\widehat{\mathbf{h}}_{m}^{\ \ \beta}$,
where                  $\widehat{\mathbf{h}}_{m}^{\                  \
\beta}\equiv\widehat{\mathbf{h}}_{m\beta}$ in the mass eigenbasis.  We
emphasize however that the resummation itself must be performed in the
mass  eigenbasis only. Further  justification for  the choice  of this
basis will  be given in  Section~\ref{sec:5.1}, where we show  that in
the degenerate  symmetry limit of  the minimal RL$_\ell$ model,  it is
important  to choose  the correct  basis  in order  to get  meaningful
results.

\subsection{Rate Equations for Decay and Inverse Decay}
\label{sec:4.1}

In  order to  obtain the  rate  equations from  the general  transport
equations         \eqref{eq:evol_lept},         \eqref{eq:evol_lept2},
\eqref{eq:evol_neu}  and  \eqref{eq:evol_neu2},   we  need  to  impose
kinetic equilibrium.  This can  be ensured throughout the evolution of
the system  by assuming that the elastic  scattering processes rapidly
change the momentum distributions on time-scales much smaller than the
statistical  evolution time  of the  particle number  densities.  This
approximation  is  valid  as  long  as  the  mass  splittings  between
different flavours inside thermal  integrals are much smaller than the
average    momentum    scale,    i.e.~$|\ve    k|    \sim    T    \gg
|m_{N_\alpha}-m_{N_\beta}|$.     In   this   regime,    the   momentum
distributions   governed  by   the  elastic   processes   are  flavour
singlets~\cite{McKellar:1994, Bell:2000kq}.  Using this approximation,
we  introduce  an average  mass  for $\mathcal{N}_N$  quasi-degenerate
heavy neutrinos:
\begin{equation}
  \label{av_mN}
  m^2_N \ =\ \frac{1}{\mathcal{N}_N} \,
  \Tr\big(\mat{M}^\dagger_N \, \mat{M}_N\big) \;,
\end{equation}
to  be used  within the  thermal integrals.  Correspondingly, we may introduce an average 
energy $E_N(\ve k) = (|\ve  k|^2 + m_N^2)^{1/2}$. Furthermore, to simplify the
general  transport equations given  in Section~\ref{sec:3.4},  we take
the   classical   statistical  limit   in   which  \eqref{Fstat}   and
\eqref{Fbarstat} can be approximated as
\begin{align}
  \Tdu{\mat{\mathcal{F}}(\ve p, \ve q, \ve k,t)}{}{}{}{} \ &
  \simeq \ n^{\Phi}(\ve q,t) \, \mat{{n}}^L(\ve p,t)  \otimes {\mat 1} \;
  -\; \mat{1} \otimes  \mat{{n}}^{N}(\ve k,t) \;,
  \label{Fstat_clas}\\[6pt]
  \Tdu{\overline{\mat{\mathcal{F}}}(\ve p, \ve q, \ve k,t)}{}{}{}{} \ &
  \simeq \  \bar{n}^{\Phi}(\ve q,t) \, \bar{\mat{n}}^L(\ve p,t) 
  \otimes {\mat 1} \;-\; \mat{1} \otimes  \bar{\mat{n}}^{N}(\ve k,t) 
  \;.  \label{Fbarstat_clas}
\end{align}
The  spinor   traces   appearing   in
\eqref{eq:Gamma1}    and    \eqref{eq:Gamma2} can be simplified if we assume 
the charged-leptons to be massless, neglecting their 
thermal  masses. In  this limit, $\mat{n}^L$ is the number density
matrix  for the  LH lepton  doublets $L_l$  and one  helicity  index  for  the  charged
leptons can be dropped in the spinor 
traces, thus yielding   (see
also~\ref{app:propagator})
\begin{align}
  \label{eq:traces}
 \ &  \sum_{r=-,+}\, \Tr\big\{u(\ve k , r) \, \bar u(\ve k , r) \,
  \mathrm{P_L} \, u(\ve p,-) \, \bar u(\ve p,-) \,
  \mathrm{P_R} \big\}\ &
  \notag\\
   \ & = \ \sum_{r=-,+}\, \Tr\big\{v(\ve k , r) \, \bar v(\ve k , r) \,
  \mathrm{P_L} \, v(\ve p,+) \, \bar v(\ve p,+) \,
  \mathrm{P_R} \big\} \ = \  2 \, k \cdot p \; ,
\end{align}
where $-  (+)$ corresponds to the  helicity of the  massless LH lepton
(RH anti-lepton).

With  these  approximations,  the  heavy-neutrino  kinetic-equilibrium
number  density,  given  by  the flavour-covariant  generalization  of
\eqref{nat} (with  the chemical potential now being  a rank-2 tensor),
can be approximated as
\begin{align}
  \n{N}{}{k}{\alpha}{\beta}{t} \ &\equiv \ \sum_{r}
  \n{N}{r r}{k}{\alpha}{\beta}{t} \ 
  = \ g_N \,
  \Big[ e^{- \big(E_N(\ve k) - \mu_N(t)\big)/T}
  \Big]_\alpha^{\phantom \alpha \beta}
  \notag \\
  & 
  \simeq \ g_N  \, 
  \big[e^{\mu_N(t)/T} \big]_\alpha^{\phantom \alpha \beta} \, e^{- E_N(\ve k)/T}
  \ = \ g_N \, \frac{[n^N(t)]_{\alpha}^{\phantom{\alpha}\beta}}
  {n^{N}_{\rm eq}} \, 
  e^{- E_N(\ve k)/T} \;,
\end{align}
where $\mat{n}^N(t)$  is the total  heavy neutrino number  density, as
defined in  \eqref{def_n_tot}, and  $n^N_{\rm eq}$ is  the equilibrium
number density  given by \eqref{nat}  setting $\mu_N=0$ and $g_N  = 2$
for the  two heavy-neutrino helicity states.   An analogous expression
can be obtained for the charged lepton number density
\begin{align}
  \n{L}{}{p}{l}{m}{t} \ \simeq \ \frac{[n^{L}(t)]_{l}^{\phantom{l}m}}
  {n^{L}_{\rm eq}} \,
  e^{- E_L(\ve p)/T} \; ,
  \label{nLeq}
\end{align}
where $E_L$ is the average energy of the lepton doublets, and 
in the massless limit, 
$n^L_{\rm eq}$  is the equilibrium number density given by  \eqref{nrel}   
with $g_L  = 2$ for  the two isospin states. Notice that the  factor $g_L$ is
not present in \eqref{nLeq}, and will appear explicitly only after the
trace        over       isospin        is        performed  
[cf.~\eqref{def_n_tot}].  Finally, for  the Higgs  number  density, we
assume an equilibrium distribution  $n^\Phi(\ve q,t) = e^{- E_\Phi(\ve
q)/T}$.  Throughout the  remainder of  this article,  we  suppress the
$t$-dependence of the number densities for notational convenience.

We  may   now  integrate   both  sides  of   \eqref{eq:evol_lept}  and
\eqref{eq:evol_lept2} over the phase space and sum over the degenerate
helicity and isospin degrees  of freedom. The resulting rate equations
for the total  number densities of the charged  lepton and anti-lepton
doublets,  accounting for  the decay  and inverse  decay of  the heavy
neutrinos, can be written in the form
\begin{align}
  \frac{\D{}{\N{L}{l}{m}}}{\D{}{t}} \ &= \ - \: i \;
  \Big[{\cal E}_L,n^L\Big]_l^{\ m} \: - \: \frac{1}{2 \, n^L_{\rm eq}} \,
  \Big\{n^L, \, \gamma(L\Phi \to N) \Big\}_l^{\phantom l m} \;
  + \; \frac{\N{N}{\beta}{\alpha}}{n^N_{\rm eq}} \,
  \Tdu{[\gamma(N \to L\Phi)]}{l}{m}{\alpha}{\beta} \;,
  \label{evol_L_tot} \\
  \frac{\D{}{\Nb{L}{l}{m}}}{\D{}{t}} \ &= \ + \: i \: \Big[{\cal E}_L,\bar{n}^L\Big]_l^{\ m}
\ - \;
  \frac{1}{2 \, n^L_{\rm eq}} \,
  \Big\{\bar{n}^L, \,
  \gamma(L^{\tilde{c}}\Phi^{\tilde{c}} \to N) \Big\}_l^{\phantom l m} \;
  + \; \frac{\Nb{N}{\beta}{\alpha}}{n^N_{\rm eq}} \,
  \Tdu{[\gamma(N \to L^{\tilde{c}}\Phi^{\tilde{c}})]}
  {l}{m}{\alpha}{\beta} \;,
  \label{evol_Lb_tot}
\end{align}
where  $\mat{\mathcal{E}}_L$  is  the   thermally-averaged  effective
lepton energy matrix 
\begin{equation}
  \mat{\mathcal{E}}_L \ \equiv \ \frac{g_L}{n^L_{\rm eq}} \, 
  \int_{\ve p} \, \mat{E}_L(\ve p) \, e^{- E_L(\ve p)/T} 
  \ = \ 
  \frac{K_1(z)}{K_2(z)} \;
  \big[\mat{M}_L^\dagger \, \mat{M}_L \big]^{1/2} \;+\; 3T \, \mat{1}
  \; ,
  \label{therm_EL}
\end{equation}
with $\mat{M}_L (T)$ being the thermal mass matrix of the lepton doublets. 
Note that  the $3T$-term  on the RHS  of \eqref{therm_EL}  is  isotropic  
in flavour  space, 
commutes  with the  number densities,  and  therefore, does  not give  any
contribution   to   the    rate   equations   \eqref{evol_L_tot}   and
\eqref{evol_Lb_tot}.    
The  $1\to2$ and  $2\to  1$
collision     rates     appearing     in    \eqref{evol_L_tot}     and
\eqref{evol_Lb_tot}  are derived  from the  rank-4  absorptive tensors
\eqref{eq:Gamma1}  and  \eqref{eq:Gamma2}.   Replacing the  tree-level
Yukawa  couplings $\h{l}{\alpha}$  appearing in  \eqref{eq:Gamma1} and
\eqref{eq:Gamma2} with  the resummed ones  $\hr{l}{\alpha}$, the $1\to
2$ collision rates can be explicitly written as
\begin{align}
  \Tdu{[\gamma(N \to L\Phi)]}{l}{m}{\alpha}{\beta} \ &
  \equiv \ \int_{N L \Phi} g_L g_\Phi (2\, p_N \cdot p_L) 
  \hrs{m}{\alpha}  \hr{l}{\beta} \
  = \ \frac{m_N^4}{\pi^2 z} \frac{K_1(z)}{16 \pi} \; 
  \hrs{m}{\alpha} \, \hr{l}{\beta} \;,
  \label{coll_nlp} \\
  \Tdu{[\gamma(N \to L^{\tilde{c}}\Phi^{\tilde{c}})]}{l}{m}{\alpha}{\beta}
  \ & \equiv \ \int_{N L \Phi} g_L g_\Phi(2\, p_N \cdot p_L) 
  \hrc{m}{\alpha}  \hrcs{l}{\beta} \ 
  = \ \frac{m_N^4}{\pi^2 z}  \frac{K_1(z)}{16 \pi} 
  \hrc{m}{\alpha}  \hrcs{l}{\beta} ,
  \label{coll_nlcpc}
\end{align}
which are the flavour-covariant generalizations of the collision rates
defined   in   \eqref{col}.   The   $2\to   1$   collision  rates   in
\eqref{evol_L_tot} and \eqref{evol_Lb_tot} are related to the $1\to 2$
rates   \eqref{coll_nlp}   and   \eqref{coll_nlcpc}   by   virtue   of
$\widetilde{C}\!P\widetilde{T}$ (= $\CP T$) invariance, i.e.
\begin{align}
 \Tdu{[\gamma(L^{\tilde{c}}\Phi^{\tilde{c}} \to N)
    ]}{l}{m}{\alpha}{\beta} \ & = \ 
  \Tdu{[\gamma(N \to L\Phi)]}{l}{m}{\alpha}{\beta} \;,\\
  \Tdu{[\gamma(L\Phi \to N)]}{l}{m}{\alpha}{\beta} \ & = \ 
  \Tdu{[\gamma(N \to L^{\tilde{c}}\Phi^{\tilde{c}})
    ]}{l}{m}{\alpha}{\beta} \; .
\end{align}
The corresponding  rank-2 collision rates  within the anti-commutators
in  \eqref{evol_L_tot} and \eqref{evol_Lb_tot}  are obtained  from the
corresponding rank-4 tensors by contracting the heavy-neutrino flavour
indices, e.g.
\begin{equation}
  \Tdu{[\gamma(L\Phi \to N)]}{l}{m}{}{} \ \equiv \
  \Tdu{[\gamma(L\Phi \to N)]}{l}{m}{\alpha}{\alpha} \;.
\end{equation}

\begin{figure}[t!]
	\centering
	\subfloat[][Inverse heavy-neutrino decay,
        $n^{\Phi}\protect{[}n^L\protect{]}_l^{\protect\phantom{l}k}
        \protect{[}\gamma(L\Phi  \to  N)
        \protect{]}_{k \protect\phantom{l} \alpha}
        ^{\protect\phantom{k} l \protect\phantom{\alpha}\beta}$. ]
                {\label{fig:feynman1}\fbox{\includegraphics[scale=0.88]
                  {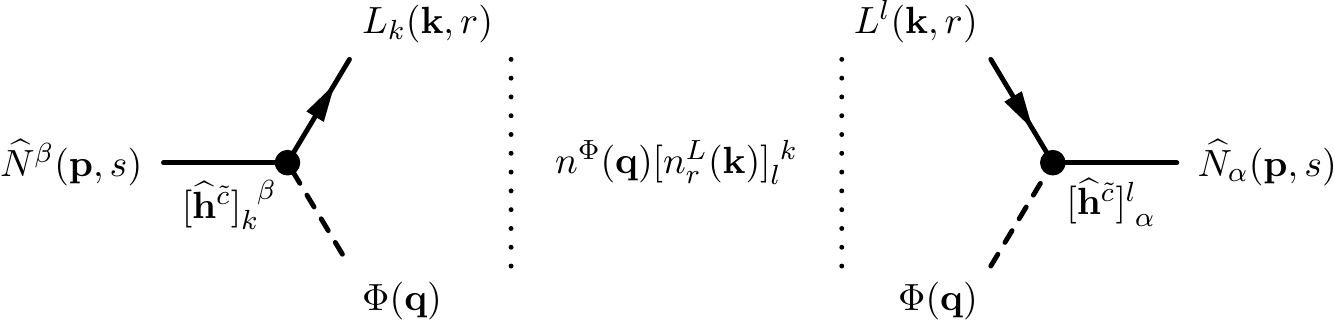}}}\\
	\subfloat[][Inverse heavy-neutrino decay,
        $\bar{n}^{\Phi}\protect{[}\bar{n}^L\protect{]}_l^{\protect\phantom{l}k}
        \protect{[}\gamma(L^{\tilde{c}}\Phi^{\tilde{c}}  \to  N)
        \protect{]}_{k \protect\phantom{l} \alpha}
        ^{\protect\phantom{k} l \protect\phantom{\alpha}\beta}$. ]
                {\fbox{\label{fig:feynman2}\includegraphics[scale=0.88]
                  {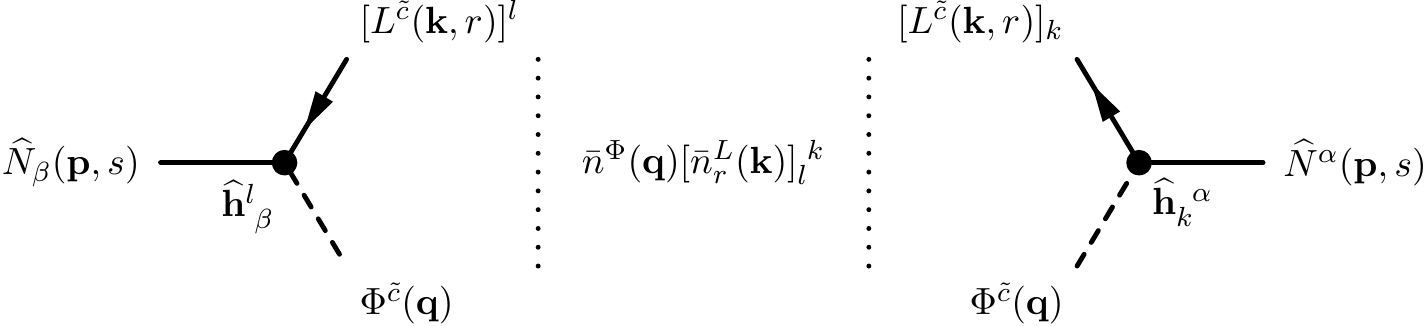}}}\vspace{.35cm}
                \caption{\emph   {Feynman   diagrams   for  $2\to 1$ inverse
        heavy-neutrino  decay, in the presence of a statistical background. The
        flavour  indices  are  shown   explicitly,  while  other  indices  are
        suppressed.} 
      \label{fig:2}
    }
\end{figure}

\begin{figure}[t!]
	\centering
	\subfloat[][Heavy-neutrino self-energy, \\ $N \to L\Phi \to N$.]
		{\fbox{\includegraphics[scale=0.8]{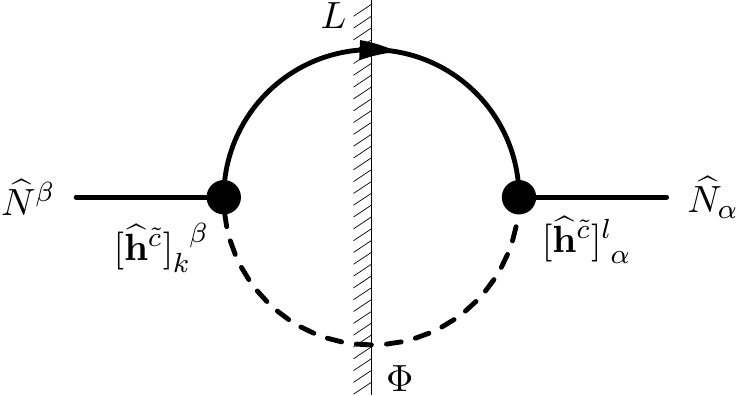}}}
        \subfloat[][Heavy-neutrino self-energy, \\ $N \to 
        L^{\widetilde{c}} \Phi^{\widetilde{c}} \to N$.]
		{\fbox{\includegraphics[scale=0.8]{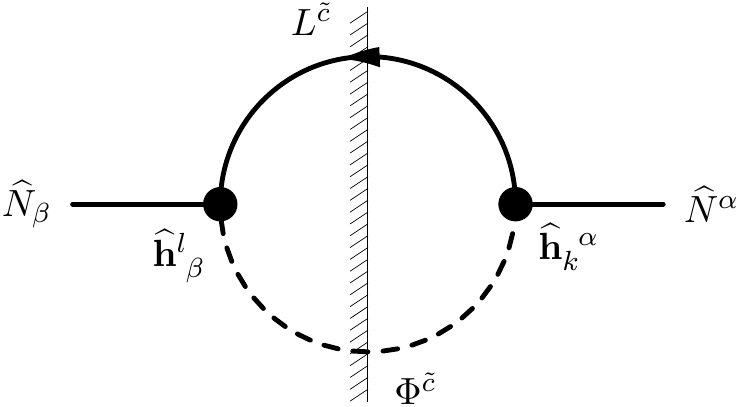}}}
                \vspace{.2cm}
	\caption{\emph {Feynman diagrams  for the self-energies of the
heavy neutrinos.  The
cut,  across  which positive  energy  flows  from  unshaded to  shaded
regions, is associated with production rates in the thermal plasma, as
described    by   the   generalized    optical   theorem    given   in
\ref{app:optical}.                       See                      also
Figure~\ref{fig:2}.}\label{fig:cuts1}}
\end{figure}

In~\ref{app:optical},  we present an  alternative derivation  of these
collision rates  by considering a  flavour-covariant generalization of
the   optical    theorem   in   the   presence    of   a   statistical
background. Therein, the necessity  of the rank-4 flavour structure of
these  collision rates is  further justified.  This is  illustrated in
Figure~\ref{fig:2} for the in-medium  production of heavy neutrinos in
a spatially-homogeneous statistical  background of lepton and
Higgs doublets. The production rates in the thermal plasma can 
be better understood from the unitarity cut 
of the partial one-loop heavy-neutrino self-energy graph, as shown in 
Figure~\ref{fig:cuts1}.  Imposing  kinetic equilibrium  as  given  by
\eqref{nLeq},  we obtain tree-level  thermally-averaged heavy-neutrino
production          rates          for          the          processes
$L^{\widetilde{c}}\Phi^{\widetilde{c}}\to   N$  and   $L\Phi   \to  N$
[cf.~\eqref{eq:NLPhi}  and \eqref{eq:LPhiN}],  which  are exactly  the
same  as those  obtained in  \eqref{coll_nlp}  and \eqref{coll_nlcpc},
respectively.


Analogous to the charged-lepton  case, we obtain the flavour-covariant
rate equations for the total  number densities of heavy neutrinos from
the    general    transport    equations    \eqref{eq:evol_neu}    and
\eqref{eq:evol_neu2}, as follows:
\begin{align}
  \frac{\D{}{\N{N}{\alpha}{\beta}}}{\D{}{t}} \ &
  = \ - \; i \, \Big[\mathcal{E}_N,\, n^{N}
  \Big]_\alpha^{\phantom \alpha \beta} \;
  + \; [\mathcal{C}^{N}]_{\alpha}^{\phantom{\alpha} \beta} \;
  + \; G_{\alpha \lambda} \,
  [\overline{\mathcal{C}}^{N}]_{\mu}^{\phantom{\mu} \lambda} \,
  G^{\mu \beta} \;, \label{evol_N_tot}\\
  \frac{\D{}{\Nb{N}{\alpha}{\beta}}}{\D{}{t}} \ &
  = \; + \; i \, \Big[\mathcal{E}_N,\, 
  \bar{n}^{N}\Big]_\alpha^{\phantom \alpha \beta} \; 
  + \; [\overline{\mathcal{C}}^{N}]_{\alpha}^{\phantom{\alpha} \beta} \; 
  + \; G_{\alpha \lambda} \,
  [\mathcal{C}^{N}]_{\mu}^{\phantom{\mu} \lambda} \,
  G^{\mu \beta} \; , \label{evol_Nb_tot}
\end{align}
where   $\mat{\mathcal{E}}_N$  is  the   thermally-averaged  effective
heavy-neutrino energy matrix, defined analogous to \eqref{therm_EL}. 
The   thermally   averaged   collision   terms
$\mat{\mathcal{C}}$      and     $\overline{\mat{\mathcal{C}}}$     in
\eqref{evol_N_tot} and \eqref{evol_Nb_tot} can be defined analogous to
$\mat{\mathcal{E}}_N$, and are explicitly given by
\begin{align}
  [\mathcal{C}^{N}]_{\alpha}^{\phantom{\alpha} \beta} 
  \ & = \  - \; \frac{1}{2 \, n^N_{\rm eq}} \,
  \Big\{n^{N}, \, \gamma(N \to L\Phi) \Big\}
  _\alpha^{\phantom \alpha \beta} \;
  + \; \frac{\N{L}{m}{l}}{n^L_{\rm eq}} \,
  \Tdu{[\gamma(L\Phi \to N)]}{l}{m}{\alpha}{\beta} \;, \label{CN1} \\
  [\overline{\mathcal{C}}^{N}]_{\alpha}^{\phantom{\alpha} \beta} \ &
  = \  - \; \frac{1}{2 \, n^N_{\rm eq}} \, 
  \Big\{\bar{n}^{N}, \, 
  \gamma(N \to L^{\tilde{c}}\Phi^{\tilde{c}})
  \Big\}_\alpha^{\phantom \alpha \beta} \; 
  + \; \frac{\Nb{L}{m}{l}}{n^L_{\rm eq}} \,
  \Tdu{[\gamma(L^{\tilde{c}}\Phi^{\tilde{c}} \to N)]}
  {l}{m}{\alpha}{\beta} \;,
  \label{CN2}
\end{align}
where  the rank-4 collision  rates are  given by  \eqref{coll_nlp} and
\eqref{coll_nlcpc}, and the  corresponding rank-2 objects appearing in
\eqref{CN1}   and  \eqref{CN2}   are  obtained   by   contracting  the
charged-lepton indices, e.g.
\begin{equation}
  \Tdu{[\gamma(N\to L\Phi)]}{\alpha}{\beta}{}{} \ \equiv \
  \Tdu{[\gamma(N\to L\Phi)]}{l}{l}{\alpha}{\beta} \;.
\end{equation}

Using the expressions~\eqref{coll_nlp}  and \eqref{coll_nlcpc}, we can
define  the flavour-covariant  generalizations of  the  $\CP$-even and
$\CP$-odd quantities  in \eqref{cpc}  and \eqref{cpv}, which  now have
definite transformation properties under $\widetilde{C}\!P$:
\begin{align}
  \Tdu{[\gamma^N_{L\Phi}]}{l}{m}{\alpha}{\beta} \ &
  = \ + \, \Tdu{[\gamma_N^{L\Phi}]}{l}{m}{\alpha}{\beta} \
  = \ {\cal O}(h^2) \;,\\
  \Tdu{[\delta \gamma^N_{L\Phi}]}{l}{m}{\alpha}{\beta} \ &
  = \ - \, \Tdu{[\delta \gamma_N^{L\Phi}]}{l}{m}{\alpha}{\beta} \
  = \ {\cal O}(h^4) \;.
\end{align}
The corresponding  rate equations for  the $\widetilde{C}\!P$-``even''
and -``odd''  number densities [cf.~\eqref{gctrans}]  are derived from
\eqref{evol_L_tot},    \eqref{evol_Lb_tot},   \eqref{evol_N_tot}   and
\eqref{evol_Nb_tot}:
\begin{align}
  \frac{\D{}{}[\underline{n}^{N}]_{\alpha}^{\phantom{\alpha}\beta}}{\D{}{t}}
  \ & = \ - \; \frac{i}{2} \,
  \Big[\mathcal{E}_N,\, \delta n^{N}\Big]_\alpha^{\phantom \alpha \beta} \;
  + \; \Tdu{\big[\widetilde{\rm Re}
    (\gamma^{N}_{L \Phi})\big]}{}{}{\alpha}{\beta} \;
  - \; \frac{1}{2 \, n^N_{\rm eq}} \,
  \Big\{\underline{n}^N, \, \widetilde{\rm Re}(\gamma^{N}_{L \Phi})
  \Big\}_{\alpha}^{\phantom{\alpha}\beta}
   \;, \label{evol_n}\\[3pt]
  \frac{\D{}{[\delta n^N]_\alpha^{\phantom \alpha \beta}}}{\D{}{t}} \ &
  = \ - \; 2 \, i \, 
  \Big[\mathcal{E}_N,\, \underline{n}^{N}\Big]
  _\alpha^{\phantom \alpha \beta} \;
  - \; 2\, i\,  \Tdu{\big[\widetilde{\rm Im}
    (\delta \gamma^{N}_{L \Phi})\big]}{}{}{\alpha}{\beta} \;-\; 
  \frac{i}{n^N_{\rm eq}} \, \Big\{\underline{n}^N, \,
  \widetilde{\rm Im}(\delta\gamma^{N}_{L \Phi}) \Big\}
  _{\alpha}^{\phantom{\alpha}\beta} \notag\\
  & \quad \;\, - \ \frac{1}{2 \, n^N_{\rm eq}}  \,
  \Big\{\delta n^N, \, \widetilde{\rm Re}(\gamma^{N}_{L \Phi})
  \Big\}_{\alpha}^{\phantom{\alpha}\beta}\;, \label{evol_dn} 
  \\
  \frac{\D{}{[\delta n^L]_l^{\phantom l m}}}{\D{}{t}} \ &
  = \ \Tdu{[\delta \gamma^{N}_{L \Phi}]}{l}{m}{}{} \;
  +\; \frac{[\underline{n}^{N}]_{\beta}^{\phantom{\beta}\alpha}}
  {n^N_{\rm eq}} \,
  \Tdu{[\delta \gamma^{N}_{L \Phi}]}{l}{m}{\alpha}{\beta} \;
  + \; \frac{[\delta n^N]_{\beta}^{\phantom\beta \alpha}}{2\,n^N_{\rm eq}} \,
  \Tdu{[\gamma^{N}_{L \Phi}]}{l}{m}{\alpha}{\beta}
  - \frac{1}{4 \, n^L_{\rm eq}} 
  \Big\{\delta n^L, \, \gamma^{N}_{L \Phi} \Big\}_{l}^{\phantom{l}m} \;,
  \label{evol_dL}
\end{align}
where we  have kept terms only  up to ${\cal O}(\mu_a/T)$ and ${\mathcal  O}(h^4)$, 
except the
last term  on the  RHS of \eqref{evol_dL},  which is the  only washout
term for the lepton asymmetry, and appears at ${\mathcal O}(h^6)$.  
In \eqref{evol_n}  and  \eqref{evol_dn}, we  have  defined, for a given Hermitian
matrix  $\mat{A}  =   \mat{A}^\dagger$,  its  generalized  real  and
imaginary parts, as follows:
\begin{align}
  \big[\widetilde{\rm Re}(A)\big]_{\alpha}^{\phantom{\alpha}\beta} \ &
  \equiv \ \frac{1}{2} \, \Big( \Tdu{A}{\alpha}{\beta}{}{} \; 
  + \; G_{\alpha \lambda} \,\Tdu{A}{\mu}{\lambda}{}{}\,
  G^{\mu \beta}\Big) \;, \label{4.26}  \\
  \big[\widetilde{\rm Im}(A)\big]_{\alpha}^{\phantom{\alpha}\beta} \ &
  \equiv \ \frac{1}{2 \, i} \, 
  \Big( \Tdu{A}{\alpha}{\beta}{}{} \; 
  - \; G_{\alpha \lambda} \,\Tdu{A}{\mu}{\lambda}{}{}\,
  G^{\mu \beta}\Big) \;  . \label{4.27}
\end{align}
Observe that in the heavy-neutrino mass eigenbasis, the definitions \eqref{4.26} and \eqref{4.27} reduce to the usual real and imaginary parts:
\begin{equation}
  \big[\widetilde{\rm Re}(\widehat{A})\big]_{\alpha}^{\phantom{\alpha}\beta} \ 
  = \ \mathrm{Re}\big(  \Tdu{\widehat{A}}{\alpha}{\beta}{}{}\big) \;,
  \qquad
  \big[\widetilde{\rm Im}(\widehat{A})\big]_{\alpha}^{\phantom{\alpha}\beta} \ 
  = \ \mathrm{Im}\big(  \Tdu{\widehat{A}}{\alpha}{\beta}{}{}\big) \;.
\end{equation}
In obtaining \eqref{evol_n} and \eqref{evol_dn}, we have used the relations
\begin{equation}
  \widetilde{\rm Re}(\underline{\mat{n}}^N) \ 
  = \ \underline{\mat{n}}^N \;, \qquad \qquad
  i\, \widetilde{\rm Im}(\mat{\delta n}^N) \ 
  = \ \mat{\delta n}^N \;,
\end{equation}
which can be  derived from \eqref{eq:Majorana_bar_G}, \eqref{eq:dN_CP}
and \eqref{eq:N_CP}. Observe that 
the commutators in \eqref{evol_L_tot} and \eqref{evol_Lb_tot} cancel to leading order in ${\cal O}(\mu_L/T)$ by virtue of \eqref{nL_tot}, even if the thermal masses are included. 
On the other hand, the commutators of the thermally-averaged effective heavy-neutrino energy 
matrix with the number densities  in \eqref{evol_n} and
\eqref{evol_dn} are non-vanishing, and describe  the  coherent   oscillations  between
different  heavy neutrino flavours.   

\begin{figure}[t!]
	\centering
        \vspace{-0.5cm}
 	\subfloat[][$\Delta L = 0$ scattering,
        $n^{\Phi}\protect{[}n^L\protect{]}_l^{\protect\phantom{l}k}
        \protect{[}\gamma(L\Phi  \to  L\Phi)
        \protect{]}_{k \protect\phantom{l} m}
        ^{\protect\phantom{k} l \protect\phantom{m}n}$. ]
	{\label{fig:feynman3}\fbox{\includegraphics[scale=0.8]{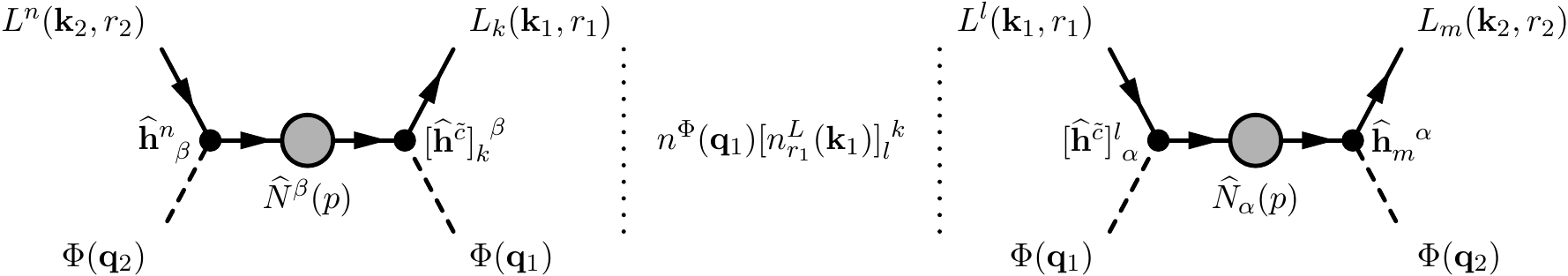}}}\\
        \subfloat[][$\Delta L = 0$ scattering,
        $\bar{n}^{\Phi}\protect{[}\bar{n}^L\protect{]}_l^{\protect\phantom{l}k}
        \protect{[}\gamma(L^{\tilde{c}}\Phi^{\tilde{c}}  \to 
        L^{\tilde{c}}\Phi^{\tilde{c}})
        \protect{]}_{k \protect\phantom{l} m}
        ^{\protect\phantom{k} l \protect\phantom{m}n}$. ]
	{\label{fig:feynman4}\fbox{\includegraphics[scale=0.8]{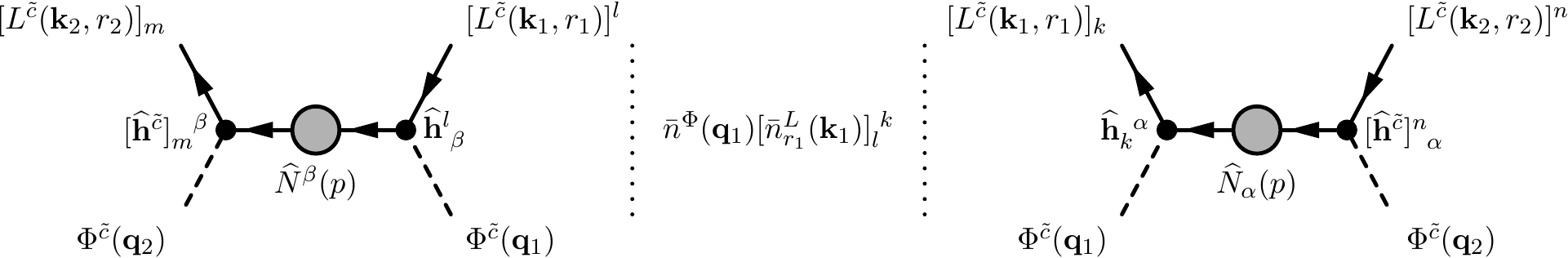}}}\\
	\subfloat[][$\Delta L  =  2$ scattering,
        $n^{\Phi}\protect{[}n^L\protect{]}_l^{\protect\phantom{l}k}
        \protect{[}\gamma(L\Phi  \to  L^{\tilde{c}} \Phi^{\tilde{c}})
        \protect{]}_{k \protect\phantom{l} m}
        ^{\protect\phantom{k} l \protect\phantom{m} n}$.]
	{\label{fig:feynman5}\fbox{\includegraphics[scale=0.8]{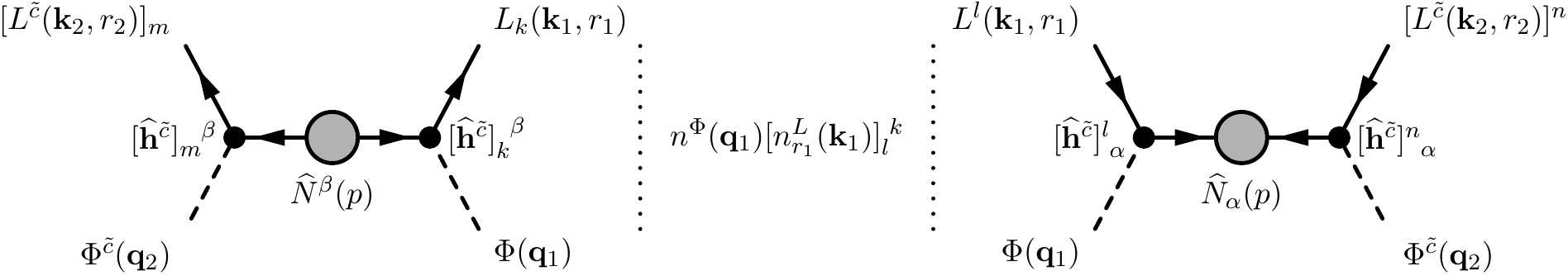}}}\\
	\subfloat[][$\Delta L  =  2$ scattering,
        $\bar{n}^{\Phi}\protect{[}\bar{n}^L\protect{]}_l^{\protect\phantom{l}k}
        \protect{[}\gamma(L^{\tilde{c}}\Phi^{\tilde{c}}  \to  L \Phi)
        \protect{]}_{k \protect\phantom{l} m}
        ^{\protect\phantom{k} l \protect\phantom{m} n}$.]
        {\label{fig:feynman6}\fbox{\includegraphics[scale=0.8]{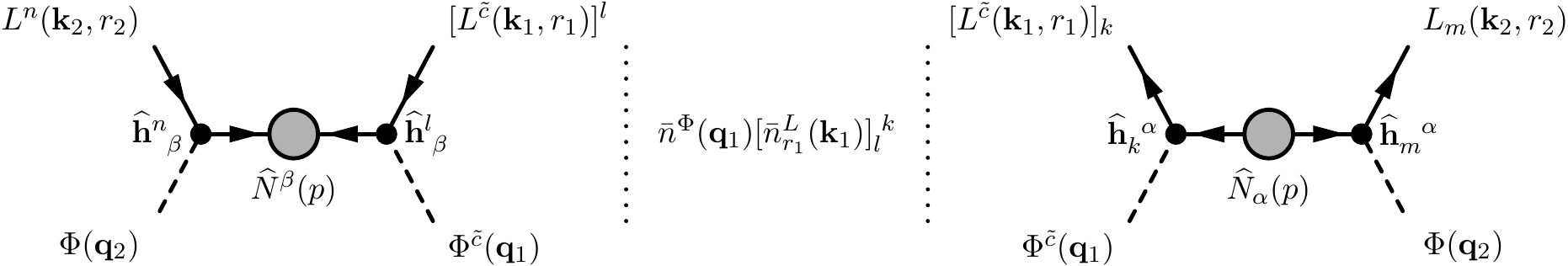}}}
        \vspace{.35cm}
	\caption{\emph {Feynman diagrams for $\Delta L = 0$ scattering
[(a), (b)] and  $\Delta L = 2$ scattering [(c),  (d)], in the presence
of a statistical background. The flavour indices are shown explicitly,
while other indices are suppressed.}\label{fig:feynman} }
\end{figure}

Note  that  the  $\widetilde{C}\!P$-``odd''  inverse  decay  terms  in
\eqref{evol_dn}       and        \eqref{evol_dL},       i.e.~$-
2i\Tdu{[\widetilde{\rm       Im}       (\delta       \gamma^{N}_{L
\Phi})]}{}{}{\alpha}{\beta}$   and   $+\Tdu{[\delta  \gamma^{N}_{L
\Phi}]}{l}{m}{}{}$, appear with the wrong  sign and do not lead to the
correct equilibrium  behaviour, since,  by construction, there  are no
washout  terms,  induced  by  the  $2\leftrightarrow  2$  scatterings,
introduced  so far.   It is  well known  that the  inclusion  of these
scattering terms  (with the RIS contribution  subtracted), changes the
sign of  these inverse decay  terms and gives the  correct equilibrium
limit~\cite{Kolb:1980,Pilaftsis:2003gt}.   In Section~\ref{sec:4.4} we
will  explicitly show  this result  in a  flavour-covariant  manner by
including in the rate equations the RIS-subtracted collision rates for
scattering in the presence of a statistical background, as illustrated
in Figure~\ref{fig:feynman}.   For the moment, we take  this result at
face  value,  and  correct  the   signs  `by  hand',  to  be  able  to
qualitatively  discuss some  important physical  phenomena,  before we
include the scattering terms.  Finally,  we also take into account the
Hubble expansion  of the Universe and change  the independent variable
$t$ in favour of $z=m_N/T$, to  write down the rate equations in terms
of the normalized number densities, introduced in Section~\ref{sec:2}:
\begin{align}
   \frac{H_{N} \, n^\gamma}{z}\,
   \frac{\D{}{[\underline{\eta}^{N}]_{\alpha}^{\phantom{\alpha}\beta}}}
   {\D{}{z}} \ &
  = \ - \, i \, \frac{n^\gamma}{2} \,
  \Big[\mathcal{E}_N,\, \delta \eta^{N}\Big]_\alpha^{\phantom \alpha \beta} \;
  + \; \Tdu{\big[\widetilde{\rm Re}(\gamma^{N}_{L \Phi})\big]}
  {}{}{\alpha}{\beta} \;
  - \; \frac{1}{2 \, \eta^N_{\rm eq}} \,
  \Big\{\underline{\eta}^N, \, \widetilde{\rm Re}(\gamma^{N}_{L \Phi})
  \Big\}_{\alpha}^{\phantom{\alpha}\beta},    \label{eq:evol_decay2} 
\\ 
  \frac{H_{N} \, n^\gamma}{z}\,
  \frac{\D{}{[\delta \eta^N]_\alpha^{\phantom \alpha \beta}}}{\D{}{z}} \ &
  = \ - \; 2 \, i \, n^\gamma \,
  \Big[\mathcal{E}_N,\, \underline{\eta}^{N}\Big]_\alpha^{\phantom \alpha \beta} \;
  + \; 2\, i\,  \Tdu{\big[\widetilde{\rm Im}
    (\delta \gamma^{N}_{L \Phi})\big]}{}{}{\alpha}{\beta} \;-\; 
  \frac{i}{\eta^N_{\rm eq}} \, \Big\{\underline{\eta}^N, \,
  \widetilde{\rm Im}
  (\delta\gamma^{N}_{L \Phi}) \Big\}_{\alpha}^{\phantom{\alpha}\beta} \notag\\
  & \quad \;\, - \ \frac{1}{2 \, \eta^N_{\rm eq}}  \,
  \Big\{\delta \eta^N, \, \widetilde{\rm Re}(\gamma^{N}_{L \Phi})
  \Big\}_{\alpha}^{\phantom{\alpha}\beta}\;, \label{eq:evol_decay3} 
\\ 
  \frac{H_{N} \, n^\gamma}{z}\,
  \frac{\D{}{[\delta \eta^L]_l^{\phantom l m}}}{\D{}{z}} \ &
  = \ - \, \Tdu{[\delta \gamma^{N}_{L \Phi}]}{l}{m}{}{} \;
  +\; \frac{[\underline{\eta}^{N}]_{\beta}^{\phantom{\beta}\alpha}}
  {\eta^N_{\rm eq}} \,
  \Tdu{[\delta \gamma^{N}_{L \Phi}]}{l}{m}{\alpha}{\beta} \;
  + \; \frac{[\delta \eta^N]_{\beta}^{\phantom\beta \alpha}}{2\,\eta^N_{\rm eq}} \,
  \Tdu{[\gamma^{N}_{L \Phi}]}{l}{m}{\alpha}{\beta} \notag\\
  &\quad\;\, - \, \frac{1}{3} \, 
  \Big\{\delta \eta^L, \, \gamma^{N}_{L \Phi} \Big\}_{l}^{\phantom{l}m} \;.
  \label{eq:evol_decay1}
\end{align}
In the  last term on the  RHS of \eqref{eq:evol_decay1},  we have used
$\eta^L_{\rm  eq}   =  3/4$,  which  follows   from  \eqref{nrel}  and
\eqref{ngamma}.

\subsection{Lepton Asymmetry via Heavy Neutrino Oscillations}
\label{sec:4.2}

The   rate  equations  \eqref{eq:evol_decay2}--\eqref{eq:evol_decay1},
contain  two  physically distinct  mechanisms  for  the generation  of
lepton  asymmetry.   One  is  the standard  $T=0$  $\varepsilon$-  and
$\varepsilon'$-type $ \CP$ violation given by \eqref{cpasy} due to the
mixing and decay  of the heavy Majorana neutrinos.  We have taken this
into  account  by means  of  the  one-loop  resummed effective  Yukawa
couplings  defined in  \eqref{resum1}, which  appear in  the collision
rates \eqref{coll_nlp} and \eqref{coll_nlcpc}.  This is the {\it only}
source of  lepton number  asymmetry in the  flavour-diagonal Boltzmann
equations \eqref{be2} and \eqref{be3}.  However, the flavour-covariant
rate   equations  \eqref{eq:evol_decay2}--\eqref{eq:evol_decay1}  also
include a second source for  the asymmetry, due to the $\CP$-violating
oscillations of the on-shell heavy neutrinos in the thermal bath. This
originates  from  the sequence  of  an  on-shell  production of  heavy
neutrinos  in the  bath due  to  inverse decays,  which can  oscillate
between  different flavours  in  the  bath and  then  decay back  into
charged-leptons. Formally, this process  has the same structure as the
scattering diagrams  in Figure~\ref{fig:feynman}. While  the $T=0$ QFT
processes  are taken  into account  by the  resummation of  the Yukawa
couplings~\cite{Pilaftsis:1997jf,  Pilaftsis:2003gt},  the oscillation
phenomenon  corresponds  to  the  thermal  part  of  the  intermediate
heavy-neutrino  propagator.    Thus,  our  flavour-covariant  approach
captures the  leading order effect  of a complete  thermal resummation
procedure      that       would      generalize      the      analysis
in~\cite{Pilaftsis:2003gt}.

In  this  section, we  will  present  a  qualitative analysis  of  the
heavy-neutrino  oscillation phenomenon  in the  RL scenario,  and show
that this mechanism contributes to the total lepton asymmetry at order
${\cal  O}(h^4)$  around  $z=1$.   Note  that,  although  conceptually
similar,  this is  qualitatively as  well as  quantitatively different
from  the  phenomenon  first proposed  in~\cite{Akhmedov:1998qx},  and
studied  in~\cite{Asaka:2005, Shaposhnikov:2008pf, Canetti:2012kh}  
for the  light  sterile neutrino  case,
where the final  lepton number asymmetry is of  order ${\cal O}(h^6)$,
as  also  recently  stressed  in~\cite{Shuve:2014zua}.   In  order  to
extract    the   heavy-neutrino    oscillation    effect   from    the
flavour-covariant                    rate                    equations
\eqref{eq:evol_decay2}--\eqref{eq:evol_decay1},    we    consider    a
simplified case  with the tree-level Yukawa couplings  (instead of the
resummed   ones),  thus   artificially  `switching   off'   all  $T=0$
$\varepsilon$-  and $\varepsilon'$-type  $\CP$  violation effects.  In
this  case, we  can drop  the  $\widetilde{C}\!P$-``odd'' $\mat{\delta
\gamma}$     collision     terms     in     the     rate     equations
\eqref{eq:evol_decay3}--\eqref{eq:evol_decay1}, which further simplify
in the mass eigenbasis to give
\begin{align}
 \frac{H_{N} \, n^\gamma}{z}\,
  \frac{\D{}{[\delta \widehat{\eta}^N]_{\alpha\beta}}}{\D{}{z}} \ &
  \supset \ - \; 2 \,i \, n^\gamma \, \Big[\widehat{\mathcal{E}}_N,\,
  \widehat{\underline{\eta}}^{N}\Big]_{\alpha\beta} \; 
  -\; \frac{1}{2\,\eta^N_{\rm eq}} \,
  \Big\{\delta\widehat{\eta}^N, \,
  \mathrm{Re}\big(\widehat\gamma^{N}_{L \Phi}\big) \Big\}_{\alpha\beta} \; ,
  \label{eq:tl_dN} \\
  \frac{H_{N} \, n^\gamma}{z}\, \frac{\D{}{[\delta\widehat{\eta}^L]_{lm}}}
  {\D{}{z}} \ & \supset \
  \frac{[\delta\widehat{\eta}^N]_{\beta \alpha}}
  {2\,\eta^N_{\rm eq}} \, 
  [\widehat{\gamma}^{N}_{L \Phi}]_{lm\alpha\beta} \; 
  -\; \frac{1}{3} \, \Big\{\delta\widehat\eta^L, \, 
  \widehat\gamma^{N}_{L \Phi} \Big\}_{lm} \; .
  \label{eq:tl_L} 
\end{align}
The  rate equation  for $\mat{\widehat{\underline{\eta}}}^N$  is still
given by  \eqref{eq:evol_decay2}, specialized to  the mass eigenbasis.
Notice  that, in the  mass eigenbasis,  the flavour  rotation matrices
$\mat{G} = \mat{\mathcal{G}} = \mat{1}$, and therefore, we do not need
to   distinguish  between   upper   and  lower   flavour  indices   in
\eqref{eq:tl_dN}  and  \eqref{eq:tl_L}.  It  is  useful  to perform  a
time-stepping analysis to see  the infinitesimal time evolution of the
total  lepton  asymmetry.   We   start  with  an  incoherent  diagonal
heavy-neutrino  number  density  matrix,  and a  zero  initial  lepton
asymmetry at some $z = z_{\rm in}$:
\begin{equation}
  [\widehat{\underline{\eta}}^N]_{\alpha \beta}  =  0
  \quad \text{for} \quad \alpha \neq \beta\, ,
  \qquad [\delta\widehat{\eta}^N]_{\alpha \beta} =  0 \, ,
  \qquad [\delta\widehat\eta^L]_{l m} =  0\;.
\end{equation}
Interference  in the  inverse  decays will  generate off-diagonals  in
$\widehat{\mat{\eta}}^{N}$ at  ${\cal O}(h^2)$.  However,  we see from
\eqref{eq:evol_decay2}, \eqref{eq:tl_dN} and \eqref{eq:tl_L}  that only
$\widehat{\underline{\mat{\eta}}}^N$,         and         not        $
\mat{\delta}\widehat{\mat{\eta}}^N$                                  or
$\mat{\delta}\widehat{\mat{\eta}}^L$,          receives          these
contributions. This is due to the fact that the processes
\begin{equation}
  L \, \Phi \ \to\ \sum_\alpha c_\alpha \, \widehat N_{\mathrm{R}, \alpha}
  \; , \qquad \qquad 
  L^{\tilde{c}} \, \Phi^{\tilde{c}} \ \to \ 
  \sum_\alpha c_\alpha^* \, \widehat N_{\mathrm{R}, \alpha}^{\tilde{c}} \;,
\end{equation}
(the $c_\alpha$ being complex  coefficients in the linear combination)
have  to proceed  at the  same rate  for $\widetilde{C}\!P$-conserving
inverse decays, if no initial  lepton asymmetry is present, and hence,
only              $\mathrm{Re}(\widehat{\mat{\eta}}^{N})             =
\widehat{\underline{\mat{\eta}}}^N$ can be generated. Therefore, after
a small time interval, at $z = z_{\rm in} + \delta z_1$ we have
\begin{equation}
  [\widehat\eta^N]_{\alpha \beta}  =  {\cal O}(h^2) \quad \text{for} 
  \quad \alpha \neq \beta\, , \qquad
  [\delta\widehat\eta^N]_{\alpha \beta} \simeq  0 \, , \qquad
  [\delta\widehat\eta^L]_{l m} \simeq  0 \;.
  \label{eq:off-diag_etaN}	
\end{equation}
Now  from \eqref{eq:tl_dN},  we see  that  heavy-neutrino oscillations
will induce imaginary parts  of $\widehat{\mat{\eta}}^N$, and hence, a
non-zero  $\mat{\delta}\widehat{\mat{\eta}}^N$  [cf.~\eqref{eq:tl_dN}]
due    to   the    off-diagonals   of    $\widehat{\mat{\eta}}^N$   in
\eqref{eq:off-diag_etaN}.  Thus,  at a  later time $z  = z_{\rm  in} +
\delta z_2$, the number densities will be
\begin{equation}
  [\widehat{\underline{\eta}}^N]_{\alpha \beta}  = 
  {\cal O}(h^2) \quad \text{for} \quad \alpha \neq \beta\, , \qquad
  [\delta\widehat\eta^N]_{\alpha \beta} =  {\cal O}(h^2) \, , \qquad
  [\delta\widehat\eta^L]_{l m} \simeq  0 \;.
\end{equation}
Finally at $z  = z_{\rm in} + \delta z_3$,  interference in the ${\cal
O}(h^2)$  decays will  create  a non-zero  lepton  asymmetry of  order
${\cal O}(h^4)$ from  the $\mat{\delta}\widehat{\mat{\eta}}^N$ term in
\eqref{eq:tl_L}:
\begin{equation}
  [\widehat{\underline{\eta}}^N]_{\alpha \beta}  = 
  {\cal O}(h^2) \quad \text{for} \quad \alpha \neq \beta\, , \qquad
  [\delta\widehat\eta^N]_{\alpha \beta} =  {\cal O}(h^2) \, , \qquad
  [\delta\widehat\eta^L]_{l m} = {\cal O}(h^4) \;.
\end{equation}
The physical reason for this is that the $ \gCP$-conjugated processes
\begin{equation}
  \sum_\alpha c_\alpha \, \widehat N_{\mathrm{R}, \alpha} \ \to\ L \, \Phi  \; ,
  \qquad \qquad 
  \sum_\alpha c_\alpha^* \, \widehat N_{\mathrm{R}, \alpha}^{\tilde{c}} \
  \to \ L^{\tilde{c}} \, \Phi^{\tilde{c}}  \;,
\end{equation}
are    respectively    proportional    to   the    number    densities
$\widehat{\mat{\eta}}^{N}$  and  $\widehat{\mat{\overline\eta}}^{N}  =
(\widehat{\mat{\eta}}^{N})^*$, which now  differ by ${\cal O}(h^2)$ in
the off-diagonal (interference) terms.

Thus, the  ${\cal O}(h^4)$ contribution to the  total lepton asymmetry
is due  to a sequence  of the coherent heavy-neutrino  inverse decays,
oscillations and decays. These effects get enhanced in the same regime
as the resonant $T=0$ $\varepsilon$-type $ \CP$ violation, namely, for
$z       \approx       1$       and       $\Delta       m_N       \sim
\Gamma_{N_\alpha}$~\cite{Pilaftsis:1997jf}.   For   $z  \ll  1$,  this
effect is suppressed by the  small mass of the Majorana neutrinos, and
for $z  \gg 1$ the  inverse decays are  frozen out and  ineffective to
create an  asymmetry. Similarly, if  the heavy-neutrino mass-splitting
$\Delta  m_N$  is  too  large  compared  to  $\Gamma_{N_\alpha}$,  the
oscillations are averaged out during  the decay time scale, whereas if
it   is  too   small   the  oscillations   proceed   too  slowly   and
$\mat{\delta}\widehat{\mat{\eta}}^N$  produced  thereof is  constantly
washed out.

In Section~\ref{sec:6.2},  we will  show quantitatively that,  for the
$\mathrm{RL_\tau}$  model   under  consideration  there,   the  lepton
asymmetry  generation via  the  heavy-neutrino oscillation  phenomenon
discussed above is of the same order as the one due to their mixing in
the vacuum, and  hence, leads to an enhanced  total lepton asymmetry
(even in  the charged-lepton flavour  diagonal case) compared  to that
predicted by the flavour-diagonal Boltzmann equations discussed in 
Section~\ref{sec:2}.

\subsection{Decoherence in the Charged Lepton Sector}
\label{sec:4.3}

In this subsection, we will include in the rate equations the effect of
the  charged-lepton  Yukawa couplings,  described  by the  interaction
Lagrangian
\begin{equation}
  \mathcal{L}_y \ = \ y_k^{\phantom{k} l} \,
  \bar{L}^k \, \Phi \, e_{R, l} \; + \; {\rm H.c.} \;,
  \label{lagl}
\end{equation}
where $e_{R,l} \equiv l_R$~(with $l=e,\mu,\tau$) and 
the Yukawa couplings are real and diagonal in the charged-lepton
Yukawa   eigenbasis,  i.e.~$\widehat{y}_k^{\phantom{k}  l}   =  y_l\,
\delta_k^{\phantom{k}l}$. The  number densities of the  RH leptons and
anti-leptons,  $\mat{n}^R$ and  $\bar{\mat{n}}^R$ respectively,  can be
defined    analogous     to    $\mat{n}^L$    and    $\bar{\mat{n}}^L$
[cf.~\eqref{eq:def_n_1}   and   \eqref{eq:def_n_2}].   The   processes
involving the charged-lepton Yukawa  couplings will be responsible for
the decoherence of the LH  leptons into the mass eigenbasis.  However,
due to the interaction of the charged-leptons with the heavy-neutrinos
[cf.~\eqref{eq:Lagr}], non-zero  off-diagonal elements are necessarily
induced  in the charged-lepton  number density  matrix, which  tend to
recreate a coherence between the charged-lepton flavours.  Thus, there
is  a  competition  between   the  coherence  effect  induced  by  the
heavy-neutrino Yukawa couplings and  the decoherence effect due to the
charged-lepton  Yukawa couplings,  and for  reasonably  large neutrino
Yukawa  couplings, the coherence  effect could  be significant,  as we
will see  explicitly in Section~\ref{sec:6.2}. In  particular, we will
find that  the decoherence  is not complete  in the  temperature range
relevant for the production of the asymmetry in the $\mathrm{RL}_\tau$
scenarios with $200~{\rm GeV}\lsim m_N \lsim 2~{\rm TeV}$. 
Hence, it is important to include
the off-diagonal  lepton number densities,  and the dynamics  of their
decoherence effects, in the rate equation for the lepton asymmetry.

Let us  first consider  the contribution of  thermal Higgs  decays and
inverse decays
\begin{equation}
  \label{4.41}
  \Phi(\ve q) \ \leftrightarrow \  L(\ve p) \, \bar{e}_{R}(\ve k) \;,
\end{equation}
and then generalize it  to other relevant processes.  The contribution
of  this   process  to   the  LH  charged-lepton   transport  equation
\eqref{eq:evol_lept}  can  be obtained  in  a  similar  manner as  the
heavy-neutrino    decays    and    inverse   decays    discussed    in
Section~\ref{sec:3.4}. Explicitly, we obtain
\begin{equation}
  \frac{\D{}{\mat{n}^L}}{\D{}{t}} \ \supset \
  \int_{\ve{p},\ve{k},\ve{q}} \left( \; - \; \frac{1}{2} \,
    \Big\{\mat{n}^L(\ve p), \, 
    \mat{\Gamma}_{\rm dec}(\ve{p}, \ve{k}, \ve{q}) \Big\} \; + \;
    \mat{\Gamma}_{\rm dec}^{\rm back}(\ve{p}, \ve{k}, \ve{q}) \right) \; . 
  \label{eq:coh_nL1}
\end{equation}
In  the above,  we  have defined  the  charged-lepton decoherence  and
back-reaction rates
\begin{align}
  [\Gamma_{\rm dec}(\ve{p}, \ve{k}, \ve{q})]_{l}^{\phantom l m} \ &
  \equiv \ A(\ve{p}, \ve{k}, \ve{q}) \; y_{l}^{\phantom l i} \, 
  y^{m}_{\phantom m j} \, [\bar{n}^R(\ve {k})]_i^{\ j} \;, \\
  [\Gamma_{\rm dec}^{\rm back}(\ve{p}, \ve{k}, \ve{q})]_{l}^{\phantom l m} \ &
  \equiv \ A(\ve{p}, \ve{k}, \ve{q}) \; y_{l}^{\phantom l i} \, 
  y^{m}_{\phantom m i} \, n^{\Phi}(\ve q) \;,
\end{align}
where  the  flavour-singlet term  $A(\ve{p},  \ve{k}, \ve{q})$,  whose
explicit form  is not needed  here, contains the  relevant kinematic
factors.

In the would-be  mass eigenbasis for the charged-leptons  in which the
charged-lepton  Yukawa  coupling  matrix  is  diagonal,  the  diagonal
entries of \eqref{eq:coh_nL1} have the form
\begin{equation}
  \frac{\D{}{[\widehat{n}^L]_{ll}}}{\D{}{t}} \ \supset \ 
  \int_{\ve{p},\ve{k},\ve{q}} A(\ve{p}, \ve{k}, \ve{q}) \; 
  y_l^2 \, \Big( n^\Phi(\ve q) \, - \, [\widehat{n}^L(\ve p)]_{ll} \,
  [\widehat{\bar {n}}^R(\ve k)]_{ll}\Big) \,  \;,
\label{coh_nL2}
\end{equation}
where  the  index  $l$  is  not  summed  over,  and  we  have  assumed
$\widehat{\bar{\mat{n}}}^R(\ve   k)$   to   be  diagonal,   neglecting
higher-order  phenomena  involving  heavy-neutrino  Yukawa  couplings.
Since  the  evolution  equations  of  $\widehat{\bar{\mat{n}}}^R$  and
$n^\Phi$ contain the same term  \eqref{coh_nL2}, and since the rate of
the process \eqref{4.41}  is much larger than the  Hubble rate for the
relevant time period,  we can safely assume that  their fast evolution
always guarantees  the chemical (as  well as kinetic)  equilibrium for
this reaction, i.e.~$\mu_\Phi  = \mu_{L,\,l} + \bar{\mu}_{R,\,l}$, in
addition   to    the   decoherence   of   $\widehat{\bar{\mat{n}}}^R$.
Therefore, the evolution equations for $\Phi$ and $\bar{e}_R$ need not
be  considered  explicitly,  and  instead,  we can  use  the  relevant
detailed  balance condition,  or  the KMS  relation~\cite{Kubo:1957mj,
Martin:1959jp,  Haag:1967sg},  to solve  the  rate  equation. For  the
process \eqref{4.41}  under consideration, the KMS  relation is simply
given by
\begin{eqnarray}
  n^\Phi \ = \ [\widehat{n}^L(\ve p)]_{ll} \,
  [\widehat{\bar {n}}^R(\ve k)]_{ll} \; ,
  \label{KMS}
\end{eqnarray}
for  all  $l$ (not  summed  over),  which  implies that  the  diagonal
contribution  \eqref{coh_nL2}  identically   vanishes,  and  only  the
off-diagonal entries  of the anti-commutator  in \eqref{eq:coh_nL1} are
responsible  for  the  decoherence  of  charged leptons  to  the  mass
eigenbasis.   Note that  at this  stage it  is inconsistent  to assume
$\Phi$ and $\bar{e}_R$ to be at equilibrium, since it violates the KMS
relation~\eqref{KMS},  and does not  lead to  the correct  approach to
equilibrium.

It  can be shown  that the  form \eqref{eq:coh_nL1}  is valid  for any
flavour-dependent process  involving one  LH charged lepton,  with $\mat{\Gamma}_{\rm dec}$ being the corresponding decoherence rate.  Since the  reactions that  cause the
decoherence in the  LH charged-lepton sector are fast  compared to the
Hubble  rate,  the  back-reaction  rate  $\mat{\Gamma}_{\rm  dec}^{\rm
back}$ in \eqref{eq:coh_nL1} can be determined from the conditions
\begin{equation} \label{4.46}
  \Big[\mat{\Gamma}_{\rm dec} \, , \;
  \mat{\Gamma}_{\rm dec}^{\rm back}\Big] \ 
  = \ \maf{0} \; , \qquad\qquad
  [\widehat{\Gamma}_{\rm dec}^{\rm back}]_{l l} \ 
  = \ [\widehat{\Gamma}_{\rm dec}]_{l l} \, [\widehat{n}^L]_{l l} \;.
\end{equation}
The  first  condition  comes  from the  fact  that  $\mat{\Gamma}_{\rm
dec}^{\rm  back}$  and  $\mat{\Gamma}_{\rm  dec}$  are  simultaneously
diagonal in the charged-lepton  mass eigenbasis.  The second condition
is the  generalized KMS relation \eqref{KMS} involving  any species in
chemical and  kinetic equilibrium with  the LH charged leptons  in the
mass  eigenbasis, whose fast  evolution ensures  the vanishing  of the
overall statistical factors that  multiply the large diagonal rates in
the rate equation \eqref{eq:coh_nL1}.
  
The  charged-lepton  Yukawa contributions  to  the  rate equation  for
anti-lepton number density $\bar{\mat{n}}^L$ will be analogous to that
given in \eqref{eq:coh_nL1}.  To obtain the corresponding contribution
to the rate equation for the  lepton asymmetry, we use the same set of
approximations  as in  Section~\ref{sec:4.1}, and  in  particular, the
kinetic-equilibrium number  density \eqref{nLeq}. Taking  into account
the expansion of the Universe, we finally obtain
\begin{equation}
  \frac{H_N \, n^\gamma}{z}\frac{\D{}{ \, \mat{\delta \eta}^L}}
  {\D{}{z}}     \ \supset \ 
  - \; \frac{1}{2 \, \eta^L_{\rm eq}} \, 
  \Big\{\mat{\delta} \mat{\eta}^L, \, 
  \mat{\gamma}_{\rm dec } \Big\} \;
  +\; \mat{\delta \gamma}_{\rm dec}^{\rm back} \;,
\end{equation}
where $\mat{\gamma}_{\rm  dec}$ and $\mat{\delta\gamma}_{\rm dec}^{\rm
back}$  are the  $\widetilde{C}\!P$-even  and -odd  thermally-averaged
decoherence  and  back-reaction  rates,  respectively.  Here  we  have
ignored      the     sub-dominant      $\{\eta^L_{\rm      eq},     \,
\mat{\delta\gamma}_{\rm dec}\}$  term, which depends  on the asymmetry
in the RH  charged-lepton sector that is assumed  to be small compared
to the asymmetry in the LH sector.  The $\widetilde{C}P$-even rate can
be  expressed  in  terms   of  the  charged-lepton  thermal  width  as
$\mat{\gamma}_{\rm dec}  = \mat{\Gamma}_{T} \, n^L_{\rm  eq}$.  In the
mass     eigenbasis,    the    thermal     width    is     given    by
$\widehat{\mat{\Gamma}}_{T} =  \text{diag}\{\Gamma_{T,\,l}\}$, and has
been calculated explicitly in~\cite{Cline:1993bd}, taking into account
the  inverse   Higgs  decays  and  the  relevant   fermion  and  gauge
scatterings:
\begin{equation}\label{eq:Gamma_T}
  \Gamma_{T,\,l} \ \simeq \ 3.8 \times 10^{-3} \, T \, y^2_l \,
  \big[(-1.1 + 3.0 x) \,+\, 1.0 \, +\, y^2_t (0.6 - 0.1 x) \big] \;,
\end{equation}
where   $y_t$  is   the  top   quark   Yukawa  coupling,   and  $x   =
M_H(T)/T=zM_H(T)/m_N$,  $M_H(T)$ being  the Higgs  thermal  mass. Note
that while calculating the final rates for the processes involving the
charged-lepton Yukawa couplings, it  is important to take into account
their thermal  masses, which control  the phase space  suppression for
the decay  and inverse  decay of the  Higgs boson~\cite{Cline:1993bd}.
Additionally,  note  that  all  the  chemical potentials  can  be  consistently
neglected in  the calculation of the  rate, as long as  we satisfy the
generalized KMS  relations given by \eqref{4.46}. After thermal
averaging, \eqref{4.46} can be written as
\begin{equation}\label{eq:det_bal}
  \Big[\mat{\gamma}_{\rm dec} \, , \;
  \mat{\delta \gamma}_{\rm dec}^{\rm back}\Big] \ 
  = \ \maf{0} \; , \qquad\qquad
  [\delta \widehat{\gamma}_{\rm dec}^{\rm back}]_{l l} \ 
  = \ [\widehat{\gamma}_{\rm dec}]_{l l} \, 
  \frac{ [\widehat{\eta}^L]_{l l} }{\eta^L_{\rm eq}} \; .
\end{equation}
These  equations  ensure  that  the detailed  balance  conditions  are
satisfied, without having to resort  to following the evolution of all
the  SM   species  involved  in   the  charged-lepton  Yukawa-mediated
processes.

\subsection{Scattering Terms}
\label{sec:4.4}

In this section, we will describe the flavour-covariant generalization
of the subtraction  of the so-called RIS contributions  present in the
$\Delta L\:=\:2$ and $\Delta L\:=\:0$ scattering terms 
(see Figure~\ref{fig:feynman}).  Specifically,
we will  show how the  sign of the  $\widetilde{C}\!P$-``odd'' inverse
decay terms in \eqref{evol_dn} and \eqref{evol_dL} is flipped, so that
the  correct approach to  equilibrium is  restored. Moreover,  we will
illustrate that it is necessary  to account for thermal corrections in
the   RIS   contributions   when  considering   off-diagonal   flavour
correlations.

Following  \cite{Pilaftsis:2003gt}, we  first consider  the case  of a
single scalar sneutrino $\widetilde{N}$,  and write down the thermally
corrected       scattering       amplitude       $\mathcal{T}(L\Phi\to
L^{\tilde{c}}\Phi^{\tilde{c}})$ as
\begin{align}
  \mathcal{T}(L\Phi\to L^{\tilde{c}}\Phi^{\tilde{c}}) 
  \ = \ \mathcal{T}(L\Phi\to\widetilde{N}^*)
  \: \frac{p^2-m_{\widetilde{N}}^2
    -i\mathrm{Im}\,\Pi^{\mathrm{eq}}_{\widetilde{N}\widetilde{N}}
    (p_0,\mathbf{p})}
  {\big(p^2-m_{\widetilde{N}}^2\big)^2
    +\big[\mathrm{Im}\, 
    \Pi^{\mathrm{eq}}_{\mathrm{ret},\widetilde{N}\widetilde{N}}
    (p_0,\mathbf{p})\big]^2} \: 
  \mathcal{T}(\widetilde{N}^*\to L^{\tilde{c}}\Phi^{\tilde{c}}) \; ,
\end{align}
where  we have  used on-shell renormalization scheme   and have neglected  
thermal dispersive
corrections.    Notice   that,  since   the   Lorentz-covariance   of
thermally-corrected self-energies  is broken, the  absorptive parts of
the   time-ordered   and   retarded  equilibrium   CTP   self-energies
$\mathrm{Im}\,   \Pi^{\mathrm{eq}}_{\widetilde{N}\widetilde{N}}  (p_0,
\mathbf{p})$                     and                    $\mathrm{Im}\,
\Pi^{\mathrm{eq}}_{\mathrm{ret},\widetilde{N}\widetilde{N}}       (p_0,
\mathbf{p})$,   respectively,  are   functions  of   both   $p_0$  and
$\mathbf{p}$; see~\ref{app:propagator} [cf.~\eqref{B101}].

By virtue  of the fluctuation-dissipation theorem  for the equilibrium
self-energies, we have the relation
\begin{equation}
  \mathrm{Im}\,\Pi^{\mathrm{eq}}_{\widetilde{N}\widetilde{N}}(p_0,\mathbf{p}) \
  = \ \varepsilon(p_0)\big[1+2\big\{\theta(p_0)
  n_{\mathrm{eq}}^{\widetilde{N}}(p_0)
  +\theta(-p_0)n_{\mathrm{eq}}^{\widetilde{N}}(-p_0)\big\}\big]
  \mathrm{Im}\,\Pi^{\mathrm{eq}}_{\mathrm{ret},\widetilde{N}\widetilde{N}}
  (p_0,\mathbf{p})\, ,
\end{equation}
where   $\varepsilon(p_0)  \equiv  \theta(p_0)-\theta(-p_0)$   is  the
generalized signum function.  In  the pole-dominance region, using the
NWA given by \eqref{nwa}, we obtain
\begin{equation}
  \label{eq:TRIS}
  |\mathcal{T}_{\mathrm{RIS}}(L\Phi\to L^{\tilde{c}}\Phi^{\tilde{c}})|^2 \ 
  = \ \frac{\pi}{m_{\widetilde{N}}\Gamma_{\widetilde{N}}(s)}
  \theta(\sqrt{s})\delta(s-m_{\widetilde{N}}^2)
  \big(1+4n_{\mathrm{eq}}^{\widetilde{N}}(\sqrt{s})\big) \, .
\end{equation}
Here,  we  have neglected  the  statistical  factors  internal to  the
thermal Breit-Wigner width
\begin{equation}
  \Gamma_{\widetilde{N}}^{\mathrm{eq}}(p_0,\mathbf{p}) \
  = \ \frac{1}{m_{\widetilde{N}}}
  \mathrm{Im}\,\Pi^{\mathrm{eq}}_{\mathrm{ret},\widetilde{N}\widetilde{N}}
  (p_0,\mathbf{p}) \ 
  \simeq\ \Gamma_{\widetilde{N}}(p^2) \, ,
\end{equation}
where   $\Gamma_{\widetilde{N}}(p^2)$   is    the   $T=0$   width   of
$\widetilde{N}$.  An expression  analogous to  \eqref{eq:TRIS}  can be
derived for chiral fermions (see \ref{app:propagator}).

Generalizing the  result in \eqref{eq:TRIS}  to fermions, we  see that
the            RIS            contribution           to            the
$L\Phi\:\to\:L^{\tilde{c}}\Phi^{\tilde{c}}$  scattering terms contains
the following combination of statistical factors:
\begin{equation}
  \mat{\mathcal{F}}_{\rm scat} \ = \ (1\:+\:n^{\Phi})
  (\mat{1}\:-\:\mat{n}^L)\:\otimes\:
  (\mat{1}-4\mat{n}^{N}_{\mathrm{eq}})\:\otimes\:
  (\mat{1}\:-\:\bar{\mat{n}}^{L})(1\:+\:\bar{n}^{\Phi})\; .
\label{4.56}
\end{equation}
In the classical statistical limit, \eqref{4.56} becomes
\begin{equation}
  \label{eq:statfacRIS}
  \mat{\mathcal F}_{\rm scat} \ \simeq \ -\: n^{\Phi}\mat{n}^L\:\otimes\:
  \mat{1} \:\otimes\:\mat{1}\:
  - \: \mat{1}\:\otimes\:\mat{1}\:\otimes\:
  \bar{n}^{\Phi}\bar{\mat{n}}^L\:-\:4\;(\mat{1}\:\otimes\:
  \mat{n}^{N}_{\mathrm{eq}}\:\otimes\:\mat{1})\;.
\end{equation}
The first two terms  in \eqref{eq:statfacRIS} arise from the $T\:=\:0$
part  of the  RIS  subtraction and  contribute  to the  charged-lepton
transport  equations  \eqref{eq:evol_lept} and  \eqref{eq:evol_lept2}.
The  third  term  arises  from  the  thermal  correction  to  the  RIS
subtraction and contributes  to the heavy-neutrino transport equations
\eqref{eq:evol_neu} and \eqref{eq:evol_neu2}.

\subsection*{Contributions to Lepton Transport Equations}
\label{sec:4.4.1}

We   begin  by   describing  the   scattering  contributions   to  the
charged-lepton  transport  equations.  The  contributions  of  $\Delta
L\:=\:2$  scattering to  the  charged-lepton number  densities can  be
obtained by a flavour-covariant generalization of the relevant part of
the flavour-diagonal rate equation \eqref{be4}:
\begin{align}
  \frac{\D{}{\N{L}{l}{m}}}{\D{}{t}} \ &\supset \
  - \; \frac{1}{2 \, n^L_{\mathrm{eq}}} \,
  \Big( \Tdu{[\gamma'(L\Phi \to L^{\tilde{c}} \Phi^{\tilde{c}})
    ]}{l}{n}{k}{k} \, \N{L}{n}{m} \; 
  + \; \N{L}{l}{n} \,
  \Tdu{[\gamma'(L\Phi \to L^{\tilde{c}} \Phi^{\tilde{c}})
    ]}{n}{m}{k}{k} \Big)  \notag\\
  & \qquad + \ \frac{\Nb{L}{k}{n}}{n^L_{\mathrm{eq}}} \,
  \Tdu{[\gamma'(L^{\tilde{c}} \Phi^{\tilde{c}} \to L \Phi)]}{n}{k}{l}{m} \; ,
  \\
  \frac{\D{}{\Nb{L}{l}{m}}}{\D{}{t}} \ &\supset \ 
  - \; \frac{1}{2 \, n^L_{\mathrm{eq}}} \,
  \Big( \Tdu{[\gamma'(L^{\tilde{c}} \Phi^{\tilde{c}} \to L\Phi)
    ]}{l}{n}{k}{k} \, \Nb{L}{n}{m} \; 
  + \; \Nb{L}{l}{n} \,
  \Tdu{[\gamma'(L^{\tilde{c}} \Phi^{\tilde{c}} \to L\Phi)
    ]}{n}{m}{k}{k} \Big) \notag\\
  & \qquad + \ \frac{\N{L}{k}{n}}{n^L_{\mathrm{eq}}} \,
  \Tdu{[\gamma'(L \Phi \to L^{\tilde{c}} \Phi^{\tilde{c}})
    ]}{n}{k}{l}{m}\; .
\end{align}
The contribution to the total lepton asymmetry is then given by 
\begin{equation}
  \frac{\D{}{[\delta \eta^{L}]_{l}^{\phantom{l}m}}}{\D{}{t}} \
  \supset \ 
  - \; 2 \, \Tdu{[\delta \gamma'^{L\Phi}_{L^{\tilde{c}} \Phi^{\tilde{c}}}
    ]}{}{}{l}{m} \; 
  - \; \frac{1}{4 \, n^L_{\mathrm{eq}}} \,
  \Big\{ \delta n^{L} , \,\gamma'^{L\Phi}_{L^{\tilde{c}} \Phi^{\tilde{c}}}
  \Big\}_l^{\phantom l m} \; 
  - \; \frac{[\delta n^{L}]_{k}^{\phantom{k}n}}
  {2 \, n^L_{\mathrm{eq}}} \, 
  \Tdu{[\gamma'^{L\Phi}_{L^{\tilde{c}} \Phi^{\tilde{c}}}]}{n}{k}{l}{m} \; , 
  \label{cov_asy_2}
\end{equation}
where,  by  virtue of  $  \gCP  \widetilde{T}$,  we have  defined  the
contractions
\begin{align}
  \Tdu{[\gamma^{L\Phi}_{L^{\tilde{c}} \Phi^{\tilde{c}}}]}{}{}{l}{m} \ &
  \equiv \ \Tdu{[\gamma^{L\Phi}_{L^{\tilde{c}} \Phi^{\tilde{c}}}
    ]}{l}{m}{k}{k} \ 
  = \ \Tdu{[\gamma^{L\Phi}_{L^{\tilde{c}} \Phi^{\tilde{c}}}
    ]}{k}{k}{l}{m} \; , \\
  \Tdu{[\delta \gamma^{L\Phi}_{L^{\tilde{c}} \Phi^{\tilde{c}}}]}{}{}{l}{m} \ &
  \equiv \ \Tdu{[\delta \gamma^{L\Phi}_{L^{\tilde{c}} \Phi^{\tilde{c}}}
    ]}{l}{m}{k}{k} \
  = \ \Tdu{[\delta \gamma^{L\Phi}_{L^{\tilde{c}} \Phi^{\tilde{c}}}
    ]}{k}{k}{l}{m}   \; .
\end{align}
The rank-4 scattering rates introduced here can be derived by means of
the  generalized  optical  theorem  given in  \ref{app:optical}.   For
consistency with the previous  calculations, only their resonant parts
must      be      kept      and      these     are      listed      in
\eqref{eq:scatrates1}--\eqref{eq:scatrates2}.   Using the NWA  for the
heavy  neutrino propagator,  and the  same approximations  as  for the
decay and inverse-decay terms  in Section~\ref{sec:4.1}, we obtain the
$\gCP$-``even'' collision rates
\begin{align} \label{cov_scat2}
  \Tdu{[\gamma^{L \Phi}_{L^{\tilde{c}} \Phi^{\tilde{c}}}]}{l}{m}{n}{k} \ &
  = \ \sum_{\alpha , \beta}
  \frac{\Big[(\widehat{\gamma}^N_{L \Phi})_{\alpha\alpha} \,
  +\, (\widehat{\gamma}^N_{L \Phi})_{\beta\beta}\Big]}{
  \left(1 \, - \, 2 i \, \frac{\widehat{M}_{N_\alpha}
      \, - \,  \widehat{M}_{N_\beta}}
    {\widehat{\Gamma}_{N_\alpha} 
     \,  + \, \widehat{\Gamma}_{N_\beta}} \right)} 
  \frac{2 \left(\chr{l}{\beta} \, \chr{n}{\beta} \, \chrs{m}{\alpha} \,
    \chrs{k}{\alpha} \;
    +\; \chrcst{l}{\alpha} \, \chrcst{n}{\alpha} \,
    \chrct{m}{\beta} \, \chrct{k}{\beta} \right) }
  {\Big[(\mathbf{\widehat{h}}^{\dagger} \,
    \mathbf{\widehat{h}})_{\alpha \alpha}  \,
    + \, (\mathbf{\widehat{h}}^{\tilde{c}\dagger} \,
    \mathbf{\widehat{h}}^{\tilde{c}})_{\alpha \alpha} \,
    + \, (\mathbf{\widehat{h}}^{\dagger} \, 
    \mathbf{\widehat{h}})_{\beta \beta}  \,
    + \, (\mathbf{\widehat{h}}^{\tilde{c}\dagger} \,
    \mathbf{\widehat{h}}^{\tilde{c}})_{\beta \beta}\Big]^2} \; ,
\end{align}
which  are the  flavour-covariant  generalizations of  those given  by
\eqref{scat2}. The RIS contribution is obtained by taking the diagonal
$\alpha  =  \beta$ elements  in  the  summation in  \eqref{cov_scat2}.
Similarly,  the  $\mat{\gamma}'$-terms   can  be  obtained  by  taking
$\alpha\neq  \beta$.   Using  the  fact that  $\Tdu{[\delta  \gamma^{L
\Phi}_{L^{\tilde{c}}  \Phi^{\tilde{c}}}]}{}{}{l}{m}=0$  up  to  ${\cal
O}(h^4)$    due     to    the    unitarity     of    the    scattering
matrix~\cite{Kolb:1980},  we obtain the  RIS-subtracted $\gCP$-``odd''
collision rates
\begin{equation}
  \Tdu{[\delta \gamma'^{L \Phi}_{L^{\tilde{c}} \Phi^{\tilde{c}}}]}{l}{m}{k}{k} \ 
  = \ - \sum_\alpha \: (\widehat{\gamma}^N_{L \Phi})_{\alpha\alpha} \,
  \frac{(\mathbf{\widehat{h}}^{\tilde{c}\dagger} \,
    \mathbf{\widehat{h}}^{\tilde{c}})_{\alpha \alpha} \,
    \chrcst{l}{\alpha} \, \chrct{m}{\alpha} \; 
    - \; (\mathbf{\widehat{h}}^{\dagger} \, 
    \mathbf{\widehat{h}})_{\alpha \alpha} \, \chr{l}{\alpha} \,
    \chrs{m}{\alpha}}{\big[(\mathbf{\widehat{h}}^{\dagger} \, 
    \mathbf{\widehat{h}})_{\alpha \alpha}  \; 
    + \; (\mathbf{\widehat{h}}^{\tilde{c}\dagger} \,
    \mathbf{\widehat{h}}^{\tilde{c}})_{\alpha \alpha}  \big]^2} \; .
\end{equation}

\begin{figure}[t!]
  \centering
	\subfloat[][Charged-lepton self-energies, with 
        $\Delta L  =  0$ internally.]
		{\fbox{\includegraphics[scale=0.8]{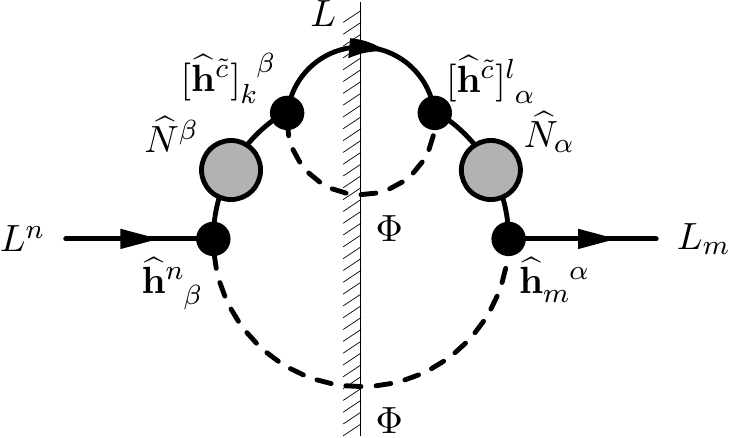}\hspace{2em}
		\includegraphics[scale=0.8]{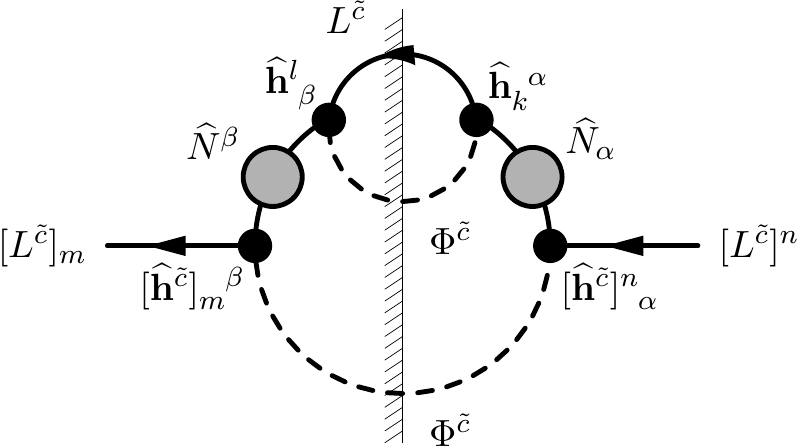}}}\\
	\subfloat[][Charged-lepton self-energies, with 
        $\Delta L  =  2$ internally.]
                {\fbox{\includegraphics[scale=0.8]{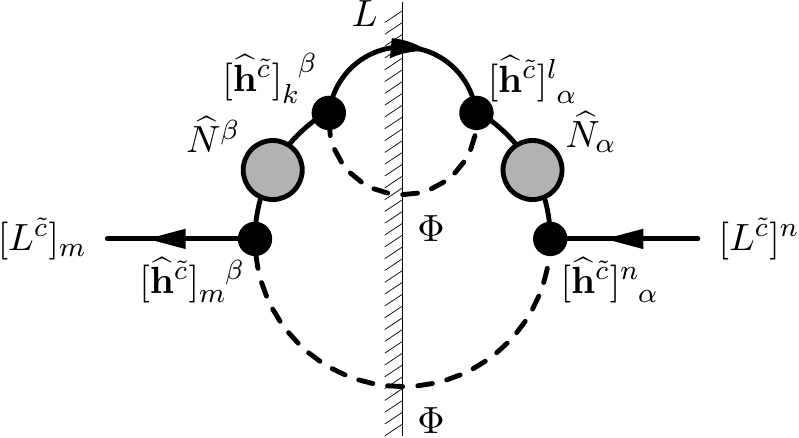}\hspace{2em}
                \includegraphics[scale=0.8]{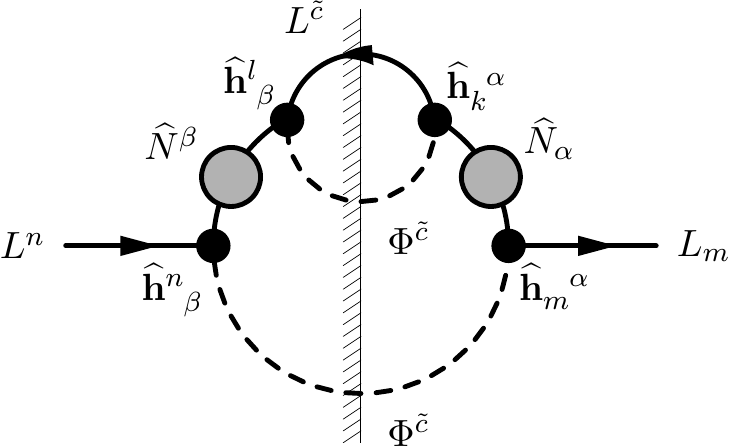}}}
            \vspace{.2cm}
	\caption{\emph {Feynman diagrams  for the self-energies of the lepton doublets. 
The cut,  across  which positive  energy  flows  from  unshaded to  shaded
regions, is associated with production rates in the thermal plasma, as
described    by   the   generalized    optical   theorem    given   in
\ref{app:optical}.                       See                      also
Figure~\ref{fig:feynman}.}\label{fig:cuts2}}
\end{figure}

The  contributions  of  the  $\Delta  L \:=\:  0$  scattering  to  the
charged-lepton transport equations are given by
\begin{align}
  \frac{\D{}{\N{L}{l}{m}}}{\D{}{t}} \ &\supset \
  - \; \frac{1}{2 \, n^L_{\mathrm{eq}}} \,
  \Big( \Tdu{[\gamma'(L\Phi \to L \Phi)]}{l}{n}{k}{k} \, \N{L}{n}{m} \; 
  + \; \N{L}{l}{n} \, \Tdu{[\gamma'(L\Phi \to L \Phi)
    ]}{n}{m}{k}{k} \Big) \notag\\
 \ & \qquad + \ \frac{\N{L}{k}{n}}{n^L_{\mathrm{eq}}} \,
  \Tdu{[\gamma'(L \Phi \to L \Phi)]}{n}{k}{l}{m} \; , \\
  \frac{\D{}{\Nb{L}{l}{m}}}{\D{}{t}} \ &\supset \
  - \; \frac{1}{2 \, n^L_{\mathrm{eq}}} \,
  \Big( \Tdu{[\gamma'(L^{\tilde{c}} \Phi^{\tilde{c}} \to
    L^{\tilde{c}}\Phi^{\tilde{c}})]}{l}{n}{k}{k} \, \Nb{L}{n}{m} \; 
  + \; \Nb{L}{l}{n} \,
  \Tdu{[\gamma'(L^{\tilde{c}} \Phi^{\tilde{c}} \to
    L^{\tilde{c}}\Phi^{\tilde{c}})]}{n}{m}{k}{k} \Big) \notag\\
 \ & \qquad + \ \frac{\Nb{L}{k}{n}}{n^L_{\mathrm{eq}}} \,
  \Tdu{[\gamma'(L^{\tilde{c}} \Phi^{\tilde{c}}
    \to L^{\tilde{c}} \Phi^{\tilde{c}})]}{n}{k}{l}{m} \; ,
\end{align}
and the corresponding contribution to the asymmetry is 
\begin{equation}
  \frac{\D{}{[\delta\eta^{L}]_{l}^{\phantom{l}m}}}{\D{}{t}} \ \supset \
  - \; 2 \, \Tdu{[\delta \gamma'^{L\Phi}_{L \Phi}]}{}{}{l}{m} \; 
  - \; \frac{1}{4 \, n^L_{\mathrm{eq}}} \,
  \Big\{ \delta n^{L} , \,\gamma'^{L\Phi}_{L \Phi} \Big\}_l^{\phantom l m} \;
  + \; \frac{[\delta n^{L}]_{k}^{\phantom{k}n}}{2 \, n^L_{\mathrm{eq}}} \,
  \Tdu{[\gamma'^{L\Phi}_{L \Phi}]}{n}{k}{l}{m} \; .
  \label{cov_asy_0}
\end{equation}
In \eqref{cov_asy_0}, using  $\gCP \widetilde{T}$-invariance,  we  have defined  the
contractions
\begin{align}
  \Tdu{[\gamma^{L\Phi}_{L \Phi}]}{}{}{l}{m} \ &\equiv \
  \Tdu{[\gamma^{L\Phi}_{L \Phi}]}{l}{m}{k}{k} \
  = \ \Tdu{[\gamma^{L\Phi}_{L \Phi}]}{k}{k}{l}{m} \; , \\
  \Tdu{[\delta \gamma^{L\Phi}_{L \Phi}]}{}{}{l}{m} \ &\equiv \
  \Tdu{[\delta \gamma^{L\Phi}_{L \Phi}]}{l}{m}{k}{k} \
  = \ -\, \ \Tdu{[\delta \gamma^{L\Phi}_{L \Phi}]}{k}{k}{l}{m}  \; .
\end{align}
Using  the  results from  \eqref{eq:scatrates1}--\eqref{eq:scatrates2}
and the same set  of approximations as in the $\Delta L  = 2$ case, we
obtain  the  flavour-covariant generalization  of  the collision  rate
given by \eqref{scat1}
\begin{align}
  \Tdu{[\gamma^{L \Phi}_{L \Phi}]}{l}{m}{n}{k} \ &
  = \ \sum_{\alpha , \beta} \,
  \frac{\Big[(\widehat{\gamma}^N_{L \Phi})_{\alpha\alpha} \,
  +\, (\widehat{\gamma}^N_{L \Phi})_{\beta\beta}\Big]}
  {\left(1 \,-\, 2 i \, \frac{\widehat{M}_{N,\alpha} - \widehat{M}_{N,\beta}}
    {\widehat{\Gamma}_{N,\alpha} + \widehat{\Gamma}_{N,\beta}} \right)}
  \frac{2\left( \chrcst{l}{\alpha} \, \chrs{k}{\alpha} \,
    \chrct{m}{\beta} \, \chr{n}{\beta} \;
    +\; \chr{l}{\beta} \, \chrct{k}{\beta} \,
    \chrs{m}{\alpha} \, \chrcst{n}{\alpha} \right) }
  {\Big[(\mathbf{\widehat{h}}^{\dagger} \, 
    \mathbf{\widehat{h}})_{\alpha \alpha}  \,
    + \, (\mathbf{\widehat{h}}^{\tilde{c}\dagger} \,
    \mathbf{\widehat{h}}^{\tilde{c}})_{\alpha \alpha} \,
    + \, (\mathbf{\widehat{h}}^{\dagger} \, 
    \mathbf{\widehat{h}})_{\beta \beta}  \, 
    + \, (\mathbf{\widehat{h}}^{\tilde{c}\dagger} \,
    \mathbf{\widehat{h}}^{\tilde{c}})_{\beta \beta}\Big]^2} \; , 
\label{cov_scat1}
\end{align}
and the corresponding RIS-subtracted $\gCP$-``odd'' quantity 
\begin{align}
  \Tdu{[\delta \gamma'^{L \Phi}_{L \Phi}]}{l}{m}{k}{k} \ &
  = \ - \sum_\alpha (\widehat{\gamma}^N_{L \Phi})_{\alpha\alpha} \,
  \frac{(\mathbf{\widehat{h}}^{\dagger} \,
    \mathbf{\widehat{h}})_{\alpha \alpha} \, \chrcst{l}{\alpha} \,
    \chrct{m}{\alpha} \;
    - \; (\mathbf{\widehat{h}}^{\tilde{c}\dagger} \,
    \mathbf{\widehat{h}}^{\tilde{c}})_{\alpha \alpha} \,
    \chr{l}{\alpha} \, \chrs{m}{\alpha}}
  {\big[(\mathbf{\widehat{h}}^{\dagger} \,
    \mathbf{\widehat{h}})_{\alpha \alpha}  \;
    + \; (\mathbf{\widehat{h}}^{\tilde{c}\dagger} \,
    \mathbf{\widehat{h}}^{\tilde{c}})_{\alpha \alpha}  \big]^2} \; .
\end{align}
The flavour structure of the $\mat{\gamma}$-terms in \eqref{cov_scat2}
and \eqref{cov_scat1} can be  understood diagrammatically, as shown in
Figures~\ref{fig:cuts1} and~\ref{fig:cuts2}, from  the unitarity cuts of
partial  self-energies, obtained  by virtue  of a  generalized optical
theorem (see ~\ref{app:optical}).

Combining   \eqref{cov_asy_2}   and   \eqref{cov_asy_0},   the   total
contribution of $2\leftrightarrow 2$  scattering to the total lepton
asymmetry can be written as
\begin{align} \label{cov_asy_tot}
  \frac{\D{}{[\delta\eta^{L}]_{l}^{\phantom{l}m}}}{\D{}{t}} \ &\supset \ 
  - \; 2 \, \Big( 
  \Tdu{[\delta \gamma'^{L\Phi}_{L^{\tilde{c}} \Phi^{\tilde{c}}}
    ]}{}{}{l}{m} \,
  +\, \Tdu{[\delta \gamma'^{L\Phi}_{L \Phi}]}{}{}{l}{m}\Big) \;  
  - \; \frac{1}{4 \, n^L_{\mathrm{eq}}} \, 
  \Big\{ \delta n^{L} , \,\gamma'^{L\Phi}_{L^{\tilde{c}} \Phi^{\tilde{c}}} \,
  +\, \gamma'^{L\Phi}_{L \Phi}\Big\}_l^{\phantom l m} \notag \\
  \ & \qquad - \ \frac{[\delta n^{L}]_{k}^{\phantom{k}n}}{2 \, n^L_{\mathrm{eq}}} \,
  \Big( \Tdu{[\gamma'^{L\Phi}_{L^{\tilde{c}} \Phi^{\tilde{c}}}]}{n}{k}{l}{m} \,
  -\, \Tdu{[\gamma'^{L\Phi}_{L \Phi}]}{n}{k}{l}{m} \Big) \; .
\end{align}

From  the results  of this  section,  we can  establish the  following
identities valid up to $\mathcal{O}(h^4)$:
\begin{align}
  \Tdu{[\delta \gamma'^{L\Phi}_{L^{\tilde{c}} \Phi^{\tilde{c}}}]}{}{}{l}{m} \,
  +\, \Tdu{[\delta \gamma'^{L\Phi}_{L \Phi}]}{}{}{l}{m} \ &
  = \  \Tdu{[\delta \gamma^{N}_{L \Phi}]}{}{}{l}{m} \; ,
  \label{identity1} \\
  \Tdu{[\gamma^{L\Phi}_{L^{\tilde{c}} \Phi^{\tilde{c}},{\rm RIS}}]}{}{}{l}{m} \,
  +\, \Tdu{[\gamma^{L\Phi}_{L \Phi,{\rm RIS}}]}{}{}{l}{m} \ &
  = \  \Tdu{[\gamma^{N}_{L \Phi}]}{}{}{l}{m} \; ,
  \label{eq:RIS_eq1} \\
  \Tdu{[\gamma^{L\Phi}_{L^{\tilde{c}} \Phi^{\tilde{c}},{\rm RIS}}]}{n}{k}{l}{m} \,
  -\, \Tdu{[\gamma^{L\Phi}_{L \Phi,{\rm RIS}}]}{n}{k}{l}{m} \ &
  = \  0 \; .
  \label{eq:RIS_eq2}
\end{align}
Using  these  identities  in  the  scattering  contribution  given  by
\eqref{cov_asy_tot},   and   including  it   in   the  rate   equation
\eqref{evol_dL} for decay and inverse decay, we obtain
\begin{align}
  \frac{\D{}{[\delta\eta^{L}]_{l}^{\phantom{l}m}}}{\D{}{t}} \ &
  = \ - \,\Tdu{[\delta \gamma^{N}_{L \Phi}]}{l}{m}{}{} \;
  +\; \frac{[\underline{n}^{N}]_{\beta}^{\phantom{\beta}\alpha}}
  {n^N_{\mathrm{eq}}} \,  
  \Tdu{[\delta \gamma^{N}_{L \Phi}]}{l}{m}{\alpha}{\beta}  \; 
  + \; \frac{[\delta n^{N}]_{\beta}^{\phantom{\beta}\alpha}}
  {2\, n^N_{\mathrm{eq}}} \, 
  \Tdu{[\gamma^{N}_{L \Phi}]}{l}{m}{\alpha}{\beta}
  \notag \\  
  &- \ \frac{1}{4 \, n^L_{\mathrm{eq}}} \,
  \Big\{ \delta n^{L} , \,\gamma^{L\Phi}_{L^{\tilde{c}} \Phi^{\tilde{c}}} \,
  +\, \gamma^{L\Phi}_{L \Phi}\Big\}_l^{\phantom l m} 
  \; - \; \ \frac{[\delta n^{L}]_{k}^{\phantom{k}n}}
  {2 \, n^L_{\mathrm{eq}}} \,
  \Big( \Tdu{[\gamma^{L\Phi}_{L^{\tilde{c}} \Phi^{\tilde{c}}}]}{n}{k}{l}{m} \,
  -\, \Tdu{[\gamma^{L\Phi}_{L \Phi}]}{n}{k}{l}{m} \Big) \; .
\end{align}
Note that, thanks to the first identity \eqref{identity1}, the sign of
the inverse-decay  term (first  term in the  RHS) is now  flipped with
respect  to that  in \eqref{evol_dL},  as  anticipated at  the end  of
Section~\ref{sec:4.1},  and  it  guarantees  the correct  approach  to
equilibrium.    The   remaining   identities  \eqref{eq:RIS_eq1}   and
\eqref{eq:RIS_eq2} are  important to guarantee the  consistency of the
formalism:  following \cite{Kolb:1980},  we  have described  processes
like   $L  \Phi   \to  N   \to  L   \Phi$  (with   an   on-shell  $N$)
\emph{statistically}, i.e.~as the successive statistical evolution of
the number density $\mat{n}^N$ first  due to an inverse decay and then
a decay  process. The RIS-subtracted scattering term  is considered in
order  to  avoid  double-counting.   However,  \eqref{eq:RIS_eq1}  and
\eqref{eq:RIS_eq2} allow  us to write the  washout term in  terms of a
complete (including  RIS) scattering rate, with no  inverse decay rate
at all,  thus describing resonant processes  like $L \Phi \to  N \to L
\Phi$ \emph{field-theoretically}. Both  these descriptions lead to the
same result, as shown in \cite{Deppisch:2010fr}.

\subsection*{Contributions to Heavy-Neutrino Transport Equations}
\label{sec:4.4.2}

The  $\Delta L    =  0$  and  $\Delta L    =  2$  scattering
contributions to the heavy-neutrino rate equations are given by
\begin{align}
  \frac{\D{}{[n^N]_{\alpha}^{\phantom{\alpha}\beta}}}{\D{}{t}}\ & 
  \supset\ [\mathcal{S}^N]_{\alpha}^{\phantom{\alpha}\beta}
  \; + \; G_{\alpha\lambda} \, [\bar{\mathcal{S}}^N]_{\mu}^{\phantom{\mu}\lambda} \,
  G^{\mu\beta} \;,\\
  \frac{\D{}{[\bar{n}^{N}]_{\alpha}^{\phantom{\alpha}\beta}}}{\D{}{t}}\ 
  & \supset\ [\bar{\mathcal{S}}^N]_{\alpha}^{\phantom{\alpha}\beta}
  \; + \; G_{\alpha\lambda} \, [\mathcal{S}^N]_{\mu}^{\phantom{\mu}\lambda} \, 
  G^{\mu\beta}\;,
\end{align}
where, for notational simplicity, we have defined 
\begin{align}
  [\mathcal{S}^{N}]_{\alpha}^{\phantom{\alpha}\beta}\ &
  = \ 2\,[\gamma'(L\Phi\to L\Phi)]_{\alpha}^{\phantom{\alpha}\beta}
  \;+\; [\gamma'(L\Phi\to L^{\tilde{c}}\Phi^{\tilde{c}})]
  _{\alpha}^{\phantom{\alpha}\beta}
  \; - \; [\gamma'(L^{\tilde{c}}\Phi^{\tilde{c}}\to L\Phi)]
  _{\alpha}^{\phantom{\alpha}\beta} \;,\\[0.2cm]
  [\bar{\mathcal{S}}^N]_{\alpha}^{\phantom{\alpha}\beta}\ &
  = \ 2\,[\gamma'(L^{\tilde{c}}\Phi^{\tilde{c}}
  \to L^{\tilde{c}}\Phi^{\tilde{c}})]_{\alpha}^{\phantom{\alpha}\beta}
  \;-\; [\gamma'(L\Phi\to L^{\tilde{c}}\Phi^{\tilde{c}})]
  _{\alpha}^{\phantom{\alpha}\beta}
  \;+\; [\gamma'(L^{\tilde{c}}\Phi^{\tilde{c}}\to L\Phi)]
  _{\alpha}^{\phantom{\alpha}\beta}\; ,
\end{align}
and  the  charged-lepton indices  are  understood  to  be traced  over
[cf.~\eqref{eq:Nscatrate}].  As  explained in  the  beginning of  this
subsection,  these  terms  arise due  to  the  last  term on  the  RHS
of~\eqref{eq:statfacRIS}.   Following  the same  procedure  as in  the
charged-lepton   case  discussed   above,  we   obtain   the  relevant
contributions  to the rate  equations for  the $\widetilde{C}\!P$-``even"
and -``odd" number densities:
\begin{align}
  \frac{\D{}{[\underline{n}^{N}
    ]_{\alpha}^{\phantom{\alpha}\beta}}}{\D{}{t}}\ & 
  \supset\ 2\, \Tdu{\big[\widetilde{\mathrm{Re}}
  (\gamma'{}^{L\Phi}_{L\Phi})\big]}{\alpha}{\beta}{}{} 
  \; , \label{scat_n_under}\\
  \frac{\D{}{[\delta n^{N}
    ]_{\alpha}^{\phantom{\alpha}\beta}}}{\D{}{t}}\ & 
  \supset\ \, 4\, i\, \Tdu{\big[\widetilde{\mathrm{Im}}
    (\delta\gamma^{N}_{L\Phi})\big]}{\alpha}{\beta}{}{}  \; .
  \label{scat_deln}
\end{align}
This contribution to $\mat{\delta  \widehat{n}}^{N}$, when added to the
decay and  inverse decay contribution given  by \eqref{evol_dn}, flips
the  sign of  the inverse  decay term  with respect  to that  given in
\eqref{evol_dn},  as  expected  in  order to  achieve  the  correct
equilibrium behaviour.

\subsection{Final Rate Equations}
\label{sec:4.4.3}

Here  we put  together the  various contributions  from heavy-neutrino
decays  and inverse  decays discussed  in  Section~\ref{sec:4.1}, from
processes     involving     charged-lepton     Yukawa     interactions
[cf.~Section~\ref{sec:4.3}], and from  $\Delta L = 0$ and  $\Delta L =
2$       scatterings      via       heavy       neutrino      exchange
[cf.~Section~\ref{sec:4.4}].   Finally,   taking   into  account   the
expansion  of the  Universe,  the following  set  of {\it  manifestly}
flavour-covariant rate equations is obtained for the $\gCP$-``even" number 
density matrix $\mat{\underline{\eta}}^{N}$ and $\gCP$-``odd" number density 
matrices $\mat{\delta \eta}^N$ and $\mat{\delta \eta}^L$:
\begin{align}
   \frac{H_{N} \, n^\gamma}{z}\,
   \frac{\D{}{[\underline{\eta}^{N}]_{\alpha}^{\phantom{\alpha}\beta}}}{\D{}{z}} \ &
  = \ - \, i \, \frac{n^\gamma}{2} \,
  \Big[\mathcal{E}_N,\, \delta \eta^{N}\Big]_\alpha^{\phantom \alpha \beta} \;
  + \; \Tdu{\big[\widetilde{\rm Re}
    (\gamma^{N}_{L \Phi})\big]}{}{}{\alpha}{\beta} \;
  - \; \frac{1}{2 \, \eta^N_{\rm eq}} \,
  \Big\{\underline{\eta}^N, \, \widetilde{\rm Re}(\gamma^{N}_{L \Phi})
  \Big\}_{\alpha}^{\phantom{\alpha}\beta} \;,
  \label{eq:evofinal2}\\
  \frac{H_{N} \, n^\gamma}{z}\,
  \frac{\D{}{[\delta \eta^N]_\alpha^{\phantom \alpha \beta}}}{\D{}{z}} \ &
  = \ - \; 2 \, i \, n^\gamma \,
  \Big[\mathcal{E}_N,\, \underline{\eta}^{N}\Big]_\alpha^{\phantom \alpha \beta} \;
  + \; 2\, i\,  \Tdu{\big[\widetilde{\rm Im}
    (\delta \gamma^{N}_{L \Phi})\big]}{}{}{\alpha}{\beta} \;-\; 
  \frac{i}{\eta^N_{\rm eq}} \, \Big\{\underline{\eta}^N, \,
  \widetilde{\rm Im}
  (\delta\gamma^{N}_{L \Phi}) \Big\}_{\alpha}^{\phantom{\alpha}\beta} \notag\\
  & \quad \;\, - \ \frac{1}{2 \, \eta^N_{\rm eq}}  \,
  \Big\{\delta \eta^N, \, \widetilde{\rm Re}(\gamma^{N}_{L \Phi})
  \Big\}_{\alpha}^{\phantom{\alpha}\beta}
  \label{eq:evofinal3}\;, \\[0.5cm] 
 \frac{H_{N} \, n^\gamma}{z}\, \frac{\D{}{[\delta \eta^L]_l^{\phantom l m}}}
  {\D{}{z}} \ &
  = \ - \, \Tdu{[\delta \gamma^{N}_{L \Phi}]}{l}{m}{}{} \;
  +\; \frac{[\underline{\eta}^{N}]_{\beta}^{\phantom{\beta}\alpha}}
  {\eta^N_{\rm eq}} \,
  \Tdu{[\delta \gamma^{N}_{L \Phi}]}{l}{m}{\alpha}{\beta} \;
  + \; \frac{[\delta \eta^N]_{\beta}^{\phantom\beta \alpha}}{2\,\eta^N_{\rm eq}} \,
  \Tdu{[\gamma^{N}_{L \Phi}]}{l}{m}{\alpha}{\beta} 
 \notag\\
  \ & \quad\;\, - \ \frac{1}{3} \,
  \Big\{ \delta {\eta}^{L} , \,
  {\gamma}^{L\Phi}_{L^{\tilde{c}} \Phi^{\tilde{c}}} \,
  +\, {\gamma}^{L\Phi}_{L \Phi}\Big\}_{l}^{\phantom l m} 
  \; - \; \ \frac{2}{3} \, \Tdu{[\delta {\eta}^L]}{k}{n}{}{} \,
  \Big( \Tdu{[{\gamma}^{L\Phi}_{L^{\tilde{c}} \Phi^{\tilde{c}}}]}{n}{k}{l}{m} \,
  -\, \Tdu{[{\gamma}^{L\Phi}_{L \Phi}]}{n}{k}{l}{m} \Big)
  \notag\\[2mm] 
  \ & \quad\;\, 
  - \; \frac{2}{3} \, 
  \Big\{\delta \eta^L, \, 
  \gamma_{\rm dec } \Big\}_{l}^{\phantom l m} \;
  +\; [\delta \gamma_{\rm dec}^{\rm back}]_{l}^{\phantom l m} \;. 
  \label{eq:evofinal1}
\end{align}
Here,   we   have   dropped  the   $\mathcal{O}(h^4)$   RIS-subtracted
contribution  [cf.~\eqref{scat_n_under}]  to  the  rate  equation  for
$\underline{\mat{\eta}}^N$,  since this  is sub-dominant,  compared to
the other terms on the  RHS of \eqref{eq:evofinal2} which are formally
of   order   $\mathcal{O}(h^2)$.    However,   the   ${\cal   O}(h^4)$
contribution  from  \eqref{scat_deln} must  be  included  in the  rate
equation for $\mat{\delta  \eta}^N$, since all the other  terms on the
RHS of \eqref{eq:evofinal3} are also of the same order.

The             flavour-covariant            rate            equations
\eqref{eq:evofinal2}--\eqref{eq:evofinal1} are the main new results of
this section. They  provide a complete and unified  description of the
RL  phenomenon,  consistently capturing  the following physically 
distinct effects in a  single framework, applicable in any temperature 
regime: 
\begin{itemize}
\item [(i)]  Resonant mixing  between heavy neutrinos, described by the resummed 
Yukawa couplings in $\gamma^N_{L\Phi}$ and $\delta\gamma^N_{L\Phi}$. This provides a 
flavour-covariant generalization of the mixing effects discussed earlier in~\cite{Pilaftsis:2003gt}.  
\item [(ii)] Coherent oscillations  between heavy neutrinos, described by the commutators 
in~\eqref{eq:evofinal2} and~\eqref{eq:evofinal3}, and transferred to the lepton asymmetry 
via the {\emph {new}} rank-4 term $\Tdu{[\gamma^{N}_{L \Phi}]}{l}{m}{\alpha}{\beta}$ 
in the first line of~\eqref{eq:evofinal1}. We should stress here that this phenomenon of 
coherent oscillations is an ${\cal O}(h^4)$ effect on the {\it total} lepton asymmetry, and so  differs from the ${\cal O}(h^6)$ mechanism proposed in~\cite{Akhmedov:1998qx} (see Section~\ref{sec:4.2}).  
\item [(iii)] Decoherence effects  due to charged-lepton Yukawa couplings, described by the 
last line of~\eqref{eq:evofinal1}. Our description of these effects goes along the lines of~\cite{Abada:2006fw}, which has been generalized here to an arbitrary flavour basis. 
\end{itemize}

As an application, we will use   the    rate    equations \eqref{eq:evofinal2}--\eqref{eq:evofinal1}  in Section~\ref{sec:6.2}  for  the  numerical  evaluation of  the  lepton
asymmetry in  the RL$_\tau$ model under consideration  there.  We will
also  derive  approximate analytic  solutions  of  these general  rate
equations in Section~\ref{sec:5.3}.  Finally,  we note that taking the
limit    in    which    the    number    densities    are    diagonal,
i.e.~$[\eta^N]_\alpha^{\    \beta}     =    \delta_\alpha^\beta    \:
\eta^N_\beta$   and   $[\delta\eta^L]_l^{\    m}   =   \delta_l^m   \:
\delta\eta^L_m$   in   \eqref{eq:evofinal2}--\eqref{eq:evofinal1},  we
recover  the  flavour-diagonal  Boltzmann  equations  \eqref{be2}  and
\eqref{be3}.

\section{Minimal Resonant \texorpdfstring{$\ell$}{l}-Genesis
  Model}
\label{sec:5}

In this section, we discuss the basic theoretical framework underlying
the  minimal  RL$_\ell$  model   in  which  the  lepton  asymmetry  is
dominantly  generated  and  stored  in an  individual  lepton  flavour
$\ell$~\cite{Pilaftsis:2004xx}.   We  start  with the  heavy  Majorana
neutrino sector Lagrangian given  by \eqref{eq:Lagr}.  Note that above
the  scale  of the  electroweak  phase  transition,  only the  singlet
neutrinos  are massive,  whose  origin must  lie  in some  ultraviolet-complete
extension of the  SM.  Within the minimal RL$_\ell$  setup, the masses
of  all these heavy  neutrinos $N_\alpha$  ($\alpha=1,...,{\cal N}_N)$
are nearly degenerate.   This can be ensured naturally  by assuming an
$O({\cal N}_N)$-symmetric  heavy neutrino  sector at some  high energy
scale $\mu_X$,  thereby imposing  the universal boundary  condition on
the  heavy-neutrino mass  matrix:  $\mat{M}_N(\mu_X)=m_N\mat{1}$.  The
corresponding  boundary values  for the Yukawa  coupling matrix
elements   $[h(\mu_X)]_{l}^{\  \alpha}$   depend  on   the  particular
RL$_\ell$   model  under   consideration~\cite{Deppisch:2010fr}.   The
Majorana neutrino  mass matrix at the  phenomenologically relevant low
energy scale can then be written down as
\begin{eqnarray}
  \mat {M}_N \ = \ m_N\mat{1} \; + \; \mat{\Delta M}_N \; , 
  \label{mnrg}
\end{eqnarray}
where  $\mat  {\Delta  M}_N$  is a  general  $O({\cal  N}_N)$-breaking
perturbation matrix induced by the  RG evolution of the heavy neutrino
mass matrix $\mat {M}_N$ from the high scale $\mu_X$ down to the scale
of $m_N$:
\begin{eqnarray}
 \mat{\Delta  M}_N \ = \ -\frac{m_N}{8\pi^2}
  \ln\left(\frac{\mu_X}{m_N}\right)
  {\rm Re}\left[\mat{h}^\dag(\mu_X) \mat{h}(\mu_X)\right] \; .
  \label{deltam}
\end{eqnarray}
The  compatibility of  the  light neutrino  masses  generated via  the
seesaw mechanism  with the solar and  atmospheric neutrino oscillation
data requires  that, for  electroweak-scale heavy neutrinos  with $m_N
\sim {\cal O}(100)$  GeV, the norm of the  Yukawa coupling matrix must
be much  smaller than unity.  Given \eqref{deltam},  this implies that
the norm  of the $O({\cal  N}_N)$-breaking matrix $\mat{  \Delta M}_N$
must  be small  compared  to $m_N$,  i.e.~$\|\mat{ \Delta  M}_N\|/m_N
\lsim 10^{-7}$.   As shown in~\cite{Deppisch:2010fr},  a $\mat{ \Delta
M}_N$ of  the required order  can indeed be generated  radiatively for
RL$_\ell$ models.

A  non-zero total  lepton asymmetry  summed over  all flavours  can be
created,      if     and      only     if      the      $     \CP$-odd
quantity~\cite{Pilaftsis:1997jf}
\begin{eqnarray}
  \Delta_{ \CP} \ \equiv \ {\rm Im}\left[{\rm Tr}
  \left(\mat{h}^\dag \mat{h} \mat{M}_N^\dag \mat{M}_N \mat{M}_N^\dag
    \mat{h}^{\sf T} \mat{h}^*\mat{M}_N\right)\right] 
  \label{deltacp}
\end{eqnarray}
does not vanish for a {\it finite} non-zero interval of RG scales.  In
general, the  total number of all non-trivial  $ \CP$-violating phases
in  a model  with  ${\cal  N}_L$ weak  iso-doublets  and ${\cal  N}_N$
neutral   iso-singlets   is    ${\cal   N}_{   \CP}={\cal   N}_L({\cal
N}_N-1)$~\cite{Branco:1986,  Korner:1992zk}.  This  results  in ${\cal
N}_{\CP}$  $\CP$-odd quantities, analogous to the one defined in 
\eqref{deltacp},  which generally  mix
under RG effects.  However, after  summing over all final state lepton
flavours  occurring   in  the  heavy  neutrino  decays,   only  one  $
\CP$-asymmetry   remains,   which  is   odd   under  the   generalized
$\CP$-transformations  discussed in Section~\ref{sec:3.2}.   Using the
definition  of  $\Delta_{  \CP}$  in  \eqref{deltacp}  and  the  heavy
neutrino mass  matrix \eqref{mnrg}, and taking into account the RG  evolution, 
one can
find the  necessary and sufficient  condition for the generation  of a
non-zero total $ \CP$-asymmetry in  the minimal RL setup. The  results  
of this exercise, including RG effects {\it except} for the charged-lepton 
Yukawa couplings, are summarized in Table~\ref{tab:cp} for different  choices of  ${\cal
N}_L$ and ${\cal N}_N$.

\begin{table}[t!]
  \centering
  \begin{tabular}{c|ccc}\hline\hline
    $\Delta_{ \CP}$ for & \multicolumn{3}{c}{${\cal N}_N$}\\ \cline{2-4}
    ${\cal N}_L$ & 1 & 2 & 3\\ \hline\hline
    1 & 0 & 0 & 0 \\ 
    2 & 0 & 0 & $\slashed{0}$ \\ 
    3 & 0 & 0 & $\slashed{0}$ \\ \hline\hline
  \end{tabular}
  \caption{\it Total $ \CP$-asymmetry in the  minimal RL  model with  ${\cal N}_L$
    lepton iso-doublets and ${\cal N}_N$ neutral iso-singlets, including the
    RG effects, except for the charged-lepton Yukawa couplings. 
Here $\slashed{0}$ means $\Delta_{ \CP}\neq 0$.}
  \label{tab:cp}
\end{table}

\subsection{Geometry of the Degeneracy Limit}
\label{sec:5.1}

In this section, we provide  a geometric and physical understanding of
the degeneracy of  the minimal RL$_\ell$ model parameter  space in the
$O({\cal N}_N)$-symmetric limit  $\mat{\Delta M}_N \to \mat{0}$ (where
$\mat{0}$   is   the  null   matrix).    Under   a  general   $O({\cal
N}_N)$-rotation,  the  heavy neutrino  mass  eigenstates transform  as
$N_\alpha \to N'_\alpha = O_{\alpha}^{\phantom{\alpha}\beta} N_\beta$,
and accordingly, the Yukawa coupling  matrix transforms as a vector in
the  heavy-neutrino  mass eigenbasis,  i.e.~$h_{l}^{\phantom{l}\alpha}
\to   h_l'^{\phantom{l}\alpha}   =  h_{l}^{\phantom{l}\beta}   [O^{\sf
T}]_\beta^{\phantom{\beta}\alpha}$.   Depending on  the dimensionality
of the  rotation space, we can  rotate away some of  the components of
$\mat{h}$, such that  some elements  of the  resummed  Yukawa coupling
matrix $\widehat{\maf{h}}$, as given by \eqref{resum1}, will vanish.

\begin{figure}[t!]
  \centering
  \includegraphics[width=11cm]{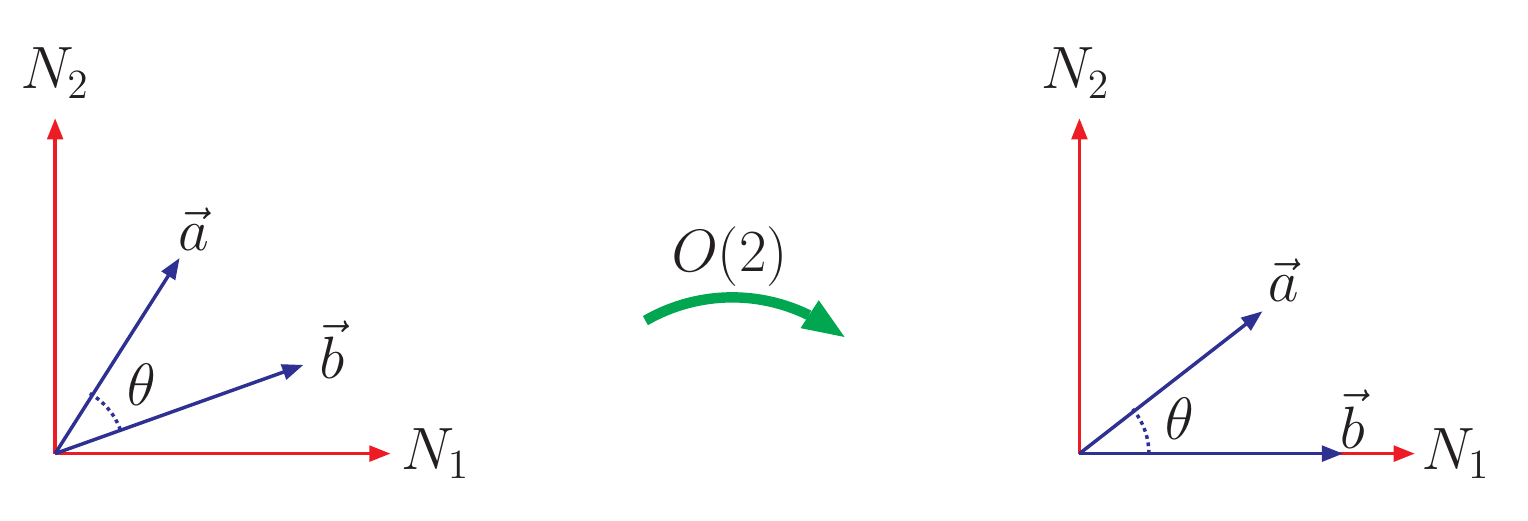}
  \caption{\emph {$O(2)$ transformation of
      $\vec{h}_l=(\widehat{h}_{l1},\widehat{h}_{l2})$.}}
  \label{rot1}
\end{figure}

To illustrate this point, let us first consider a simple case with one
charged-lepton  flavour  (${\cal N}_L  =  1$)  and two  heavy-neutrino
flavours (${\cal N}_N = 2$).  In this case, the tree-level heavy-neutrino
mass matrix  ${\bm M}_N$ is  symmetric under $O(2)$. Let  us therefore
define  the tree-level  Yukawa coupling  as a  two-dimensional complex
vector in the heavy  neutrino mass eigenbasis $\{N_1, N_2\}$: $\vec{h}
\equiv  (\widehat{h}_{l1},\widehat{h}_{l  2})$,   which  can  also  be
written  in   terms  of  two  real  vectors,   i.e.~$\vec{h}  =  {\rm
Re}(\vec{h})+i{\rm Im}(\vec{h}) \equiv \vec{a} + i\vec{b}$.  Using the
$O(2)$-invariance  of   the  $\{N_1,N_2\}$  parameter   space  in  the
degenerate limit,  one can  always rotate the  vectors $\vec  {a}$ and
$\vec{b}$,  such that  ${\vec b}$  is along  the  $N_1$-axis, i.e.~the
$N_2$-component  of ${\vec b}$  vanishes, which  in turn  implies ${\rm
Im}({\widehat{h}}_{l 2}) = 0$ (see Figure~\ref{rot1}).

For  ${\cal  N}_N  =  2$,  the  $R_{\alpha\beta}$-dependent  terms  in
\eqref{resum1}  are  absent [cf.~\eqref{resum11}].  Thus,  in  the
degenerate limit $m_{N_\alpha}  = m_{N_\beta}$, the resummed Yukawa 
couplings given in \eqref{resum1} become 
\begin{eqnarray}
  \widehat{\maf{h}}_{l 1} &=& \widehat{h}_{l 1}
  -\widehat{h}_{l2}\frac{A_{12}+A_{21}}{2A_{22}}
  = \frac{1}{2}\left(\widehat{h}_{l 1}
    -\widehat{h}^*_{l 1}
    \frac{\widehat{h}_{l 2}}{\widehat{h}^*_{l 2}}\right) \; ,
  \nonumber\\
  \widehat{\maf{h}}_{l 2} &=& \widehat{h}_{l 2}
  -\widehat{h}_{l1}\frac{A_{21}+A_{12}}{2A_{11}}
  = \frac{1}{2}\left(\widehat{h}_{l 2}
    -\widehat{h}^*_{l 2}
    \frac{\widehat{h}_{l 1}}{\widehat{h}^*_{l 1}}\right) \; .
  \label{resum2}
\end{eqnarray}
Note that  in the  $O({\cal N}_N)$-symmetric limit,  any basis  in the
heavy-neutrino  flavour space  is a  mass eigenbasis,  and  hence, the
resummed Yukawa couplings can  be defined consistently in any $O({\cal
N}_N)$-rotated basis.   From \eqref{resum2}, we find  that in general,
$\widehat{\maf{h}}_{l  1,2}  \neq  0$  for $\widehat{h}_{l  1,2}  \in
\mathbb{C}$.  However, if both $\widehat{h}_{l 1}$ and $\widehat{h}_{l
2}$ are real, i.e.~$\vec{b}=\vec{0}$, we can rotate $\vec{a}$ to align
with either  $N_1$ or $N_2$ direction in  Figure~\ref{rot1}, such that
either $\widehat{\maf{h}}_{l 1} = 0$ or $\widehat{\maf{h}}_{l 2} = 0$.
Thus,  for $\widehat{h}_{l  1,2} \in  \mathbb{R}$, the  resummed heavy
neutrino    Yukawa    couplings  flow to  the exact $O(2)$-symmetric limit of
the  theory, i.e.~$\widehat{\maf{h}}_{l1,2}=0$.   However, as  we will
see  below, the  RG effects  play  an instrumental  role in  consistently
lifting this $O(2)$ degeneracy.

\begin{figure}[t!]
  \centering
  \includegraphics[width=12cm]{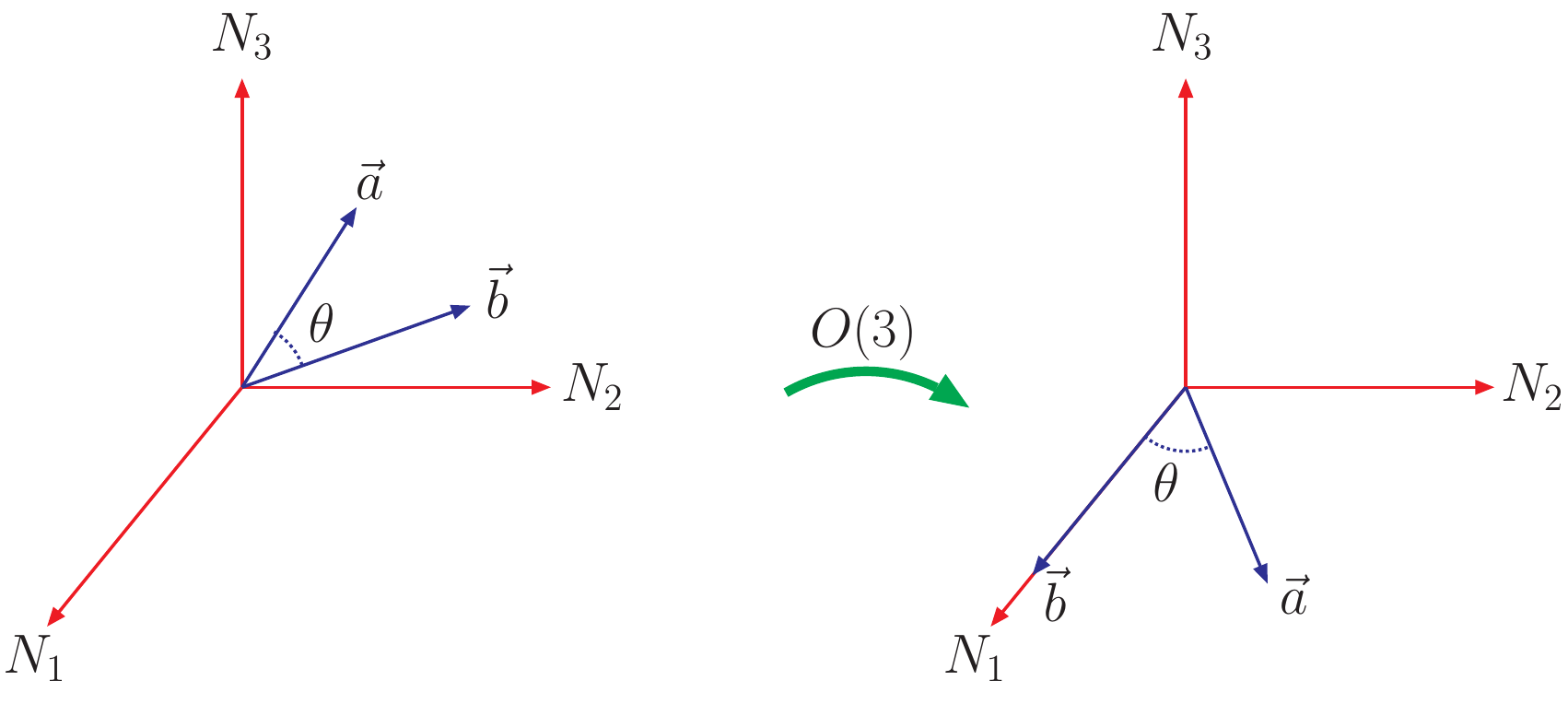}
  \caption{{\it $O(3)$ transformation of
      $\vec{h}_l=(\widehat{h}_{l1},\widehat{h}_{l2},\widehat{h}_{l3})$.}}
  \label{rot2}
\end{figure}

Similarly  for three heavy-neutrino flavours (${\cal  N}_N =  3$), 
we  can define  a three-dimensional
complex vector in the $\{N_1,N_2,N_3\}$ mass eigenbasis: ${\vec h} =
(\widehat {h}_{l 1},\widehat{h}_{l 2},\widehat{h}_{l  3}) = {\vec a} +
i{\vec  b}$.   In  this  case,  using  the  $O(3)$-invariance  of  the
parameter space  in the  degenerate limit, one  can always  rotate the
vectors ${\vec a}$ and ${\vec b}$ in such a way that ${\vec b}$ points
in the  $N_1$-direction and ${\vec a}$ lies  on the $(N_1,N_2)$-plane,
as  shown in  Figure~\ref{rot2}.  Thus,  the $N_3$-components  of both
${\vec   a}$   and   ${\vec   b}$  identically   vanish,   i.e.~${\rm
Re}(\widehat{h}_{l  3}) =  0 =  {\rm Im}(\widehat{h}_{l3})$.   For the
simple case  with ${\cal  N}_L = 1$,  as considered above, 
the resummed     Yukawa    couplings 
flow to the $O(3)$-symmetric limit, i.e.     
$\widehat{\maf{h}}_{l  \alpha} =  0$ for  $\widehat{h}_{l  \alpha} \in
\mathbb{C}$.   However, for  $\widehat{h}_{l \alpha}  \in \mathbb{R}$,
$\widehat{\maf{h}}_{l  \alpha}$  is  undetermined  (0/0 form)  in  the
degenerate limit. One can similarly work out the degeneracy limits for
other  values  of ${\cal  N}_L$;  the  results  of this  analysis  are
summarized in Table~\ref{tab:resum}.

\begin{table}[t!]
  \centering
  \begin{tabular}{c|ccc}\hline\hline
    $\widehat{\maf{h}}$ for & \multicolumn{3}{c}{${\cal N}_N$} \\
    \cline{2-4}
    ${\cal N}_L$ & 1 & 2 & 3 \\ 
    \hline\hline
    1 & $\slashed{0}$ & $\slashed{0}^*$ & $0^*$ \\ 
    2 & $\slashed{0}$ & $\slashed{0}$ & $\slashed{0}^*$ \\ 
    3 & $\slashed{0}$ & $\slashed{0}$ & $\slashed{0}$ \\
    \hline\hline
  \end{tabular}
  \caption{{\it The   resummed   Yukawa   couplings   in   the   degeneracy
    limit, without including RG effects. Here 
$\slashed{0}$  means $\widehat{\maf h}_{l  \alpha}  \neq
     0$ for $\widehat{h}_{l\alpha}   \neq  0$; $\slashed{0}^*$ means
    $\widehat{\maf h}_{l \alpha}  \neq  0$ for $\widehat{h}_{l \alpha}
     \in  \mathbb{C}$, but $\widehat{\maf h}_{l \alpha}  =  0$ for
    $\widehat{h}_{l\alpha}     \in   \mathbb{R}$;   and  $0^*$  means
    $\widehat{\maf h}_{l \alpha}  =  0$ for $\widehat{h}_{l \alpha} 
    \in  \mathbb{C}$, but undetermined for $\widehat{h}_{l\alpha}  \in
     \mathbb{R}$.}}
\label{tab:resum}
\end{table} 

In  a  realistic  situation,  the degeneracy  of  the  
heavy-neutrino parameter space in the $O({\cal N}_N)$-symmetric limit   
$\mat{\Delta M}_N \to \mat{0}$ will be broken by RG effects. Specifically, the
RG evolution from a high scale $\mu_X$, at which  the heavy neutrino masses
are degenerate, i.e.~$m_{N_\alpha} = m_{N_\beta}$, to the scale $m_N$
induces  a  non-zero  mass-splitting  given  by  \eqref{deltam}.   For
instance,  for the case $({\cal  N}_L,{\cal N}_N)  = (1,3)$ discussed
above, the  inclusion of  RG effects  yields $\widehat{\maf{h}}_{l
\alpha} \neq 0$ for  $\widehat{h}_{l\alpha} \in \mathbb{C}$ ($\alpha =
1,2$) and $\widehat{\maf{h}}_{l 3} =  0$.  This is consistent with the
fact  that $\widehat{h}_{l3}$  does not  run in  the  mass eigenbasis,
since $\widehat{h}_{l 3}(\mu_X)$  can be rotated to zero,  as shown in
Figure~\ref{rot2}. As a consequence, the mass parameter $m_{N_3}$ does 
not evolve under the action of the RG.  

It  is  worth  noting  that  the  degeneracies  encountered  above  are
reminiscent of the singular  degeneracies occurring in ordinary Quantum
Mechanics,   where    one   has   to    apply   carefully   degenerate
time-independent  perturbation theory  in order  to  obtain meaningful
results. For illustration, let  us consider a three-state system whose
time-evolution  is  governed by  the  following perturbed  Hamiltonian
(see~\cite{sakurai}, p.~348):
\begin{eqnarray}
  H \ = \ H_0\:+\:\Delta H\ =\ \left(
    \begin{array}{ccc}
      E_1 & 0 & 0\\
      0 & E_1 & 0\\
      0 & 0 & E_2
    \end{array}
  \right)\:+\:\left(
    \begin{array}{ccc}
      0 & 0 & a\\
      0 & 0 & b\\
      a^* & b^* & 0
    \end{array}
  \right)\;,
\label{5.5}
\end{eqnarray}
where the unperturbed  energy levels are $E_1, E_1,  E_2$ (with $E_2 >
E_1$), and $a, b \ll (E_2-E_1)$ are  treated as small perturbations.  
Since the perturbation matrix $\Delta H$ is off-diagonal, the
first-order  perturbation  does  not  change the  energy  eigenvalues,
i.e.~does  not remove  the  degeneracy between  the  first two  energy
eigenstates.   At second-order,  applying  non-degenerate perturbation
theory would lead to an undetermined result for the corrections to the
degenerate eigenvalues, i.e.
\begin{eqnarray}
  \Delta_1^{(2)} \ = \ \frac{0}{0} \:
  - \: \frac{|a|^2}{E_2-E_1}\;,
  \qquad \Delta_2^{(2)} \ = \ \frac{0}{0} \:
  - \:\frac{|b|^2}{E_2-E_1}\;,
  \qquad \Delta_3^{(2)} \ = \ \frac{|a|^2+|b|^2}{E_2-E_1}\;.
\end{eqnarray}
Instead,  applying  degenerate  perturbation  theory,  in  a  suitable
$O(2)$-rotated  basis,  leads   to  well-defined  second-order  energy
shifts, thus lifting the degeneracy:
\begin{eqnarray}
  \Delta_1^{(2)} \ = \ 0\;,
  \qquad \Delta_2^{(2)} \ = \ -\frac{|a|^2+|b|^2}{E_2-E_1}\;,
  \qquad \Delta_3^{(2)} \ = \ \frac{|a|^2+|b|^2}{E_2-E_1}\;.
\end{eqnarray}
Note that  the first energy
eigenvalue  will remain unperturbed to all orders. This  is  in close analogy  
with the  $({\cal
N}_L,{\cal    N}_N)   =   (1,3)$    case   discussed    above,   where
$m_{N_3}$ remains invariant, irrespective of the RG effects.

Consequently, the  resummed Yukawa couplings in  the minimal RL$_\ell$
model are  finite and consistently flow to  the $O(N)$-symmetric limit
of the theory.  The role  of RG
flow in  lifting the  degeneracy of heavy  neutrino masses is  akin to
that  of  degenerate time-independent  perturbation  theory in
ordinary  Quantum Mechanics (e.g.~electric field  in  a
linear Stark effect). Just as one  needs to  choose carefully a
basis in which to apply perturbation theory, one must define the 
resummed Yukawa couplings only in the mass eigenbasis, in which  the  RG  effects  consistently break  the
degeneracies  of the heavy-neutrino  parameter space.  
\subsection{A Model of Resonant \texorpdfstring{$\tau$}{tau}-Genesis}
\label{sec:5.2}

As  an explicit  example of  the  RL$_\ell$ scenario,  we consider  an
RL$_\tau$  model with  $O(3)$  symmetry imposed  on  the heavy-neutrino
sector at  the GUT  scale, $\mu_X\sim 2\times  10^{16}$ GeV,  which is
explicitly   broken  to  the   $U(1)_{L_e+L_\mu}\times  U(1)_{L_\tau}$
subgroup of  lepton-flavour symmetries  by a neutrino  Yukawa coupling
matrix of the following form~\cite{Deppisch:2010fr}:
\begin{eqnarray}
  \mat{h} \ = \ \left(\begin{array}{ccc}
      0 & ae^{-i\pi/4} & ae^{i\pi/4}\\
      0 & be^{-i\pi/4} & be^{i\pi/4}\\
      0 & 0 & 0
    \end{array}\right) \: + \: \mat{\delta h} \; , 
  \label{yuk}
\end{eqnarray}
where  $\mat{\delta h}$  vanishes in  the flavour  symmetric limit.  In this
symmetric limit, the light neutrinos  remain massless to all orders in
perturbation  theory~\cite{Pilaftsis:1991ug}, while  $a$  and $b$  are
arbitrary complex  parameters. In  order to give  masses to  the light
neutrinos, we consider the following form of $\mat {\delta h}$ as a minimal
departure from the flavour-symmetric limit~\cite{Deppisch:2010fr}:
\begin{eqnarray}
  \mat{\delta h} \ = \ \left(\begin{array}{ccc}
      \epsilon_e & 0 & 0\\
      \epsilon_\mu & 0 & 0\\
      \epsilon_\tau & \kappa_1 e^{-i(\pi/4-\gamma_1)} &
      \kappa_2 e^{i(\pi/4-\gamma_2)}
    \end{array}\right) \; ,
\label{delta_h}
\end{eqnarray}
where  $|\epsilon_l|,\kappa_{1,2}\ll |a|,|b|$, and  $\gamma_{1,2}$ are
arbitrary phases. To leading order in the symmetry-breaking parameters
$\Delta \mat{ M}_N$ and  $\mat{\delta h}$,  the tree-level  light neutrino
mass matrix is given by the seesaw formula
\begin{eqnarray}
  \bm M_\nu \ \simeq \ -\frac{v^2}{2}\mat{h} \bm M_N^{-1} \mat{h}^{\sf T} \ 
  \simeq \ \frac{v^2}{2m_N}\left(
    \begin{array}{ccc}
      \kappa_N a^2+\epsilon_e^2 &
      \kappa_N ab + \epsilon_e\epsilon_\mu &
      \epsilon_e\epsilon_\tau\\
      \kappa_N ab+\epsilon_e\epsilon_\mu &
      \kappa_N b^2+\epsilon_\mu^2 &
      \epsilon_\mu\epsilon_\tau\\
      \epsilon_e\epsilon_\tau &
      \epsilon_\mu\epsilon_\tau &
      \epsilon_\tau^2
    \end{array}\right) \; .
  \label{mnu}
\end{eqnarray}
In deriving this expression,  we     have     assumed     that
$\kappa_{1,2}\sqrt{|\kappa_N|}\ll  \epsilon_l$, where 
\begin{eqnarray}
  \kappa_N \ \equiv \ \frac{1}{8\pi^2}\ln
  \left(\frac{\mu_X}{m_N}\right)
  \left[2\kappa_1\kappa_2\sin(\gamma_1+\gamma_2)
    +i(\kappa_2^2-\kappa_1^2)\right]. 
\end{eqnarray}
The light  neutrino mass
matrix in \eqref{mnu} is diagonalized by the usual PMNS mixing matrix
\begin{eqnarray}
  \bm M_\nu \ = \ U_{\rm PMNS}\ {\rm diag}(m_{\nu_1},m_{\nu_2},m_{\nu_3}) \
  U_{\rm PMNS}^{\sf T} \; ,
  \label{mnu2}
\end{eqnarray}
where the $m_{\nu_i}$'s are the light neutrino mass eigenvalues. As we
will see in Section~\ref{sec:6.1}, for the benchmark points considered
therein, the non-unitarity  of the $3\times 3$ PMNS  mixing matrix due
to  the  light-heavy  neutrino mixing~\cite{Antusch:2006vwa}  is  very
small.  Hence,  we can assume  that $U_{\rm PMNS}$ in  \eqref{mnu2} is
unitary, and  express it  in terms of  the three  experimentally known
light neutrino mixing angles  $\theta_{ij}$, and the yet unconstrained
Dirac phase $\delta$ and Majorana phases $\varphi_{1,2}$:
\begin{align}  
  U_{\rm PMNS} \ = \ \left(\begin{array}{ccc}
      c_{12}c_{13} & s_{12}c_{13} & s_{13}e^{-i\delta}\\
      -s_{12}c_{23}-c_{12}s_{23}s_{13}e^{i\delta} &
      c_{12}c_{23}-s_{12}s_{23}s_{13}e^{i\delta} & s_{23}c_{13}\\ 
      s_{12}s_{23}-c_{12}c_{23}s_{13}e^{i\delta} &
      -c_{12}s_{23}-s_{12}c_{23}s_{13}e^{i\delta} & c_{23}c_{13} 
    \end{array}\right)
  {\rm diag}(e^{i\varphi_1/2},e^{i\varphi_2/2},1)\; , \label{PMNS}
\end{align}
with   $c_{ij}\equiv   \cos\theta_{ij},s_{ij}\equiv  \sin\theta_{ij}$.
Assuming a particular mass hierarchy between the light neutrino masses
$m_{\nu_i}$'s   and   for   given   values  of   the   $   \CP$-phases
$\delta,\varphi_{1,2}$,  we can fully  reconstruct the  light neutrino
mass   matrix  using   \eqref{mnu2}.   Substituting   \eqref{mnu2}  in
\eqref{mnu}, we can determine the following model parameters appearing
in the Yukawa coupling matrix \eqref{yuk}:
\begin{align}
  a^2 \ = \ \frac{2m_N}{v^2\kappa_N}
  \left(M_{\nu,{11}}-\frac{M^2_{\nu,{13}}}{M_{\nu,{33}}}\right)\; , 
\qquad  \qquad 
  b^2 \ = \ \frac{2m_N}{v^2\kappa_N}
  \left(M_{\nu,{22}}-\frac{M^2_{\nu,{23}}}{M_{\nu,{33}}}\right)\; ,
  \nonumber \\
  \epsilon_e^2 \  = \ \frac{2m_N}{v^2}\frac{M^2_{\nu,{13}}}{M_{\nu,{33}}} \; ,
\qquad \qquad 
  \epsilon_\mu^2 \ = \ \frac{2m_N}{v^2}\frac{M^2_{\nu,{23}}}{M_{\nu,{33}}}\; ,
\qquad \qquad 
  \epsilon_\tau^2 \ = \ \frac{2m_N}{v^2}M_{\nu,{33}}\; . 
  \label{esq} 
\end{align}
In  this way, the Yukawa  coupling matrix  \eqref{yuk} in  the RL$_\tau$
model can be  completely fixed in terms of the  heavy neutrino mass scale
$m_N$   and  the   symmetry-breaking  parameters   $\kappa_{1,2}$  and
$\gamma_{1,2}$.  Similar  models can  be  constructed  for RL$_e$  and
RL$_\mu$ scenarios~\cite{Deppisch:2010fr}.

\subsection{Approximate Analytic Solutions}
\label{sec:5.3}

In this section, we will find some analytic solutions of the evolution
equations   in   different   regimes.    In   the   first   part,   we
\emph{qualitatively} study the  role of the heavy-neutrino coherences,
and  the  generation of  lepton  number  asymmetry via  heavy-neutrino
oscillations (see also Section~\ref{sec:4.2}). Here, we will obtain an
approximate analytic solution  for the rate equations with  the 
$\varepsilon$- and  $\varepsilon'$-type $\CP$ asymmetries artificially
switched  off,  and  also  neglecting  the $\gamma'$  effects and charged-lepton off-diagonal number densities,  for  a
simplified case with two heavy neutrinos.  Even
though this is  a rough approximation, this will  allow us to estimate
the  relative  magnitude  of  the  different effects  present  in  the
statistical  evolution of  the system.   In  the second  part, we  will
obtain a \emph{quantitatively} accurate analytic solution for the case
of diagonal  heavy-neutrino number densities  (hence no oscillations),
but  retaining the  full  off-diagonal number-density  matrix for  the
charged  leptons,  thereby  capturing the  charged-lepton  decoherence
effects.   In Section~\ref{sec:6.2},  we will  show that  the analytic
solution  presented   here  reproduces  quite   accurately  the  exact
numerical   solution  of   the  full   charged-lepton   rate  equation
\eqref{eq:evofinal1} in the attractor limit.

\subsection*{Qualitative Estimate of the Asymmetry via Oscillations}
\label{sec:5.3.1}

Let  us  find  the  approximate  solution to  the  simplified  set  of
equations       \eqref{eq:evol_decay2},      \eqref{eq:tl_dN}      and
\eqref{eq:tl_L}, with the tree-level  Yukawa couplings, instead of the
resummed ones, where  the $\varepsilon$- and $\varepsilon'$-type $\CP$
violating   sources  have   been  artificially   switched   off.   For
simplicity, we take the $\mathcal{N}_N  = 2$ case and neglect the charged-lepton off-diagonal number densities.     Promoting    the    vector    $\upeta^N$    in
Section~\ref{sec:2.2} to a matrix in the heavy-neutrino flavour space:
\begin{equation}
  \widehat{\mat{\upeta}}^N \ \equiv \ \frac{\widehat{\mat{\eta}}^N}
  {\eta^N_{\rm eq}} \, - \, \mat{1} \ 
  = \ \frac{1}{\eta^N_{\rm eq}}
  \left( \underline{\widehat{\mat{\eta}}}^N \, 
    + \, \frac{\mat{\delta} \widehat{\mat{\eta}}^N}{2} \right) \, 
  - \, \mat{1} \;,
\label{upeta}
\end{equation}
and combining \eqref{eq:evol_decay2} and \eqref{eq:tl_dN}, we find
\begin{equation}\label{eq:upeta_N}
  \frac{\D{}{\widehat{\mat{\upeta}}^N}}{\D{}{z}} \ 
  = \ \frac{K_1(z)}{K_2(z)} \, \bigg( \mat{1} \; 
  + \; \widehat{\mat{\upeta}}^N \; 
  - \; iz \, \Big[\frac{\widehat{\mat{M}}_N}{\zeta(3)H_N}, \, 
  \widehat{\mat{\upeta}}^N \Big] \, 
  - \, \frac{z}{2} \, \Big\{ {\rm Re}(\widehat{\mathbf{K}}^N), \, 
  \widehat{\mat{\upeta}}^N \Big\} \bigg) \;,
\end{equation}
where the  matrix $\widehat{\maf{K}}^N$  is defined as  the tree-level
generalization of ${\rm K}_\alpha$'s appearing in \eqref{be5}:
\begin{equation}
[\widehat{\mathrm{K}}]_{lm \alpha \beta} \ = \ \frac{1}{\zeta(3) H_N} \, \frac{m_N}{8 \pi} \, \widehat{h}^*_{m \alpha} \widehat{h}_{l \beta} \; , \qquad
  \widehat{\mathrm{K}}^N_{\alpha\beta}  
\ \equiv \ \sum_l \:  [\widehat{\mathrm{K}}]_{ll \alpha \beta} 
\ = \ \frac{1}{\zeta(3)H_N}\frac{m_N}{8\pi}(\widehat{h}^\dag \widehat{h})_{\alpha\beta} \;.
\label{wash-gen}
\end{equation}

In the  strong washout regime,  i.e. for~$[\widehat{\mathrm{K}}^N]_{\alpha
\beta}  \gg 1$,  the system  evolves towards  the  attractor solution,
obtained by setting the RHS of \eqref{eq:upeta_N} to zero:
\begin{equation}\label{eq:upeta_N_attractor}
  i \, \Big[\frac{\widehat{\mat{M}}_N}
  {\zeta(3)H_N}, \, \widehat{\mat{\upeta}}^N \Big] \, 
  + \, \frac{1}{2} \, \Big\{ {\rm Re}(\widehat{\mathbf{K}}^N), \,
  \widehat{\mat{\upeta}}^N \Big\} \ \simeq \ \frac{\mat{1}}{z} \;,
\end{equation}
where  we   have  neglected  $\widehat{\mat{\upeta}}^N$   compared  to
$\mat{1}$.   From \eqref{eq:upeta_N_attractor}, it  is clear  that all
the elements  of $\widehat{\mat{\upeta}}^N$ will have  the usual $1/z$
behaviour,     as     expected      in     the     attractor     limit
[cf.~\eqref{eq:N_anal_diag}].   The exact  numerical  solution of  the
fully            flavour-covariant            rate           equations
\eqref{eq:evofinal2}--\eqref{eq:evofinal1}     also    exhibit    this
behaviour,   as  shown   explicitly   in  Section~\ref{sec:6.2}   (see
Figure~\ref{figN}).

To compute the charged-lepton asymmetry we are interested in the value
of  $[\delta  \widehat{\eta}^N]_{12}  =  2  \eta^N_{\rm eq}  \,  i  \,
\mathrm{Im}([\widehat{\upeta}^N]_{12})$   (see   the   discussion   in
Section~\ref{sec:4.2}).  From \eqref{eq:upeta_N_attractor}, we get
\begin{equation}
  \label{eq:upeta_N_attr_sol}
  \mathrm{Im}([\widehat{\upeta}^N]_{12}) \ \simeq \ 
  \frac{\zeta(3) H_N}{z} \; 
  \frac{\mathrm{Re}([\widehat{\Gamma}_N^{(0)}]_{12})}
  {[\widehat{\Gamma}_N^{(0)}]_{11}\,[\widehat{\Gamma}_N^{(0)}]_{22}} \; 
  \frac{\Delta m_N\widetilde{\Gamma}_N}
  {\Delta m_N^2 \; 
    + \; \frac{\widetilde{\Gamma}_N^2 \,
        {\rm det}[{\rm Re}(\widehat{\mat{\Gamma}}_N^{(0)})]}
    {[\widehat{\Gamma}_N^{(0)}]_{11}\,
      [\widehat{\Gamma}_N^{(0)}]_{22}}} \;,
\end{equation}
where      we      have      defined   $\Delta m_N = m_{N_1}-m_{N_2}$ and   
$\widetilde{\Gamma}_N      =
([\widehat{\Gamma}_N^{(0)}]_{11}                                      +
[\widehat{\Gamma}_N^{(0)}]_{22})/2$. Neglecting the charged-lepton off-diagonal coherences, the rate  equation for the lepton
asymmetry \eqref{eq:tl_L} takes the form
\begin{equation}
  \frac{\D{}{[\delta \widehat{\eta}^L}]_{ll}}{\D{}{z}}   \ \supset \ z^3 K_1(z) \, 
  \left[ - \, \frac{1}{3} \,[\widehat{\mathrm{K}}^L]_{ll} \, [\delta \widehat{\eta}^L]_{ll} \; 
    + \; \mathrm{Im}([\widehat{\upeta}^N]_{12}) \, 
    \mathrm{Im}([\widehat{\mathrm{K}}]_{ll12}) \right] \;,
\label{5.20}
\end{equation}
where the index $l$ is not summed over and $[\widehat{\rm K}^L]_{lm} \equiv  \sum_\alpha \: [\widehat{\mathrm{K}}]_{lm \alpha \alpha}  =  (m_N/8\pi)(\widehat{h} \widehat{h}^\dag)_{lm}/(\zeta(3) H_N)$ [cf.~\eqref{wash-gen}]. 
The attractor solution is obtained  by setting the RHS of \eqref{5.20}
to  zero:
\begin{align}\label{eq:delta_eta_L_osc_anal}
  \delta \widehat{\eta}^L \ \supset \ \delta \widehat{\eta}^L_{\rm osc} \ &\simeq \ 
\frac{3}{2z} \sum_l \, \frac{1}{[\widehat{\mathrm{K}}^L]_{ll}} \; 
  \frac{2 \, \mathrm{Im}\big(\widehat{h}^*_{l1} \widehat{h}_{l2} \big)\mathrm{Re} \big[(\widehat{h}^\dagger \widehat{h})_{12}\big]}{(\widehat{h}^\dagger \widehat{h})_{11} \, 
    (\widehat{h}^\dagger \widehat{h})_{22}} \; \notag\\
    & \qquad  \times \; \frac{2 \, (m_{N_1}^2 - m_{N_2}^2) \, m_N \, 
    \widetilde{\Gamma}_N}{(m_{N_1}^2 - m_{N_2}^2)^2 \; 
    + \; \frac{4 \, m_N^2 \widetilde{\Gamma}_N^2 \, {\rm det}[{\rm Re}(\widehat{\mat{h}}^\dagger 
\widehat{\mat{h}})]}{(\widehat{h}^\dagger \widehat{h})_{11} \, 
    (\widehat{h}^\dagger \widehat{h})_{22}}} \; ,
\end{align}
which is valid only for $|\Delta m_N| \ll m_N$. Notice that in the single charged-lepton flavour limit, we have ${\rm det}(\widehat{\mat{h}}^\dagger 
\widehat{\mat{h}})=0$, from which it follows that ${\rm det}[{\rm Re}(\widehat{\mat{h}}^\dagger 
\widehat{\mat{h}})]={\rm Im}[(\widehat{h}^\dagger \widehat{h})_{12}]^2$ in the denominator of \eqref{eq:delta_eta_L_osc_anal}. 

Comparing~\eqref{eq:delta_eta_L_osc_anal}  
 with the lepton asymmetry due to mixing effects [cf.~\eqref{eq:anal_diag}], 
we  see that the total lepton  asymmetry due to
heavy-neutrino oscillations around  $z\sim 1$ is of the  same sign and
of the same order in magnitude,  as compared to that obtained from the
standard  $\varepsilon$-type  $\CP$  asymmetry due  to  heavy-neutrino
mixing.  Even though some of the approximations leading to this result
may    not    be   realistic    in    certain    cases,   we    expect
\eqref{eq:delta_eta_L_osc_anal}  to be  qualitatively correct,  and in
Section~\ref{sec:6.2}   we   will   numerically   verify   this   (see
Figures~\ref{fig5}--\ref{fig7}) for the $\mathrm{RL}_\tau$ model under
consideration.

\subsection*{Analytic Results for the Charged Lepton
  Decoherence Effect} \label{sec:5.3.2}

We  will   now  obtain  the   attractor  analytic  solution   for  the
charged-lepton  asymmetry neglecting  the  heavy-neutrino off-diagonal
number densities  and performing a number of  approximations valid for
the  $\mathrm{RL}_\tau$ scenario  discussed  in Section~\ref{sec:5.2}.
Neglecting  the  heavy-neutrino  coherences and the       sub-dominant
$\widehat{\mat{\gamma}}^{L\Phi}_{L^{\tilde{c}}   \Phi^{\tilde{c}}}   -
\widehat{\mat{\gamma}}^{L\Phi}_{L  \Phi}$ term,  but retaining  the  full
flavour structure for  the charged leptons, the rate  equation for the
asymmetry \eqref{eq:evofinal1}  can be written in  the would-be mass
eigenbasis for charged-leptons as follows: 
\begin{align}\label{eq:eta_L_evol_simpl}
  \frac{\D{}{}}{\D{}{z}} [\delta \widehat{\eta}^L]_{lm} \ & = \ 
  \frac{z^3 K_1(z)}{2} \bigg(\sum_\alpha \: [\widehat{\upeta}^N]_{\alpha \alpha} \,
  [\delta \widehat{\mathrm{K}}^N_{L \Phi}]_{l m \alpha \alpha} \
  - \ \frac{1}{3} \, \Big\{\delta \widehat{\eta}^{L} , \,
  \widehat{\mathrm{K}}^{\rm eff}\Big\}_{lm} \notag\\
  \ & \qquad \qquad - \; \frac{2}{3} \, 
  \Big\{\delta \widehat{\eta}^L, \, 
  \widehat{\mathrm{K}}_{\rm dec } \Big\}_{l m} \;
  +\; [\delta \widehat{\mathrm{K}}_{\rm dec}^{\rm back}]_{ l m}  \bigg) \;,
\end{align}
where    the  various effective
$\mathrm{K}$-factors are defined as
\begin{align}
  \widehat{\mathbf{K}}^{\rm eff} \ & = \  \upkappa  
  \Big( \widehat{\mat{\gamma}}^{L\Phi}_{L^{\tilde{c}} \Phi^{\tilde{c}}} 
  + \widehat{\mat{\gamma}}^{L\Phi}_{L \Phi} \Big)\;, \, &
  \mat{\delta}\widehat{\maf{K}}^N_{L\Phi} \ & = \ \upkappa  
  \mat{\delta} \widehat{\mat{\gamma}}^N_{L\Phi} \; , \nonumber \\
  \widehat{\mathbf{K}}_{\rm dec} \ & = \ \upkappa  
  \widehat{\mat{\gamma}}_{\rm dec} \; , \, &
  \mat{\delta} \widehat{\mathbf{K}}_{\rm dec}^{\rm back} \ &
  = \ \upkappa 
  \mat{\delta} \widehat{\mat{\gamma}}_{\rm dec}^{\rm back} \; , 
\end{align}
with  $\upkappa =  \pi^2 z/(\zeta(3)  H_N m_N^3  K_1(z))$.   

The $\CP$
asymmetries in the charged-lepton flavour space can be defined as
\begin{equation}
 \widehat {\varepsilon}_{lm} \ \equiv \ \sum_\alpha 
  \frac{[\delta \widehat{\mathrm{K}}^{N}_{L \Phi}]_{l m \alpha \alpha}}
  {[\widehat{\mathrm{K}}^{N}]_{\alpha \alpha}}\;.
\end{equation}
Notice  that, in  general,  this  is a  tensor  in the  charged-lepton
flavour space, even though here we are working in the would-be mass eigenbasis for charged leptons. In the 2 heavy-neutrino mixing case, it can be approximated as
\begin{equation}
  \widehat{\varepsilon}_{lm} \ \approx \ \sum_{\alpha \neq \beta} \; \frac{ - i \, \big(\widehat{h}^*_{m\alpha} \widehat{h}_{l\beta} \: - \, \widehat{h}_{l\alpha} \widehat{h}^*_{m \beta}\big) \; \mathrm{Re}
    \big[(\widehat{h}^\dag \widehat{h})_{\alpha\beta}\big]}
  {(\widehat{h}^\dag \widehat{h})_{\alpha\alpha} \, (\widehat{h}^\dag  \widehat{h})_{\beta\beta}} \; \frac{(m^2_{N_\alpha}-m^2_{N_\beta})
   \, m_{N_\alpha} \Gamma_{N_\beta}^{(0)}}
  {(m^2_{N_\alpha}-m^2_{N_\beta})^2
    \, + \, \big(m_{N_\alpha}\Gamma_{N_\beta}^{(0)} \big)^2}
  \;.
\label{eps-lm}
\end{equation}
This expression is analogous to that of the $\varepsilon$-type $\CP$-asymmetry 
$\varepsilon_{l\alpha}$ (see~\ref{app:cp}) in the quasi-degenerate heavy neutrino limit. 
More precisely, for $\Delta m_N\ll m_{N_\alpha}$, we have the relation $\widehat{\varepsilon}_{ll} = \sum_\alpha \varepsilon_{l\alpha}$, where $\varepsilon_{l\alpha}$ is given in~\eqref{eps22}.  

Using the analytic solution  for the diagonal heavy-neutrino evolution
equation  \eqref{eq:N_anal_diag} $[\widehat{\upeta}^N]_{\alpha \alpha}
\simeq 1/([\widehat{\mathrm{K}}^{N}]_{\alpha \alpha}  z)$, we find the
attractor solution in the strong  washout regime by setting the RHS of
\eqref{eq:eta_L_evol_simpl} to zero, thus obtaining
\begin{equation}
  \frac{1}{3} \, \Big\{ \mat{\delta} \widehat{\mat{\eta}}^{L} , \,
  \widehat{\mathbf{K}}^{\rm eff} + 2 \,\widehat{\mathbf{K}}_{\rm dec}
  \Big\} \;
  -\; \mat{\delta} \widehat{\mathbf{K}}_{\rm dec}^{\rm back} \ 
  \simeq \ \frac{\widehat{\mat{\varepsilon}}}{z} \; ,
  \label{attractor_lm}
\end{equation}
In  the  RL$_\tau$   model  discussed  in  Section~\ref{sec:5.2},  the
dominant  contribution to the  total lepton  asymmetry comes  from the
$\tau$-sector    involving   $[\delta    \widehat{\eta}^L]_{k   \tau}$
(with $k=e,\mu,\tau$)  for which the  third column  of \eqref{attractor_lm}
provides a  closed set of  equations.  Imposing  the   detailed  
balance   condition \eqref{eq:det_bal}, the third column of
\eqref{attractor_lm} can be explicitly written as
\begin{align}
  \label{eq:eta_L_attr_1}
  \left(
    \begin{array}{c}
      [\delta \widehat{\eta}^L]_{e k} 
      [\widehat{\mathrm{K}}^{\rm eff}]_{k \tau} \,
      +\, [\widehat{\mathrm{K}}^{\rm eff}]_{e k} 
      [\delta \widehat{\eta}^L]_{k \tau} \;
      +\; 2  \big([\widehat{\mathrm{K}}_{\rm dec}]_{e e} 
      + [\widehat{\mathrm{K}}_{\rm dec}]_{\tau \tau}\big) 
      [\delta \widehat{\eta}^L]_{e \tau} \\[0pt]
      [\delta \widehat{\eta}^L]_{\mu k} 
      [\widehat{\mathrm{K}}^{\rm eff}]_{k \tau} \,
      +\, [\widehat{\mathrm{K}}^{\rm eff}]_{\mu k} 
      [\delta \widehat{\eta}^L]_{k \tau} \;
      +\; 2  \big([\widehat{\mathrm{K}}_{\rm dec}]_{\mu \mu} 
      + [\widehat{\mathrm{K}}_{\rm dec}]_{\tau \tau}\big)
      [\delta \widehat{\eta}^L]_{\mu \tau} \\[0pt]
       [\delta \widehat{\eta}^L]_{\tau k} 
      [\widehat{\mathrm{K}}^{\rm eff}]_{k \tau} \,
      +\, [\widehat{\mathrm{K}}^{\rm eff}]_{\tau k}
      [\delta \widehat{\eta}^L]_{k \tau}
      \end{array} \right)
      \ \simeq \ 
  \frac{3}{z} \begin{pmatrix}
    \widehat{\varepsilon}_{e \tau} \\
    \widehat{\varepsilon}_{\mu \tau} \\
    \widehat{\varepsilon}_{\tau \tau} 
  \end{pmatrix} \; .
\end{align}
We  can safely neglect $[\widehat{\mathrm{K}}_{\rm
dec}]_{e  e}$ and  $[\widehat{\mathrm{K}}_{\rm dec}]_{\mu  \mu}$ which
are much  smaller compared to 
\begin{equation}
[\widehat{\mathrm{K}}_{\rm dec}]_{\tau \tau} \ = \ \frac{3}{2 \, H_N z^2 K_1(z)} \; \Gamma_{T,\,\tau} \;,
\end{equation}
with $\Gamma_{T,\,\tau}$ given by \eqref{eq:Gamma_T}. Moreover,   $[\delta
\widehat{\eta}^L]_{\tau \tau}$ is much  larger than the other entries,
whereas  $[\widehat{\mathrm{K}}^{\rm eff}]_{k  \tau}$ (with $k=e,\mu,\tau$)
are   much  smaller   than  the   entries  $[\widehat{\mathrm{K}}^{\rm
eff}]_{ee,   e\mu,   \mu\mu}$. This   allows   us   to   further 
approximate \eqref{eq:eta_L_attr_1} as
\begin{equation}
  \label{eq:eta_L_attr_2}
  \left(
    \begin{array}{c}
      [\widehat{\mathrm{K}}^{\rm eff}]_{e k} 
      [\delta \widehat{\eta}^L]_{k \tau} \;
      +\; 2 \, [\widehat{\mathrm{K}}_{\rm dec}]_{\tau \tau} 
      [\delta \widehat{\eta}^L]_{e \tau} \\[0pt]  
      [\widehat{\mathrm{K}}^{\rm eff}]_{\mu k} 
      [\delta \widehat{\eta}^L]_{k \tau} \;
      +\; 2 \, [\widehat{\mathrm{K}}_{\rm dec}]_{\tau \tau} 
      [\delta \widehat{\eta}^L]_{\mu \tau} \\    
       2 \, \mathrm{Re} \big([\widehat{\mathrm{K}}^{\rm eff}]_{\tau k}
      [\delta \widehat{\eta}^L]_{k \tau}\big) 
    \end{array}\right) \ \simeq \ 
  \frac{3}{z} \begin{pmatrix}
    \widehat{\varepsilon}_{e \tau} \\
    \widehat{\varepsilon}_{\mu \tau} \\
    \widehat{\varepsilon}_{\tau \tau} 
  \end{pmatrix} \;.
\end{equation}
Assuming  that  the   imaginary  part  of  $[\widehat{\mathrm{K}}^{\rm
eff}]_{\tau k}  [\delta \widehat{\eta}^L]_{k \tau}$  is small compared
to its  real part,  \eqref{eq:eta_L_attr_2} has the  form of  a closed
linear system of equations for $[\delta \widehat{\eta}^L]_{k \tau}$:
\begin{equation}
  \label{eq:eta_L_attr_3}
  \left( \widehat{\mathbf{K}}^{\rm eff} \; 
    + \; 2 \,  \mathrm {diag} \left\{
      [\widehat{\mathrm{K}}_{\rm dec}]_{\tau \tau},\ 
      [\widehat{\mathrm{K}}_{\rm dec}]_{\tau \tau}, \ 0 \right\} \right)
  \begin{pmatrix}
    [\delta \widehat{\eta}^L]_{e \tau} \\
    [\delta \widehat{\eta}^L]_{\mu \tau} \\
    [\delta \widehat{\eta}^L]_{\tau \tau} 
  \end{pmatrix}
  \ \simeq \ \frac{3}{2 z} \begin{pmatrix}
    2 \widehat{\varepsilon}_{e \tau} \\
    2 \widehat{\varepsilon}_{\mu \tau} \\
    \widehat{\varepsilon}_{\tau \tau} 
  \end{pmatrix}\;.
\end{equation}
Notice  that,  in  the limit  $[\widehat{\mathrm{K}}_{\rm  dec}]_{\tau
\tau}  \to \infty$,  we  recover  the solutions of the  diagonal  rate
equations     in     the
$\mathrm{RL}_\tau$  model,~i.e.
\begin{equation}
[\delta    \widehat{\eta}^L]_{e    \tau} \ = \   [\delta
\widehat{\eta}^L]_{\mu  \tau} \ \to \ 0 \;, \qquad  
[\delta \widehat{\eta}^L]_{\tau
\tau} \  \to \ [\delta \widehat{\eta}^L_{\rm mix}]_{\tau\tau} \simeq  \frac{3}{2z} \frac{\widehat{\varepsilon}_{\tau   \tau}}{[\widehat{\mathrm{K}}^{\rm     eff}]_{\tau  \tau}} \;,
\end{equation}   
as  expected [cf.~\eqref{eq:anal_diag}].
However, we have checked numerically that for a realistic value of the
$\tau$-Yukawa      coupling     in     the      RL$_{\tau}$     model,
$[\widehat{\mathrm{K}}_{\rm  dec}]_{\tau \tau}$ has  to be  treated as
finite,  and  the   discrepancy  from  the  flavour-diagonal  solution
\eqref{eq:anal_diag} can  be of  one order of  magnitude, as  shown in
Section~\ref{sec:6.2}.        In       this      case,       inverting
\eqref{eq:eta_L_attr_3},  we  finally   find  an  approximate  analytic
solution for the $\tau$-lepton asymmetry:
\begin{align}
  \label{eq:eta_L_attr_anal}
  \delta \widehat{\eta}^L \ &\supset \ \delta \widehat{\eta}^L_{\rm mix} \;+\; \delta \widehat{\eta}^L_{\rm dec} \ \simeq \ 
  [\delta \widehat{\eta}^L]_{\tau \tau} \notag\\
  &\simeq \ \frac{3}{2 z} \; \mathrm{Re}\Bigg( 
  \left[\left( \widehat{\mathbf{K}}^{\rm eff} \; 
      + \; 2 \, [\widehat{\mathrm{K}}_{\rm dec}]_{\tau \tau} \, 
      \mathrm{diag}(1,1,0)  \right)^{-1}\right]_{\tau k} 
  (2 - \delta_{k \tau}) \, \widehat{\varepsilon}_{k \tau} \Bigg) \; ,
\end{align}
where we  have only  taken the  real part of  the solution,  since the
imaginary  parts  in the  third  row  of \eqref{eq:eta_L_attr_2}  were
assumed to be small in this derivation.

In  the next  section,  we  will show  that  the approximate  analytic
solution  \eqref{eq:eta_L_attr_anal}  reproduces  the exact  numerical
solution  of the  rate equations  remarkably  well for  the case  with
diagonal heavy-neutrino  number densities, and provides  a fairly good
estimate for the total  lepton number asymmetry predicted by the
fully flavour-covariant rate equations.

\section{Numerical Examples}
\label{sec:6}

In this section,  we present some numerical results  for the evolution
of  the  lepton  asymmetry  governed  by  the  flavour-covariant  rate
equations  given  by \eqref{eq:evofinal2}--\eqref{eq:evofinal1}.   For
definiteness,  we choose  to  work within  a  minimal RL$_\tau$  model
presented in Section~\ref{sec:5.2}. For illustration, we take a set of
neutrino Yukawa couplings satisfying the neutrino oscillation data for
a normal hierarchy of light neutrino masses with the lightest neutrino
mass $m_{\nu_1}=0$.   We use the best-fit values of  the light neutrino
oscillation   parameters   from   a   recent   three-neutrino   global
analysis~\cite{Capozzi:2013csa}:
\begin{align}
  & \Delta m^2_{\rm sol} \ = \  7.54\times 10^{-5}~{\rm eV}^2\;, \qquad \qquad 
  \Delta m^2_{\rm atm} \ = \ 2.44\times 10^{-3}~{\rm eV}^2\;,
  \nonumber\\
  &   \sin^2\theta_{12} \ = \ 0.308 \; , \qquad 
  \sin^2\theta_{23} \ = \ 0.425 \; , \qquad 
  \sin^2\theta_{13} \ = \ 0.0234 \; .
  \label{global}
\end{align}
For definiteness, we choose  the leptonic  $ \CP$ phases  $\delta = 0$,  $\varphi_1 =
\pi$  and $\varphi_2  = 0$,  and  reconstruct the  light neutrino  mass
matrix using the definition \eqref{mnu2}.  From this, we can determine
the parameters $a, b,  \epsilon_{e,\mu,\tau}$ of the RL$_\tau$ model
using  the  relations \eqref{esq}  for  a  given  value of  the  heavy
neutrino  mass  scale  $m_N$,  after  taking  into  account  the  mass
splitting  between the  three heavy  neutrinos due  to the  RG effects
given by  \eqref{deltam}. Note  that for a  given light  neutrino mass
matrix  $\mat{M}_\nu$, the  solutions for  $a$ and  $b$  obtained using
\eqref{esq} are  unique up to a  sign factor, and  the sign discrepancy
could be eliminated only if the  sign of Re($a$) were known.  There is
a  similar  sign  freedom  for $\epsilon_{e,\mu,\tau}$,  but  this  is
irrelevant since it  applies to a whole column  of the Yukawa coupling
matrix $\mat{\delta h}$  [cf.~\eqref{delta_h}] and  can be rotated away.   Thus, the
only  free  parameters   we  have  in  this  model,   apart  from  the
neutrino-sector $ \CP$ phases, are $\kappa_{1,2},\ \gamma_{1,2},\ {\rm
sign}[{\rm Re}(a)]$ and $m_N$.  Below we present some benchmark values
for these free parameters.

\subsection{Benchmark Points}
\label{sec:6.1}

\begin{table}[t]
  \begin{center}
    \begin{tabular}{c|c|c|c}\hline\hline
      Parameters & BP1 & BP2 & BP3\\ \hline\hline
      $m_N$ & 120 GeV & 400 GeV & 5 TeV \\
      $\gamma_1$ & $\pi/4$ & $\pi/3$ & $3\pi/8$ \\
      $\gamma_2$ & 0 & 0 & $\pi/2$ \\
      $\kappa_1$ & $4\times 10^{-5}$ & $2.4\times 10^{-5}$ &
      $2\times 10^{-4}$ \\ 
      $\kappa_2$ & $2\times 10^{-4}$ & $6\times 10^{-5}$ &
      $2\times 10^{-5}$ \\ \hline
      $a$ & $(7.41 - 5.54 \, i)\times 10^{-4}$ &
      $(4.93-2.32 \, i)\times 10^{-3}$ &
      $(4.67+4.33 \, i)\times 10^{-3}$\\
      $b$ & $(1.19-0.89 \, i)\times 10^{-3}$ &
      $(8.04 - 3.79 \, i)\times 10^{-3}$ &
      $(7.53+6.97 \, i)\times 10^{-3}$\\
      $\epsilon_e$ & $3.31\times 10^{-8}$ &
      $5.73\times 10^{-8}$ & $2.14\times 10^{-7}$ \\
      $\epsilon_\mu$ & $2.33\times 10^{-7}$ &
      $4.3\times 10^{-7}$ & $1.5\times 10^{-6}$ \\
      $\epsilon_\tau$ & $3.5\times 10^{-7}$ &
      $6.39\times 10^{-7}$ & $2.26\times 10^{-6}$ \\  
      \hline\hline
    \end{tabular}
  \end{center}
  \caption{{\it The   numerical    values   of   the   free   ($m_N$,
$\gamma_{1,2}$,   $\kappa_{1,2}$)  and   derived
parameters ($a$,  $b$, $\epsilon_{e,\mu,\tau}$), with $\mathrm{Re}(a)>0$,  
in  the RL$_\tau$ model
for three chosen benchmark points.}}
  \label{tab3}
\end{table}  

We choose three benchmark scenarios with the heavy neutrino mass scales
$m_N = 120$ GeV, 400 GeV and  5 TeV, in order to illustrate the 
flavourdynamics of the RL$_\ell$ model in different temperature regimes. 
The other free model parameters are
chosen  such   that  they   satisfy  all  the   relevant  experimental
constraints, as discussed below. The values
of the  free model parameters, viz.  $m_N$, $\gamma_{1,2}$, $\kappa_{1,2}$ 
for  our  benchmark scenarios  are given  in
Table~\ref{tab3}, along with the corresponding values of the 
derived model parameters, viz. $a$, $b$, $\epsilon_{e,\mu,\tau}$, 
with  $\mathrm{Re}(a)>0$.   
The low-energy  lepton flavour violating (LFV) and 
lepton number violating (LNV) observables are  briefly discussed
below for  completeness, and their predicted values  for the benchmark
points  are   shown  in  Table~\ref{tab4},  along   with  the  current
experimental limits.
\subsection*{LFV Observables}
\label{6.1.1}
The mixing between the light and ${\cal N}_N$ heavy Majorana neutrinos 
induces LFV processes such  as $\ell\to \ell'\gamma$~\cite{seesaw1, Petcov:1976ff,    
Marciano:1977wx, Bilenky:1977du, Cheng:1980tp, Lim:1981kv, Langacker:1988up}, 
$\ell\to \ell'\ell_1\bar{\ell}_2$~\cite{Lim:1981kv, Garcia:1991be, Ilakovac:1994kj} and 
$\mu \to e$ conversion in nuclei~\cite{Weinberg:1959,        
Marciano:1977cj, Shanker:1979ap,
Ilakovac:1995km, Alonso:2012ji}, through loops
involving  the heavy neutrinos. In general, the light-heavy neutrino 
mixing is parametrized in terms of an arbitrary $3 \times {\cal N}_N$ matrix $\mat{\xi}$~\cite{Korner:1992zk}, which depends on the Yukawa coupling matrix $\mat{h}$ and the heavy Majorana neutrino mass matrix $\mat{M}_N$. In the mass eigenbasis, and to leading order in $\|\mat{\xi}\|$, the mixing is given by\footnote{An all-order expression for the mixing in terms of $\mat{\xi}$ may be found in~\cite{Pilaftsis:2008qt, Dev:2012sg, Dev:2012bd}. Some approximate seesaw expressions are studied in~\cite{Grimus:2000vj, Hettmansperger:2011bt}.}  
\begin{eqnarray}
B_{l\alpha} \ \simeq \ \xi_{l\alpha} \ \equiv \ \frac{v}{\sqrt 2}\: \widehat{h}_{l\alpha} \: \widehat{M}_{N_\alpha}^{-1} \; ,
\label{mixing}
\end{eqnarray}  
which governs the rare LFV decay rates, as discussed below.   

The branching  ratio for  the $\mu\to
e\gamma$   process   is   given   by~\cite{Cheng:1980tp}
\begin{eqnarray}
  {\rm BR}(\mu\to e\gamma) \ = \ \frac{\alpha_w^3 s_w^2}
  {256\pi^2}\: \frac{m_\mu^4}{M_W^4}\: \frac{m_\mu}{\Gamma_\mu} \:
  \left|G_\gamma^{\mu e}\right|^2 \; ,
\label{brmue}
\end{eqnarray}
where $m_\mu$ and $\Gamma_\mu$ are  respectively the mass and width of
the muon,  $s_w\equiv \sin\theta_w$ is the weak  mixing parameter, and
$\alpha_w\equiv  g_w^2/(4\pi)$  is  the  weak coupling  strength,  all
evaluated at the weak scale $M_Z$.  The form factor $G_\gamma^{\mu e}$
is  defined in~\ref{app:loop}.  The other  kinematically  allowed rare
decay modes  of this type,  namely, $\tau\to \mu \gamma$  and $\tau\to
e\gamma$, can  be defined  similar to \eqref{brmue},  in terms  of the
mass  and width of  the $\tau$-lepton.  The branching ratio predictions  
of these rare  decay modes for our  chosen benchmark points  are given in
Table~\ref{tab4}.   For   comparison,  we   also   give  the   current
experimental upper  limits~\cite{pdg, Adam:2013mnn}. The  MEG limit on
$\mu\to  e\gamma$  branching  ratio~\cite{Adam:2013mnn}  is  the  most
stringent one, and the model prediction for BP2 is within reach of the
upgraded MEG sensitivity~\cite{Baldini:2013ke}.

The   branching  ratio  for   the  $\mu^- \to   e^-e^+e^-$  process   is  given
by~\cite{Ilakovac:1994kj}
\begin{align}
  {\rm BR}(\mu\to eee) \  =  & \   \frac{\alpha_w^4}{24576 \pi^3}\:
  \frac{m_\mu^4}{M_W^4} \: \frac{m_\mu}{\Gamma_\mu}
  \left[2 \left|\frac{1}{2}F_{\rm Box}^{\mu eee}
      +F_Z^{\mu e}-2s_w^2\left(F_Z^{\mu e}-F_\gamma^{\mu e}\right)
    \right|^2 \right. \nonumber \\
  \ & \left. + \ 4s_w^4\left|F_Z^{\mu e}-F_\gamma^{\mu e}\right|^2 
    \; + \; 16s_w^2 \; {\rm Re}\left[\left(F_Z^{\mu e}
        +\frac{1}{2}F_{\rm Box}^{\mu eee}\right)
      \left(G_\gamma^{\mu e}\right)^*\right] \right. \label{brmu3e}\\
  \ & \left.
    - \ 48s_w^4 \; {\rm Re}\left[\left(F_Z^{\mu e}
        -F_\gamma^{\mu e}\right)\left(G_\gamma^{\mu e}\right)^*\right] 
    \; + \; 32 s_w^4\left|G_\gamma^{\mu e}\right|^2
    \left\{\ln\left(\frac{m_\mu^2}{m^2_e}\right)-\frac{11}{4}\right\}
  \right] \; , \nonumber
\end{align}  
where  the various  form  factors are  defined in~\ref{app:loop}.  The
predictions for BR($\mu\to eee$) for the three benchmark points chosen
here are shown  in Table~\ref{tab4} and, for comparison,  we have also
shown the  current experimental  upper limit~\cite{pdg}. We  find that
the  model predictions  are well  within  the current  limit. One  can
similarly    define    the    LFV    decay   rates    involving    the
$\tau$-lepton~\cite{Ilakovac:1994kj};  however,  the numerical  values
for these rates  turn out to be several orders  of magnitude smaller than
the current  experimental limits and hence we  do not show  them in
Table~\ref{tab4}.

The $\mu\to e$ conversion rate in  an atomic nucleus $_Z^A X$ is given
by~\cite{Alonso:2012ji}
\begin{eqnarray}
  R_{\mu\to e} \ = \ \frac{2G_F^2\alpha_w^2m_\mu^5}
  {16\pi^2\Gamma_{\rm capt}} \ \left|4V^{(p)} 
    \left(2\tilde{F}_u^{\mu e} + \tilde{F}_d^{\mu e}\right) 
    \; + \; 4V^{(n)}\left(\tilde{F}_u^{\mu e} 
      + 2\tilde{F}_d^{\mu e}\right) \; 
    + \; \frac{s_w^2}{2e}G_\gamma^{\mu e}D\right|^2 \; , 
  \label{ratemue}
\end{eqnarray}
where $e=g_ws_w$ is the magnitude  of the electron charge, 
$\Gamma_{\rm capt}$ is the nuclear capture  rate and $V^{(p), (n)},~D$ 
are various nuclear form factors, whose numerical values for some typical nuclei 
of  interest are  given in~\ref{app:loop} [cf. Table~\ref{tabn}].  
The  electroweak form  factors
$\tilde{F}_q^{\mu e}$ $(q=u,d)$ in \eqref{ratemue} are defined as
\begin{eqnarray}
  \tilde{F}_q^{\mu e} \ = \ Q_qs_w^2F_\gamma^{\mu e} \;
  + \; \left(\frac{I^q_{3L}}{2}-Q_qs_w^2\right)F_Z^{\mu e} \;
  + \; \frac{1}{4}F_{\rm Box}^{\mu e qq} \; ,
\end{eqnarray}
where  $Q_q$  is   the  electric  charge  of  the quark $q$  in  units  of  $e$
($Q_u=2/3,~Q_d=-1/3$), $I^q_{3L}$  is the third component  of the weak
isospin  ($I^u_{3L}=1/2,~I^d_{3L}=-1/2$),   and  the  individual  form
factors  $F_\gamma^{\mu e},~F_Z^{\mu  e},~F_{\rm  Box}^{\mu eqq}$  are
defined  in~\ref{app:loop}. The  predictions for the $\mu\to e$  conversion 
rate for the
three  benchmark points  are given  in Table~\ref{tab4}  for certain 
isotopes of titanium,
gold and lead nuclei, along  with their experimental upper limits from
SINDRUM-II~\cite{mueTi,  mueAu, muePb}.  It  is worth  mentioning here
that the  next generation experiments, such  as COMET~\cite{comet} and
Mu2e~\cite{mu2e} have planned sensitivities around $10^{-16}$, which could 
easily test the first two benchmark points. 
The  distant  future  proposal  PRISM/PRIME~\cite{prism}  
could  probe $\mu\to e$ conversion rates down to $10^{-18}$, thus 
testing the third benchmark point as well. 

\begin{table}[t]
  \begin{center}
    \begin{tabular}{c|c|c|c|c}\hline\hline
      Low-energy &  &  &  & Experimental\\ 
      observables & BP1 & BP2 & BP3 & Limit \\ \hline\hline
      BR($\mu\to e\gamma$) & $4.5\times 10^{-15}$ & $1.9\times 10^{-13}$
      & $2.3\times 10^{-17}$ & $<5.7\times 10^{-13}$~\cite{Adam:2013mnn} \\
      BR($\tau \to \mu\gamma$) & $1.2\times 10^{-17}$ & $1.6\times 10^{-18}$ 
      & $8.1\times 10^{-22}$ & $<4.4\times 10^{-8}$~\cite{pdg} \\
      BR($\tau \to e \gamma$) & $4.6\times 10^{-18}$ & $5.9\times 10^{-19}$ 
      & $3.1\times 10^{-22}$ & $<3.3\times 10^{-8}$~\cite{pdg} \\ \hline
      BR($\mu \to 3e$) &  $1.5\times 10^{-16}$ & $9.3\times 10^{-15}$ 
      & $4.9\times 10^{-18}$ & $<1.0\times 10^{-12}$~\cite{pdg}\\ \hline
      R$_{\mu\to e}^{\text{Ti}}$ & $2.4\times 10^{-14}$ & $2.9\times 10^{-13}$ 
      & $2.3\times 10^{-20}$ & $< 6.1 \times 10^{-13}$~\cite{mueTi}\\
      R$_{\mu\to e}^{\text{Au}}$ & $3.1\times 10^{-14}$ & $3.2\times 10^{-13}$ 
      & $5.0\times 10^{-18}$ & $< 7.0 \times 10^{-13}$~\cite{mueAu} \\
      R$_{\mu\to e}^{\text{Pb}}$ & $2.3\times 10^{-14}$ & $2.2\times 10^{-13}$ 
      & $4.3\times 10^{-18}$ &  $< 4.6 \times 10^{-11}$~\cite{muePb}\\ \hline
      $|\Omega|_{e\mu} $ & $5.8\times 10^{-6}$ & $1.8\times 10^{-5}$ &
      $1.6\times 10^{-7}$ & $< 7.0 \times 10^{-5}$~\cite{Antusch:2006vwa}
      \\ \hline
      $\langle m \rangle$ [eV] & $3.8\times 10^{-3} $ 
      & $3.8\times 10^{-3}$ & $3.8\times 10^{-3}$ & $<$ (0.11--0.25) ~\cite{kamland}
      \\
      \hline\hline
    \end{tabular}
  \end{center}
\caption{{\it The model predictions for  the low-energy observables for the
three chosen  benchmark points and  their comparison with  the current
experimental limits.}}
\label{tab4}
\end{table} 

Apart  from these  LFV  observables, a  non-zero light-heavy  neutrino
mixing also  leads to a non-unitary  PMNS mixing matrix,  which can be
parametrized as~\cite{Korner:1992zk, Dev:2012sg, Dev:2012bd}
\begin{eqnarray}
  \widetilde{U}_{\rm PMNS} \ = \ \left(\mat{1}  +  \mat{\xi}^* \mat{\xi}^{\sf T} 
\right)^{-1/2} U_{\rm PMNS} 
 \ \equiv \  
\bigg(\mat{1} - \frac{1}{2}\: \mat{\Omega}\bigg)
  U_{\rm PMNS} \, ,
\label{nuty}
\end{eqnarray} 
where $U_{\rm PMNS}$ is the  unitary matrix given by \eqref{PMNS}, and
the  non-unitarity  effects  are  captured  by  the  Hermitian  matrix
$\mat{\Omega}$, which is a function of the light-heavy neutrino mixing 
parameter $\mat{\xi}$ [cf.~\eqref{mixing}].  To leading order  in $\|\mat{\xi}\|^2$,
the non-unitarity parameters are given by $\Omega_{\ell\ell'} \simeq \sum_{\alpha} B^*_{\ell\alpha} B_{\ell'\alpha}$.   
For the  benchmark points  given in  Table~\ref{tab3}, the
predictions for $|\Omega|_{e\mu}$ are shown in Table~\ref{tab4}, along
with the  current experimental limit at  90\% CL from a  global fit of
neutrino oscillation data,  electroweak decays, universality tests and
rare  charged-lepton  decays~\cite{Antusch:2006vwa}.  The  predictions
for  other  elements of  $\mat{\Omega}$  are  much  below the  current
experimental  limits, and  hence, are  not  shown here.   Note that  an
upgraded MEG  limit on BR$(\mu\to e\gamma)$ could  reach a sensitivity
of  $|\Omega|_{e\mu}<10^{-6}$ which includes  the first  two benchmark
points  in   Table~\ref{tab4}.   Since  the   non-unitarity  parameter
$\mat{\Omega}$ is very small for all the benchmark points chosen here,
we use $U_{\rm PMNS}$ [cf.~\eqref{PMNS}] instead  of $\widetilde{U}_{\rm PMNS}$ 
[cf.~\eqref{nuty}] 
as the diagonalizing
matrix in \eqref{mnu2} for our phenomenological purposes.

\subsection*{LNV Observables}
\label{sec:6.1.2}

The Majorana  nature of  the light and  heavy neutrinos in  the type-I
seesaw models violate  lepton number by two units,  which can manifest
in  the   neutrinoless double beta decay ($0\nu\beta\beta$)  process  at  low-energy   (for  a  review, see e.g.~\cite{rode}).  In the minimal  seesaw model,  the $0\nu\beta\beta$
process gets  contributions from diagrams  mediated by both  light and
heavy Majorana neutrinos, and the corresponding half-life is given by
\begin{eqnarray}
  \frac{1}{T_{1/2}^{0\nu}} \ = \ G_{01}^{0\nu}
  \left|{\cal M}_\nu^{0\nu}A_{\nu}
    \; + \; {\cal M}_{N}^{0\nu}A_{N}\right|^2 \; ,
  \label{half}
\end{eqnarray}
where  $G_{01}^{0\nu}$  is  the   decay  phase  space  factor,  ${\cal
M}_{\nu,N}^{0\nu}$'s  are  the  nuclear  matrix  elements  (NMEs)  for
$0\nu\beta\beta$ mediated  by light and  heavy neutrinos respectively,
and the dimensionless parameters $A_{\nu,N}$ are defined as
\begin{eqnarray}
  A_\nu \ = \ \frac{1}{m_e}\sum_i (U_{\rm PMNS})_{ei}^2 \; m_{\nu_i},\qquad 
  A_{N} \ = \ m_p\sum_\alpha \frac{B_{e\alpha}^2}{m_{N_\alpha}}\; ,
  \label{eta_nN}
\end{eqnarray}
where   $m_e$  and  $m_p$   are  the   electron  and   proton  masses,
respectively. For all the  benchmark points given in Table~\ref{tab3},
the heavy neutrino contribution $A_N$ to the half-life \eqref{half}
turns out to be negligible compared to the light neutrino contribution
$A_\nu$, and  hence, we can  ignore the second  term on the  RHS of
\eqref{half}, to rewrite it in the canonical form
\begin{eqnarray}
  \frac{1}{T_{1/2}^{0\nu}} \ = \ G_{01}^{0\nu}
  \left|{\cal M}_\nu^{0\nu}\right|^2
  \frac{\langle m\rangle^2}{m_e^2} \; ,
  \label{half2}
\end{eqnarray}
where   $\langle  m\rangle   \equiv   |\sum_i  (U_{\rm   PMNS})_{ei}^2
m_{\nu_i}|$  is known  as the  effective neutrino  mass. For  a normal
hierarchy  of  neutrino  masses  with  $m_{\nu_1}=0$,  and  using  the
three-neutrino oscillation parameter  values given in \eqref{global},
we obtain $\langle m\rangle=3.8$ meV. For comparison, we note that the
current  90\%  CL  experimental  upper limits  are  $\langle  m\rangle
<$ (0.3--0.9)      eV      from       the      NEMO-3      limit      on
$T_{1/2}^{0\nu}(^{100}$Mo)~\cite{nemo},   $<$ (0.2--0.4)  eV   from  the
GERDA+Heidelberg-Moscow+IGEX         combined         limit         on
$T_{1/2}^{0\nu}(^{76}$Ge)~\cite{gerda},  and $<$  (0.12--0.25)  eV from
the        KamLAND-Zen+EXO-200         combined        limit        on
$T_{1/2}^{0\nu}(^{136}$Xe)~\cite{kamland},  where the range  of limits
is   due   to  the   NME   uncertainties
involved.\footnote{After   taking  into   account   the  updated   NME
uncertainties,  it  was   found~\cite{Dev:2013vxa}  that  the  current
$^{136}$Xe limit on $\langle m \rangle$ is the strongest for any given
NME calculation. Therefore, we have only shown the experimental limit from
$^{136}$Xe in Table~\ref{tab4}.}

Apart from  the low-energy  observables discussed above,  the Majorana
nature of the  heavy neutrinos as well as their  mixing with the light
neutrinos  could manifest simultaneously  via their  `smoking gun' 
signature of same-sign dilepton plus two jets with no missing 
energy~\cite{Keung:1983uu} at the 
LHC~\cite{Pilaftsis:1991ug, Datta:1993nm, Almeida:2000pz, Panella:2001wq, 
Han:2006ip, Aguila:2007em, Aguila:2008cj}. 
Note that even  for quasi-degenerate heavy Majorana neutrinos
(as    occurs    in    the    RL$_{\ell}$    models    discussed    in
Section~\ref{sec:5}),  the LNV  signal can  be sizable  when  the mass
splitting  $\Delta  m_N\equiv |m_{N_\alpha}-m_{N_\beta}|$  is  comparable to  the
average width  $\Gamma_N\equiv  (\Gamma_{N_\alpha}+\Gamma_{N_\beta})/2$~\cite{BLP}.  
Within the minimal seesaw framework, both CMS~\cite{CMS-LL}
and  ATLAS~\cite{ATLAS-LL}  experiments  have  derived limits  on  
the mixing parameters $|B_{l\alpha}|^2$ (for $l=e,\mu$) 
between 0.01--0.1  for  $m_N$ = 100--300  GeV.  Including  infrared
enhancement  effects due to  $t$-channel processes  involving photons,
these limits can be improved (by at least a factor of 5) and extended to 
higher heavy neutrino masses at $\sqrt s=14$
TeV LHC~\cite{Dev:2013wba}. The improved limits will encompass the  
first two benchmark
points in Table~\ref{tab3}. 

\subsection{Results for Lepton Flavour Asymmetries}
\label{sec:6.2}

\begin{figure}[p!]
  \centering
\vspace{-1.5cm}
 \includegraphics[width=15cm]{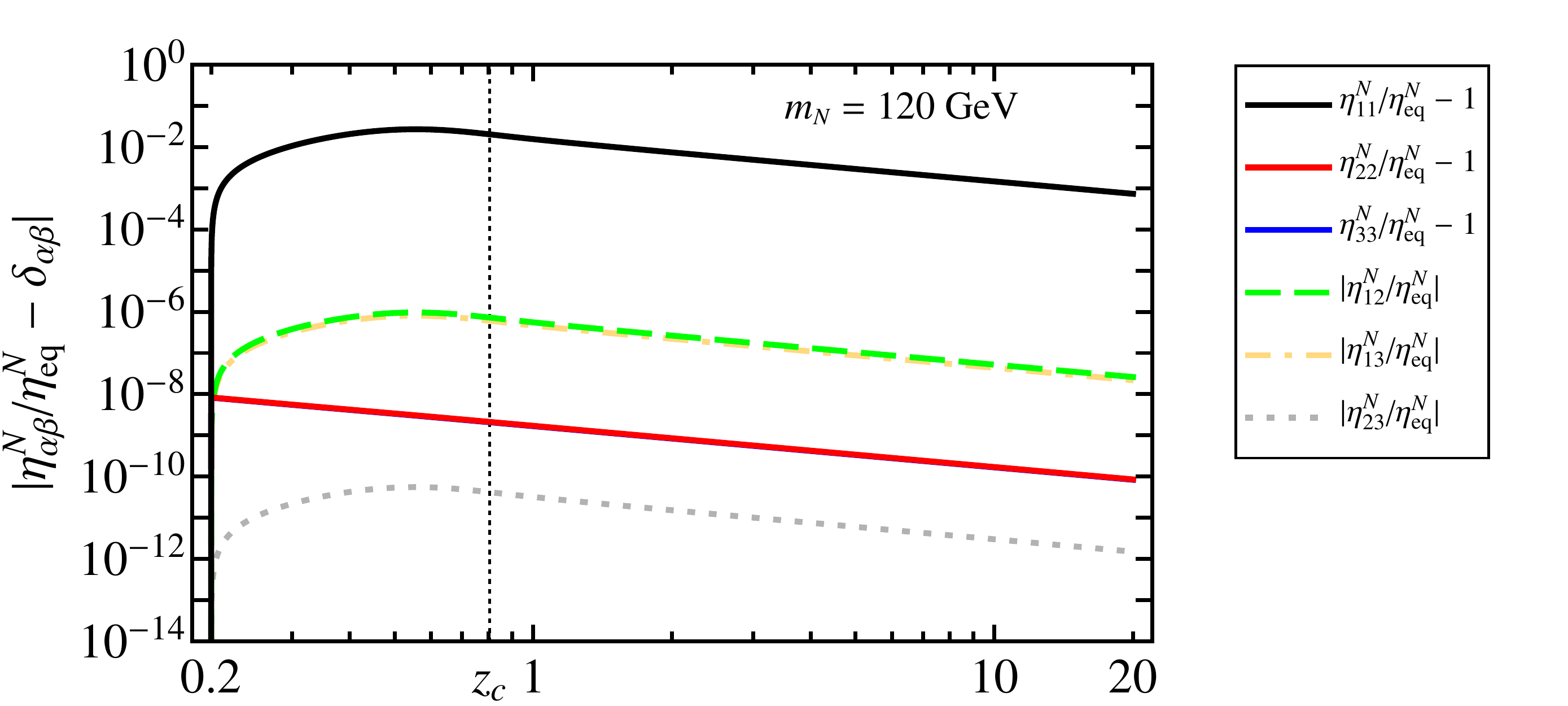}
\\
 \includegraphics[width=15cm]{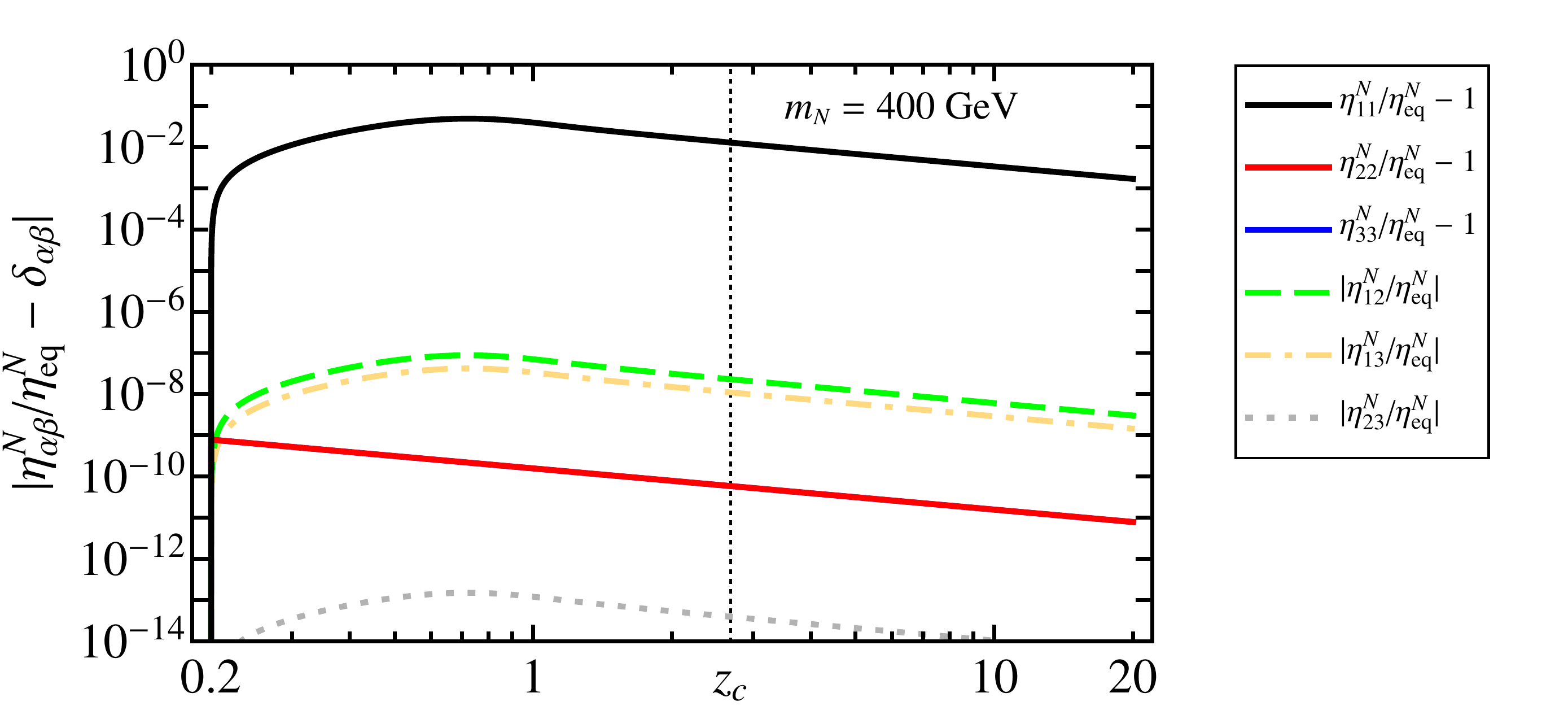} \\
\includegraphics[width=15cm]{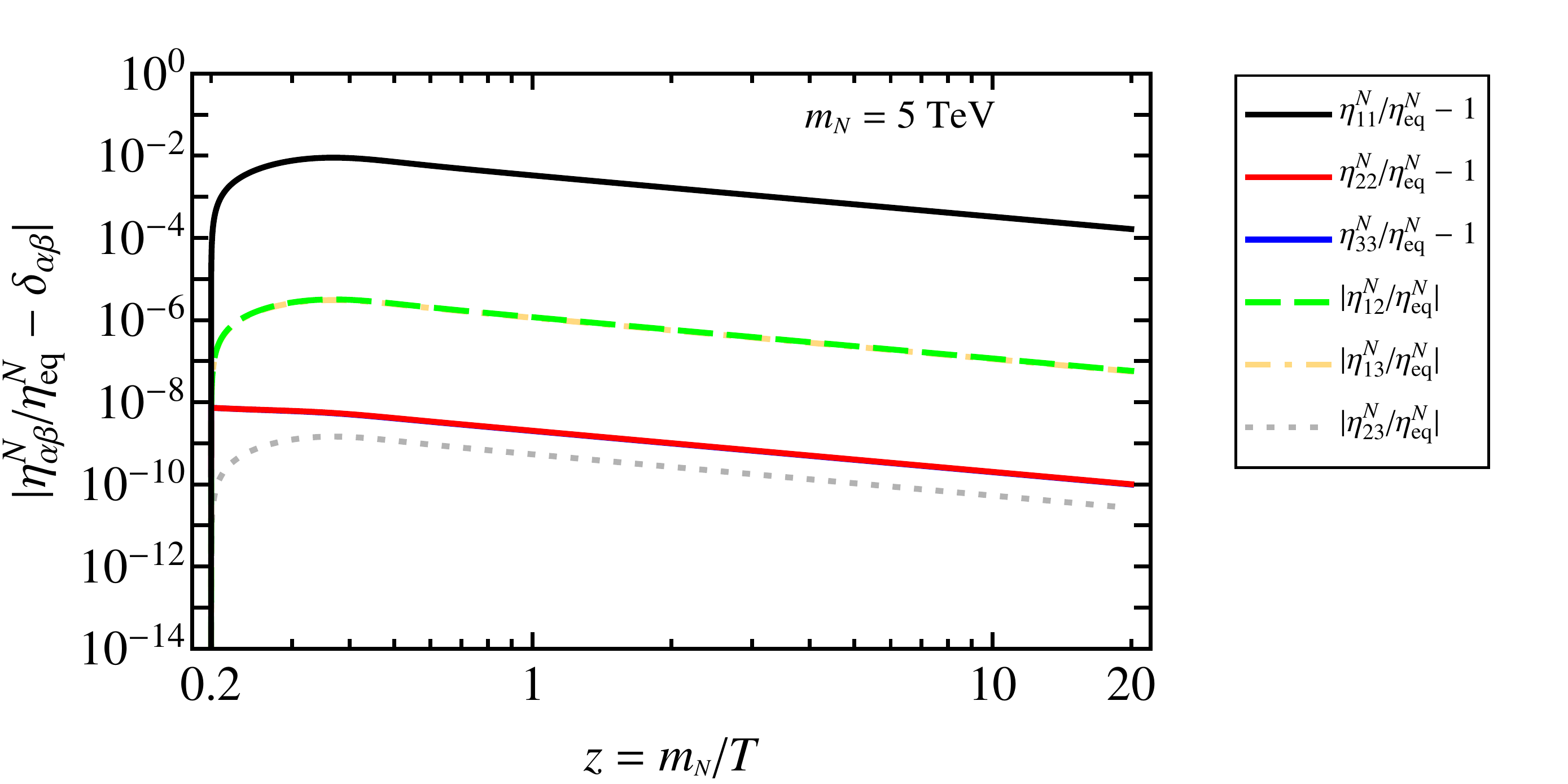} 
  \caption{\emph {The deviation of the heavy-neutrino number densities
$\upeta^N_{\alpha\beta}=\eta^N_{\alpha\beta}/\eta^N_{\mathrm{eq}}-\delta_{\alpha\beta}$
from their equilibrium values for  the three benchmark points given in
Table~\ref{tab3}.   The  different lines  show  the  evolution of  the
diagonal  (solid   lines)  and  off-diagonal   (dashed  lines)  number
densities  in  the fully  flavour-covariant  formalism. The  numerical
values of $\upeta^N_{22}$ and $\upeta^N_{33}$ coincide with each other
in all three cases.  }}
\label{figN}
\end{figure}

Using the  parameter values given in  Table~\ref{tab3}, we numerically
solve        the        flavour-covariant        rate        equations
\eqref{eq:evofinal2}--\eqref{eq:evofinal1}  for the  evolution  of the
charged-lepton and heavy-neutrino  number densities. For definiteness,
in this section we work in  the basis in which the heavy-neutrino mass
matrix  as  well as  the  charged-lepton  Yukawa  coupling matrix  are
diagonal. First,  we discuss the results for  the heavy-neutrino number
densities,  as  shown in  Figure~\ref{figN}  for  the three  benchmark
points  given in Table~\ref{tab3}.   Here we  have chosen  the initial
conditions  with zero lepton  asymmetry, i.e.~$\mat{\delta\eta}^L_{\rm
in}=\mat{0}$,  and   the  heavy  neutrinos   in  thermal  equilibrium,
i.e.~$\mat{\eta}^N_{\rm in} = \eta^N_{\rm  eq}\mat{1}$. As we will see
below, other  choices of initial  conditions lead to  similar results.
The vertical  dotted line indicates the  critical value $z_c=m_N/T_c$,
where    $T_c$    is    the    critical    temperature [cf.~\eqref{Tcritnum}]. 
The  number densities are  shown in terms  of the
deviation         from        their         equilibrium        values:
$\upeta^N_{\alpha\beta}=\eta^{N}_{\alpha\beta}/\eta^N_{\mathrm{eq}}   -
\delta_{\alpha\beta}$  [cf.~\eqref{upeta}].    The  evolution  of  the
diagonal  elements  are   shown  as  solid  lines  and   that  of  the
off-diagonal   elements   as    dashed   lines.    As   discussed   in
Section~\ref{sec:5.3}     [cf.~\eqref{eq:upeta_N_attractor}],     both
diagonal  and  off-diagonal  heavy-neutrino number  densities  rapidly
follow   the   attractor   solution.    Numerically,  the   value   of
$\upeta^N_{11}$ is  several orders of magnitude larger  than the other
elements,  whereas the values  of $\upeta^N_{22}$  and $\upeta^N_{33}$
overlap  for all  the benchmark  points.  In  all three  cases, unlike
$\upeta^N_{23}$,   the  off-diagonal   elements   $\upeta^N_{12}$  and
$\upeta^N_{13}$ are larger than the diagonal elements $\upeta^N_{22}$,
$\upeta^N_{33}$.   Therefore,   the   effect   of   the   off-diagonal
contributions of $\eta^N_{\alpha\beta}$ to the lepton asymmetry cannot
be neglected, as we will illustrate below.

\begin{figure}[p!]
  \centering
\vspace{-1.5cm}
 \includegraphics[width=14.5cm]{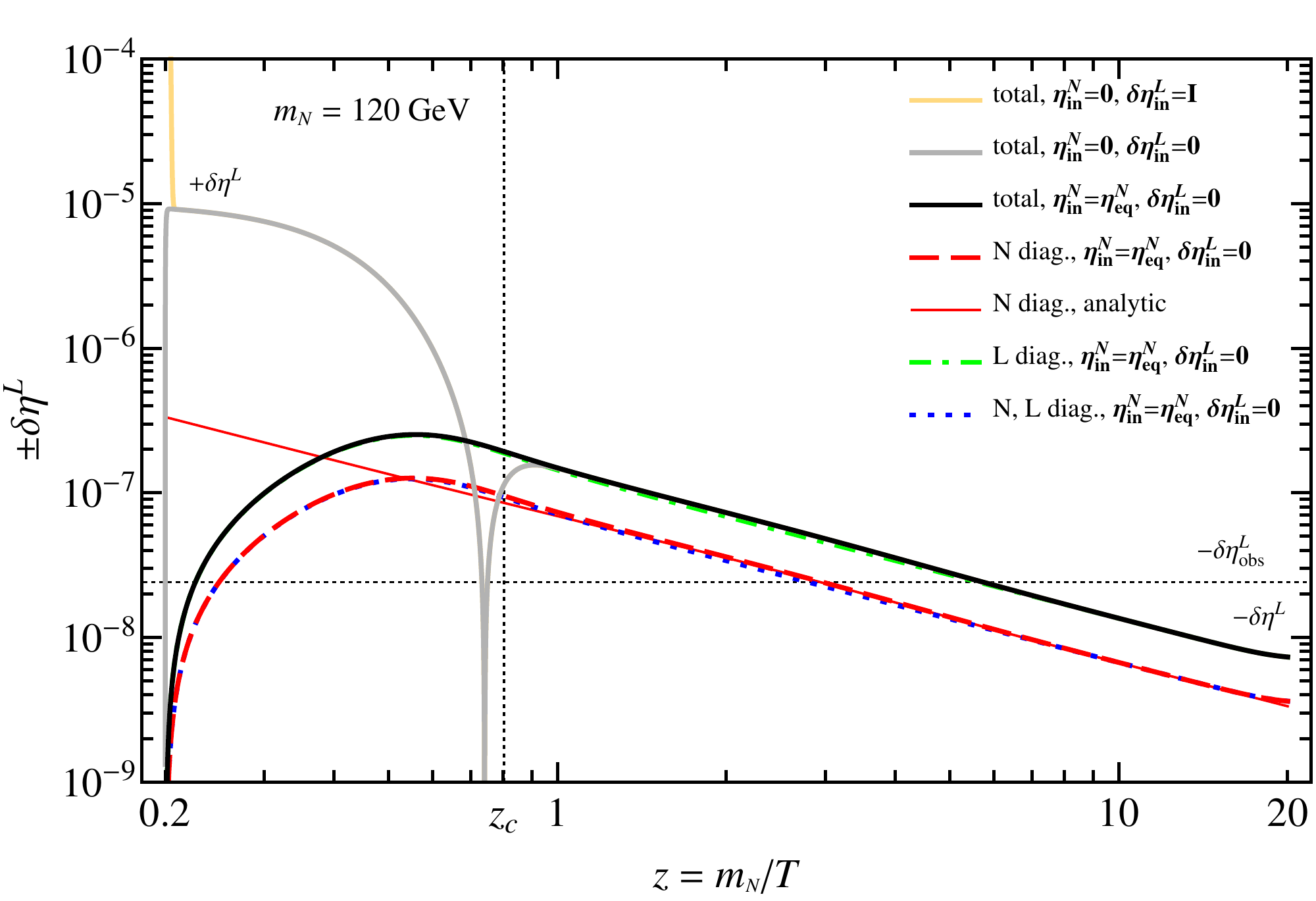}
\\
 \includegraphics[width=14.5cm]{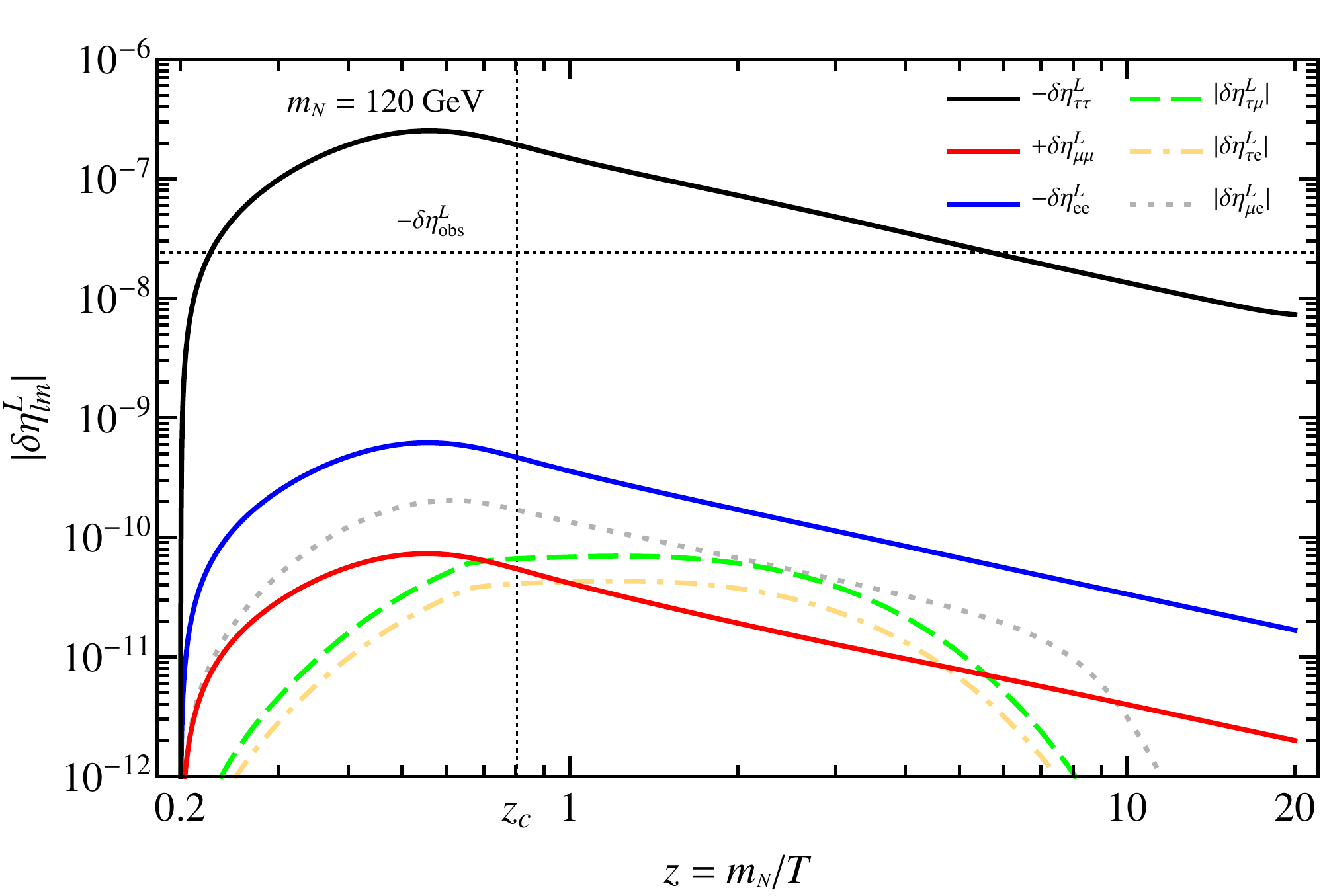}
  \caption{\emph {Lepton  flavour asymmetries as predicted  by the BP1
parameters  given  in  Table~\ref{tab3}.   The  top  panel  shows  the
comparison  between  the  total  asymmetry obtained  using  the  fully
flavour-covariant formalism (thick solid lines, with different initial
conditions) with  those obtained using  the flavour-diagonal formalism
(dashed lines).  Also shown (thin  solid line) is the  analytic result
discussed  in  Section~\ref{sec:5.3}.   The  bottom  panel  shows  the
diagonal (solid lines) and off-diagonal (dashed lines) elements of the
total lepton  number asymmetry  matrix in the  fully flavour-covariant
formalism. For details, see text.}}
\label{fig5}
\end{figure}

\begin{figure}[p!]
  \centering
  \includegraphics[width=14.5cm]{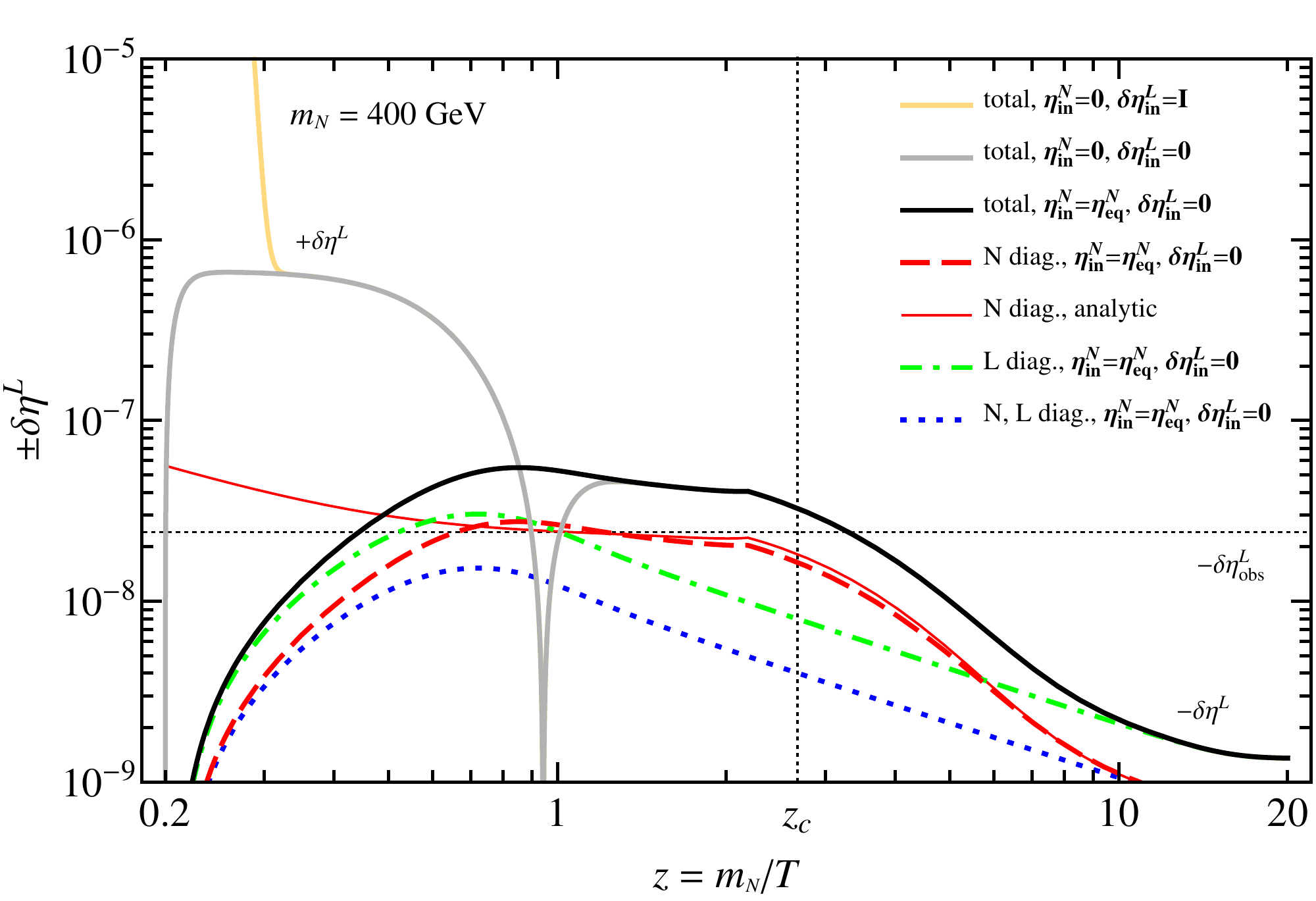}\\
  \includegraphics[width=14.5cm]{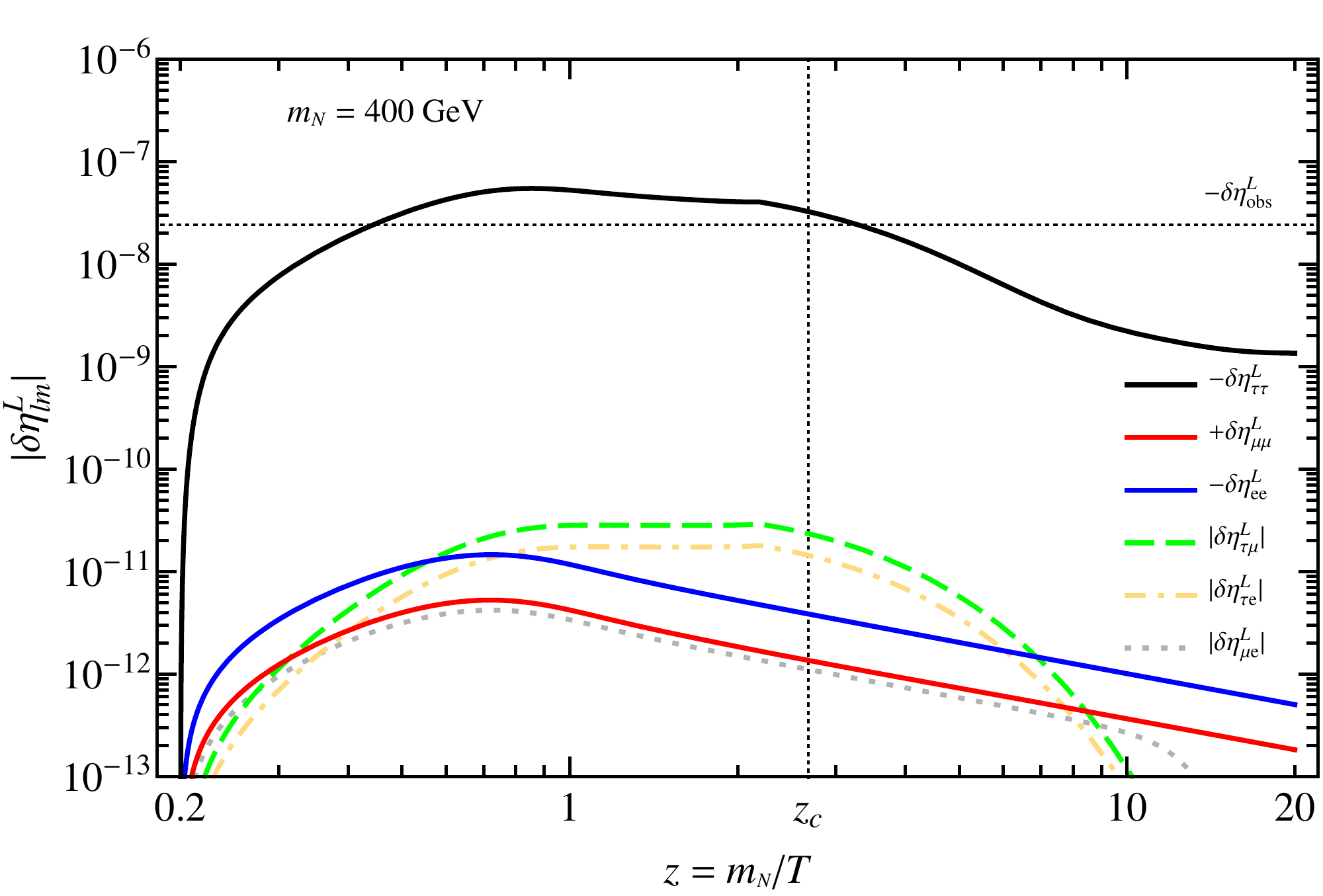}
    \caption{\emph  {Lepton flavour asymmetries as  predicted by
the BP2 minimal RL$_\tau$  model parameters given in Table~\ref{tab3}.
The labels are the same as in Figure~\ref{fig5}. \vspace{2em}}}
  \label{fig6}
\end{figure}

\begin{figure}[p!]
  \centering
  \includegraphics[width=14.5cm]{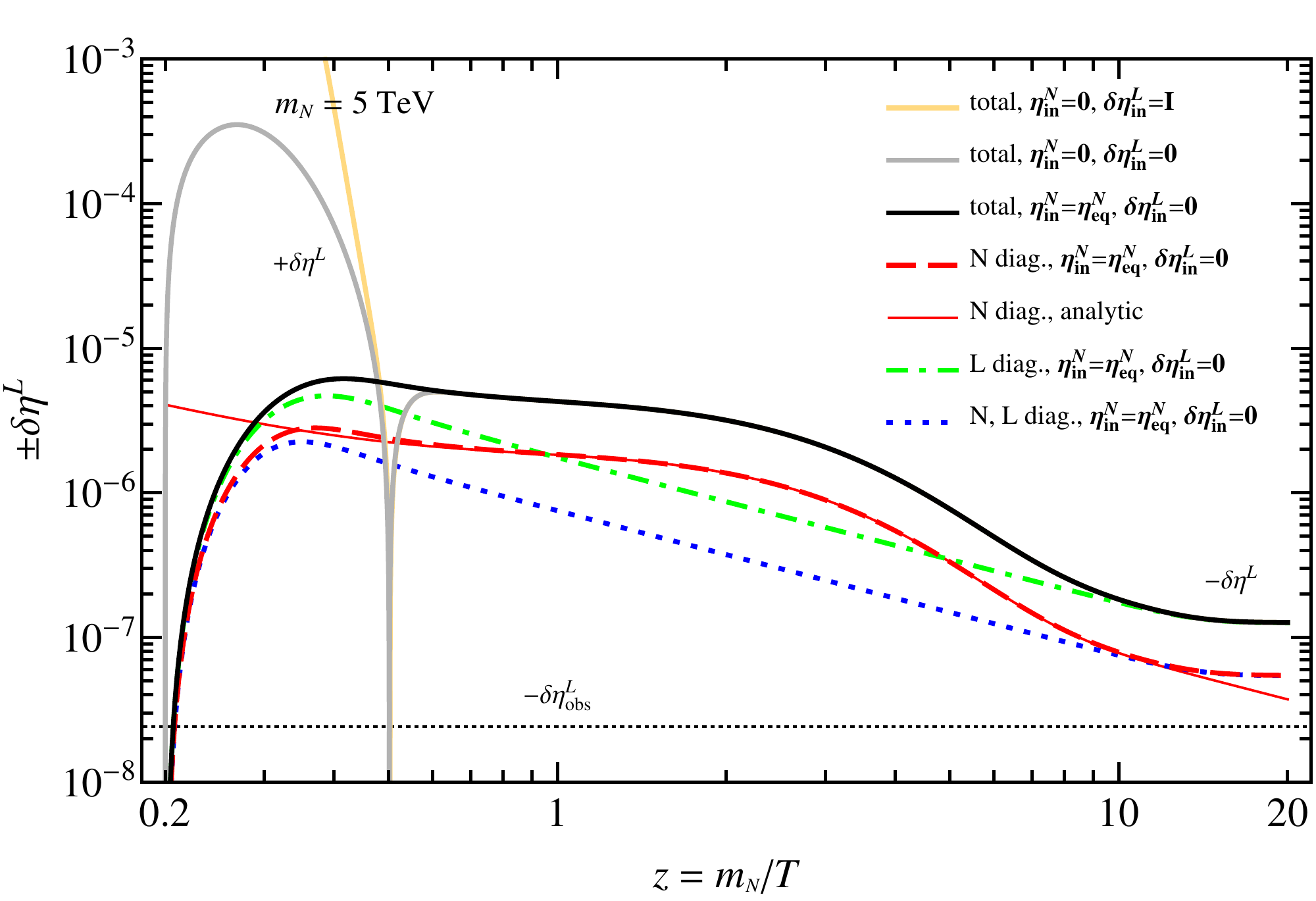}\\
  \includegraphics[width=14.5cm]{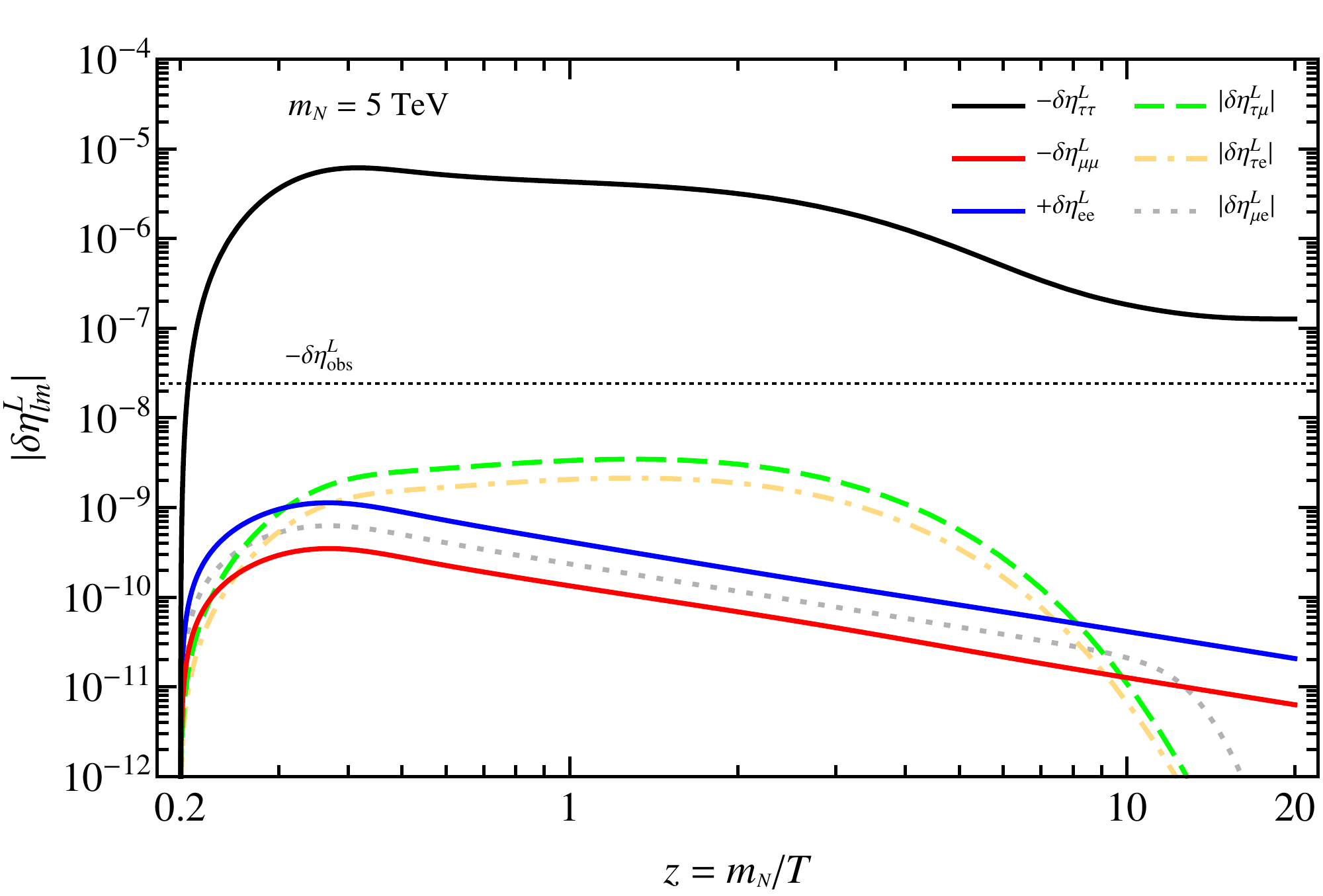}
  \caption{\emph {Lepton flavour asymmetries as predicted by the
BP3 minimal RL$_\tau$ model parameters given in Table~\ref{tab3}.  The
labels are the same as in Figure~\ref{fig5}.\vspace{2em}}}
  \label{fig7}
\end{figure}

Our results  for the asymmetries  in the SM lepton-doublet  sector for
the  three benchmark  points given  in Table~\ref{tab3}  are  shown in
Figures~\ref{fig5}--\ref{fig7}.  In each figure, the horizontal dotted
line shows the  value of $\delta\eta^L$ [cf.~\eqref{eta_obs}] required
to  explain  the  observed  baryon  asymmetry in  the  Universe.   The
conversion of the asymmetry  stops for $z>z_c$ (vertical dotted line),
since the sphaleron  processes are no longer in  action.  As such, the
observed value  for $\delta\eta^L$ should  be compared with  the model
prediction      at      $z=z_c$.       The     top      panels      in
Figures~\ref{fig5}--\ref{fig7} show the  evolution of the total lepton
asymmetry,  $\delta  \eta^L  \equiv  {\rm  Tr}(\mat{\delta  \eta}^L)$,
obtained   using    the   fully   flavour-covariant    rate   equation
\eqref{eq:evofinal1},  for three  different initial  conditions (thick
solid lines):
\begin{enumerate}
\item[(i)]    zero     lepton    asymmetry    $\mat{\delta\eta}^L_{\rm
in}=\mat{0}$,     heavy    neutrinos     in     thermal    equilibrium
$\mat{\eta}^N_{\rm in} = \eta^N_{\rm eq}\mat{1}$ (thick black line);
\item[(ii)]    zero    lepton    asymmetry    $\mat{\delta\eta}^L_{\rm
in}=\mat{0}$,    heavy     neutrinos    strongly    out-of-equilibrium
$\mat{\eta}^N_{\rm in} = \mat{0}$ (thick grey line);
\item[(iii)]  extremely  large  lepton  asymmetry with  opposite  sign
compared  to the  observed one  $\mat{\delta\eta}^L_{\rm in}=\mat{1}$;
heavy neutrinos  strongly out-of-equilibrium $\mat{\eta}^N_{\rm  in} =
\mat{0}$ (thick yellow line).
\end{enumerate} It is clear  that the final lepton asymmetry $\delta
\eta^L (z\gg 1)$ is independent  of the initial conditions, which is a
general consequence of the RL mechanism in the strong washout regime.
Even starting  with extremely large initial  lepton asymmetry and/or
with the wrong  sign, this primordial asymmetry is  rapidly washed out
and  the  final  asymmetry,  with   the  right  sign  as  required  by
\eqref{eta_obs}, is set by the RL mechanism itself for $z \sim 1$.

We may  now compare  the results of  our fully  flavour-covariant rate
equations with their diagonal \cite{Pilaftsis:2003gt,Pilaftsis:2005rv}
and partially  flavour-dependent limits.  To  this end, we show in the
top  panels of  Figures~\ref{fig5}--\ref{fig7}  limiting cases,  where
either  the heavy-neutrino  number density  (red dashed  line)  or the
charged-lepton number  density (green dash-dotted line)  or both (blue
dotted  line) are  treated as  diagonal in  flavour space.   For these
cases, we have  chosen the initial conditions $\mat{\delta\eta}^L_{\rm
in}=\mat{0}$ and $\mat{\eta}^N_{\rm  in} = \eta^N_{\rm eq}\mat{1}$ for
concreteness.   As mentioned  above,  the final  lepton asymmetry  is
insensitive to the initial conditions  in each case.  We also show the
analytic  solution, discussed in  Section~\ref{sec:5.3}, for  the case
with diagonal heavy-neutrino number  densities, which agrees well with
the asymptotic limit of the corresponding exact numerical solution.

From Figures~\ref{fig5}--\ref{fig7},  we see that  the final asymmetry
including all  flavour effects is  significantly enhanced in  the {\it
fully}  flavour-covariant formalism,  as compared  to  the predictions
from  the partially  flavour-dependent  limits.  The  enhanced $  \CP$
asymmetry  in the  flavour-covariant  formalism can  be understood  as
arising predominantly from two physically-distinct phenomena:
\begin{itemize}
\item[(i)]  coherent  oscillations  between  different  heavy-neutrino
flavours,   which  create   an  ${\cal   O}(h^4)$  asymmetry   in  the
charged-lepton   sector  in   the   RL  scenario,   as  discussed   in
Sections~\ref{sec:4.2}   and   \ref{sec:5.3}.    This  leads   to   an
enhancement by a factor of  two over the flavour-diagonal limit in all
the benchmark points shown in Figures~\ref{fig5}--\ref{fig7}.
\item[(ii)] the evolution of  flavour coherences in the charged-lepton
sector,  which   are  generated  through   the  heavy-neutrino  Yukawa
couplings and  destroyed through the  charged-lepton Yukawa couplings,
as discussed in Sections~\ref{sec:4.3} and \ref{sec:5.3}.
\end{itemize} 
The latter  charged-lepton decoherence effects give rise
to the  distinctive `plateau' in  the off-diagonal $\tau e$  and $\tau
\mu$ elements of the lepton asymmetry matrix $\mat{\delta\eta}^{L}$ in
the  RL$_\tau$ model.   This can  be seen  from the  bottom  panels of
Figures~\ref{fig5}--\ref{fig7},    which     show    the    individual
charged-lepton flavour contributions to the total asymmetry.  Since we
are  considering an  RL$_\tau$  model, the  asymmetry produced  around
$z=1$ is dominantly in the $\tau$-flavour ($\delta\eta^L_{\tau\tau}$),
which has  relatively {\it smaller}  couplings to the  heavy neutrinos
[cf.~\eqref{yuk}], and hence, a  smaller washout factor.  On the other
hand, the asymmetries  generated in electron ($\delta\eta^L_{ee}$) and
muon   ($\delta\eta^L_{\mu\mu}$)  flavours,  with   relatively  larger
couplings to the heavy neutrinos, are suppressed due to larger washout
rates.  In addition to the lepton asymmetry in the diagonal 
element $\delta\eta^L_{\tau\tau}$, the coherence effect in the
charged-lepton sector generates an extra asymmetry in the off-diagonal
number  densities   involving the $\tau$-flavour, i.e. 
$\delta\eta^L_{\tau  e}$  and  $\delta\eta^L_{\tau
\mu}$.   These  could  be  large  compared to  those  involving  other
flavours,   i.e.    $\delta\eta^L_{ee}$,   $\delta\eta^L_{e\mu}$   and
$\delta\eta^L_{\mu\mu}$,  depending   on  the  values   of  the  input
parameters.   This effect is more prominent around  $z=1$, since
with increasing $z$ values, the off-diagonal lepton-flavour coherences
decay,  leading   to  a complete  decoherence  of  the   system  to  the
charged-lepton  Yukawa  eigenbasis.   This  explains  the  distinctive
plateau   at   intermediate   $z$   values   in   the   evolution   of
$\delta\eta_{\tau     e}$    and     $\delta\eta_{\tau     \mu}$    in
Figures~\ref{fig5}--\ref{fig7}   (bottom  panels).    This  additional
source contributes to  the total lepton asymmetry 
$\delta\eta^L$, which  can  exhibit a similar feature, depending on the 
model parameters. In the case  where the sphalerons freeze out   
within the `plateau' region, which occurs for $200~{\rm GeV}\lsim m_N\lsim 2~{\rm TeV}$, 
an  additional enhancement of a factor $\sim 5$ 
in the final asymmetry
is  observed,  as  can  be  seen for  BP2  in  Figure~\ref{fig6}  (top
panel). For BP3 with a much higher heavy-neutrino mass scale, 
the coherence effect is already subdued, and the system has completely decohered 
to the charged-lepton Yukawa eigenbasis,  well above the critical temperature, 
thus giving no additional enhancement, 
as shown in Figure~\ref{fig7} (top panel). On the other hand,  for BP1, the neutrino Yukawa  couplings are smaller
than those  in BP2 and BP3,  and hence, the coherence effects are 
not pronounced in the total asymmetry, as can be seen from Figure~\ref{fig5} 
(top panel). 

The impact of the additional enhancements, discussed above, is exemplified 
in BP2 (see Figure~\ref{fig6}), where the total lepton asymmetry obtained in the fully flavour-covariant 
formalism is above the observed value, whereas the predictions obtained in various 
partially flavour-dependent limits all fall below the observed asymmetry. Note that 
the total lepton asymmetries predicted for all of the benchmark points exceed the observed asymmetry. Nevertheless, due to the freedom in the choice of the $\CP$ phases $\gamma_{1,2}$, these benchmark points represent viable choices of model parameters for successful leptogenesis.

Before concluding this section, we note  that there  is no {\it decoherence}  
effect  in the heavy-neutrino  sector,  and hence, 
both diagonal and off-diagonal 
heavy-neutrino number densities decay {\it coherently}, as shown  
in  Figure~\ref{figN}. This is due to the fact that the 
evolution of the heavy-neutrino number densities are  entirely   
governed   by   the heavy-neutrino Yukawa couplings [cf.~\eqref{eq:evofinal2} and  
\eqref{eq:evofinal3}], ignoring sub-dominant collision terms, such as $\Delta L=1$ scattering processes. In the heavy-neutrino mass eigenbasis, the corresponding 
decay rates are not diagonal, which leads to the occurrence of coherences in the 
heavy-neutrino sector. 

\section{Conclusions}
\label{sec:7}

We have  presented a  {\it fully} flavour-covariant formalism  for transport
phenomena, in which we have derived Markovian master equations describing the
time-evolution  of  particle   number  densities in  a
quantum-statistical ensemble with  arbitrary flavour degrees of freedom.  
In particular, we have obtained a flavour-covariant  generalization  of  the
semi-classical flavour-diagonal Boltzmann   equations.

In   order   to  explicitly   demonstrate   the   importance  of  the effects captured 
{\it only} in the flavour-covariant  formalism,  we  have  discussed  a
particular  application  to the  phenomenon  of resonant  
leptogenesis (RL). It is known that the   RL   scenario 
offers  a unique opportunity for testing the connection between the origin of neutrino mass and matter-antimatter asymmetry by the  ongoing LHC experiments as well
as by various low-energy experiments probing lepton flavour  and number 
violation.  For this reason,  
it is essential to  capture all  the flavour  effects due  to the
heavy neutrinos  as well as SM leptons {\it  in a consistent
manner}, in  order to obtain a  more accurate prediction for the baryon
asymmetry in this scenario. As  we have shown in this paper, including
{\it all} flavour off-diagonal effects could  enhance the predicted lepton 
asymmetry as much as one order of magnitude in certain RL models, 
as compared to predictions obtained from partially flavour-dependent 
treatments.  Thus, our flavour-covariant  formalism allows us to access 
an enlarged parameter space of the RL models, which could be tested 
in ongoing and planned experiments at both the high energy and intensity frontiers. 

The main {\it new} results of our fully flavour-covariant formalism for
RL scenarios, as contained in the final rate equations~\eqref{eq:evofinal2}--\eqref{eq:evofinal1},  are the following:

\begin{itemize}
\item[(i)] The  appearance of new  rank-4 tensors in flavour  space in
   transport  equations   (see  Section~\ref{sec:3.4}).    These are 
necessary  to  describe  the  time-evolution  of  the  number  density
matrices   for   leptons    and   heavy   neutrinos   in   a
flavour-covariant manner. One can extend this formalism, by introducing 
even higher rank rate tensors, to describe sub-dominant processes, such as 
$LN\leftrightarrow Le_R$, involving more flavour degrees of freedom. 
The existence  of the tensorial structure in
the rate equations  is firmly supported by an  explicit calculation of
the transition  matrix elements, by virtue of a generalization of the  optical theorem
(see~\ref{app:optical}).  To further  elucidate the consistency of our
treatment,  we   develop a  flavour-covariant
generalization of  the helicity amplitude  technique, applied to 
spinorial  fields in the  presence of time-dependent  and 
spatially-inhomogeneous backgrounds (see~\ref{app:propagator}).

\item[(ii)]  A systematic treatment  of two  intrinsically  quantum
effects, i.e. oscillations between different
heavy  neutrino  flavours  (see  Section~\ref{sec:4.2})  and  quantum
decoherence     between    the     charged-lepton     flavours    (see
Section~\ref{sec:4.3}).  Numerical studies for a particular RL$_\tau$ model 
reveal that these flavour off-diagonal effects could enhance  the total lepton asymmetry 
by up to one order of magnitude, as compared  to  the  flavour-diagonal case  (see
Figures~\ref{fig5}--\ref{fig7}).

\item[(iii)]     The    approximate     analytic     solutions    (see
Section~\ref{sec:5.3}) to  the  fully   flavour-covariant  transport
equations, which  capture  the two relevant flavour effects
discussed  above. Taking this into account 
in the strong washout regime, the total lepton asymmetry at $z\gsim 1$ 
may be estimated by the sum of the contributions from flavour mixing, oscillation and 
decoherence effects:
\begin{eqnarray}
\delta \eta^L_{\rm tot} \ \simeq \ \delta \eta^L_{\rm mix} \: + \: \delta \eta^L_{\rm osc} 
\: + \: \delta \eta^L_{\rm dec} \; ,
\end{eqnarray}
as given by \eqref{eq:anal_diag}, \eqref{eq:delta_eta_L_osc_anal} and 
\eqref{eq:eta_L_attr_anal}, respectively.  The  quantitative  predictions obtained  from  the
analytic solutions for all our  benchmark points agree well with
the exact numerical results obtained from the full flavour-covariant transport 
equations.  The analytic expressions are
presented with the aim to facilitate phenomenological
studies for a given model,  without necessarily having to solve the full
flavour-covariant rate equations. 

\end{itemize}

Aside  from  these  main  results specific  to  the  flavour-covariant
formalism, we  have also given a  geometric and physical understanding of
the  degeneracy  limit in  the  heavy  neutrino  parameter space  (see
Section~\ref{sec:5.1})  for  the  minimal  RL scenario  in  which  the
quasi-degeneracy of the heavy neutrino  masses at low scale can be 
naturally explained as a small deviation from  the $O({\cal N_N})$-symmetric limit at some
high scale through RG effects. We also point out that the role of RG effects in lifting the degeneracies encountered here is reminiscent of the role of time-independent perturbations in the degenerate perturbation theory of ordinary  Quantum Mechanics.

We  also  comment on  the  various  existing  forms of  the self-energy regulator
used to calculate the  $\varepsilon$-type $\CP$-asymmetry 
and  make a  comparative study  in  a simple  toy model,  in order  to
demonstrate their behaviour in certain lepton number conserving limits
(see~\ref{app:cp}).   We  find  that   only  the  regulator  given  by
\eqref{fpu}  gives a  valid  and well-defined  $\CP$-asymmetry in  the
entire parameter space possessing  the correct $L$-conserving limit, whereas
the other  regulators are not  well-defined in certain regions  of the
parameter space.

In conclusion, our flavour-covariant formalism provides a complete and
unified  description of  transport phenomena  in RL  models, capturing
three relevant physical phenomena: (i)~the resonant mixing between the
heavy  neutrino states, (ii)~coherent oscillations  between different
heavy neutrino flavours, and  (iii)~quantum decoherence effects in the
charged-lepton sector.  The  formalism developed  here is
rather general  and may also find applications in  various other 
transport phenomena involving  flavour effects. 

\section*{Acknowledgments}

The work  of P.S.B.D., P.M.  (in part) and  A.P.  is supported  by the
Lancaster-Manchester-Sheffield  Consortium   for  Fundamental  Physics
under  STFC   grant  ST/J000418/1.  P.M.  is also supported in part by the 
IPPP through STFC grant ST/G000905/1 and would like  to
acknowledge  the  conferment  of  visiting researcher  status  at  the
University of Sheffield. 
The work of D.T. has been supported by a fellowship of the
EPS  Faculty of  the  University  of Manchester.  


\appendix

\section{\texorpdfstring{$\CP$}{CP} Asymmetry in the
  \texorpdfstring{$L$}{L}-conserving Limit}
\label{app:cp}

The  analytic results  for the  leptonic $  \CP$-asymmetries  given by
\eqref{cpasy} simplify  considerably in the  two heavy-neutrino mixing
limit  ($\alpha=1,2$).  In this  case,  the $R_{\alpha\beta}$-dependent terms 
in the expression \eqref{resum1} for the effective   Yukawa   
couplings can be set to zero, since they always involve a sum over
more than two heavy neutrinos.  Thus,  in   this   limit,
\eqref{resum1} reduces to
\begin{eqnarray}
  \widehat{\maf{h}}_{l\alpha} \ = \  \widehat{h}_{l\alpha} \: - \: 
i\: \widehat{h}_{l\beta} \: 
  \frac{m_{N_\alpha}\left(m_{N_\alpha}A_{\alpha\beta}
      +m_{N_\beta} A_{\beta\alpha}\right)}
  {m_{N_\alpha}^2-m_{N_\beta}^2+2im^2_{N_\alpha}A_{\beta\beta}}  \qquad 
(\alpha\neq \beta)\; .
  \label{resum11}
\end{eqnarray}
Using~\eqref{resum11} in~\eqref{cpasy} and neglecting higher-order Yukawa couplings  of ${\cal  O}(h_{l\alpha}^4)$ at  the amplitude  
level, we obtain the following expression for the $\varepsilon$-type $ \CP$-asymmetry:\footnote{The Yukawa structure in~\eqref{eps22} agrees with~\cite{Covi:1996wh}, but differs from that given in~\cite{Branco:2009by, Deppisch:2010fr}, which do not have the second term in the numerator on the RHS of~\eqref{eps22}. This additional term can be dropped only after taking the sum over $l$.}  
\begin{eqnarray}
\varepsilon_{l\alpha} \ \approx \ \frac{{\rm Im}\left[\widehat{h}^*_{l\alpha} \widehat{h}_{l\beta}(\widehat{h}^\dag \widehat{h})_{\alpha\beta}\right]+\frac{m_\alpha}{m_\beta}\:{\rm Im}\left[\widehat{h}^*_{l\alpha} \widehat{h}_{l\beta}(\widehat{h}^\dag \widehat{h})_{\beta\alpha}\right]}
  {(\widehat{h}^\dag \widehat{h})_{\alpha\alpha}(\widehat{h}^\dag \widehat{h})_{\beta\beta}}\: f_{\rm reg}\; , 
\label{eps22}
\end{eqnarray}
where $\alpha,\beta=1,2$ $(\alpha\neq \beta)$, and the  self-energy regulator is
given by~\cite{Pilaftsis:1997jf, Pilaftsis:2003gt}
\begin{eqnarray}
  f_{\rm reg} \ = \ \frac{\left(m^2_{N_\alpha}-m^2_{N_\beta}\right)
    m_{N_\alpha}\Gamma_{N_\beta}^{(0)}}
  {\left(m^2_{N_\alpha}-m^2_{N_\beta}\right)^2
    +\left(m_{N_\alpha}\Gamma_{N_\beta}^{(0)}\right)^2}\; .
  \label{fpu}
\end{eqnarray}
Note that  $\widehat{h}$ and  $\Gamma_{N_\beta}^{(0)}$ are the  {\it tree-level}
Yukawa couplings and decay width, respectively.  
In  the  degenerate  heavy   neutrino  mass limit,  $\Delta  m_{N}  \equiv 
(m_{N_1}-m_{N_2})\to 0$, 
the would-be singular behaviour
of the $ \CP$-asymmetry is        regularized         by        the        absorptive        term
$(m_{N_\alpha}\Gamma_{N_\beta}^{(0)})^2$  in the  denominator on
the  RHS   of  \eqref{fpu}. 

Based  on  the   simplified  expression
\eqref{eps22}, the  following two necessary conditions for a resonant 
enhancement of the leptonic  $\CP$-asymmetry may be derived~\cite{Pilaftsis:1997jf}:
\begin{eqnarray}
  {\rm (i)}~~\Delta m_N \ \sim \ \frac{\Gamma_{N_{1,2}}}{2} \ \ll \ m_{N_{1,2}}\; ,
\qquad \qquad
  {\rm (ii)}~~\frac{\left|{\rm Im}
      \left[(\widehat{h}^\dag \widehat{h})^2_{\alpha\beta}\right]
    \right|}{(\widehat{h}^\dag \widehat{h})_{\alpha\alpha}(\widehat{h}^\dag \widehat{h})_{\beta\beta}} \ \sim \ 1 \; .
\end{eqnarray}
The generic feature of these  
resonant conditions  remains valid even in the presence of flavour 
effects~\cite{Pilaftsis:2005rv}. 
The  condition (i)  is exactly  met when  the unitarity  limit  on the
resummed            heavy-neutrino           propagator           gets
saturated~\cite{Pilaftsis:1997dr}, i.e.~when the total $\CP$-asymmetry
$\varepsilon_l=\sum_\alpha  \varepsilon_{l\alpha}$  takes its  maximum
possible value  of unity. Note  that the limit  $\varepsilon_l\leq 1$,
similar     to      the     Lee-Wolfenstein     bound      for     the
$K^0\overline{K}^0$-system~\cite{Lee:1965},  gets  saturated when  the
regulator  in  \eqref{eps22}  takes  its  maximum  possible  value  of
$|f^{\mathrm{max}}_{\mathrm{reg}}|  = 1/2$. 
 
Since the exact location of  the pole of the propagator determines the
maximum enhancement of  the $ \CP$ asymmetry as  well as its behaviour
in some limiting cases, it is worth commenting on other analytic forms
of the regulator for the  resonant part of the lepton asymmetries in
\eqref{eps22},   existing    in   the   literature~\cite{Flanz:1996fb,
Buchmuller:1997yu, Garny:2011hg}. In  particular, their results differ
by the way in which the singularity $\Delta m_N\to 0$ in \eqref{eps22}
is regulated  once the  heavy neutrino mixing  effects are  taken into
account.   For  instance,   using  a  perturbative  quantum-mechanical
approach,~\cite{Flanz:1996fb} obtained a regulator of the form
\begin{eqnarray}
  f_{\rm reg}^{\rm I} \ = \ \frac{\Delta m_N \Gamma_{N_\alpha}^{(0)}/2}
  {(\Delta m_N)^2+m_{N_\alpha}^2
    [{\rm Re}(A_{\alpha\beta})]^2} \; ,
\end{eqnarray}
which has  a pathological  behaviour for certain  flavour-dependent RL
scenarios, for which    ${\rm    Re}(A_{\alpha\beta})=0$,    but    ${\rm
Re}(A^l_{\alpha\beta})\neq  0$.  In this  limit,  the unitarity  upper
bound   $\varepsilon_l\leq   1$  is   violated   in  the   degenerate
heavy-neutrino mass  limit $\Delta m_N\to 0$ and  the individual $\CP$
asymmetries $\varepsilon_{l}$ become singular.

Following a  modified version of the  quantum field-theoretic approach
introduced  in~\cite{Pilaftsis:1997jf},   a  different  regulator  was
obtained by~\cite{Buchmuller:1997yu, Anisimov:2005hr}:
\begin{eqnarray}
  f_{\rm reg}^{\rm II} \ = \ \frac{\left(m^2_{N_\alpha}-m^2_{N_\beta}\right)
    m_{N_\alpha}\Gamma_{N_\beta}^{(0)}}
  {\left(m^2_{N_\alpha}-m^2_{N_\beta}\right)^2
    +\left(m_{N_\alpha}\Gamma_{N_\alpha}^{(0)}
      -m_{N_\beta}\Gamma_{N_\beta}^{(0)}\right)^2} \; .
  \label{fbp}
\end{eqnarray} 
This regulator diverges in  the doubly degenerate limit $\Delta m_N\to
0$               and               $\Delta              \Gamma_N\equiv
|\Gamma_{N_1}^{(0)}-\Gamma_{N_2}^{(0)}|\to  0$. In  RL  scenarios with
small   $\Delta  m_N$,   one   could  have   $\Gamma^{(0)}_{N_1}\simeq
\Gamma^{(0)}_{N_2}$  even though $A^l_{11}\neq  A^l_{22}$ for  a given
lepton flavour $l$.  For instance, such a scenario can occur naturally
in   approximate   $L$-conserving   RL   models~\cite{Blanchet:2009kk,
Blanchet:2010kw}.   In these  cases, using  the  regulator \eqref{fbp}
might lead to an overestimation  of the leptonic $ \CP$ asymmetries by
several  orders   of  magnitude~\cite{Deppisch:2010fr},  as   we  will
explicitly demonstrate below for a toy model.

Finally, using  an effective Kadanoff-Baym approach  with a particular
quasi-particle  ansatz,~\cite{Garny:2011hg,  Iso:2013lba}  derived  an
effective regulator of the form
\begin{eqnarray}
  f_{\rm reg}^{\rm III} \ = \ \frac{\left(m^2_{N_\alpha}-m^2_{N_\beta}\right)
    m_{N_\alpha}\Gamma_{N_\beta}^{(0)}}
  {\left(m^2_{N_\alpha}-m^2_{N_\beta}\right)^2
    +\left(m_{N_\alpha}\Gamma_{N_\alpha}^{(0)}
      +m_{N_\beta}\Gamma_{N_\beta}^{(0)}\right)^2}\; .
  \label{fKB}
\end{eqnarray}
The  predictions   for  the   $  \CP$-asymmetry  for   the  regulators
\eqref{fbp}  and \eqref{fKB}  are  comparable so  long  as the  widths
$\Gamma_{N_{1,2}}$   are    hierarchical.   Although   the   regulator
\eqref{fKB}  does not have  any pathological  behaviour in  the doubly
degenerate  limit  $\Delta  m_N\to  0$,  $\Delta  \Gamma_N\to  0$,  it
predicts a $ \CP$-asymmetry 4 times smaller than that predicted by the
regulator  $f_{\rm   reg}$  in  \eqref{fpu}.   As  a  result,   the  $
\CP$-asymmetry  $\varepsilon\leq 1/2$ never  gets saturated  to unity,
unlike  the   general  expectations  based   on  unitarity  arguments~\cite{Pilaftsis:1997dr}.
Moreover,  as  shown  in~\ref{app:propagator},  there  are  additional
subtleties  in the  existing treatment  of flavour  mixing  within the
Kadanoff-Baym  approach  and  their   resolution  might  be  crucial  in
determining which of the two regulating expressions \eqref{fpu} and
\eqref{fKB} captures properly the resonant dynamics.

To illustrate the different behaviours of the $ \CP$-asymmetry for the
regulators in \eqref{fpu} and \eqref{fbp} in the $L$-conserving limit,
let us  consider a toy model  with two heavy  neutrinos $N_{1,2}$ with
opposite lepton  numbers, i.e.~$L(N_1) = -L(N_2)  = 1$.  The relevant
Yukawa Lagrangian is given by~\cite{Gavela:2009cd, Blanchet:2009kk}
\begin{eqnarray}
  -{\cal L}_Y \ = \ Y_{l}\overline{L}_l\widetilde{\Phi}N_1
  +Y'_{l}\overline{L}_l\widetilde{\Phi}N_2+
  \frac{1}{2}\left(M\overline{N}_1N_2^C
    +\mu_1\overline{N}_1N_1^C+\mu_2e^{i\theta}
    \overline{N}_2N_2^C\right)+{\rm H.c.} \; ,
  \label{lagtoy}
\end{eqnarray}
where  $M,\mu_{1,2}$  are  real  parameters. The  full  neutrino  mass
matrix, in the basis $\{\nu_{L,l}^C,N_1,N_2^C\}$, is given by
\begin{eqnarray}
  {\cal M}_{\nu N} \ = \  \left(\begin{array}{ccc}
      0 & \frac{v}{\sqrt 2}Y & \frac{v}{\sqrt 2}Y'\\
      \frac{v}{\sqrt 2}Y^{\sf T} & \mu_1 & M\\
      \frac{v}{\sqrt 2}Y'^{\sf T} & M & \mu_2 e^{i\theta}
    \end{array}\right) \; .
  \label{massmtx}
\end{eqnarray}
To first order  in the lepton number breaking  parameters $Y'_{l}$ and
$\mu_{1,2}$, the light neutrino mass matrix is given by
\begin{eqnarray}
  \mat{M}_\nu \ = \  \frac{v^2}{2M}\left(YY'^{\sf T}+Y'Y^{\sf T}
    -\frac{\mu_{\rm eff}}{M}YY^{\sf T}\right) \; .
  \label{mnuloop}
\end{eqnarray}
Note that, although at tree-level  the light neutrino mass matrix does
not depend  on $\mu_1$, it  receives a one-loop contribution  from the
electroweak radiative corrections~\cite{Pilaftsis:1991ug, Dev:2012sg},
which  is  directly  proportional  to $\mu_1$.  This is taken into account 
by the effective $\mu$-parameter          in          \eqref{mnuloop}: $\mu_{\rm
eff}=\mu_2+x_Nf(x_N)\mu_1$,   where  $x_N=M^2/M_W^2$   and   the  loop
function $f(x_N)$ is given by~\cite{Dev:2012sg}
\begin{eqnarray}
  f(x_N) \ = \ \frac{\alpha_w}{16\pi}\left[\frac{x_H}{x_N-x_H}
    \ln\left(\frac{x_N}{x_H}\right) 
    + \frac{3x_Z}{x_N-x_Z}\ln\left({x_N}{x_Z}\right)\right] \; ,
\end{eqnarray}
with     $x_H=M_H^2/M_W^2,~x_Z=M_Z^2/M_W^2\equiv    1/\cos^2\theta_w$.
Therefore,  in order  to  satisfy the  neutrino  mass constraints,  we
require   the   lepton-number  breaking   parameters   to  be   small,
i.e.~$|Y'_l|\ll |Y_l|$ and $\mu_{1,2}\ll M$.

\begin{figure}[t]
\centering
\includegraphics[width=8cm]{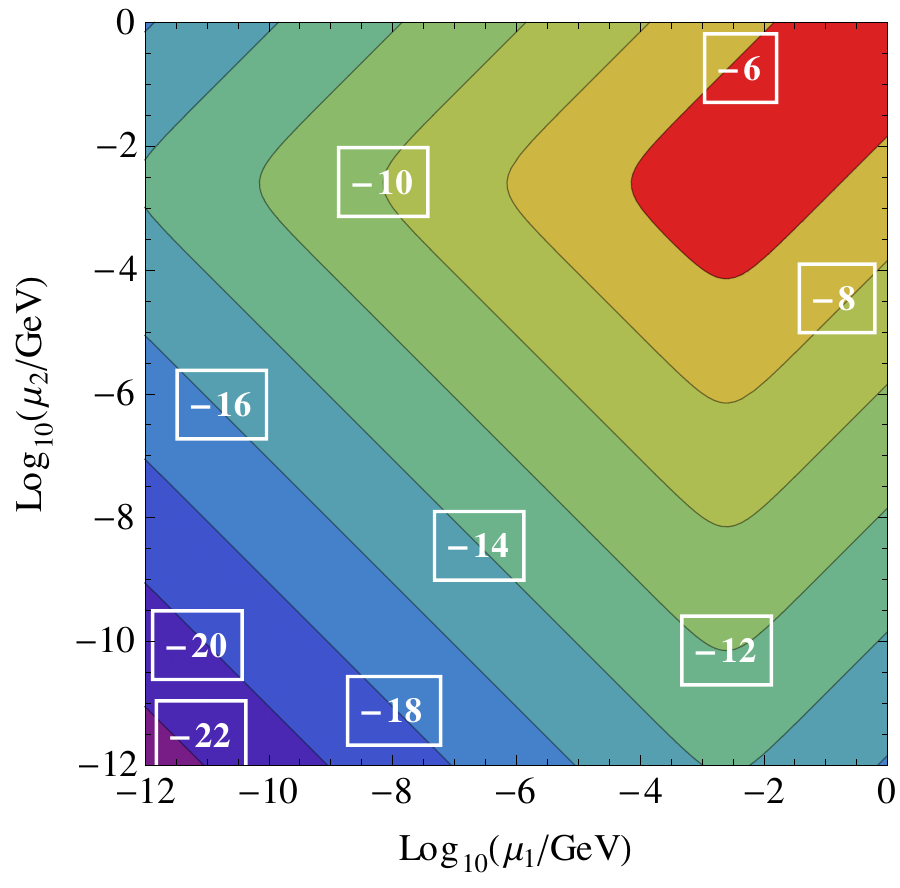}
\hspace{0.2cm}
\includegraphics[width=8cm]{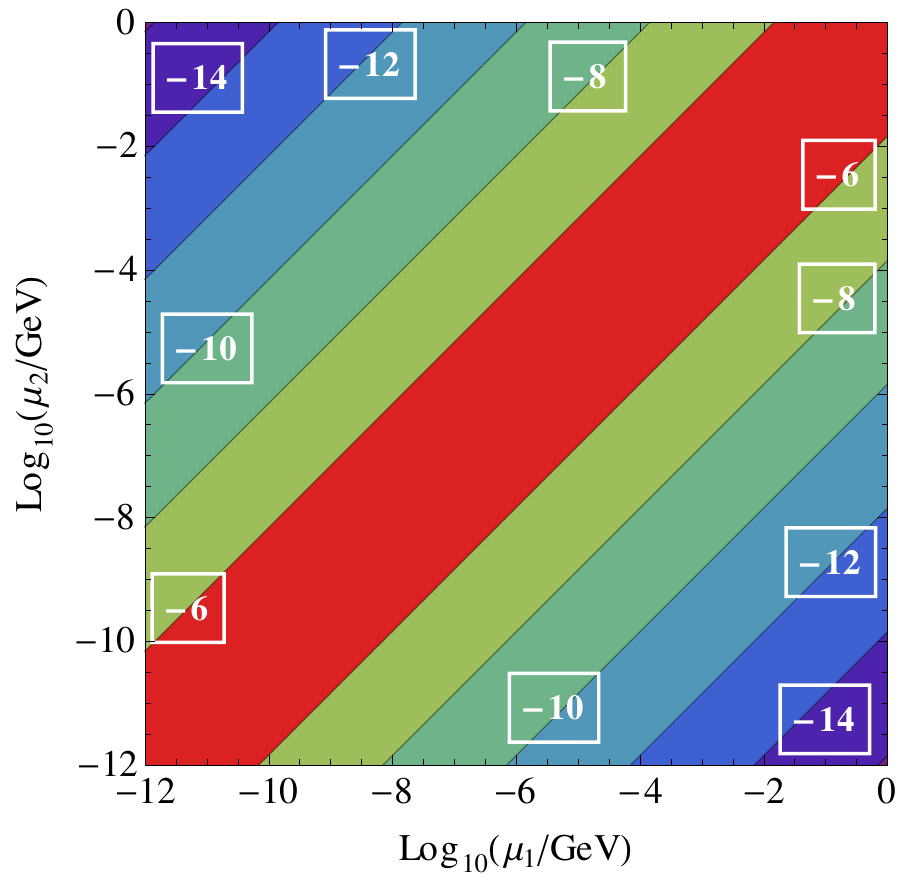}
\caption{\emph  {The  magnitude  of  the  total  $\varepsilon$-type  $
\CP$-asymmetry       in      a       single       lepton      flavour,
$\log_{10}(|\varepsilon_l|)$,  as  a  function  of  the  $L$-violating
parameters $\mu_{1,2}$, as obtained  with the regulators $f_{\rm reg}$
(left panel) and $f^{\rm II}_{\rm reg}$ (right panel).}}
\label{fig:reg}
\end{figure}

In the basis in which the  heavy neutrino mass matrix is diagonal with
real and positive eigenvalues, the Lagrangian \eqref{lagtoy} can
be  recast into  the  general  form given  by  \eqref{eq:Lagr},  with
$\alpha = 1,2$  and  the  heavy  neutrino masses  $M_{1,2}  \simeq  M\mp
\mu/2$. Here,  we have  defined $\mu e^{i\phi}  = \mu_1+\mu_2
e^{i\theta}$.   To  leading order  in  $\mu/M$  and  $Y'$, the  Yukawa
couplings in this mass eigenbasis are given by~\cite{Blanchet:2009kk}
\begin{eqnarray}
  \widehat{h}_{l1} &\simeq &\frac{i}{\sqrt{2}}e^{-i(\phi-\lambda)/2}
  \left[\left(1+\frac{\mu_2^2-\mu_1^2}{4M\mu}\right)
    e^{i\phi}Y_{l}-Y'_l\right],\nonumber \\
  \widehat{h}_{l2} &\simeq &\frac{1}{\sqrt{2}}e^{-i(\phi+\lambda)/2}
  \left[\left(1-\frac{\mu_2^2-\mu_1^2}{4M\mu}\right)
    e^{i\phi}Y_{l}+Y'_l\right],\label{h2}
\end{eqnarray}
where $\lambda=\sin\theta(\mu_1\mu_2/\mu M)$.
Using  the tree-level  Yukawa couplings  given by  \eqref{h2}  and the
corresponding  decay widths given  by \eqref{gammatree},  we calculate
the $\varepsilon$-type $ \CP$-asymmetry in a particular lepton flavour
$\varepsilon_l=\sum_\alpha         \varepsilon_{l\alpha}$,        with
$\varepsilon_{l\alpha}$  given  by   \eqref{eps22},  and  compare  the
magnitudes of $\varepsilon_l$ obtained  using the two regulators given
by   \eqref{fpu}   and  \eqref{fbp}.   Our   results   are  shown   in
Figure~\ref{fig:reg},  where  the  different  contours  show  constant
values       of        $\log_{10}(|\varepsilon_l|)$       in       the
$(\mu_1,\mu_2)$-plane.  Here,  we  have chosen  $Y'_l=0$,  $Y_l = 0.05$, 
$M=100~{\rm GeV}$ and $\theta=\pi/4$ for illustration.  From
Figure~\ref{fig:reg}  (left  panel),  it  can  be seen  that  for  the
regulator  \eqref{fpu}, the  total  $\varepsilon$-type $\CP$-asymmetry
goes  to  zero  in  the  $L$-conserving  limit  $\mu_{1,2}\to  0$,  as
expected.  On the  other hand,  for the  regulator \eqref{fbp},  the $
\CP$-asymmetry gets  enhanced with  smaller $\mu_{1,2}$ and  does {\em
not}   vanish  in   the   limit  in   which   $\mu_{1,2}\to  0$,   see
Figure~\ref{fig:reg}  (right  panel).  In addition,  for  $\mu_1\simeq
\mu_2$, it  approaches a constant value of  ${\cal O}(10^{-5})$, which
clearly demonstrates the pathological behaviour of \eqref{fbp}.

\begin{figure}[t]
\centering
\includegraphics[width=8cm]{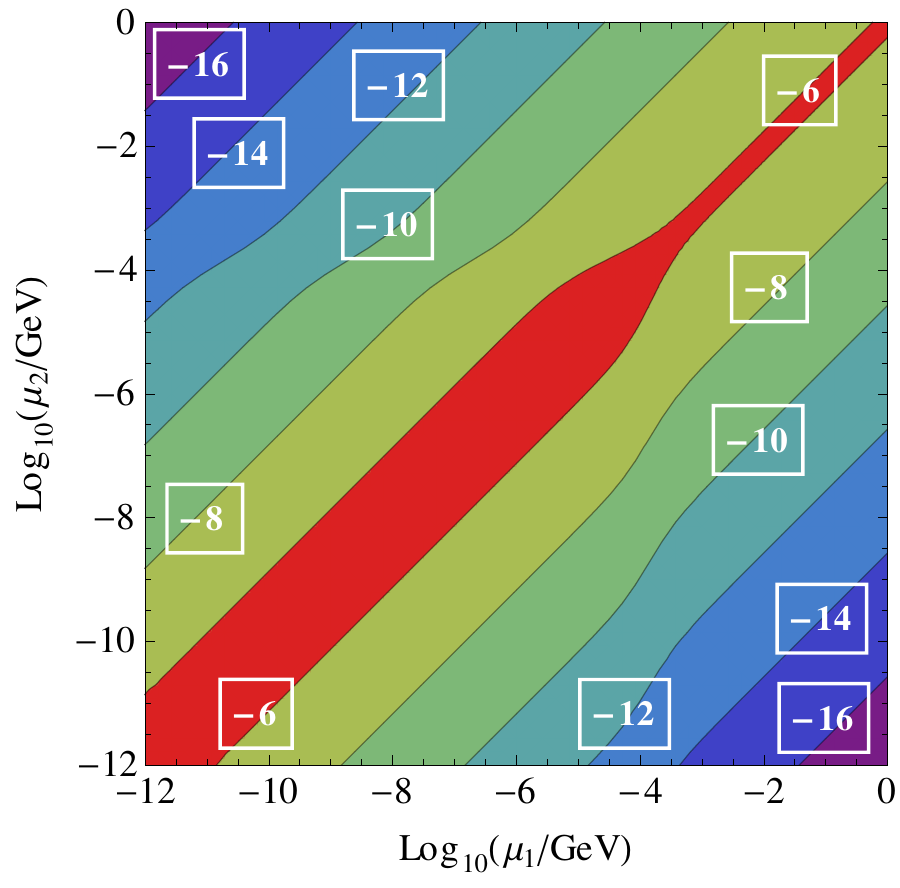}
\hspace{0.2cm}
\includegraphics[width=8cm]{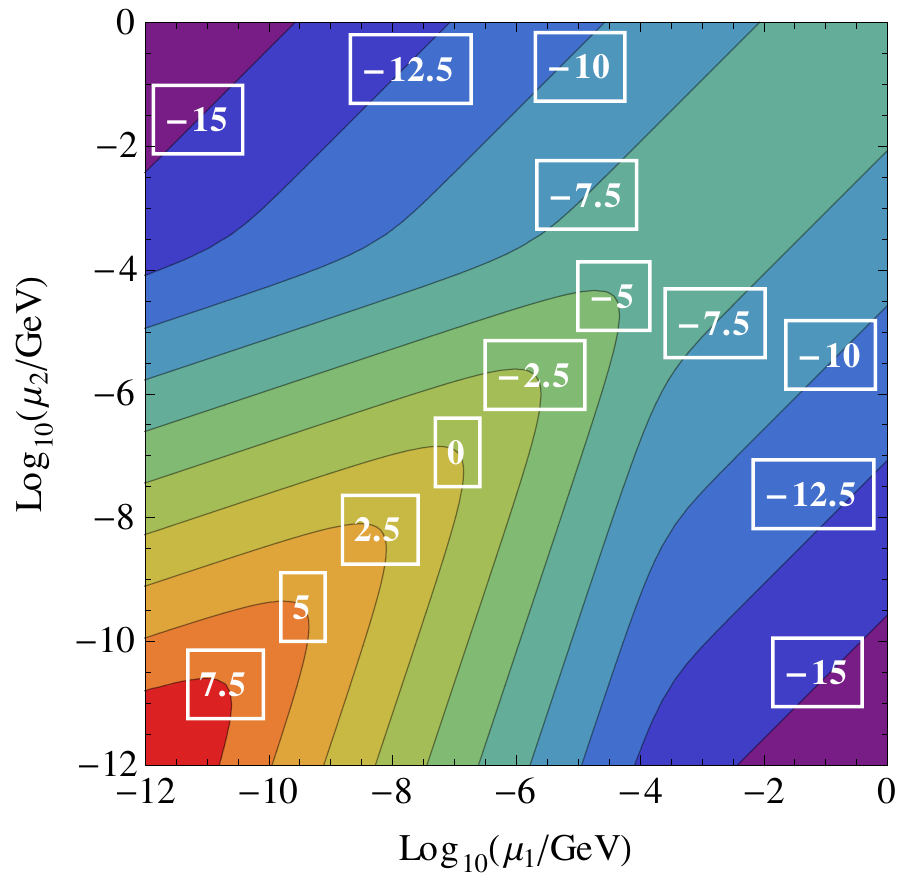}
\caption{\emph {The magnitude  of $\log_{10}(\varepsilon_l/\kappa_l)$ in a
single lepton  flavour as a  function of the  $L$-violating parameters
$\mu_{1,2}$,  as  obtained with  the  regulators  $f_{\rm reg}$  (left
panel) and  $f^{\rm II}_{\rm reg}$  (right panel). The  expression for
$\kappa_l$ is given in \eqref{kappal}. }}
\label{fig:reg2}
\end{figure}

The  singular behaviour  of the  regulator is  \eqref{fbp}  is further
illustrated   in  Figure~\ref{fig:reg2}  where   we  plot   the  ratio
$\varepsilon_l/\kappa_l$,  with $\kappa_l$ defined  by \eqref{kappal}.
Note  that  in  the   $L$-conserving  limit,  $\kappa_l$  vanishes  by
construction,    since     in    this    limit,     there    are    no
$\gamma^{L\Phi}_{L^c\Phi^c}$ terms in  \eqref{kappal}. Now in the same
$L$-conserving limit, the $\CP$-asymmetry obtained using the regulator
\eqref{fpu}  also  vanishes, as  shown  in Figure~\ref{fig:reg}  (left
panel). For  the toy  model under consideration,  both $\varepsilon_l$
and  $\kappa_l$ go  to  zero at  the  {\it same}  rate, while  keeping
$\varepsilon_l/\kappa_l$ constant,  in the limit  $\mu_1\to \mu_2$, as
shown in Figure~\ref{fig:reg2} (left  panel). Note that this does {\it
not}   mean    the   final   lepton   number    asymmetry   given   by
\eqref{eq:anal_diag}  is non-zero in  the $L$-conserving  limit, since
the  expression  \eqref{eq:anal_diag}  is  valid only  in  the  strong
washout regime. In the weak  washout regime with small $\kappa_l$, the
final  lepton  number  asymmetry   will  instead  be  proportional  to
$\varepsilon_l$ alone [cf.~\eqref{be7}], and therefore, vanishes as long as
$\varepsilon_l\to 0$.  However, for the  regulator given by
\eqref{fbp}, due to  the fact that $\varepsilon_l$ does  not vanish in
the $\mu_{1,2}\to  0$ limit, the  ratio $\varepsilon_l/\kappa_l$ blows
up, as shown in Figure~\ref{fig:reg2} (right panel).

It was  argued in~\cite{Anisimov:2005hr} that, due  to the limitations
of the perturbative approach used  to derive \eqref{fbp}, the range of
validity  of  the  analytic  expression  \eqref{eps22}  with  the
regulator  structure  given  by   \eqref{fbp}  is  restricted  to  the
parameter space in  which the degree of degeneracy  of heavy neutrinos
must be much  larger than the pole expansion  parameter, determined by
the neutrino Yukawa couplings. In  the toy model presented above, this
validity condition can be written as
\begin{eqnarray}
  \frac{1}{16\pi^2}\left|{\rm Re}[(\widehat{h}^\dag \widehat{h})_{12}]\right| \ \ll \ 
  \left|\frac{M_2-M_1}{M_1}\right|\; .\label{val1}
\end{eqnarray}
Here $(M_2-M_1)/M_1\sim \mu/M$  in the  heavy neutrino
mass eigenbasis, whereas from \eqref{h2} with $Y'_l=0$ we obtain
\begin{eqnarray}
  {\rm Re}[(\widehat{h}^\dag \widehat{h})_{12}] \ = \ \frac{1}{2}\sin\lambda
  \left[1-\left(\frac{\mu_2^2-\mu_1^2}{4M\mu}\right)^2\right]|Y_l|^2 \; ,
  \label{val11}
\end{eqnarray}
In the pathological limit $\mu_1\to  \mu_2$ for the regulator given by
\eqref{fbp}, the LHS and RHS of the condition \eqref{val1} reduce to
\begin{eqnarray}
  \lim_{\mu_1\to \mu_2} \: \frac{1}{16\pi^2}{\rm Re}[(\widehat{h}^\dag \widehat{h})_{12}]  &
  =& \frac{|Y_l|^2}{32\pi^2}\sin\left(\frac{\mu_2}{M}
    \sin{\frac{\theta}{2}}\right) \; ,\\
  \lim_{\mu_1\to \mu_2} \:  \frac{M_2-M_1}{M_1}  &
  =& \frac{2\mu_2}{M}\cos{\frac{\theta}{2}}\; , 
\end{eqnarray} 
from which it is clear that for any choice of $\theta\neq (2n+1)\pi/2$
(with $n=0,1,2,...$),  the condition \eqref{val1}  is always satisfied, 
as long  as $|Y_l|^2\leq 1$.  Thus, the pathological behaviour  of the
regulator  \eqref{fbp}  cannot  be  avoided  simply  by  imposing  the
validity condition \eqref{val1}.

On the other hand, for  the  analytic solution  \eqref{fKB}  obtained in  the
Kadanoff-Baym  approach,  the expansion  for  small Yukawa  couplings
requires an additional validity condition~\cite{Garny:2011hg}
\begin{eqnarray}     
  {\rm Re}[(\widehat{h}^\dag \widehat{h})_{12}] \ \ll \ |(\widehat{h}^\dag \widehat{h})_{22}-(\widehat{h}^\dag \widehat{h})_{11}| \; ,
  \label{val2}
\end{eqnarray}
while  the  condition \eqref{val1}  is  somewhat  relaxed, i.e.~${\rm
Re}[(\widehat{h}^\dag \widehat{h})_{12}]  \lesssim 8\pi(M_2-M_1)/M_1$. For  our toy model,
the RHS of \eqref{val2} is given by
\begin{eqnarray}
  |(\widehat{h}^\dag \widehat{h})_{22}-(\widehat{h}^\dag \widehat{h})_{11}| \ = \ 
  \left|\frac{\mu_2^2-\mu_1^2}{2M\mu}\right||Y_l|^2 \; .
  \label{val22}
\end{eqnarray}
Comparing \eqref{val11}  and \eqref{val22}, we see  that the condition
\eqref{val2} is {\it not} satisfied  in the limit $\mu_1\to \mu_2$ and
hence, the regulator \eqref{fKB} is also not valid in the pathological
limit. These issues do not  arise for the regulator \eqref{fpu}, which
is  well-defined  in  the   {\it  entire}  parameter  space  shown  in
Figure~\ref{fig:reg}.

For comparison  with the $\varepsilon$-type  $\CP$-asymmetry discussed
above, let us also compute the $\varepsilon'$-type $\CP$-asymmetry due
to one-loop vertex corrections.  For the two heavy neutrino case, this
is given by~\cite{Covi:1996wh}
\begin{eqnarray}
  \varepsilon'_{l\alpha} \ = \ \frac{{\rm Im}
    \left[(\widehat{h}^*_{l\alpha} \widehat{h}_{l\beta})(\widehat{h}^\dag \widehat{h})_{\alpha\beta}\right]}
  {(\widehat{h}^\dag \widehat{h})_{\alpha\alpha}(\widehat{h}^\dag \widehat{h})_{\beta\beta}} 
\frac{\Gamma_{N_\beta}^{(0)}}{m_{N_\beta}}  
f\left(\frac{m_{N_\beta}^2}{m^2_{N_\alpha}}\right) \; ,
\end{eqnarray}
where     $f(x)=\sqrt{x}\left[1-(1+x)\ln(1+1/x)\right]$     is     the
Fukugita-Yanagida  loop function~\cite{Fukugita:1986hr}. 
\begin{figure}[t!]
\centering
\includegraphics[width=8cm]{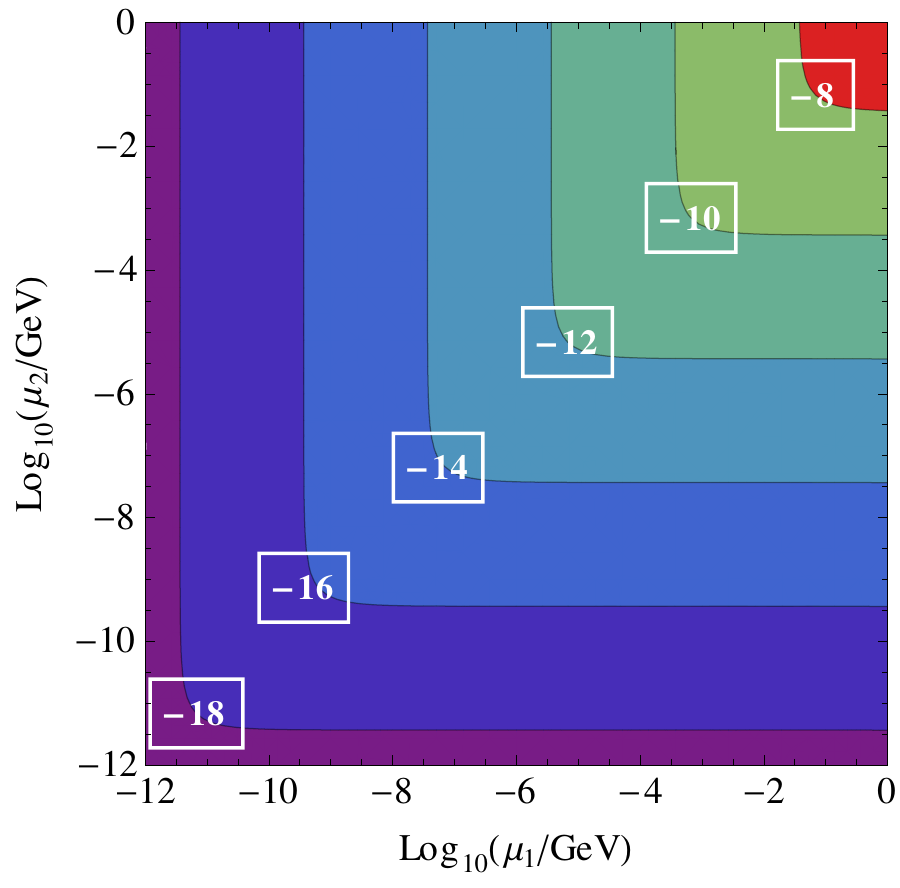}
\caption{\emph {The magnitude  of the individual $\varepsilon'$-type $
\CP$-asymmetries      in       a      single      lepton      flavour,
$\log_{10}(|\varepsilon'_{l\alpha}|)$, 
as  a  function  of   the  $L$-violating  parameters $\mu_{1,2}$.}}
\label{fig:epsp}
\end{figure}  
The   magnitude  of   the   $\varepsilon'_{l\alpha}$  contribution  
is shown numerically in  Figure~\ref{fig:epsp} for our toy model  with the same
parameter values as chosen for Figure~\ref{fig:reg}. 
We find that, as expected, the  $\varepsilon'$-type $\CP$-asymmetry vanishes in the  $L$-conserving  limit. Moreover, the  $\varepsilon'$-part  of  the  transition  amplitude
squared for the $N_1$ decay becomes equal and opposite in sign to that
for the  $N_2$ decay,  and thus, these  two contributions  cancel each
other to give a  vanishing {\it total} $\CP$-asymmetry $\varepsilon'_l
=  \sum_\alpha  \varepsilon'_{l\alpha}$, which goes to zero faster than the 
individual contributions  in  the   limit  $\mu_{1,2}\to   0$,  similar   to  the
$\varepsilon$-case with  the regulator \eqref{fpu}.  
We  also note that 
the $\varepsilon'$-type $\CP$ violation given by
Figure~\ref{fig:epsp} is  smaller than the
$\varepsilon$-type $\CP$ violation given by Figure~\ref{fig:reg} (left
panel) for all values of $\mu_{1,2}$.  
This is consistent with the general expectation that the  $\varepsilon'$-type  
contribution  can become  comparable  to  the
$\varepsilon$-type term only in the hierarchical limit $m_{N_2}\gg m_{N_1}$, 
when it approaches the asymptotic relation $\sum_l \varepsilon'_{l\alpha} = (1/2)\sum_l \varepsilon_{l\alpha}$~\cite{Covi:1996wh}. 

\section{Flavour Covariant Helicity Amplitude Formalism}
\label{app:propagator}

In  this appendix,  we describe  the  pertinent details  of the  fully
flavour-covariant quantization of spinorial  fields in the presence of
time-dependent and  spatially inhomogeneous backgrounds.   Thereby, we
exemplify the  consistency of our  treatment of flavour mixing  in the
transport equations  derived in Sections~\ref{sec:3}  and \ref{sec:4}.
To   this  end,   we   begin  by   describing  the   flavour-covariant
generalization  of the  helicity  amplitude formalism  \cite{Bouchiat,
Michel},  in the context  of an  $\mathcal{N}$-flavour model  of Dirac
fermions.   After  highlighting   the  generalized  discrete  symmetry
transformations of the Dirac  helicity four-spinors, we illustrate the
inter-dependence of the spinorial and flavour structure.\footnote{Previously, the issue of spin coherence in quantum kinetic equations has been considered using a truncated gradient expansion of Wigner functions \cite{Vlasenko:2013fja}, which we do not follow here.} Subsequently,
we   derive    the   flavour-covariant   propagators    of   our   toy
$\mathcal{N}$-flavour model,  thereby generalizing the non-homogeneous
propagators  described in  \cite{Millington:2012pf}.  In addition,  we
introduce    the    spatially    inhomogeneous   and    time-dependent
flavour-covariant  statistical distribution  functions  relevant to  a
complete  quantum field  theoretic treatment  of  flavour-coherent and
Gaussian statistical backgrounds.  Finally, we highlight a consequence
of   this  treatment   that  is   anticipated  to   impact   upon  the
flavour-dependent quasi-particle  approximations currently employed in
the literature in  the application of the CTP  formalism and resulting
Kadanoff-Baym equations to transport phenomena.  Specifically, we show
how the time-translational invariance of flavour-covariant propagators
is necessarily broken in  the presence of flavour-coherent statistical
backgrounds.

\subsection*{B.1\ Flavour Covariant Spinor Algebra}

We begin by introducing the four-component Dirac spinor $\psi_{k}$ and
its Dirac conjugate $\bar{\psi}^k$ in the Weyl basis
\begin{equation}
  \psi_{\smrm{W},\,k}(x)\ =\ \begin{pmatrix} \xi_{\mathfrak{a},\, k}(x) \\
  \bar{\eta}_{k}^{\dot{\mathfrak{a}}}(x) \end{pmatrix}\;,\quad
  \bar{\psi}_{\smrm{W}}^k(x)\ =\
  \begin{pmatrix} \eta^{\mathfrak{a},\, k}(x) &
  \bar{\xi}^k_{\dot{\mathfrak{a}}}(x)\end{pmatrix}\;,
\end{equation}
where  $\mathfrak{a},   \dot{\mathfrak{a}},  \dots$  are   the  spinor
indices  and $k, l,  \dots$ the  flavour indices.   The two-component
Weyl        spinors        $\xi_{\mathfrak{a},\,        k}$        and
$\bar{\eta}_k^{\dot{\mathfrak{a}}}$    are   operator-valued   complex
vectors in the  flavour space $\mathcal{V}$, transforming respectively
as   $(\tfrac{1}{2},0)$  and  $(0,\tfrac{1}{2})$   representations  of
$SL(2,\mathbb{C})$  and covariant  vectors  of $U(\mathcal{N})$.   The
contravariant  Weyl   spinors  $\bar{\xi}^k_{\dot{\mathfrak{a}}}$  and
$\eta^{\mathfrak{a},\,   k}$   transform   as   complex   vectors   of
$U(\mathcal{N})$ in  the dual space $\mathcal{V}^*$.   Notice that the
left      and     right      spinors      $\xi_{\mathfrak{a}}$     and
$\bar{\eta}^{\dot{\mathfrak{a}}}$ transform in the same representation
of $U(\mathcal{N})$.

The $U(\mathcal{N})$-symmetric Dirac Lagrangian may be written in the
following form:
\begin{equation} \label{Dirac_lag}
  \mathcal{L}_{\mathrm{D}}(x)\ =\
  \bar{\psi}_{\smrm{W}}^{k}(x)
  \big(i\gamma_{\smrm{W}}^{\mu}\partial_{\mu}\delta_{k}^{\ l}
  \:-\:m_{k}^{\ l}\big)\psi_{\smrm{W},\,l}(x)\;,
\end{equation}
where the gamma matrices are defined in the Weyl basis:
\begin{equation}
  \gamma^{\mu}_{\smrm{W}}\ =\
  \begin{pmatrix}
    0 & (\sigma^{\mu})_{\mathfrak{a}\dot{\mathfrak{b}}} \\
    (\bar{\sigma}^{\mu})^{\dot{\mathfrak{a}}\mathfrak{b}} & 0
  \end{pmatrix}\;, 
\end{equation}
with $\sigma^\mu\equiv  (1,\mat{\sigma})$ and $\bar{\sigma}^\mu \equiv
(1,-\mat{\sigma})$,  $\sigma^i$'s being  the usual  $2\times  2$ Pauli
matrices.  The mass  matrix $m_k^{\ l}$ transforms as  a rank-2 tensor
under $\mathrm{U}(\mathcal{N})$.   We may  rotate to the  Dirac basis,
independent  of the  flavour  structure, by  means  of the  orthogonal
transformation
\begin{equation}
  \gamma^{\mu}\ =\ O^{\mathsf{T}}\gamma^{\mu}_{\smrm{W}}O\;,\qquad
  \psi_k\ =\ O^{\mathsf{T}}\psi_{\smrm{W},\, k}\;,\qquad 
  O\ =\ \frac{1}{\sqrt{2}}
  \begin{pmatrix}
    \mat{1}_2 & -\mat{1}_2 \\ 
    \mat{1}_2 & \mat{1}_2
  \end{pmatrix}\;,
\end{equation}
where the gamma matrices in the Dirac representation are 
\begin{equation}
  \gamma^0\ =\ 
  \begin{pmatrix}
    \mat{1}_2 & \mat{0}_2 \\
    \mat{0}_2 & -\mat{1}_2
  \end{pmatrix}
  \;, \qquad
  \gamma^i\ =\
  \begin{pmatrix}
    \mat{0}_2 & \sigma^i \\
    -\sigma^i & \mat{0}_2
  \end{pmatrix}
  \;,\qquad
  \gamma^5\ =\
  \begin{pmatrix}
    \mat{0}_2 & \mat{1}_2 \\
    \mat{1}_2 & \mat{0}_2
  \end{pmatrix}\; ,
\end{equation} 
with $\mat{0}_2$ and $\mat{1}_2$ being the $2\times 2$ null and identity matrices, respectively. 
The relevant flavour transformations for the Dirac fields and the mass
matrix are
\begin{equation}
  \psi'_k(x)\ =\ U_{k}^{\ l}\psi_{l}(x)\;,\qquad 
  \psi'{}^k(x)\ =\ U^{k}_{\ l}\psi^{l}(x)\;,\qquad
  m'{}_k^{\ l}\ =\ U_{k}^{\ m}U^{l}_{\ n}m_m^{\ \ n}\;,
\end{equation}
where $U^{k}_{\  l} \equiv (U_{k}^{\ l})^*$ and  $ U_{k}^{\ l}U^{k}_{\
m}  = U^{l}_{\  k}U_{m}^{\  \  k} =  \delta^{l}_{\  m}$.  Varying  the
Lagrangian with respect to the fields, we obtain the flavour-covariant
Dirac equations
\begin{equation} \label{dirac}
  \big(i\gamma^{\mu}\overrightarrow{\partial}_{\!\!\mu}
  \delta_{k}^{\phantom{k}l} \:
  - \: m_{k}^{\phantom{k}l}\big)\psi_{l}(x) \ = \ 0\;,
  \qquad
  \bar{\psi}^k(x)\big(i\gamma^{\mu}\overleftarrow{\partial}_{\!\!\mu}
  \delta_{k}^{\phantom{k}l}\:
  + \: m_{k}^{\phantom{k}l}\big) \ = \ 0\;,
\end{equation}
where,  in   the  latter,  the   derivative  acts  to  the   left  and
$\bar{\psi}^k(x)   \:   =   \:  [\psi_k(x)]^{\dag}\gamma^0$   is   the
Dirac-conjugate spinor.

The Dirac field operators in the interaction picture may be written as
follows:
\begin{align}
  \psi_k(x)\ &=\ \sum_{s}\int_{\ve{p}}
  [(2E(\mathbf{p}))^{-1}]_{k}^{\phantom{k}l}
  \Big([e^{-ip\cdot x}]_{l}^{\phantom{l}m}
  [u(\mathbf{p},s)]_{m}^{\phantom{m}n}
  \underline{b}_{n}(\mathbf{p},s,0)
  \nonumber\\&\qquad
  +\:[e^{ip\cdot x}]_l^{\phantom{l}m}
  [v(\mathbf{p},s)]_{m}^{\phantom{m}n}
  \underline{d}^{\dag}_{n}(\mathbf{p},s,0)\Big)\;,\\
  \bar{\psi}^k(x)\ &=\ \sum_{s}\int_{\ve{p}} 
[(2E(\mathbf{p}))^{-1}]^{k}_{\phantom{k}l}
  \Big([e^{-i p\cdot x}]^{l}_{\ m}
  [\bar{v}(\mathbf{p},s)]^{m}_{\phantom{m}n}
  \underline{d}^{\dag\,n}(\mathbf{p},s,0)
  \nonumber\\&\qquad 
  +\:[e^{i p\cdot x}]^{l}_{\ m}
  [\bar{u}(\mathbf{p},s)]^{m}_{\phantom{m}n}
  \underline{b}^{n}(\mathbf{p},s,0)\Big)\;,
\end{align}
where  $s =  \pm$ is  the helicity  index, denoting  the  two helicity
states  with  the unit  spin  vector  $\maf{n}  = s\maf{s}$   aligned
parallel  and  anti-parallel   to  the  three  momentum  $\mathbf{p}$,
respectively, i.e.
\begin{equation}
  \maf{s} \ = \ \maf{p}/|\maf{p}| \ = \
  (\sin\theta\cos\phi,\    \sin\theta\sin\phi,\   \cos\theta)\;.
\end{equation}
The three momentum $\maf{p}$ is  obtained by boosting from the rest
frame along the direction  specified by $\maf{s}$.  Notice that the
four-component Dirac  spinors $u$ and $v$ transform  as rank-2 tensors
under $U(\mathcal{N})$.   In addition, we  draw attention to  the fact
that $\underline{b}_k$ and $\underline{d}^{\dag}_k$ ($\underline{b}^k$
and  $\underline{d}^{\dag,k}$) transform  under  the same  fundamental
(anti-fundamental)  representation of  $U(\mathcal{N})$.   The particle
and  anti-particle  creation and  annihilation  operators satisfy  the
anti-commutation relations
\begin{align}
  \label{eq:bdcom}
  \{\underline{b}_{k}(\mathbf{p},s,\tilde{t}),\
  \underline{b}^{l}(\mathbf{p}',s',\tilde{t})\}
  \ &=\ 
  \{\underline{d}^{\dag\,l}(\mathbf{p},s,\tilde{t}),\
  \underline{d}^{\dag}_{k}(\mathbf{p}',s',\tilde{t})\}
  \ &=\ (2\pi)^3[2E(\mathbf{p})]_{k}^{\phantom{k}l}\,
  \delta^{(3)}(\mathbf{p}-\mathbf{p}')\delta_{ss'}\;.
\end{align}
where
\begin{equation}
  [|E(\mathbf{p})|^2]_{k}^{\phantom{k}l} \ = \ 
  [E(\mathbf{p})]^{m}_{\phantom{m}k}[E(\mathbf{p})]_{m}^{\phantom{m}l} \ = \
  \mathbf{p}^2\delta_{k}^{\phantom{k}l} \: + \: [m^{\dag}m]_{k}^{\phantom{k}l}\;.
\end{equation}
The  creation and  annihilation operators  of mass  dimension  $-1$ in
\eqref{eq:bdcom}  are  related  to  the  corresponding  ones  of  mass
dimension $-\tfrac{3}{2}$ in  \eqref{eq:b_d_anticomm} by the following
Bogoliubov transformations:
\begin{align}
  \label{eq:bog1}
  \underline{b}_{k}(\mathbf{p},s,\tilde{t})\ &=\
  [(2E(\mathbf{p}))^{1/2}]_{k}^{\phantom{k}l}\,
  \Big(\cos\omega_b\,b_{l}(\mathbf{p},s,\tilde{t})\:
  +\:\sin\omega_b\,\mathcal{G}_{lm}b^{m}(\mathbf{p},s,\tilde{t})
  \Big)\;,\\
  \label{eq:bog2}
  \underline{b}^{k}(\mathbf{p},s,\tilde{t})\ &=\
  [(2E(\mathbf{p})^{1/2}]^{k}_{\phantom{k}l}\,
  \Big(\cos\omega_b\,b^{l}(\mathbf{p},s,\tilde{t})\:
  +\:\sin\omega_b\,\mathcal{G}^{lm}b_{m}(\mathbf{p},s,\tilde{t})
  \Big)\;,
\end{align}
with   analogous  expressions  for   the  antiparticle   creation  and
annihilation  operators, obtained  by  the replacement  $b  \: \to  \:
d^{\dag}$  and  $\omega_b  \:   \to  \:  \omega_d$  of  the  arbitrary
Bogoliubov  angles.   The  matrix  $\mathcal{G}_{lm}$  is  defined  in
\eqref{eq:C_trans}.

In momentum space, the Dirac field operators take the forms
\begin{align}
  \psi_k(p;\tilde{t}_i)\ &=\ \sum_{s}2\pi
  [\delta(p^2 \: - \: m^2)]_{k}^{\phantom{k}l}
  \Big(\theta(p_0)[u(\mathbf{p},s)]_{l}^{\phantom{l}m}
  \underline{b}_{m}(\mathbf{p},s,0)
  \nonumber\\& \qquad
  +\:\theta(-p_0)[v(-\mathbf{p},s)]_{l}^{\phantom{l}m}
  \underline{d}^{\dag}_{m}(-\mathbf{p},s,0)\Big)\;,\\
  \bar{\psi}^k(p;\tilde{t}_i)\ &=\ \sum_{s}2\pi
  [\delta(p^2 \: - \: m^2)]^{k}_{\phantom{k}l}
  \Big(\theta(p_0)[\bar{u}(\mathbf{p},s)]^{l}_{\phantom{l}m}
  \underline{b}^{m}(\mathbf{p},s,0)
  \nonumber\\& \qquad
  + \: \theta(-p_0)[\bar{v}(-\mathbf{p},s)]^{l}_{\phantom{l}m}
  \underline{d}^{\dag\, m}(-\mathbf{p},s,0)\Big)\;,
\end{align}
where  the   rank-2  tensor  delta  function  $[\delta(p^2   \:  -  \:
m^2)]_{k}^{\phantom{k}l}$ is understood in the following sense:
\begin{equation}
  \int\!\D{}{p_0}\;2[p_0]_k^{\phantom{k}m}\,\theta(\pm p_0)\,
  [\delta(p^2 \: - \: m^2)]_{m}^{\phantom{m}l} \ 
  = \ \: \pm \, \delta_{k}^{\phantom{k}l}\;.
\end{equation}
Finally, the  Dirac equations \eqref{dirac} in momentum  space read as
follows:
\begin{equation}
  [\slashed{p} \: - \: m]_{k}^{\phantom{k}l}
  \psi_{l}(p;\tilde{t}_i) \ = \ 0\;,\qquad 
  \bar{\psi}^k(p;\tilde{t}_i)[\slashed{p} \: + \: m]
  _{k}^{\phantom{k}l}\ =\ 0\;,
\end{equation}
where
\begin{equation}
  \slashed{p}_{k}^{\phantom{k}l}\ = \ \gamma^{\mu}[p_{\mu}]
  _{k}^{\phantom{k}l}\ = \
  \begin{pmatrix}
    [E(\mathbf{p})]_{k}^{\phantom{k}l}\,\mat{1}_2 &
    -\delta_{k}^{\phantom{k}l}\,\bm{\sigma}\cdot\mathbf{p} \\
    \delta_{k}^{\phantom{k}l}\,\bm{\sigma}\cdot\mathbf{p} & 
    -[E(\mathbf{p})]_{k}^{\phantom{k}l}\,\mat{1}_2
  \end{pmatrix}\!\;.
\end{equation}

The four-component helicity spinors may be written explicitly as
  \begin{align}
  \label{eq:4spinors1}
  [u(\mathbf{p},s)]_{k}^{\phantom{k}l} \ &
  = \ [\slashed{p} \: + \: m]_{k}^{\phantom{k}m}
  \{[(E(\ve{p}) \: + \: m)2m]^{-1/2}\}_{m}^{\phantom{m}n}
  [u(\mathbf{0},s)]_{n}^{\phantom{n}l}\;,\\
  \label{eq:4spinors2}
  [v(\mathbf{p},s)]_{k}^{\phantom{k}l} \ &
  = \ [-\slashed{p} \: + \: m]_{k}^{\phantom{k}m}
  \{[(E(\ve{p}) \: + \: m)2m]^{-1/2}\}_{m}^{\phantom{m}n}
  [v(\mathbf{0},s)]_{n}^{\phantom{n}l}\;,
  \end{align} 
where the rest-frame four-spinors are given by
\begin{equation} \label{rest-spinor}
  [u(\mathbf{0},s)]_{k}^{\phantom{k}l} \ = \
  [m^{1/2}]_{k}^{\phantom{k}l}e^{i\varphi^s_{u}}\!
  \begin{pmatrix}
    u_s(\mathbf{s}) \\ \mat{0}_2 
  \end{pmatrix}\;,\qquad
  [v(\mathbf{0},s)]_{k}^{\phantom{k}l} \ = \ 
  [m^{1/2}]_{k}^{\phantom{k}l}e^{i\varphi^{-s}_{v}}\!
  \begin{pmatrix}
    \mat{0}_2 \\ -u_{s}(-\mathbf{s})
  \end{pmatrix}\;,
\end{equation}
with the two-component spinors 
\begin{equation}
  \label{eq:uDdef}
  u_s(\mathbf{s}) \ = \
  \begin{cases}
\sqrt{2}\,
  \begin{pmatrix}
    \cos\tfrac{\theta}{2}e^{-i\frac{\phi}{2}} \\ 
    \sin\tfrac{\theta}{2}e^{i\frac{\phi}{2}}
  \end{pmatrix}
  \;,\quad s\ =\ +\;;\\
  i\sqrt{2}\,
  \begin{pmatrix}
    -\sin\tfrac{\theta}{2}e^{-i\frac{\phi}{2}} \\ 
    \cos\tfrac{\theta}{2}e^{i\frac{\phi}{2}}
  \end{pmatrix}
  \;,\quad s\ =\ -\;;
\end{cases}
\end{equation}
and the corresponding phases $\varphi^\pm_{u(v)}$.  We note the useful
identities\footnote{Throughout  this  appendix,  the  conventions  and
notation are based on \cite{Haber:1994pe, Pokorski} (see also \cite{Pal}) with the exception
that   the   azimuthal   phase   of   the   two-spinors   defined   in
\eqref{eq:uDdef} differs.  The latter will  impact on the $C$, $P$ and
$T$  transformations of  the spinors,  as we  will see  later  in this
appendix.}
\begin{align}
  \label{eq:twospinident1}
  u_s(-\mathbf{s}) \  = \ su_{-s}(\mathbf{s})\;, \qquad \qquad
  u_{s}^*(-\mathbf{s}) \  = \ (-i\sigma^1\sigma^3)u_s(\mathbf{s})\;.
\end{align}
Using \eqref{rest-spinor} and \eqref{eq:twospinident1}, the Dirac four
spinors  in  \eqref{eq:4spinors1}   and  \eqref{eq:4spinors2}  may  be
rewritten as
\begin{align} \label{dirac4u}
 [u(\mathbf{p},s)]_{k}^{\phantom{k}l} \ &
 = \ \frac{e^{i\varphi^s_{u}}}{\sqrt{2}}
  \begin{pmatrix}
    [(E(\ve{p}) \: + \: m)^{1/2}]_{k}^{\phantom{k}l}\,u_s(\mathbf{s}) \\
    s\,[(E(\ve{p}) \: - \: m)^{1/2}]_{k}^{\phantom{k}l}\,u_s(\mathbf{s})
  \end{pmatrix}\;,\\
  [v(\mathbf{p},s)]_{k}^{\phantom{k}l} \ &
  = \ \frac{e^{i\varphi^{-s}_{v}}}{\sqrt{2}}
  \begin{pmatrix}
    [(E(\ve{p}) \: - \: m)^{1/2}]_{k}^{\phantom{k}l}\,u_{-s}(\mathbf{s}) \\
    -s\,[(E(\ve{p}) \: + \: m)^{1/2}]_{k}^{\phantom{k}l}\,u_{-s}(\mathbf{s})
  \end{pmatrix}\;. \label{dirac4v}
\end{align}
With the  aid of \eqref{eq:4spinors1} and  \eqref{eq:4spinors2} we may
verify that these four-spinors are solutions to the Dirac equations
\begin{equation}
  [\slashed{p} \: - \: m]_k^{\phantom{k}l}
  [u(\mathbf{p},s)]_{l}^{\phantom{l}m}\ =\ 0\;,\qquad 
  [\slashed{p} \: + \: m]_{k}^{\phantom{k}l}
  [v(\mathbf{p},s)]_{l}^{\phantom{l}m}\ =\ 0\;, 
\end{equation}
and helicity eigenstates, satisfying
\begin{equation}
  \label{eq:heleigen}
  \frac{\bm{\Sigma}\cdot\mathbf{p}}{|\mathbf{p}|}
  [u(\mathbf{p},s)]_{k}^{\phantom{k}l} \
   = \frac{1}{2}s[u(\mathbf{p},s)]_{k}^{\phantom{k}l}\;,\qquad
  \frac{\bm{\Sigma}\cdot\mathbf{p}}{|\mathbf{p}|}
  [v(\mathbf{p},s)]_{k}^{\phantom{k}l} \
   = -\frac{1}{2}s[v(\mathbf{p},s)]_{k}^{\phantom{k}l}\;,
\end{equation}
where
\begin{equation}
  \bm{\Sigma} \ = \ \frac{1}{2}
  \begin{pmatrix}
    \bm{\sigma} & 0 \\
    0 & \bm{\sigma}
  \end{pmatrix}
\end{equation}
is the spin operator. The expressions~\eqref{eq:heleigen} follow immediately from the
fact that
\begin{equation}
  \frac{\bm{\sigma}\cdot\mathbf{p}}{|\mathbf{p}|}
  u_{s}(\pm \; \mathbf{s})\ =\ \pm s \; 
  u_s(\pm \;  \mathbf{s})\;.
\end{equation} 

Boosting to the frame in  which the particle momentum is $\mathbf{p}$,
the  rest-frame spin four-vector  $s^{\mu} \:  = \:  (0,\ \mathbf{s})$
transforms as
\begin{equation}
  [s'{}^{\mu}]_{k}^{\phantom{k}l} \
  = \ \Lambda^{\mu}_{\phantom{\mu}\nu}(\beta_{k}^{\phantom{k}l})s^{\nu} \
  = \ \Big(|\mathbf{p}|[m^{-1}]_{k}^{\phantom{k}l}\;,\
  [E(\ve{p})]_{k}^{\phantom{k}m}[m^{-1}]_{m}^{\phantom{m}l}
  \frac{\mathbf{p}}{|\mathbf{p}|}\Big)\;,
\end{equation}
satisfying
\begin{equation}
  [s'_{\mu}]_{k}^{\phantom{k}m}[s'^{\mu}]_{m}^{\phantom{m}l} \
  = \ -\delta_{k}^{\phantom{k}l}\;, \qquad
  [s'_{\mu}]_{k}^{\phantom{k}m}[p^{\mu}]_{m}^{\phantom{m}l} \ =\ 0\;.
\end{equation}
Notice  that  since   the  boost  factor  $\beta_{k}^{\phantom{k}l}  =
|\mathbf{p}|[E^{-1}(\mathbf{p})]_{k}^{\phantom{k}l}$  depends  on  the
mass              matrix,              the              transformation
$\Lambda^{\mu}_{\phantom{\mu}\nu}(\beta_{k}^{\phantom{k}l})$  is, in a
general basis,  a rank-2 tensor  in flavour space.   Alternatively, we
can consider the Lorentz-factor for this boost, which takes the form
\begin{equation}
  [\gamma^2]_{k}^{\phantom{k}l} \
  = \ \delta_{k}^{\phantom{k}l} \:
  + \: |\mathbf{p}|^2[m^{-2}]_{k}^{\phantom{k}l}\;,
\end{equation}
as one may readily justify by rotating from the mass eigenbasis
\begin{equation}
  [\gamma^2]_{l}^{\phantom{l}m} \
  = \ U_{l}^{\phantom{l}k}U^{m}_{\phantom{m}k}\widehat{\gamma}^2_k\;, 
\end{equation}
where
\begin{equation}
\widehat{\gamma}^{2}_k\ =\ 1 \: + \: \frac{|\mathbf{p}|^2}{m^2_k} \;.
\end{equation}
In this case, we obtain the flavour-covariant generalization of the Einstein 
mass-energy relation 
\begin{equation}
E_{k}^{\phantom{k}l}\ =\ \gamma_{k}^{\phantom{k}m}m_m^{\phantom{m}l}\; .
\end{equation}
Thus, we see
that boosting to  the rest frame of a  decaying flavour coherence with
respect  to a definite  three-momentum, all  Lorentz-covariant objects
will naturally become rank-2 tensors in this frame.

The Dirac-conjugate  spinors can  be written explicitly,  analogous to
\eqref{eq:4spinors1} and \eqref{eq:4spinors2}:
\begin{align}
  [\bar{u}(\mathbf{p},s)]^{k}_{\phantom{k}l}\ &=\ 
  [u(\mathbf{0},s)]^{k}_{\phantom{k}m}\gamma^0
  \{[2m(E(\ve{p}) \: + \: m)]^{-1/2}\}^{m}_{\phantom{m}n}
  [\slashed{p} \: + \:m]^{n}_{\phantom{n}l}\;,\\
  [\bar{v}(\mathbf{p},s)]^{k}_{\phantom{k}l}\ & = \
  [v(\mathbf{0},s)]^{k}_{\phantom{k}m}\gamma^0
  \{[2m(E(\ve{p}) \: + \: m)]^{-1/2}\}^{m}_{\phantom{m}n}
  [-\slashed{p} \: + \: m]^{n}_{\phantom{n}l}\;,
\end{align}
thus yielding
\begin{align}
  [\bar{u}(\mathbf{p},s)]^{k}_{\phantom{k}l} \ &
  = \ \frac{e^{-i\varphi^s_{u}}}{\sqrt{2}}
  \Big(
    [(E(\ve{p}) \: + \: m)^{1/2}]^{k}_{\phantom{k}l}\,u^{\dag}_s(\mathbf{s})\;,\ 
    -s\,[(E(\ve{p}) \: - \: m)^{1/2}]^{k}_{\phantom{k}l}\,u^{\dag}_s(\mathbf{s})
  \Big)\;,\\
  [\bar{v}(\mathbf{p},s)]^{k}_{\phantom{k}l}\ &
  = \ \frac{e^{-i\varphi^{-s}_{v}}}{\sqrt{2}}
  \Big(
    [(E(\ve{p}) \: - \: m)^{1/2}]^{k}_{\phantom{k}l}\,u^{\dag}_{-s}(\mathbf{s})\;,\
    s\,[(E(\ve{p}) \: + \: m)^{1/2}]^{k}_{\phantom{k}l}\,u^{\dag}_{-s}(\mathbf{s})
  \Big)\;.
\end{align}

The matrix product of two spinors is given by 
\begin{equation}
  u_s(\mathbf{s})u_{s'}^{\dag}(\mathbf{s}')\ \equiv \ 
  A_{ss'}(\mathbf{s},\mathbf{s}')\\
  =\ A_{s's}^{\dag}(\mathbf{s}',\mathbf{s})\;.
\end{equation}
In  the   homogeneous  limit  $\mathbf{s}\:=\:\mathbf{s}'$,   we  have
$A_{ss}(\mathbf{s},\mathbf{s})         =         (\sigma^0        \pm
\bm{\sigma}\cdot\mathbf{s})$ for $s=\pm$ and
\begin{equation}
  \label{eq:Asum}
  \sum_{s\:=\:s'} A_{ss'}(\mathbf{s},\mathbf{s})\ =\
  2\sigma^0\;,\qquad \sum_{s\:=\:s'} s A_{ss'}(\mathbf{s},\mathbf{s})\ =\
  2\bm{\sigma}\cdot\mathbf{s}\;.
\end{equation}
The scalar product of two-spinors is
\begin{equation}
  \label{eq:spinfac}
  u_s^{\dag}(\mathbf{s})u_{s'}(\mathbf{s}')\
  \equiv \ \Theta_{ss'}(\mathbf{s},\mathbf{s}')\ 
  =\ \mathrm{Tr}[A_{ss'}^{\dag}(\mathbf{s},\mathbf{s}')]\;.
\end{equation}
Hence, in the homogeneous limit $\mathbf{s}\:=\:\mathbf{s}'$, we have 
$\Theta_{ss'}(\mathbf{s},\mathbf{s}) = 2\delta_{ss'}$.

The $16$ possible contractions of the Dirac four-spinors are
summarized as follows:
\begin{align}
  [\bar{u}(\mathbf{p},s)]^k_{\phantom{k}l}
  [u(\mathbf{p}',s')]_m^{\phantom{m}n}\ &=\
  [\mathscr{N}(p,s;p',s')]^{k\phantom{l}\phantom{m}n}_{\phantom{k} l m}
  \exp[-i(\varphi^s_{u}\:-\:\varphi^{s'}_{u})]\;,\\
  [\bar{v}(\mathbf{p},s)]^k_{\phantom{k}l}
  [v(\mathbf{p}',s')]_m^{\phantom{m}n}\ &=\
  -\:[\mathscr{N}(-p,-s;-p',-s')]^{k\phantom{l}\phantom{m}n}_{\phantom{k} l m}
  \exp[-i(\varphi^{-s}_{v}\:-\:\varphi^{-s'}_{v})]\;,\\
  [\bar{u}(\mathbf{p},s)]^k_{\phantom{k}l}
  [v(\mathbf{p}',s')]_{m}^{\phantom{m}n}\ &=\
  i[\mathscr{N}(p,s;-p',-s')]^{k\phantom{l}\phantom{m}n}_{\phantom{k} l m}
  \exp[-i(\varphi^s_{u}\:-\:\varphi^{-s}_{v})]\;,\\
  [\bar{v}(\mathbf{p},s)]^k_{\phantom{k}l}
  [u(\mathbf{p}',s')]_m^{\phantom{m}n}\ &=\
  i[\mathscr{N}(-p,-s;p',s')]^{k\phantom{l}\phantom{m}n}_{\phantom{k} l m}
  \exp[-i(\varphi^{-s}_{v}\:-\:\varphi^{s'}_{u})]\;,
\end{align}
where $p_0 = E(\ve{p})$ is understood to be on-shell and we have defined
\begin{align}
\label{eq:norm}
  [\mathscr{N}(p,s;p',s')]^{k\phantom{l}\phantom{m}n}_{\phantom{k} l m} \ & = \
  \frac{1}{2}\,\Big\{\,[(p_0 \: + \: m)^{1/2}]^{k}_{\phantom{k}l}
  [(p_0' \: + \: m)^{1/2}]_{m}^{\phantom{m}n}
  \nonumber\\&\qquad 
    - \: ss'\,[(p_0 \: - \: m)^{1/2}]^{k}_{\phantom{k}l}
    [(p_0' \: - \: m)^{1/2}]_{m}^{\phantom{m}n}\,\Big\}\,
    \Theta_{ss'}(\mathbf{s},\mathbf{s}')\;.
\end{align}
In the homogeneous limit $\mathbf{p}  =  \mathbf{p}'$, we have
\begin{equation}
  [\mathscr{N}(p,s;p,s')]^{m\phantom{k}\phantom{m}l}_{\phantom{m}k m} \
  = \ 2m_{k}^{\phantom{k}l}\delta_{ss'}\;,\qquad
  [\mathscr{N}(p,s;-p,s')]^{m\phantom{k}\phantom{m}l}_{\phantom{m}k m}\ =\ 0\;.
\end{equation}
We then recover the familiar identities
\begin{align}
  [\bar{u}(\mathbf{p},s)]^{m}_{\phantom{m}k}
  [u(\mathbf{p},s')]_{m}^{\phantom{m}l} \ & = \
  -[\bar{v}(\mathbf{p},s)]^{m}_{\phantom{m}k}
  [v(\mathbf{p},s')]_{m}^{\phantom{m}l}\ 
  =\ 2m_{k}^{\phantom{k}l}\delta_{ss'}\;,\\
  [\bar{u}(\mathbf{p},s)]^{m}_{\phantom{m}k}
  [v(\mathbf{p},s')]_{m}^{\phantom{m}l}\ &=\ 
  [\bar{v}(\mathbf{p},s)]^{m}_{\phantom{m}k}
  [u(\mathbf{p},s')]_{m}^{\phantom{m}l}\ =\ 0\;,
\end{align}
as we would expect.

The 16 possible matrix products of the four spinors are:
\begin{align}
  [u(\mathbf{p},s)]_k^{\phantom{k}l}
  [\bar{u}(\mathbf{p}',s')]^{m}_{\phantom{m}n}\ &=\
  [\mathscr{P}(p,s;p',s')]_{k\phantom{l}\phantom{m}n}^{\phantom{k}lm}
  \exp[i(\varphi^s_{u}\:-\:\varphi^{s'}_{u})]\;,\nonumber\\
  [v(\mathbf{p},s)]_k^{\phantom{k}l}
  [\bar{v}(\mathbf{p}',s')]^{m}_{\phantom{m}n}\ &=\
  -[\mathscr{P}(-p,-s;-p,-s')]_{k\phantom{l}\phantom{m}n}^{\phantom{k}lm}
  \exp[i(\varphi^{-s}_{v}\:-\:\varphi^{-s'}_{v})]\;,\nonumber\\
  [u(\mathbf{p},s)]_k^{\phantom{k}l}
  [\bar{v}(\mathbf{p}',s')]^{m}_{\phantom{m}n}\ &=\
  i[\mathscr{P}(p,s;-p',-s')]_{k\phantom{l}\phantom{m}n}^{\phantom{k}lm}
  \exp[i(\varphi^{s}_{u}\:-\:\varphi^{s'}_{v})]\;,\nonumber\\
  [v(\mathbf{p},s)]_k^{\phantom{k}l}
  [\bar{u}(\mathbf{p}',s')]^{m}_{\phantom{m}n}\ &=\
  i[\mathscr{P}(-p,-s;p',s')]_{k\phantom{l}\phantom{m}n}^{\phantom{k}lm}
  \exp[i(\varphi^{-s}_{v}\:-\:\varphi^{s'}_{u})]\;,
  \label{eq:dyads}
\end{align}
where we have introduced
\begin{align} \label{Pps}
  &[\mathscr{P}(p,s;p',s')]_{k\phantom{l}\phantom{m}n}^{\phantom{k}lm} \ = \
 \\ & 
  \frac{1}{2} \begin{pmatrix} [(p_0 \: + \: m)^{1/2}]_{k}^{\phantom{k}l}
    [(p_0' \: + \: m)^{1/2}]^{m}_{\phantom{m}n} &
    - \: s'\,[(p_0 \: + \: m)^{1/2}]_{k}^{\phantom{k}l}
    [(p_0' \: - \: m)^{1/2}]^{m}_{\phantom{m}n} \\
    s\,[(p_0 \: - \: m)^{1/2}]_{k}^{\phantom{k}l}
    [(p_0' \: + \: m)^{1/2}]^{m}_{\phantom{m}n} &
    - \: ss'\,[(p_0 \: - \: m)^{1/2}]_{k}^{\phantom{k}l}
    [(p_0' \: - \: m)^{1/2}]^{m}_{\phantom{m}n}
  \end{pmatrix}
  \otimes \: A_{ss'}(\mathbf{s},\mathbf{s}')\;, \nonumber
\end{align}
which, in the homogeneous limit $\mathbf{p} = \mathbf{p}'$, reduces to 
\begin{align}
  [\mathscr{P}(p,s;p,s')]_{k\phantom{m}\phantom{l}m}^{\phantom{k}ml}\
  = \ \frac{1}{2}
  \begin{pmatrix} [p_0 \: + \: m]_{k}^{\phantom{k}l} &
    - \: s'\,|\mathbf{p}|\,\delta_{k}^{\phantom{k}l} \\
    s\,|\mathbf{p}|\,\delta_{k}^{\phantom{k}l} &
    - \: ss' \, [p_0 \: - \: m]_{k}^{\phantom{k}l}
  \end{pmatrix} \: \otimes \: A_{ss'}(\mathbf{s},\mathbf{s})\;.
\end{align}
In this case, we obtain the familiar helicity sums
\begin{equation}
  \sum_{s\,=\,s'}
  [u(\mathbf{p},s)]_{k}^{\phantom{k}m}
  [\bar{u}(\mathbf{p},s')]^{l}_{\phantom{l}m} \
  = \ [\slashed{p} \: + \: m]_{k}^{\phantom{k}l}\;,\qquad
  \sum_{s\,=\,s'}
  [v(\mathbf{p},s)]_{k}^{\phantom{k}m}
  [\bar{v}(\mathbf{p},s')]^{l}_{\phantom{l}m} \
  = \ [\slashed{p} \: - \: m]_{k}^{\phantom{k}l}\;,
\end{equation}
using the summations in \eqref{eq:Asum}.

\subsection*{B.2\ Discrete Symmetry Transformations}

In this  section, we summarize  the spinor identities relevant  to the
$C$,  $P$ and  $T$  transformations.  In  particular,  we justify  the
phases appearing in  the generalized discrete symmetry transformations
described in  Section~\ref{sec:3.2} for the  creation and annihilation
operators       [cf.~\eqref{eq:b_C},      \eqref{eq:P_trans1}      and
\eqref{eq:b_T}].

 Under $C$-transformations, we have
\begin{align}
 \big( [u(\mathbf{p},s)]_{k}^{\phantom{k}l}\big)^C \
  & = \ iC\exp[i(\varphi_u^s \: + \: \varphi_{v}^{-s})]
  \mathcal{G}_{kn}[\bar{v}^{\mathsf{T}}(\mathbf{p},s)]_{m}^{\phantom{m}n}
  \mathcal{G}^{ml}\;,\\
 \big( [v(\mathbf{p},s)]_{k}^{\phantom{k}l} \big)^C \
  & = \ iC\exp[i(\varphi_{v}^{-s}\:+\:\varphi_{u}^{s})]
  \mathcal{G}_{kn}[\bar{u}^{\mathsf{T}}(\mathbf{p},s)]_{m}^{\phantom{m}n}
  \mathcal{G}^{ml}\;,
\end{align}
where $C   = i\gamma^0\gamma^2$, and the $\mathcal{G}$  matrix is
defined   in  \eqref{eq:C_trans}. Under   parity,
\begin{align}
  \big([u(\mathbf{p},s)]_{k}^{\phantom{k}l}\big)^{P} \
  & = \ [u(-\mathbf{p},-s)]_{k}^{\phantom{k}l}
  = \ -sP\exp[-i(\varphi_u^s-\varphi_u^{-s})]
  [u(\mathbf{p},s)]_{k}^{\phantom{k}l}\;,\\
  \big([v(\mathbf{p},s)]_{k}^{\phantom{k}l}\big)^{P} \
  & = \ [v(-\mathbf{p},-s)]_{k}^{\phantom{k}l} \ = \
  +sP\exp[-i(\varphi_v^{-s}-\varphi_{v}^s)]
  [v(\mathbf{p},s)]_{k}^{\phantom{k}l}\;,
\end{align}
where  $P  = \gamma^0$.  Finally, under   $T$-transformations, we
have
\begin{align}
  \big([u(\mathbf{p},s)]_{k}^{\phantom{k}l}\big)^{T} \
  & = \ [u(-\mathbf{p},s)]^{k}_{\phantom{k}l}
  = \ e^{-2i\varphi_u^s}T
  \mathcal{G}^{km}[u(\mathbf{p},s)]_{m}^{\phantom{m}n}\mathcal{G}_{nl}\;,\\
  \big([v(\mathbf{p},s)]_{k}^{\phantom{k}l}\big)^{T} \
  & = \ [v(-\mathbf{p},s)]^{k}_{\phantom{k}l} \ = \
  e^{-2i\varphi_v^{-s}}T\mathcal{G}^{km}
  [v(\mathbf{p},s)]_{m}^{\phantom{m}n}\mathcal{G}_{nl}\;,
\end{align}
where $T = i\gamma^1\gamma^3$. All  of these identities may readily be
verified using those in \eqref{eq:twospinident1} for the two component
spinors.  Notice that in  the mass eigenbasis, where the $\mathcal{G}$
matrices  are proportional to  the identity,  we recover  the expected
results.

\subsection*{B.3\ Chiral Field Operators}

We write the chiral projection operators
\begin{equation}
  \rm{P}_{\chi}\ =\
  \frac{1}{2}\big(\,\maf{1}_4 + \chi\gamma^5\,\big)\;,
\end{equation}
by introducing a dummy index  $\chi = \pm$, such that $\rm{P}_+ \equiv
\rm{P_{R}}$  and   $\rm{P}_-  \equiv  \rm{P_{L}}$,   as  appearing  in
\eqref{self_LR}.  With the chiral field definitions
\begin{equation} \label{chiral-psi}
  \rm{P}_{\chi}\psi_{k}(x)\ \equiv \ \psi_{\chi,\,k}(x)\;,\qquad
  \bar{\psi}^{k}(x)\rm{P}_{\chi}\ \equiv \ \bar{\psi}_{\chi}^{k}(x)\;,
\end{equation}
 the Dirac Lagrangian \eqref{Dirac_lag} takes the form
\begin{equation}
  \mathcal{L}_{\mathrm{D}}(x) \
  = \sum_{\chi} \bar{\psi}_{\chi}^k(x) (
  i\gamma^{\mu}\partial_{\mu}  \psi_{\chi,\,k}(x) \:-\:  m_{k}^{\phantom{k}l}
  \psi_{-\chi,\,l}(x) \big)\;.
\end{equation}
In   terms   of   the   Dirac   four   spinors   \eqref{dirac4u}   and
\eqref{dirac4v}, we get
\begin{align} \label{chiral4u}
  {\rm P}_{\chi}[u(\mathbf{p},s)]_{k}^{\phantom{k}l} \ &
      = \ \frac{e^{i\varphi^s_{u}}}{\sqrt{2}}\,
      [\mathscr{E}({p},\chi s)]_{k}^{\phantom{k}l}
      \begin{pmatrix}
        u_s(\mathbf{s}) \\
        \chi u_s(\mathbf{s})
      \end{pmatrix}\;,\\
      {\rm P}_{\chi}[v(\mathbf{p},s)]_{k}^{\phantom{k}l} \ & = \
      -s\,\frac{e^{i\varphi^{-s}_{v}}}{\sqrt{2}}\,
      [\mathscr{E}({p},-\chi s)]_{k}^{\phantom{k}l}
      \begin{pmatrix}
        \chi u_{-s}(\mathbf{s}) \\
        u_{-s}(\mathbf{s})
      \end{pmatrix}\;, \label{chiral4v}
    \end{align}
where we have introduced
\begin{equation}
  [\mathscr{E}(p,\chi s)]_{k}^{\phantom{k}l} \
  = \ \frac{1}{2}\{[(p_0 + m)^{1/2}]_{k}^{\phantom{k}l}
    \: + \: \chi s\,[(p_0 -  m)^{1/2}]_{k}^{\phantom{k}l}\}\;.
\end{equation}
From \eqref{chiral4u} and \eqref{chiral4v}, we find that 
\begin{equation}
  \label{utov}
  {\rm P}_{\chi}[u(\mathbf{p},s)]_{k}^{\phantom{k}l} \
  = \ s\chi e^{i(\varphi^s_{u}-\varphi^s_{v})}
  {\rm P}_{\chi}[v(\mathbf{p},-s)]_{k}^{\phantom{k}l}\;.
\end{equation}
Hence, we can define four independent chiral four-spinors, as follows:
\begin{equation}
  [\xi_{\chi}(\mathbf{p})]_{k}^{\phantom{k}l} \
  = \ \frac{1}{\sqrt{2}}[\mathscr{E}({p},\chi)]_{k}^{\phantom{k}l}
  e^{i\varphi_u^+}
  \begin{pmatrix}
    u_+(\mathbf{s}) \\
    \chi u_+(\mathbf{s})
  \end{pmatrix}\;,\qquad
  [\eta_{\chi}(\mathbf{p})]_{k}^{\phantom{k}l} \
  = \ \frac{1}{\sqrt{2}}
  [\mathscr{E}({p},-\chi)]_{k}^{\phantom{k}l}
  e^{i\varphi_u^-}\begin{pmatrix}
    u_-(\mathbf{s}) \\
    \chi u_-(\mathbf{s})
  \end{pmatrix}\;.
\end{equation}
We may then expand the chiral field operators \eqref{chiral-psi} 
in terms of the following chiral four-spinors: 
\begin{align}
 \psi_{\chi,\,k}(x) & \ = \ \int_{\ve{p}}
 [(2E(\mathbf{p}))^{-1}]_{k}^{\phantom{k}l}\Big\{
  [e^{-ip\cdot x}]_{l}^{\phantom{l}m}\Big(
  [\xi_{\chi}(\mathbf{p})]_{m}^{\phantom{m}n}
  \,\underline{b}_{n}(\mathbf{p},+,0) + 
  [\eta_{\chi}(\mathbf{p})]_{m}^{\phantom{m}n}
  \,\underline{b}_{n}(\mathbf{p},-,0)\Big)
 \nonumber\\  &   
  -\chi [e^{ip\cdot x}]_l^{\phantom{l}m}
  \Big(e^{-i(\varphi^-_{u}-\varphi^-_{v})}
  [\eta_{\chi}(\mathbf{p})]_{m}^{\phantom{m}n}\,
  \underline{d}^{\dag}_{n}(\mathbf{p},+,0) - 
   e^{-i(\varphi^+_{u}-\varphi^+_{v})}
   [\xi_{\chi}(\mathbf{p})]_{m}^{\phantom{m}n}\,
   \underline{d}^{\dag}_{n}(\mathbf{p},-,0)\Big)\Big\},\\
\label{chiral-psi2a}
  \bar{\psi}_{\chi}^{k}(x) \ & =\ \int_{\ve{p}}
[(2E(\mathbf{p}))^{-1}]^{k}_{\phantom{k}l} 
\Big\{ [e^{-ip\cdot x}]^{l}_{\phantom{l}m}
 \Big([\bar{\xi}_{\chi}(\mathbf{p})]^{m}_{\phantom{m}n}\,
  \underline{b}^n(\mathbf{p},+,0)+
  [\bar{\eta}_{\chi}(\mathbf{p})]^{m}_{\phantom{m}n}\,
  \underline{b}^{n}(\mathbf{p},-,0)\,\Big) 
  \nonumber\\  & 
   - \chi [e^{ip\cdot x}]^{l}_{\phantom{l}m}
   \Big(\,e^{i(\varphi^-_{u}-\varphi^-_{v})}
  [\bar{\eta}_{\chi}(\mathbf{p})]^{m}_{\phantom{m}n}\,
  \underline{d}^{\dag\, n}(\mathbf{p},+,0)-
  e^{i(\varphi_{u}^+-\varphi_{v}^+)}
  [\bar{\xi}_{\chi}(\mathbf{p})]^{m}_{\phantom{m}n}\,
  \underline{d}^{\dag\, n}(\mathbf{p},-,0)\,\Big)\Big\}.
\end{align}
Notice that in the relativistic  limit $E \gg m$, the helicity and
chirality states coincide and only the $\xi_{\chi}$ spinors survive. Hereafter, we neglect the $\varphi^s_{u(v)}$-dependent phases for notational convenience.

\subsection*{B.4\ Spinor Traces}

We may  now define  the following spinor  trace involving  the objects
defined in \eqref{Pps}:
\begin{align}
  \label{eq:quadtrace}
  &\mathrm{Tr}\left\{[\mathscr{P}(p,s;p',s')]
    _{k\phantom{l}\phantom{m}n}^{\phantom{k}l m}
  {\rm P}_{\chi}
  [\mathscr{P}(q,r;q',r')]_{{k'}\phantom{l'}\phantom{m'}n'}^{\phantom{k'}l' m'}
  {\rm P}_{\chi'}\right\} 
\nonumber\\ & \quad 
\ = \ 
  [\mathscr{E}(p,\chi' s)]_{k}^{\phantom{k}l}
  [\mathscr{E}(p',-\chi s')]^{m}_{\phantom{m}n}
  [\mathscr{E}(q,\chi r)]_{{k'}}^{\phantom{k'}l'}
  [\mathscr{E}(q',-\chi' r')]^{{m'}}_{\phantom{m'}n'}
  \Theta_{sr'}^*(\mathbf{s},\mathbf{r}')
  \Theta_{rs'}^*(\mathbf{r},\mathbf{s}')\;.
\end{align}
where we have used the properties of matrix products and
the cyclicity of the trace to write 
\begin{equation} \label{theta-rs}
  \mathrm{Tr}\left\{A_{ss'}(\mathbf{s},\mathbf{s}')
    A_{rr'}(\mathbf{r},\mathbf{r}')\right\} 
  \ =\ \Theta_{sr'}^*(\mathbf{s},\mathbf{r}')
  \Theta_{rs'}^*(\mathbf{r},\mathbf{s}')\;.
\end{equation}
Using \eqref{eq:dyads} and \eqref{utov},  we may relate \eqref{eq:quadtrace}  to the 16
possible traces of the Dirac four-spinors, as follows:
\begin{align} \label{uv-trace}
  &\mathrm{Tr}\,\Big\{[u(\mathbf{p},s)]_{k}^{\phantom{k}l}
  [\bar{u}(\mathbf{p}',s')]^{m}_{\phantom{m}n}{\rm P}_{\chi}
  [u(\mathbf{q},r)]_{k'}^{\phantom{k'}l'}
  [\bar{u}(\mathbf{q}',r')]^{m'}_{\phantom{m'}n'}{\rm P}_{\chi'}\Big\}
  \nonumber\\ & \quad =\
  ss'rr'\,\mathrm{Tr}\,\Big\{[v(\mathbf{p},-s)]_{k}^{\phantom{k}l}
  [\bar{v}(\mathbf{p}',-s')]^{m}_{\phantom{m}n}{\rm P}_{\chi}
  [v(\mathbf{q},-r)]_{k'}^{\phantom{k'}l'}
  [\bar{v}(\mathbf{q}',-r')]^{m'}_{\phantom{m'}n'}{\rm P}_{\chi'}\Big\}
  \nonumber\\ & \quad =\
  -\:rr'\chi\chi'\,\mathrm{Tr}\,\Big\{[u(\mathbf{p},s)]_{k}^{\phantom{k}l}
  [\bar{u}(\mathbf{p}',s')]^{m}_{\phantom{m}n}{\rm P}_{\chi}
  [v(\mathbf{q},-r)]_{k'}^{\phantom{k'}l'}
  [\bar{v}(\mathbf{q}',-r')]^{m'}_{\phantom{m'}n'}{\rm P}_{\chi'}\Big\}
  \nonumber \\&\quad =\
  -\:ss'\chi\chi'\,\mathrm{Tr}\,\Big\{[v(\mathbf{p},-s)]_{k}^{\phantom{k}l}
  [\bar{v}(\mathbf{p}',-s')]^{m}_{\phantom{m}n}{\rm P}_{\chi}
  [u(\mathbf{q},r)]_{k'}^{\phantom{k'}l'}
  [\bar{u}(\mathbf{q}',r')]^{m'}_{\phantom{m'}n'}{\rm P}_{\chi'}\Big\}\;,
\end{align}
generalizing the result used in \eqref{eq:traces}.  In the homogeneous
limit $p = p'$ and $q = q'$, \eqref{theta-rs} is given by
\begin{equation}
  \Theta_{sr}^{*}(\mathbf{s},\mathbf{r})
  \Theta_{rs}^{*}(\mathbf{r},\mathbf{s})
  \ =\ |\Theta_{sr}(\mathbf{s},\mathbf{r})|^2\ =\
  2\,\bigg(1\:+\:sr\,
  \frac{\mathbf{p}\cdot \mathbf{q}}{|\mathbf{p}||\mathbf{q}|}\bigg)\;.
\end{equation}
and \eqref{eq:quadtrace} is given by 
\begin{align}
  \label{eq:fulltrace}
  & \mathrm{Tr}\left\{[\mathscr{P}(p,s;p,s)]
    _{k\phantom{m}\phantom{l}m}^{\phantom{k} m l}
    {\rm P}_{\chi}
    [\mathscr{P}(q,r;q,r)]_{{k'}\phantom{m'}\phantom{l'}m'}^{\phantom{k'} m' l'}
    {\rm P}_{\chi'} \right\} 
  \nonumber\\  
  & \qquad \ = \ 
  \frac{1}{8}
  \bigg(1+ sr\,\frac{\mathbf{p}\cdot \mathbf{q}}
  {|\mathbf{p}||\mathbf{q}|}\bigg)
  \: [p_0+ m - (\chi - \chi')s|\mathbf{p}|
   - \chi\chi'(p_0 - m)]_{k}^{\phantom{k}l} 
  \nonumber\\&\qquad \qquad \qquad \qquad
  \times\:  
  [q_0 + m + (\chi - \chi')r|\mathbf{q}| 
  - \chi\chi'(q_0 - m)]_{k'}^{\phantom{k'} l'} .
\end{align}
Summing over the helicities $s$ and $r$, we find
\begin{align}
  &\sum_{s,\,r}\mathrm{Tr}\left\{
  [\mathscr{P}(p,s;p,s)]
  _{k\phantom{m}\phantom{l}m}^{\phantom{k} m l}
  {\rm P}_{\chi}
  [\mathscr{P}(q,r;q,r)]
  _{{k'}\phantom{m'}\phantom{l'}m'}^{\phantom{k'} m' l'}
  {\rm P}_{\chi'} \right\} 
  \nonumber\\ & = \ 
  \frac{1}{2} 
  \Big([p_0 + m - \chi\chi'(p_0 - m)]_{k}^{\phantom{k} l}
  [q_0 + m  - \chi\chi'(q_0  - m)]_{k'}^{\phantom{k'}l'}
  - (\chi  -  \chi')^2\,\mathbf{p}\cdot\mathbf{q}\,
  \delta_{k}^{\phantom{k}l}\delta_{k'}^{\phantom{k'}l'}\Big)  .
\end{align}
The spinor  traces \eqref{uv-trace} can  be summed similarly  over $s$
and $r$, e.g.
\begin{align}
  \label{eq:fintrace}
  &\sum_{s,\,r}\mathrm{Tr}\left\{[u(\mathbf{p},s)]_{k}^{\phantom{k}m}
    [\bar{u}(\mathbf{p},s)]^{l}_{\phantom{l}m}{\rm P}_{\chi}
    [u(\mathbf{q},r)]_{k'}^{\phantom{k'}m'}
    [\bar{u}(\mathbf{q},r)]^{l'}_{\phantom{l'}m'}P_{\chi'}\right\}
 \nonumber\\ &\qquad
 =\ -\:\chi\chi'\,\sum_{s,\,r}\mathrm{Tr}\left\{[u(\mathbf{p},s)]_{k}^{\phantom{k}m}
  [\bar{u}(\mathbf{p},s)]^{l}_{\phantom{l}m}{\rm P}_{\chi}
  [v(\mathbf{q},r)]_{k'}^{\phantom{k'}m'}
  [\bar{v}(\mathbf{q},r)]^{l'}_{\phantom{l'}m'}{\rm P}_{\chi'}\right\}
  \nonumber\\&\qquad \qquad
  =\ \begin{cases}
  2p_{k}^{\phantom{k}l}\cdot q_{k'}^{\phantom{k'}l'},\quad &
  \chi = -\chi'\;;\\
  2m_k^{\phantom{k}l}m_{k'}^{\phantom{k'}l'},\quad &
  \chi = +\chi'\; .
  \end{cases}
\end{align}  
Note that  in the limit  when one pair  of Dirac spinors  is massless,
\eqref{eq:fintrace} reduces to the result quoted in \eqref{eq:traces}:
\begin{align}
\label{eq:fintrace2}
  &\sum_{s}\mathrm{Tr}\Big\{[u(\mathbf{p},s)]_{k}^{\phantom{k}m}
    [\bar{u}(\mathbf{p},s)]^{l}_{\phantom{l}m}{\rm P}_{\chi}
    u(\mathbf{q},\chi)
    \bar{u}(\mathbf{q},\chi){\rm P}_{\chi'}\Big\}
  \nonumber\\ & \qquad
  = -\:\chi\chi'\,\sum_{s}\mathrm{Tr}\Big\{[u(\mathbf{p},s)]_{k}^{\phantom{k}m}
  [\bar{u}(\mathbf{p},s)]^{l}_{\phantom{l}m}{\rm P}_{\chi}
  v(\mathbf{q},-\chi)
  \bar{v}(\mathbf{q},-\chi){\rm P}_{\chi'}\Big\}
 \nonumber\\&\qquad \qquad = \ \begin{cases}
  2p_{k}^{\phantom{k}l}\cdot q\;,\quad &
  \chi\ =\ -\chi'\;;\\
  0\;,\quad &
  \chi\ =\ +\chi'\;,
  \end{cases}
\end{align}
where  the massless  spinors are  denoted  by the  absence of  flavour
indices.   In  \eqref{eq:fintrace2},  the  factor of  $2$  for  $\chi=
-\chi'$ remains  relative to  \eqref{eq:fintrace} in spite  of summing
only over the helicity $s$, due to the action of the chiral projection
operators ${\rm P}_{\chi}$ and  ${\rm P}_{-\chi}$. This may be readily
confirmed from the general expression in \eqref{eq:fulltrace}.

\subsection*{B.5\ Flavour Covariant Free Propagators}
\label{sec:props}

A full  quantum-field theoretic description of  transport phenomena in
the  context of  flavour oscillations,  mixing and  coherences  may be
provided  by   master  equations  derived   by  \cite{Calzetta:1986ey,
Calzetta:1986cq} from the  CJT effective action \cite{Cornwall:1974vz}
as       applied        to       the       Schwinger-Keldysh       CTP
formalism~\cite{Schwinger:1961,  Keldysh:1964}.  A  full  treatment of
such a  derivation is  beyond the  scope of this  article and  will be
deferred to a future publication.   However, in order to exemplify the
consistency  of  the   flavour-covariant  Markovian  master  equations
derived in  Section~\ref{sec:4}, we highlight  below pertinent details
of the flavour-covariant propagators  that would necessarily appear in
such a treatment.

In order to define the relevant free flavour-covariant non-homogeneous
propagators  in the  CTP formalism,  we define  the  bilinear ensemble
expectation  values  (EEVs) of  creation  and annihilation  operators,
generalizing    those   introduced   for    the   scalar    field   in
\cite{Millington:2012pf} to an $\mathcal{N}$-flavour fermion model:
\begin{align}
  \label{eq:bilins}
  \braket{\underline{b}^{l}(\mathbf{p}',s',\tilde{t})
    \underline{b}_{k}(\mathbf{p},s,\tilde{t})}_t\ &=\ 
  \big[\big(2E(\mathbf{p})\big)^{1/2}\big]_{k}^{\phantom{k}m}
  [f_{ss'}(\mathbf{p},\mathbf{p}',t)]_{m}^{\phantom{m}n}
  \big[\big(2E(\mathbf{p}')\big)^{1/2}\big]^{l}_{\phantom{l}n}\;,
  \nonumber\\
  \braket{\underline{d}_{k}^{\dag}(\mathbf{p},s,\tilde{t})
  \underline{d}^{\dag\,l}(\mathbf{p}',s',\tilde{t})}_t\ &=\
  \big[\big(2E(\mathbf{p})\big)^{1/2}\big]_{k}^{\phantom{k}m}
  [\bar{f}_{ss'}(\mathbf{p},\mathbf{p}',t)]_{m}^{\phantom{m}n}
  \big[\big(2E(\mathbf{p}')\big)^{1/2}\big]^{l}_{\phantom{l}n}\;,
  \nonumber\\
  \braket{\underline{d}^{\dag\, l}(\mathbf{p}',s',\tilde{t})
    \underline{b}_{k}(\mathbf{p},s,\tilde{t})}_t\ &=\
  \big[\big(2E(\mathbf{p})\big)^{1/2}\big]_{k}^{\phantom{k}m}
  [g_{ss'}(\mathbf{p},\mathbf{p}',t)]_{m}^{\phantom{m}n}
  \big[\big(2E(\mathbf{p}')\big)^{1/2}\big]^{l}_{\phantom{l}n}\;,
  \nonumber\\
  \braket{\underline{d}^{\dag}_{k}(\mathbf{p},s,\tilde{t})
  \underline{b}^{l}(\mathbf{p}',s',\tilde{t})}_t\ &=\
  \big[\big(2E(\mathbf{p})\big)^{1/2}\big]_{k}^{\phantom{k}m}
 [\bar{g}_{ss'}(\mathbf{p},\mathbf{p}',t)]_{m}^{\phantom{m}n}
 \big[\big(2E(\mathbf{p}')\big)^{1/2}\big]^{l}_{\phantom{l}n}\;.
\end{align}
The form of  the energy factors appearing in  \eqref{eq:bilins} may be
justified   by    inverting   the   Bogoliubov    transformations   in
\eqref{eq:bog1}  and  \eqref{eq:bog2}.  Note  that  we  have  set  the
arbitrary phases $\varphi_{u,v}^{\pm}$ to zero.

The    statistical   distribution   functions    $f$   and    $g$   in
\eqref{eq:bilins} satisfy the following properties:
\begin{align}
 &  [f_{ss'}(\mathbf{p},\mathbf{p}',t)]_{k}^{\phantom{k}l}\ 
  =\ ([f_{s's}(\mathbf{p}',\mathbf{p},t)]_{l}^{\phantom{l}k})^*\ =\ ([\bar{f}_{ss'}(\mathbf{p},\mathbf{p}',t)]_{k}^{\phantom{k}l})^{\tilde{C}*}\;, \\
 &  [g_{ss'}(\mathbf{p},\mathbf{p}',t)]_{k}^{\phantom{k}l}\ 
  =\ ([\bar{g}_{ss'}(\mathbf{p},\mathbf{p}',t)]_{k}^{\phantom{k}l})^{\tilde{C}*}\;.
\end{align}
We assume a Gaussian density operator,  so that we must specify only the
bilinear EEVs  of operators.  Notice that  in this case  only the four
combinations   in   \eqref{eq:bilins}   are   permitted   by   the associated spinorial structure. In the homogeneous limit, we have the correspondence
\begin{align}
 & [f_{ss'}(\mathbf{p},\mathbf{p}',t)]_{k}^{\phantom{k}l}\ 
  \to \
  (2\pi)^3\delta^{(3)}(\mathbf{p}  - \mathbf{p}')\delta_{ss'}
  [f_s(\mathbf{p},t)]_{k}^{\phantom{k}l}\;, 
\qquad 
  [g_{ss'}(\mathbf{p},\mathbf{p}',t)]_{k}^{\phantom{k}l}\ 
   \to \ 0\;.
\end{align}

The positive and negative-frequency fermionic Wightman propagators are
defined as
  \begin{align}
    [iS_{>}(p,p',\tilde{t})]_{k}^{\phantom{k}l}\
    &= \ \braket{\psi_k(p)\bar{\psi}^{l}(p')}_t\;,\\
    [iS_{<}(p,p',\tilde{t})]_{k}^{\phantom{k}l}\
    &= \ -\braket{\bar{\psi}^{l}(p')\psi_k(p)}_t\;.
  \end{align}
By  evaluating the  EEV of  field  operators directly,  we obtain  the
explicit forms
  \begin{align}
  &[iS_{\gtrless}(p,p',\tilde{t})]_{k}^{\phantom{k}l}\
  =\ \sum_{s,\,s'}
  2\pi|2p_0|^{1/2}[\delta(p^2 - m^2)]_k^{\phantom{i}i}\: 
  2\pi|2p_0'|^{1/2}[\delta(p'^2 \: - \: m^2)]^l_{\phantom{l}m} \: 
e^{i(p_0 \: - \: p_0')\tilde{t}}
  \nonumber \\ & 
 \quad  \times\:
  [\mathscr{P}(p,s;p',s')]_{i\phantom{j}\phantom{m}n}^{\phantom{i}jm}
  \Big(\theta(\pm p_0)\theta(\pm p_0')(2\pi)^3
  \delta^{(3)}(\mathbf{p}  - \mathbf{p}')\delta_{ss'}
  \delta_{j}^{\phantom{n}n}
  -  [\tilde{f}_{ss'}(p,p',t)]_{j}^{\phantom{j}n}\Big)\;,
\label{wightman}
  \end{align}
where we have defined the ensemble function
\begin{align} 
  &[\tilde{f}_{ss'}(p,p',t)]_{k}^{\phantom{k}l}\
  =\ \theta(p_0)\theta(p_0')[f_{ss'}(\mathbf{p},\mathbf{p}',t)]_{k}^{\phantom{k}l}
  \: + \:
  i\,\theta(p_0)\theta(-p_0')
  [g_{s,-s'}(\mathbf{p},-\mathbf{p}',t)]_{k}^{\phantom{k}l}
  \nonumber\\&\qquad \:
  - \: i\,\theta(-p_0)\theta(p_0')
  [\bar{g}_{-s,s'}(-\mathbf{p},\mathbf{p}',t)]_{k}^{\phantom{k}l}
  \: + \:\theta(-p_0)\theta(-p_0')
  [\bar{f}_{-s,-s'}(-\mathbf{p},-\mathbf{p}',t)]_{k}^{\phantom{k}l}\;,
\label{fensemb}
\end{align}
satisfying the relation
\begin{equation}
  [\tilde{f}_{ss'}(p,p',t)]_{k}^{\phantom{k}l}\
  = \ ([\tilde{f}_{s',s}(p',p,t)]_{l}^{\phantom{l}k})^*\
  = \ ([\tilde{f}_{-s,-s'}(-p,-p',t)]
  _{k}^{\phantom{k}l})^{\tilde{C}*}\;.
\end{equation}

At   this  point,   we   make  an   important   observation.  In   the
spatially-homogeneous   limit   $\mathbf{p}\:   =\:\mathbf{p}'$,   the
ensemble function \eqref{fensemb} becomes
\begin{equation}
  [\tilde{f}_{ss'}(p,p',t)]_{k}^{\phantom{k}l}\
  \to \ 
  [\tilde{f}_{\mathrm{hom},\,ss'}(p_0,p_0',\mathbf{p},t)]_{k}^{\phantom{k}l}
  (2\pi)^3\delta^{(3)}(\mathbf{p}  -  \mathbf{p}')\;.
\end{equation}
In this case, the Wightman propagators \eqref{wightman} reduce to
\begin{align}
  \label{eq:prophom}
  [iS_{\gtrless}(p,p',\tilde{t})]_{k}^{\phantom{k}l}\
  & = \ \sum_{s,\,s'}
  2\pi|2p_0|^{1/2}[\delta(p_0^2 - E^2)]_k^{\phantom{i}i}
 \: 2\pi|2p_0'|^{1/2}[\delta(p_0'^2  -  E^2)]^l_{\phantom{l}m}
\: e^{i(p_0 \: - \: p_0')\tilde{t}} 
  \nonumber \\
  & \quad \times\:
  [\mathscr{P}(p,s;p',s')]_{i\phantom{j}\phantom{m}n}^{\phantom{i}jm}
  \Big(\theta(\pm p_0)\theta(\pm p_0')\delta_{ss'}
  \delta_{j}^{\phantom{n}n}\:
  - \: [\tilde{f}_{\mathrm{hom},ss'}(p_0,p_0',\mathbf{p},t)]
  _{j}^{\phantom{j}n}\Big)
\nonumber \\
  & \qquad \qquad \times\:
  (2\pi)^3\delta^{(3)}(\mathbf{p} - \mathbf{p}')\;,
\end{align}
Rotating to the mass eigenbasis, we obtain
\begin{align}
  \label{eq:prophomhat}
  &[i\widehat{S}_{\gtrless}(p,p',\tilde{t})]_{k}^{\phantom{k}l}\
   =\ \sum_{s,\,s'}
  2\pi|2p_0|^{1/2}\delta(p_0^2 - \widehat{E}_k^2)
  \: 2\pi|2p_0'|^{1/2}\delta(p_0'^2 - \widehat{E}_l^2)
  e^{i(p_0 \: - \: p_0')\tilde{t}}
  \nonumber \\ & \qquad
  \times\:
  [\mathscr{P}(p,s;p',s')]_{k\phantom{k}\phantom{l}l}^{\phantom{k}kl}
  \Big(\theta(\pm p_0)\theta(\pm p_0')\delta_{ss'}
  \delta_{l}^{\phantom{l}k}\:
  - \: [\tilde{f}_{\mathrm{hom},\,ss'}
  (p_0,p_0',\mathbf{p},t)]_{l}^{\phantom{l}k}\Big) 
  (2\pi)^3\delta^{(3)}(\mathbf{p}  -  \mathbf{p}')\;.
\end{align}
If  the system  is  out-of-equilibrium, flavour  coherences cannot  be
neglected.  As a  consequence, the  statistical  distribution function
$[f]_{k}^{\phantom{k}l}$  will in  general have  non-zero off-diagonal
entries.   In this case,  we see  from \eqref{eq:prophomhat}  that the
Wightman propagator depends explicitly on two zeroth component momenta
$p_0$  and  $p_0'$,  which  need  not  be equal.  In  this  case,  the
time-translational invariance of  the flavour-covariant propagators is
necessarily broken.

Assuming that $\tilde{f}$ is diagonal in flavour space and helicities in thermodynamic equilibrium, we have
\begin{equation}
 [\tilde{f}_{\mathrm{hom},\,ss'}(p_0,p_0',\mathbf{p},t)]_{k}^{\phantom{k}l}\
  \ \to \
  \tilde{f}_{\mathrm{eq},\, k}(p_0,p_0',\mathbf{p},t)\delta_k^{\phantom{k}l}
  \delta_{ss'}\;, \label{eq-f} 
\end{equation}
with
\begin{equation}
  \tilde{f}_{\mathrm{eq},\, k}(p_0,p_0',\mathbf{p},t)\ =\ 
  \theta(p_0)\theta(p_0')f_{\mathrm{F}}\big(E_k(\mathbf{p})\big) 
  + \theta(-p_0)\theta(-p_0')\bar{f}_{\mathrm{F}}\big(E_k(\mathbf{p})\big)\;,
\end{equation}
where $f_{\mathrm{F}}(E) =  [e^{(E-\mu)/T}+1]^{-1}$ is the Fermi-Dirac
particle distribution, and  $\bar{f}_{\mathrm{F}}(E)$ is the corresponding
anti-particle  distribution with $\mu \to -\mu$.  In  this equilibrium
limit \eqref{eq-f}, the Wightman propagators \eqref{eq:prophom} reduce
to
\begin{align}
  \label{eq:wighteq}
  &[iS_{\mathrm{eq},\,\gtrless}(p,p',\tilde{t})]_{k}^{\phantom{k}l}\
  =\  2\pi|2p_0|^{1/2}[\delta(p_0^2 \: - \: E^2)]_k^{\phantom{k}i}
  \: 2\pi|2p_0'|^{1/2}[\delta(p_0'^2 \: - \: E^2)]^l_{\phantom{l}m}
  \:  e^{i(p_0 \: - \: p_0')\tilde{t}} 
  \nonumber \\ 
  &\qquad\times\:  
  [\mathscr{P}(p_0,p_0',\mathbf{p})]_{i\phantom{j}\phantom{m}j}^{\phantom{i}jm}
  \Big[\theta(\pm p_0)\theta(\pm p_0')  -  
  \tilde{f}_{\mathrm{eq},\,j}(p_0,p_0',\mathbf{p},t)\Big]
  (2\pi)^3\delta^{(3)}(\mathbf{p}  - \mathbf{p}')\;,
\end{align}
where, following \eqref{Pps}, we have defined
\begin{align}
  &[\mathscr{P}(p_0,p_0',\mathbf{p})]_{k\phantom{l}\phantom{m}n}^{\phantom{k}lm} 
  \ = \ 
  \nonumber\\ & \quad
  \frac{1}{2}  \begin{pmatrix}
    \sigma_0\,[(p_0 + m)^{1/2}]_{k}^{\phantom{k}l}
    [(p_0'  + m)^{1/2}]^{m}_{\phantom{m}n} &
    -\bm{\sigma}\cdot\mathbf{s}\,
    [(p_0  + m)^{1/2}]_{k}^{\phantom{k}l}
    [(p_0'  -  m)^{1/2}]^{m}_{\phantom{m}n} \\
    \bm{\sigma}\cdot\mathbf{s}\,
    [(p_0  -  m)^{1/2}]_{k}^{\phantom{k}l}
    [(p_0'  +  m)^{1/2}]^{m}_{\phantom{m}n} &
    -\sigma_0\,[(p_0 \: - \: m)^{1/2}]_{k}^{\phantom{k}l}
    [(p_0'  - m)^{1/2}]^{m}_{\phantom{m}n}
  \end{pmatrix}\;.
\end{align}
In         the        mass        eigenbasis         the        factor
$[\mathscr{P}(p_0,p_0',\mathbf{p})]_{i\phantom{j}\phantom{m}j}^{\phantom{i}jm}$
in \eqref{eq:wighteq} is non-zero only for $i=m$ and thus the Wightman
propagator   becomes    proportional   to   $\delta(p_0    -   p_0')$,
i.e.~time-translational  invariance is restored. This  result is valid
in any basis, by virtue of flavour covariance.

Shifting the boundary time $\tilde{t}_i \to 0$, so that $\tilde{t} \to
\tilde{t} - \tilde{t}_i = t$ in \eqref{eq:prophomhat}, we see that the
degree of violation of  time-translational invariance is controlled by
the phase  $e^{i(p_0 -  p_0')t}$. Working in  the mass  eigenbasis, we
find that the violation of translational invariance is maximal when
\begin{equation}
  (p_0   \:  -  \:  p_0')t   \  \approx\  \frac{\Delta
    m^2_{kl}}{2|\mathbf{p}|}\;,
\end{equation} 
where  we  have assumed  $|\mathbf{p}|  \gg  \Delta m^2_{kl}$  and
$\Delta  m^2_{kl} =  m^2_k - m^2_l$ is  the mass  splitting.  This
violation of time-translational invariance is expected in the presence
of flavour coherences,  since the non-equilibrium propagators describe
correlations   of  coherent  superpositions   of  states,   and  these
superpositions are  {\it not} eigenstates of the  Hamiltonian.  In the
CTP formalism, these coherence effects  will be captured in the memory
integrals  occurring in  the collision  terms of  the  resulting master
equations.   We anticipate that  this violation  of time-translational
invariance due  to flavour mixing  may have significant impact  on the
quasi-particle resummations currently  employed in the applications of
Kadanoff-Baym  equations  to  such  phenomena~\cite{Garny:2011hg}.   A
detailed discussion may be given elsewhere.

Before   concluding   \ref{app:propagator},   let  us   consider   the
Schwinger-Dyson equation  of the fermion propagators.   Working in the
double momentum representation discussed above, this reads as follows:
\begin{equation}
  [S^{-1}_{ab}(p,p',\tilde{t})]_{k}^{\phantom{k}l}\ 
  =\ [S^{0^ {-1}}_{ab}(p,p',\tilde{t})]_{k}^{\phantom{k}l}\:
  +\:[\Sigma_{ab}(p,p',\tilde{t})]_{k}^{\phantom{k}l}\;,
\label{SchDys}
\end{equation}
where               $[S^{0}_{ab}]_{k}^{\phantom{k}l}$              and
$[S_{ab}]_{k}^{\phantom{k}l}$ (with $a,b=1,2$ being the CTP indices) 
are  respectively the free  and resummed
$2 \times 2$  matrix CTP propagators in the  doublet notation employed
by         \cite{Calzetta:1986ey,         Calzetta:1986cq},        and
$[\Sigma_{ab}]_{k}^{\phantom{k}l}$   is  the   $2  \times   2   $  CTP
self-energy matrix. Equation \eqref{SchDys} can be inverted to obtain
\begin{align}
  [S^{ab}(p,p',\tilde{t})]_{k}^{\phantom{k}l}\
  & =\ [S^{0,\, ab}(p,p',\tilde{t})]_{k}^{\phantom{k}l}
  \nonumber\\ &
  - \int_{q,\,q'}[S^{0,\, ac}(p,q,\tilde{t})]_{k}^{\phantom{k}m}
  [\Sigma_{cd}(q,q',\tilde{t})]_{m}^{\phantom{m}n}
  [S^{db}(q',p',\tilde{t})]_{n}^{\phantom{n}l}\;.
\end{align}
Due to  the violation of time-translational  invariance, the resulting
Feynman-Dyson  series will  contain  an infinite  nesting of  momentum
integrals, which will not collapse  to the usual algebraic equation of
resummation,  i.e.~for the  time-ordered $(1,1)$   component  of  the
equilibrium CTP propagator for a single flavour:
\begin{align}
  \label{eq:SF}
   S_{\mathrm{F}}(p_0,\mathbf{p}) \ & \equiv \ S^{11}(p_0,\mathbf{p})\ 
    =\ S^{0,\,1a}(p_0,\mathbf{p})\!\sum_{n\:=\:0}^{\infty}
  \!\big[\big(\Sigma(p_0,\mathbf{p})\,\cdot\, S^0(p_0,\mathbf{p})\big)^n
  \big]_{a}^{\phantom{a}1} 
  \nonumber\\ 
  & = \
  \big[\slashed{p}-m+\bar{\Sigma}(p_0,\mathbf{p})\big]^{-1}
  \big[\slashed{p}-m+\Sigma^*(p_0,\mathbf{p})\big]
  \big[\slashed{p}-m+\bar{\Sigma}^*(p_0,\mathbf{p})\big]^{-1}\;,
\end{align}
where $\Sigma(p_0,\mathbf{p})$  is the time-ordered  self-energy. This
result can readily  be verified by rotating to  the CTP eigenbasis (or
so-called       Feynman       basis)~\cite{vanEijck:1994rw}       (see
also~\cite{LeBellac}).     In      \eqref{eq:SF},     the     function
$\bar{\Sigma}(p_0,  \mathbf{p})  = \mathrm{Re}[  \Sigma_{\mathrm{ret}}
(p_0,         \mathbf{p})]         +        i         \varepsilon(p_0)
\mathrm{Im}[\Sigma_{\mathrm{ret}}(p_0,  \mathbf{p})]$ and  its complex
conjugate $\bar{\Sigma}^*(p_0,  \mathbf{p})$, written in  terms of the
retarded self-energy  $\Sigma_{\mathrm{ret}}(p_0,\mathbf{p})$, are the
eigenvalues  of  the  CTP   self-energy  matrix,  having  no  physical
significance      at      finite      temperature      (see
e.g.~\cite{Millington:2012pf}).       The     self-energy     function
$\bar{\Sigma}(p_0,\mathbf{p})$   (and    likewise   the   time-ordered
self-energy    $\Sigma(p_0,\mathbf{p})$)    permits   the    following
decomposition    in    terms    of    the   Dirac    gamma    matrices
$\gamma^{\mu}=(\gamma^0,\bm{\gamma})$ [cf.~\eqref{self_LR}]:
\begin{align}
  \bar{\Sigma}(p_0,\mathbf{p})\ &=\
  \bar{\Sigma}_{\mathrm{L}}^0(p_0,\mathbf{p})\,\gamma^0p_0\,\mathrm{P}_{\mathrm{L}}\:
  +\:\bar{\Sigma}^V_{\mathrm{L}}(p_0,\mathbf{p})\,\bm{\gamma}\cdot\mathbf{p}\,
  \mathrm{P}_{\mathrm{L}}\:
  +\:\bar{\Sigma}_{\mathrm{R}}^0(p_0,\mathbf{p})\,\gamma^0p_0\,
  \mathrm{P}_{\mathrm{R}} \nonumber\\&\qquad +\:
  \bar{\Sigma}^V_{\mathrm{R}}(p_0,\mathbf{p})\,\bm{\gamma}\cdot\mathbf{p}\,
  \mathrm{P}_{\mathrm{R}} \: + \: 
  \bar{\Sigma}^S_{\rm L}(p_0,\mathbf{p})\,\mathrm{P}_{\mathrm{L}}\:
  +\:\bar{\Sigma}^S_{\rm R}(p_0,\mathbf{p})\,\mathrm{P}_{\mathrm{R}} \; ,
\label{B101}
\end{align}
where  we emphasize the violation of  Lorentz-covariance due to
thermal effects.  In \eqref{B101}, the form  factors $\bar{\Sigma}^S_{\rm {L,R}}$ 
correspond to the
scalar      components      of       the      self-energy      and
$\bar{\Sigma}^{0,\,V}_{\rm{L,R}}$ the chiral components. In
the   zero-temperature  limit,   $\bar{\Sigma}$  coincides   with  the
time-ordered  self-energy $\Sigma$  and \eqref{eq:SF}  reduces  to the
expected form
\begin{equation}
  S_{\mathrm{F}}(\slashed{p})\ =\ \frac{1}
  {\slashed{p}-m+\Sigma(\slashed{p})} \; ,
\end{equation}
in  which  Lorentz-covariance  is   restored,  since the  form  
factors are functions only of $p^2$ and 
$\bar{\Sigma}_{\mathrm{L}(\mathrm{R})} (p^2)       
\equiv
\bar{\Sigma}^0_{\mathrm{L}(\mathrm{R})}  (p^2)                             =
\bar{\Sigma}^V_{\mathrm{L}(\mathrm{R})} (p^2)$.

Maintaining  the thermodynamic  equilibrium in \eqref{eq-f}   and  rotating  to  the  mass
eigenbasis  simultaneously,  we can  reduce  the Wightman  propagators
\eqref{eq:wighteq} to
\begin{align}
  &[i\widehat{S}_{\mathrm{eq},\,\gtrless}(p,p',\tilde{t})]
  _{k}^{\phantom{k}l}\
  =\ 2\pi\delta(p_0^2 \: - \: \widehat{E}_k^2)
  (\slashed{p} \: + \: \widehat{m}_k)
  \nonumber\\&\qquad\times\:
  \Big[\theta(\pm p_0)\:
  -\:\theta(p_0)f_{\mathrm{F}}\big(\widehat{E}_k(\mathbf{p})\big)\:
  -\:\theta(-p_0)\bar{f}_{\mathrm{F}}\big(\widehat{E}_k(\mathbf{p})\big)\Big]
  \delta_k^{\phantom{k}l}(2\pi)^4\delta^{(4)}(p \: - \: p') \; ,
\end{align}
in  which  time-translational  invariance  is  restored  in  the  free
propagators. If one also makes the Markovian approximation that energy
is conserved  in the interaction vertices,  i.e.~the interactions take
place over an infinite time domain, time-translational invariance
is restored  globally and the  Schwinger-Dyson equation \eqref{SchDys}
reduces to the relatively simpler form
\begin{equation}
  [\widehat{S}^{-1}_{ab}(p_0, \ve{p})]_{k}^{\phantom{k}l}\ 
  =\ [\widehat{S}^{0^{ -1}}_{ab}(p_0, \ve{p})]_{k}^{\phantom{k}l}\:
  +\:[\widehat{\Sigma}_{ab}(p_0, \ve{p})]_{k}^{\phantom{k}l}\;,
\end{equation}
for  which an  exact  matrix inversion  both  in the  CTP and  flavour
structure is possible.

In summary, we  have presented a fully flavour  covariant formalism of
helicity amplitudes.  In addition, we have shown  that the resummation
of  self-energy corrections  in  the  case of  flavour  mixing may  be
performed in closed algebraic  form only in a flavour-diagonal thermodynamic equilibrium.
Thus, we justify the resummation  scheme employed in the derivation of
the  Markovian master  equations of  Section~\ref{sec:4} in  which the
resummed Yukawa couplings are obtained at zero-temperature, whilst the
aforementioned  flavour coherence  effects discussed  above  have been
included at the level of the quantum statistics.

\section{Generalized Optical Theorem}
\label{app:optical}

In this  appendix, we  justify the tensorial  flavour structure  of the
rates introduced  in Sections~\ref{sec:3} and \ref{sec:4}  by means of
an explicit  calculation of transition matrix elements.   To this end,
we derive a background-dependent  analogue of the optical theorem that
is able to account for off-diagonal flavour coherences.

We  begin by  writing  the scattering operator  $S$ in  terms of  the
transition operator  $\mathcal{T}$ as  $S =  \mat{1}  + 
i\mathcal{T}$.   Subsequently, using the  unitarity of  the scattering
operator $S^{\dag}  S  =  S  S^{\dag}  =   \mat{1}$, one can 
easily show that
\begin{equation}
  \label{eq:transop}
  2\,\mathrm{Im}\, \mathcal{T}
  \ =\ \mathcal{T}^{\dag}\mathcal{T}\;.
\end{equation}
In the usual  derivation of the optical theorem,  we would now proceed
by multiplying \eqref{eq:transop}  from the right and left  by a given
initial Fock state  and insert a complete set  of final states between
the transition  operators on the  RHS of \eqref{eq:transop}.   For our
purposes, the  completeness of  the Fock space  can be written  in the
short-hand notation
\begin{equation}
  \label{eq:completesimp}
 \mat{1} \ =\ \sum_{A}\ket{A^{\{\{\alpha\},\,\{l\},\,\{\bar{l}\},\,...\}}}\!
  \bra{A_{\{\{\alpha\},\,\{l\},\,\{\bar{l}\},\,...\}}} \;,
\end{equation}
where the sum  over $A$ runs over all  possible multi-particle states,
e.g.~Higgs, heavy-neutrinos and charged-leptons  in our case, and also
contains  the   helicity  summations,  isospin   traces  and  momentum
integrals.  Writing explicitly, \eqref{eq:completesimp} has the form
\begin{align}
  \label{eq:complete}
  &\mat{1}\ =\ \sum_{f',\:g',\:h',\:i',\:j'\:=\:0}^{\infty}
  \frac{1}{f'!g'!h'!i'!j'!}\:
  \Bigg[\:\bigwedge_{f\:=\:0}^{f'}\sum_{r_f}\int_{\mathbf{k}_f}
    \ket{\mathbf{k}_f,r_f,N^{\alpha_f}}\!
    \bra{\mathbf{k}_f,r_f,N_{\alpha_f}}
    \nonumber \\ & \quad
    \otimes
    \: \bigwedge_{g\:=\:0}^{g'}\sum_{s_g,\,I_g}\int_{\mathbf{p}_g}
    \ket{\mathbf{p}_g,s_g,I_g,L^{l_g}}\!
    \bra{\mathbf{p}_g,s_g,I_g,L_{l_g}} \:
    \otimes \: \bigwedge_{h\:=\:0}^{h'}\sum_{\bar{s}_h,\, \bar{I}_h}
    \int_{\bar{\mathbf{p}}_h}
    \ket{\bar{\mathbf{p}}_h,\bar{s}_h,\bar{I}_h,\bar{L}_{\bar{l}_h}}\!
    \bra{\bar{\mathbf{p}}_h,\bar{s}_h,\bar{I}_h,\bar{L}^{\bar{l}_h}}
    \nonumber \\ & \quad
    \: \otimes \: \bigotimes_{i\:=\:0}^{i'} \sum_{J_i}\int_{\mathbf{q}_i}
    \ket{\mathbf{q}_i,J_i,\Phi^{\dag}}\!\bra{\mathbf{q}_i,J_i,\Phi^{\dag}} \:
    \otimes \: \bigotimes_{j\:=\:0}^{j'}\sum_{\bar{J}_j} \int_{\bar{\mathbf{q}}_j}
    \ket{\bar{\mathbf{q}}_j,\bar{J}_j,\Phi}\!
    \bra{\bar{\mathbf{q}}_j,\bar{J}_j,\Phi}
    \Bigg]\;,
\end{align}
where  $f'!g'!h'!i'!j'!$ are  the  symmetry factors  arising from  the
integration  over all  three-momenta, $\{r\}$,  $\{s\}$, $\{\bar{s}\}$
are the heavy-neutrino, charged-lepton and anti-lepton helicities, and
$\{I\}$,  $\{\bar{I}\}$,  $\{J\}$, $\{\bar{J}\}$  are  the lepton  and
Higgs isospin indices.   Hereafter, the dimension $-3/2$ charged-lepton, anti-lepton, Higgs
and         anti-Higgs         states        $\ket{\mathbf{p},s,L^l}$,
$\ket{\bar{\mathbf{p}},\bar{s},\bar{L}_{\bar{l}}}$,
$\ket{\mathbf{q},\Phi^{\dag}}$    and    $\ket{\bar{\mathbf{q}},\Phi}$
without  isospin  indices are  understood  to  be  vectors of  $SU(2)$
isospin. The heavy-neutrino states $\ket{\mathbf{k},r,N^{\alpha}}$ are
isospin   singlets.   In   \eqref{eq:complete},  the   wedge   product
$\bigwedge$ denotes the anti-symmetrized tensor product, e.g.
\begin{equation}
  \bigwedge_{j\:=\:1}^{2}\ket{\mathbf{p}_j}\!\bra{\mathbf{p}_j} \
  \equiv \ \ket{\mathbf{p}_1}\!\bra{\mathbf{p}_1} \: \wedge \: 
  \ket{\mathbf{p}_2}\!\bra{\mathbf{p}_2}\ =\
  \Big(\ket{\mathbf{p}_1} \: \wedge \: \ket{\mathbf{p}_2}\Big)
  \Big(\bra{\mathbf{p}_1} \: \wedge \: \bra{\mathbf{p}_2}\Big)\;,
\end{equation}
with the Slater determinant 
\begin{equation}
  \ket{\mathbf{p}_1} \: \wedge \: \ket{\mathbf{p}_2} \: \wedge \:
  \cdots \: \wedge \: \ket{\mathbf{p}_n} \
  \equiv \ \frac{1}{\sqrt{n!}}\epsilon_{i_1 i_2 \cdots i_n}
  \ket{\mathbf{p}_{i_1}}\ket{\mathbf{p}_{i_2}}\cdots\ket{\mathbf{p}_{i_n}}\;.
\end{equation}
The zeroth term in each  of the summations over $g$, $h$, $i$
and $j$ in  \eqref{eq:complete} is understood to be  the outer product
of vacuum states, e.g.
\begin{equation}
  \ket{\mathbf{p}_0,N^{\alpha_0}}\!\bra{\mathbf{p}_0,N_{\alpha_0}}
  \ \equiv\ \mathcal{V}_3\ket{0}\!\bra{0} \; ,
\end{equation}
in   which  $\mathcal{V}_3=(2\pi)^3\delta^{(3)}(\mathbf{0})$   is  the
coordinate  space   three-volume.  We  note  that   the  leptonic  and
anti-leptonic ket-states transform  under different representations of
$U(\mathcal{N}_L)$.   In  \eqref{eq:complete},  the  set  of  internal
flavour indices  $\{\{\alpha\},\ \{l\},\ \{\bar{l}\}\}$  is contracted
and the completeness of the Fock space is a singlet under the combined
heavy-neutrino  and  lepton  flavour  rotations  $U(\mathcal{N}_N)  
\otimes   U(\mathcal{N}_L)$.  For this reason,  the usual derivation
of  the  optical  theorem,  as  outlined  above,  cannot  account  for
transition amplitudes that are off-diagonal in both heavy-neutrino and
lepton flavour  indices.  In  order to incorporate  these off-diagonal
flavour coherences, we must take  into account the contribution of the
background ensemble, which we assume to be described by a factorizable 
density operator $\rho
 =  \rho_{N} \otimes  \rho_{L}   \otimes  \rho_{\Phi}$, where 
coherences between different particle species have been neglected. 
Note  that  the  density  operator  itself is  also  a  singlet  under
$U(\mathcal{N}_N)   \otimes   U(\mathcal{N}_L)$.   Throughout this
section,  we suppress the  time-dependence of  the 
density  operator   $\rho     \equiv    \rho(t)$   for  notational
convenience.

In the  presence of  a background ensemble,  we proceed by  taking the
EEV of the  unitarity relation in
\eqref{eq:transop}, yielding
\begin{equation}
  2\braket{\,\mathrm{Im}\,\mathcal{T}\,}
  \ =\ 2\,\mathrm{Tr}[\rho\,\mathrm{Im}\,\mathcal{T}]
  \ =\ \mathrm{Tr}[\rho\,\mathcal{T}^{\dag}\mathcal{T}]\;.
\end{equation}
Inserting    the     completeness    of    the     Fock    space    in
\eqref{eq:completesimp}   and  using  the   cyclicity  of   the  trace
operation, we obtain
\begin{align}
  \label{eq:optstart}
  2\sum_B\braket{B|\,\rho\,\mathrm{Im}\,\mathcal{T}\,|B}\
  & = \ 2\sum_{A,\,B}\braket{B|A}\,
  \braket{A|\,\rho\,\mathrm{Im}\,\mathcal{T}\,|B}
  \nonumber\\
  & = \ 
 \sum_{A,\,B,\,C,\,D}\braket{B|A}
  \braket{A|\,\mathcal{T}\,|C}
  \braket{C|\,\rho\,|D}
  \braket{D|\,\mathcal{T}^{\dag}\,|B}\;,
\end{align}
where we have suppressed the flavour indices on the sets of states $A$
to $D$ for the time-being.
We now isolate from the summations  over $A$ and $B$ two sets of final
states    $F^{\alpha}_r(\mathbf{k})    \:    \subset   \:    A$    and
$F^{\beta}_{r'}(\mathbf{k}') \: \subset \: B$, each containing at least one
heavy neutrino with  flavours $\alpha$, $\beta$, three-momenta
$\mathbf{k}$, $\mathbf{k}'$   and  helicities   $r$, $r'$, i.e.
\begin{equation}
  \ket{A}\ \supset\ \ket{F_r^{\alpha}(\mathbf{k})}\ =\
  \ket{\mathbf{k},r,N^{\alpha}}\:\wedge\:\ket{A'}\;,\qquad
  \ket{B}\ \supset\ \ket{F^{\beta}_{r'}(\mathbf{k}')}\ =\
  \ket{\mathbf{k}',r',N^{\beta}}\:\wedge\:\ket{B'}\;.
\end{equation}
The  sets  of states  $A'$  and $B'$  consist  of  all other  possible
multi-particle   heavy-neutrino,   lepton,   anti-lepton   and   Higgs
\emph{spectator} states.\footnote{These spectator \emph{states} should
not  be confused with  the SM  spectator \emph{processes}  which could
enhance   the   washout   of   the   lepton   asymmetry   in   thermal
leptogenesis~\cite{Buchmuller:2001sr}.}   Using the  orthonormality of
the Fock states, we have
\begin{align}
  \braket{F_{r',\,\beta}(\mathbf{k}')|F_r^{\alpha}(\mathbf{k})} \
  & = \ \bra{B'} \: \wedge \:  
  \braket{\mathbf{k}',r',N_{\beta}|\mathbf{k},r,N^{\alpha}}
  \: \wedge \: \ket{A'} 
\nonumber \\ 
& = \ (2\pi)^3\delta_{\beta}^{\ \alpha}\,\delta_{rr'}\,
  \delta^{(3)}(\mathbf{k}-\mathbf{k}')\,\delta_{A'B'} \; .
\end{align}
Consequently,   we  obtain   from  \eqref{eq:optstart}  the
following equality:
\begin{align}
  \label{eq:optmid}
  &2\braket{\,\mathrm{Im}\,\mathcal{T}\,
    a^{\beta}(\mathbf{k}',r')\,a_{\alpha}(\mathbf{k},r)\,}\ =\ 2\sum_{A'}
  \bra{A'}\:\wedge\:\bra{\mathbf{k},r,N_{\alpha}}\,\rho\,
    \mathrm{Im}\,\mathcal{T}\,\ket{\mathbf{k}',r',N^{\beta}}
    \:\wedge\:\ket{A'}
  \nonumber\\&\qquad
  =\ \sum_{A',\,C,\,D} \bra{A'}\:\wedge\:
  \bra{\mathbf{k},r,N_{\alpha}}\,\mathcal{T}\,\ket{C}
  \,\braket{C|\,\rho\,|D}
  \bra{D}\,\mathcal{T}^{\dag}\,\ket{\mathbf{k}',r',N^{\beta}}
  \:\wedge\:\ket{A'}\;.
\end{align}
The creation and annihilation operators appearing in this appendix are
understood to be asymptotic `in' operators, see \eqref{eq:asymphase}.

By inspection, we  see that the LHS of  \eqref{eq:optmid} defines the
total in-medium heavy-neutrino production rate
\begin{align}
  \label{eq:prodrate}
  &\Tdu{[\Gamma_{rr'}(\{X\} \: \to \: N;
    \mathbf{k},\mathbf{k}')]}{\alpha}{\beta}{}{}
  \ \equiv \ 2\braket{\,\mathrm{Im}\,\mathcal{T}\,
    a^{\beta}(\mathbf{k}',r')\,a_{\alpha}(\mathbf{k},r)\,} \; ,
\end{align}
in which $\{X\}$ represents the set of all possible initial states and
at least one  heavy-neutrino appears in the final  state. All internal
degrees of  freedom in  the set of  initial states $\{X\}$  are summed
over, except for isospin, i.e.~the general rate
\begin{equation}
  \Tdu{[\Gamma_{\{r\}\{s\}}(\{X\}\: \to
    \:  \{Y\})]}{\{k\}}{\{l\}}{\{\alpha\}}{\{\beta\}}
\end{equation}
is  a  tensor  in  isospin  space.  We  may  convince  ourselves  that
\eqref{eq:prodrate} necessarily gives the production rate (and not the
decay  rate),  since in  the  zero-temperature  limit,  where $\rho  =
\ket{0}\!\bra{0}$,  the  EEV  on  the RHS  of  \eqref{eq:prodrate}  is
identically  zero. Similarly, the  total in-medium  $L\Phi$ production
rate is given by
\begin{equation}
  \label{eq:scatrate}
  \Tdu{[\Gamma_{s_1s_1'}
  (\{X\}\: \to \: L\Phi; 
  \mathbf{p}_1,\mathbf{p}_1',\mathbf{q}_1,\mathbf{q}_1')]}{k}{l}{}{}
  \ \equiv\ 2\braket{\,\mathrm{Im}\,\mathcal{T}\,b^l(\mathbf{p}_1',s_1')\,
  b_k(\mathbf{p}_1,s_1)\,c^{\dag}(\mathbf{q}_1')\,c(\mathbf{q}_1)\,}\; ,
\end{equation}
in which at least one lepton and one Higgs appears in the final state.
In order  to study specific contributions to  the total heavy-neutrino
and  $L\Phi$  production  rates,  defined in  \eqref{eq:prodrate}  and
\eqref{eq:scatrate},  we must  isolate particular  initial  states and
truncate  the transition  operator to  a given  order  in perturbation
theory.

Truncating the  transition operator to first order  in the interaction
Hamiltonian density,    $\mathcal{T}     \to      \mathcal{T}^0   =
\int_x\mathcal{H}_{\mathrm{int}}(x)$, we find from \eqref{eq:prodrate} that 
only  states in $\{X\}$ that  contain  at  least  one  lepton  and  one  
Higgs  will contribute to the heavy-neutrino production rate.         
Isolating         the        initial        states
$\ket{\mathbf{p}',s',L^{l};\mathbf{q}',\Phi^{\dag}}   \wedge \ket{C'} 
\subset   \ket{C}$ and  $\ket{\mathbf{p},s,L^{k};\mathbf{q},\Phi^{\dag}} 
\wedge     \ket{D'}     \subset    \ket{D}$,  we   obtain  from
\eqref{eq:optmid}     and    \eqref{eq:prodrate}     the    tree-level
heavy-neutrino production rate
\begin{align}
  \label{eq:optmid2}
  &\Tdu{[\Gamma^{0}_{rr'}(L\Phi \: \to \: N;\mathbf{k},\mathbf{k}')]}
  {\alpha}{\beta}{}{}
  \ =\ \sum_{\,s,\,s',\,A'}
  \int_{\mathbf{p},\,\mathbf{q},\,\mathbf{p}',\,\mathbf{q}'}
  \bra{A'}\:\wedge\:
    \braket{\mathbf{p}',s',L_{l};\mathbf{q}',\Phi^{\dag}|\,\rho\,|
      \mathbf{p},s,L^{k};\mathbf{q},\Phi^{\dag}}\:\wedge\:\ket{A'}
    \nonumber\\&\qquad \qquad\times\:
    \braket{\mathbf{k},r,N_{\alpha}|\,\mathcal{T}^{0}\,|
    \mathbf{p}',s',L^{l};\mathbf{q}',\Phi^{\dag}}
  \braket{\mathbf{p},s,L_{k};\mathbf{q},\Phi^{\dag}|\
    \mathcal{T}^{0\,\dag}\,|\mathbf{k}',r',N^{\beta}} \; .
\end{align}
In \eqref{eq:optmid2}, the spectator states
$A'$   do   not  contribute   to   the   transition  matrix   elements
$\braket{\mathbf{k},   r,   N_{\alpha}|   \,  \mathcal{T}^{0}   \,   |
\mathbf{p}', s',  L^{l}; \mathbf{q}', \Phi^{\dag}}$  and $\braket{\mathbf{p},
s, L_{k};  \mathbf{q}, \Phi^{\dag}| \  \mathcal{T}^{0\,\dag} \, |\mathbf{k}',
r', N^{\beta}}$ at first order in perturbation theory.  The $\rho$-dependent term in \eqref{eq:optmid2} is
\begin{align}
  &\sum_{A'}\bra{A'}\:\wedge\:
    \braket{\mathbf{p}',s',L_l;\mathbf{q}',\Phi^{\dag}|\,\rho\,|
      \mathbf{p},s,L^k;\mathbf{q},\Phi^{\dag}}
    \:\wedge\:\ket{A'}
  \nonumber\\&\qquad
  =\ \mathrm{Tr}\left[\rho\, b^k(\mathbf{p},s)\, b_l(\mathbf{p}',s')\,
    c^{\dag}(\mathbf{q})\,c(\mathbf{q'})\right]
  \ =\ [n^{L}_{s's}(\mathbf{p}',\mathbf{p})]_{l}^{\phantom{l}k}\,
  n^{\Phi}(\mathbf{q}',\mathbf{q})\;,
\end{align}
where we have  taken the leptonic and Higgs  ensembles to be spatially
inhomogeneous in  general, depending explicitly  on two three-momenta.
The  time-dependence of  the  distribution functions  $n^X(\mathbf{p},
\mathbf{p}')$  is assumed implicitly.   In  the  spatially-homogeneous
limit,  the  lepton  and  Higgs  distribution  functions  satisfy  the
correspondence
\begin{align}
  [n^{L}_{ss'}(\mathbf{p},\mathbf{p}')]_{k}^{\phantom{k}l}\ & \to \
  [n^{L}_{s}(\mathbf{p})]_{k}^{\phantom{k}l}(2\pi)^3\delta_{ss'}
  \delta^{(3)}(\mathbf{p}  -  \mathbf{p}')\;, 
  \\ 
  n^{\Phi}(\mathbf{q},\mathbf{q}')\ & \to \
  n^{\Phi}(\mathbf{q})(2\pi)^3\delta^{(3)}(\mathbf{q}-\mathbf{q}')\;.
\end{align}
The  general  inhomogeneous  distribution  functions  $n^X(\mathbf{p},
\mathbf{p}')$    are   related    to   the    number    densities   defined in 
Section~\ref{sec:3.1} by a Wigner transformation~\cite{Winter:1986da} and
integration over all coordinate space:
\begin{equation}
  n^X(\mathbf{p})\ =\ \frac{1}{\mathcal{V}_3}\int\!\mathrm{d}^3\mathbf{x}
  \int\!\mathrm{d}^3\mathbf{q}\;
  e^{i\mathbf{q}\cdot \mathbf{x}}\;
  n^X(\mathbf{p}+\tfrac{\mathbf{q}}{2},\mathbf{p}-\tfrac{\mathbf{q}}{2}) \
  = \ \frac{1}{\mathcal{V}_3}\,n^X(\mathbf{p},\mathbf{p})\;. 
\end{equation}

The Wick  contraction of  field operators may  be performed  using the
following set of flavour-covariant field-particle duality relations:
\begin{align}
  \braket{0|\,\widetilde{\Phi}^{\dag}(x)c^{\dag}(\mathbf{p})\,|0}\ &=\
  \big(2E_{\Phi}(\mathbf{p})\big)^{-1/2}e^{-iE_{\Phi}(\mathbf{p})x_0}
  e^{i\mathbf{p}\cdot\mathbf{x}}\;,\\
  \braket{0|\,c(\mathbf{p})\widetilde{\Phi}(x)\,|0}\ &=\
  \big(2E_{\Phi}(\mathbf{p})\big)^{-1/2}e^{iE_{\Phi}(\mathbf{p})x_0}
  e^{-i\mathbf{p}\cdot\mathbf{x}}\;,\\
  \braket{0|\,\widetilde{\Phi}(x)\bar{c}^{\dag}(\mathbf{p})\,|0}\ &=\
  \big(2E_{\Phi}(\mathbf{p})\big)^{-1/2}e^{-iE_{\Phi}(\mathbf{p})x_0}
  e^{i\mathbf{p}\cdot\mathbf{x}}\;,\\
  \braket{0|\,\bar{c}(\mathbf{p})\widetilde{\Phi}^{\dag}(x)\,|0}\ &=\
  \big(2E_{\Phi}(\mathbf{p})\big)^{-1/2}e^{iE_{\Phi}(\mathbf{p})x_0}
  e^{-i\mathbf{p}\cdot\mathbf{x}}\;,\\
  \braket{0|\,L_k(x)b^l(\mathbf{p},s)\,|0}\ &=\
  \big[\big(2E_{L}(\mathbf{p})\big)^{-1/2}\big]_{k}^{\phantom{k}m}
  \big[e^{-iE_{L}(\mathbf{p})x_0}\big]_{m}^{\phantom{m}n}
  e^{i\mathbf{p}\cdot\mathbf{x}}[u(\mathbf{p},s)]_{n}^{\phantom{n}l}\;,\\
  \braket{0|\,d^{\dag\, l}(\mathbf{p},s)L_k(x)\,|0}\ &=\
  \big[\big(2E_{L}(\mathbf{p})\big)^{-1/2}\big]_{k}^{\phantom{k}m}
  \big[e^{iE_{L}(\mathbf{p})x_0}\big]_{m}^{\phantom{m}n}
  e^{-i\mathbf{p}\cdot\mathbf{x}}[v(\mathbf{p},s)]_{n}^{\phantom{n}l}\;,\\
  \braket{0|\,b_{l}(\mathbf{p},s)\bar{L}^k(x)\,|0}\ &=\
  \big[\big(2E_{L}(\mathbf{p})\big)^{-1/2}\big]^{k}_{\phantom{k}m}
  \big[e^{iE_{L}(\mathbf{p})x_0}\big]^{m}_{\phantom{m}n}e^{-i\mathbf{p}\cdot\mathbf{x}}
  [\bar{u}(\mathbf{p},s)]^{n}_{\phantom{n}l}\;,\\
  \braket{0|\,\bar{L}^k(x)d^{\dag}_{l}(\mathbf{p},s)\,|0}\ &=\
  \big[\big(2E_{L}(\mathbf{p})\big)^{-1/2}\big]^{k}_{\phantom{k}m}
  \big[e^{-iE_{L}(\mathbf{p})x_0}\big]^{m}_{\phantom{m}n}e^{i\mathbf{p}\cdot\mathbf{x}}
  [\bar{v}(\mathbf{p},s)]^{n}_{\phantom{n}l}\;,\\
  \braket{0|\,N_{R,\,\alpha}(x)a^{\beta}(\mathbf{k},r)\,|0}\ &=\
  \big[\big(2E_N(\mathbf{k})\big)^{-1/2}\big]_{\alpha}^{\phantom{\alpha}\gamma}
  \big[e^{-iE_N(\mathbf{p})x_0}\big]_{\gamma}^{\phantom{\gamma}\delta}
  e^{i\mathbf{p}\cdot\mathbf{x}}{\mathrm{P_R}}
  [u(\mathbf{k},r)]_{\delta}^{\phantom{\delta}\beta}\;,\\
  \braket{0|\,a_{\beta}(\mathbf{k},r)N_{R,\, \alpha}(x)\,|0}\ &=\
  \big[\big(2E_N(\mathbf{k})\big)^{-1/2}\big]_{\alpha}^{\phantom{\alpha}\gamma}
  \big[e^{iE_N(\mathbf{p})x_0}\big]_{\gamma}^{\phantom{\gamma}\delta}
  e^{-i\mathbf{p}\cdot\mathbf{x}}{\mathrm{P_R}}
  [v(\mathbf{k},r)]_{\delta}^{\phantom{\delta}\epsilon}G_{\epsilon\beta}\;,\\
  \braket{0|\,a_{\beta}(\mathbf{k},r)\overline{N}_R^{\alpha}(x)\,|0}\ &=\
  \big[\big(2E_N(\mathbf{k})\big)^{-1/2}\big]^{\alpha}_{\phantom{\alpha}\gamma}
  \big[e^{iE_N(\mathbf{p})x_0}\big]^{\gamma}_{\phantom{\gamma}\delta}
  e^{-i\mathbf{p}\cdot\mathbf{x}}
  [\bar{u}(\mathbf{k},r)]^{\delta}_{\phantom{\delta} \beta}{\mathrm{P_L}}\;,\\
  \braket{0|\,\overline{N}_R^{\alpha}(x)a^{\beta}(\mathbf{k},r)\,|0}\ &=\
  \big[\big(2E_N(\mathbf{k})\big)^{-1/2}\big]^{\alpha}_{\phantom{\alpha} \gamma}
  \big[e^{-iE_N(\mathbf{p})x_0}\big]^{\gamma}_{\phantom{\gamma} \delta}
  e^{i\mathbf{p}\cdot\mathbf{x}}
  [\bar{v}(\mathbf{k},r)]^{\delta}_{\phantom{\delta}\epsilon}
  {\mathrm{P_L}}G^{\epsilon\beta}\;,
\end{align}
where the creation and  annihilation operators have the mass dimension $-3/2$
and satisfy the commutator algebra given in Section~\ref{sec:3.1}.  We
emphasize that the operators appearing in this appendix are understood
to be asymptotic `in' operators.

In  the case  of  the tree-level  heavy-neutrino  production rate,  we
obtain the Hermitian conjugate pair of transition matrix elements
\begin{align}
  &\Tdu{[\mathcal{T}^{0}_{r's}(L\Phi  \to  N; \mathbf{k}',
    \mathbf{p},\mathbf{q})]}{k}{\beta}{}{}
  \ \equiv\
  \braket{\mathbf{p},s,L_k;\mathbf{q},\Phi^{\dag}|\,\mathcal{T}^{0\,\dag}\,|
    \mathbf{k}',r',N^{\beta}}
  \nonumber\\ &\qquad \qquad 
  =\ [\mathcal{M}^{0}(L\Phi  \to  N)]_i^{\phantom{i}\gamma}
  [(2E_{L}(\mathbf{p}))^{-1/2}]^{i}_{\phantom{i}j}
  [(2E_N(\mathbf{k}'))^{-1/2}]_{\gamma}^{\phantom{\gamma}\delta}
  (2E_{\Phi}(\mathbf{q}))^{-1/2}
  \nonumber\\&\qquad \qquad \qquad \qquad\times\:
  \DiFud{k'  -  p  -  q}{j}{m}{\delta}{\epsilon}
  [\bar{u}(\mathbf{p},s)]^{m}_{\phantom{m}k}
  {\mathrm{P_R}}[u(\mathbf{k}',r')]_{\epsilon}^{\phantom{\epsilon}\beta}\;, 
\label{C29}
\\
  &\Tud{[\mathcal{T}^{0}_{rs'}
     (L\Phi  \to  N;\mathbf{k},\mathbf{p}',\mathbf{q}')]}{l}{\alpha}{}{}
  \ \equiv\
  \braket{\mathbf{k},r,N_{\alpha}|\,\mathcal{T}^{0}\,|
    \mathbf{p}',s',L^{l};\mathbf{q}',\Phi^{\dag}}
  \nonumber\\ &\qquad \qquad
  =\ [\mathcal{M}^{0}(L\Phi  \to  N)]^{i'}_{\phantom{i'}\gamma'}
  [(2E_{L}(\mathbf{p}'))^{-1/2}]_{i'}^{\phantom{i'}j'}
  [(2E_N(\mathbf{k}))^{-1/2}]^{\gamma'}_{\phantom{\gamma'}\delta'}
  (2E_{\Phi}(\mathbf{q}'))^{-1/2}
  \nonumber\\&\qquad \qquad \qquad \qquad\times\:
  \DiFdu{k  -  p'  -  q'}{{j'}}{m'}{\delta'}{\epsilon'}
  [\bar{u}(\mathbf{k},r)]^{\epsilon'}_{\phantom{\epsilon'}\alpha}
  {\mathrm{P_L}}[u(\mathbf{p}',s')]_{m'}^{\phantom{m'}l}\;.
\label{C30}
\end{align}
The Dirac spinors  $u$ and $v$ of the  heavy-neutrinos and leptons are
distinguished by the character-set  of their indices: lower-case Greek
characters are  for heavy-neutrinos and lower-case  Latin for leptons.
For  a detailed  discussion  of  the flavour  structure  of the  Dirac
spinors, see~\ref{app:propagator}.

Finally, we may recast \eqref{eq:optmid2} in the following form:
\begin{align}
  \Tdu{[\Gamma^{0}_{rr'}(L\Phi  \to N; \mathbf{k},\mathbf{k}')]}
  {\alpha}{\beta}{}{}
  \ &=\ \sum_{s,\,s'}\int_{\mathbf{p},\,\mathbf{q},\,\mathbf{p}',\,\mathbf{q}'}
  \Tdu{[\Gamma^{0}_{rr'ss'}
    (L\Phi  \to N;
    \mathbf{k},\mathbf{k}',\mathbf{p},\mathbf{p}',\mathbf{q},\mathbf{q}')
      ]}{k}{l}{\alpha}{\beta}
\nonumber\\& \qquad \qquad \times \:
  [n^{L}_{s's}(\mathbf{p}',\mathbf{p})]_{l}^{\phantom{l}k}
  n^{\Phi}(\mathbf{q}',\mathbf{q})\;,
\label{C31}
\end{align}
where we have defined the spatially-inhomogeneous rank-4 tensor rate
\begin{align}
  \Tdu{[\Gamma^{0}_{rr'ss'}(L\Phi  \to N;
  \mathbf{k},\mathbf{k}',\mathbf{p},\mathbf{p}',\mathbf{q},\mathbf{q}')
      ]}{k}{l}{\alpha}{\beta} \
  &=\ \Tud{[\mathcal{T}^{0}_{rs'}
    (L\Phi \to N;\mathbf{k},\mathbf{p}',\mathbf{q}')]}{l}{\alpha}{}{}
  \nonumber\\&\qquad\times\:
       \Tdu{[\mathcal{T}^{0}_{r's}
    (L\Phi  \to  N; \mathbf{k}',\mathbf{p},\mathbf{q})]}{k}{\beta}{}{},
\label{C32}
\end{align}
carrying  two pairs  of helicity  indices, two  heavy-neutrino flavour
indices, two lepton flavour indices and three pairs of three-momenta.

For  the tree-level $\Delta  L=0$ scattering  ($L\Phi \to  L\Phi$), we
need the  transition operator  up to second  order in  the interaction
Hamiltonian  density,  $\mathcal{T}  \to \mathcal{T}^1  =  \tfrac{i}{2!}\int_{x,x'}
\mathrm{T}[                  \mathcal{H}_{\mathrm{int}}(x)
\mathcal{H}_{\mathrm{int}}(x')]$ in which $\mathrm{T}$ is
the time-ordering  operator.  Ignoring the disconnected  contributions, we
obtain the tree-level in-medium scattering rate
\begin{align}
 & \Tdu{[\Gamma^{0}_{s_1s_1'}
     (L\Phi  \to L\Phi;
     \mathbf{p}_1,\mathbf{p}_1',\mathbf{q}_1,\mathbf{q}_1')]}{k}{k'}{}{}
   \nonumber\\ & \qquad 
   = \   \sum_{s_2,\, s_2'}
   \int_{\mathbf{p}_2,\,\mathbf{q}_2,\,\mathbf{p}_2',\,\mathbf{q}_2'} 
   \Tdu{[\Gamma^{0}_{s_1s_1's_2s_2'}
        (L\Phi  \to L\Phi;
        \mathbf{p}_1,\mathbf{p}_1',\mathbf{p}_2,\mathbf{p}_2',
        \mathbf{q}_1,\mathbf{q}_1',\mathbf{q}_2,\mathbf{q}_2')
          ]}{l}{l'}{k}{k'}
\nonumber\\&\quad \qquad \qquad \times \:
    [n^{L}_{s_2's_2}(\mathbf{p}_2',\mathbf{p}_2)]_{l'}^{\phantom{l'}l}
    n^{\Phi}(\mathbf{q}_2',\mathbf{q}_2)\;,
\label{C33}
\end{align}
where the rank-4 scattering rate is
\begin{align}
  &\Tdu{[\Gamma^{0}_{s_1s_1's_2s_2'}
    (L\Phi  \to  L\Phi;
    \mathbf{p}_1,\mathbf{p}_1',\mathbf{p}_2,\mathbf{p}_2',\mathbf{q}_1,
    \mathbf{q}_1',\mathbf{q}_2,\mathbf{q}_2')]}{l}{l'}{k}{k'}
  \nonumber\\&\qquad =\ \Tud{[\mathcal{T}^{0}_{s_2s_1'}
    (L\Phi  \to  L\Phi;
    \mathbf{p}_2,\mathbf{p}_1',\mathbf{q}_2,\mathbf{q}_1')]}{k'}{l}{}{}
    \Tdu{[\mathcal{T}^{0}_{s_1s_2'}
    (L\Phi  \to  L\Phi;
    \mathbf{p}_1,\mathbf{p}_2',\mathbf{q}_1,\mathbf{q}_2')]}{k}{l'}{}{}\;.
\label{C34}
\end{align}
The Hermitian conjugate pair of scattering transition matrix elements are
\begin{align}
  \label{eq:scatT1}
  &\Tdu{[\mathcal{T}^{0}_{s_1s_2'}
     (L\Phi  \to  L\Phi;
     \mathbf{p}_1,\mathbf{p}_2',\mathbf{q}_1,\mathbf{q}_2')]}{k}{l'}{}{}
  \ =\ \braket{\mathbf{p}_1,s_1,L_k;\mathbf{q}_1,\Phi^{\dag}|\,
    \mathcal{T}^{1\,\dag}\,|
    \mathbf{p}_2',s_2',L^{l'};\mathbf{q}_2',\Phi^{\dag}}
  \nonumber\\ & \quad
  = -i\int_{k_1,\,k_1'}[(2E_{L}(\mathbf{p}_1))^{-1/2}]^{i}_{\phantom{i}j}
  [(2E_{L}(\mathbf{p}_2'))^{-1/2}]_{a}^{\phantom{a}b}
  (2E_{\Phi}(\mathbf{q}_1))^{-1/2}
  (2E_{\Phi}(\mathbf{q}_2'))^{-1/2}
  \nonumber\\& \quad \times\:
  \DiFud{k_1  -  p_1  - q_1}{j}{m}{\gamma}{\epsilon}
  \DiFdu{k_1'  -  p_2'  -  q_2'}{b}{c}{\delta}{\sigma}
  \nonumber\\& \quad \times \:
  [\bar{u}(\mathbf{p}_1,s_1)]^m_{\phantom{m}k}
  [i\Delta^{0}_{\mathrm{F},\,N}(k_1,k_1')]_{\epsilon}^{\phantom{\epsilon}\sigma}
  [u(\mathbf{p}_2',s_2')]_{c}^{\phantom{c}l'}
  [\mathcal{M}^{0}(L\Phi  \to  N)]^{a}_{\phantom{a}\delta}
  [\mathcal{M}^{0}(N \to L\Phi)]_i^{\phantom{i}\gamma}, \displaybreak 
\\
  \label{eq:scatT2}
  &\Tud{[\mathcal{T}^{0}_{s_2s_1'}
     (L\Phi \: \to \: L\Phi;
     \mathbf{p}_2,\mathbf{p}_1',\mathbf{q}_2,\mathbf{q}_1')]}{k'}{l}{}{}
  \ =\ \braket{\mathbf{p}_2,s_2,L_{l};\mathbf{q}_2,\Phi^{\dag}|\,
    \mathcal{T}^{1}\,|
    \mathbf{p}_1',s_1',L^{k'};\mathbf{q}_1',\Phi^{\dag}}
  \nonumber\\&\quad
  = +i\int_{k_2,\,k_2'}
  [(2E_{L}(\mathbf{p}_1'))^{-1/2}]_{i'}^{\phantom{i'}j'}
  [(2E_{L}(\mathbf{p}_2))^{-1/2}]^{a'}_{\phantom{a'}b'}
  (2E_{\Phi}(\mathbf{q}_2))^{-1/2}
  (2E_{\Phi}(\mathbf{q}_1'))^{-1/2}
  \nonumber\\& \quad \times\:
  \DiFdu{k_2'  -  p_1'  -  q_1'}{{j'}}{m'}{\gamma'}{\epsilon'}
  \DiFud{k_2  -  p_2 -  q_2}{{b'}}{c'}{\delta'}{\sigma'}
  \nonumber\\& \quad \times \:
  [\bar{u}(\mathbf{p}_2,s_2)]^{c'}_{\phantom{c'}l}
  [i\Delta^{0}_{\mathrm{F},\,N}(k_2',k_2)]^{\epsilon'}_{\phantom{\epsilon'}\sigma'}
  [u(\mathbf{p}_1',s_1')]_{m'}^{\phantom{m'}k'}
  [\mathcal{M}^{0}(N \to L\Phi)]^{i'}_{\phantom{i'}\gamma'}
  [\mathcal{M}^{0}(L\Phi  \to  N)]_{a'}^{\phantom{a'}\delta'},
\end{align}
where   $\Delta^0_{{\rm    F},N}$   is   the    free   non-homogeneous
heavy-neutrino Feynman propagator
\begin{equation}
  [i\Delta_{\mathrm{F},\,N}^{0}(k,k')]_{\alpha}^{\phantom{\alpha}\beta}\
  =\ \mathrm{Tr}\{\rho_N\mathrm{T}[N_{\mathrm{R},\,\alpha}(k)
  \overline{N}_{\mathrm{R}}^{\beta}(k')]\} \; .
\label{C37}
\end{equation}
The   structure  of   these  flavour-covariant   non-homogeneous  free
propagators  in   the  case  of   Dirac  fermions  is   illustrated  in
\ref{app:propagator}.  Notice  that, by  means of the  Gaussian moment
theorem   (Wick's  theorem),   the  heavy-neutrino   spectator  states
contribute the thermal part of the Feynman propagator.

\subsection*{C.1\ Decays}

Imposing  kinetic equilibrium as  described in  Section \ref{sec:4.1},
the tree-level heavy-neutrino production rate \eqref{C31} becomes
\begin{equation}
  \Tdu{[\Gamma^{0}_{r}(L\Phi \: \to \: N; \mathbf{k})]}{\alpha}{\beta}{}{}
  \ =\ \sum_s\frac{[n^{L}_s]_{l}^{\phantom{l}k}}
  {n^{L}_{\mathrm{eq}}}\,
  \int_{\mathbf{p},\,\mathbf{q}}
  e^{-\,(|\mathbf{p}|+|\mathbf{q}|)/T}\,\Tdu{[\Gamma^{0}_{rs}
    (L\Phi \: \to \: N; \mathbf{k},\mathbf{p},\mathbf{q})
      ]}{k}{l}{\alpha}{\beta}\;,
\label{C38}
\end{equation}
where, following \eqref{C32}, we have defined 
\begin{equation}
  \label{eq:twodeltas}
  \Tdu{[\Gamma^{0}_{rs}
    (L\Phi  \to  N; \mathbf{k},\mathbf{p},\mathbf{q})
      ]}{k}{l}{\alpha}{\beta}
  \ =\ \Tud{[\mathcal{T}^{0}_{rs}(L\Phi  \to  N;
    \mathbf{k},\mathbf{p},\mathbf{q})]}{l}{\alpha}{}{}
       \Tdu{[\mathcal{T}^{0}_{rs}
    (L\Phi \to  N;
    \mathbf{k},\mathbf{p},\mathbf{q})]}{k}{\beta}{}{}\;.
\end{equation}
In  this case,  the  transition amplitudes  given  in \eqref{C29}  and
\eqref{C30} simplify to
\begin{align}
  &\Tdu{[\mathcal{T}^{0}_{rs}
     (L\Phi  \to  N;
     \mathbf{k},\mathbf{p},\mathbf{q})]}{k}{\beta}{}{}
  \ \equiv\
  \braket{\mathbf{p},s,L_k;\mathbf{q},\Phi^{\dag}|\,\mathcal{T}^{0\,\dag}\,|
    \mathbf{k},r,N^{\beta}} \ = \ 
[\mathcal{M}^{0}(L\Phi  \to  N)]_{k}^{\phantom{k}\beta}
  \nonumber\\ & \qquad
  \times\: 
  (2|\mathbf{p}|)^{-1/2}
  (2E_N(\mathbf{k}))^{-1/2}
  (2|\mathbf{q}|)^{-1/2}
  (2\pi)^4\delta^{(4)}(k-p-q)
  \bar{u}(\mathbf{p},s)
  {\mathrm{P_R}}u(\mathbf{k},r)\;,\\
  &\Tud{[\mathcal{T}^{0}_{rs}
     (L\Phi  \to  N;
     \mathbf{k},\mathbf{p},\mathbf{q})]}{l}{\alpha}{}{}
  \ \equiv\
  \braket{\mathbf{k},r,N_{\alpha}|\,\mathcal{T}^{0}\,|
    \mathbf{p},s,L^{l};\mathbf{q},\Phi^{\dag}}
\  =\ [\mathcal{M}^{0}(L\Phi \: \to \: N)]_{\phantom{l}\alpha}^{l}
  \nonumber\\ &\qquad
 \times \: 
  (2|\mathbf{p}|)^{-1/2}
  (2E_N(\mathbf{k}))^{-1/2}
  (2|\mathbf{q}|)^{-1/2}
  (2\pi)^4\delta^{(4)}(p+q-k)
  \bar{u}(\mathbf{k},r)
  {\mathrm{P_L}}u(\mathbf{p},s) \; ,
\end{align}
in which the tree-level matrix elements are
\begin{equation}
  [\mathcal{M}^{0}(L\Phi  \to N)]_k^{\phantom{k}\alpha} \ 
  = \ h_{k}^{\phantom{k}\alpha}\;,\quad
  [\mathcal{M}^{0}(L\Phi  \to  N)]^{k}_{\phantom{k}\alpha}\ = \
  ([\mathcal{M}^{0}(L\Phi  \to N)]_k^{\phantom{k}\alpha})^* \ 
  = \ h^{k}_{\phantom{k}\alpha}\;.
\end{equation}
The one-loop  resummation effects due to heavy-neutrino  mixing can be
included by promoting the  tree-level Yukawa couplings to the one-loop
resummed ones,  as discussed in  Section~\ref{sec:2.1}.  Specifically,
in Section~\ref{sec:4}, we have used
\begin{equation}
  \label{eq:resyuk1}
  [\mathcal{M}(N  \to  L\Phi)]_{k}^{\phantom{k}\alpha}\
  = \ \mathbf{h}_{k}^{\phantom{k}\alpha}\;,
\end{equation}
corresponding   to   the   process    $N   \to   L\Phi$. For  its 
$\widetilde{T}$-conjugate      process      $L\Phi \to N$,  
we have by $\CP T$ invariance (see Section~\ref{sec:3.2}),
\begin{equation}
  \label{eq:resyuk2}
  [\mathcal{M}(L\Phi \: \to \: N)]^{k}_{\phantom{k}\alpha}\ =\
  \big([\mathcal{M}(N \: \to \: L\Phi)]_{k}^{\phantom{k}\alpha}
  \big)^{\widetilde{T}}
  \ =\ 
  [\mathbf{h}^{\tilde{c}}]^{k}_{\phantom{k}\alpha} \; ,
\end{equation}
where  we recall that  $\tilde{c} \  \equiv  \ \widetilde{C}P$
denotes the  generalized-$\CP$ conjugate.   The matrix element  for the
process  $N \to L^{\tilde{c}}\Phi^{\tilde{c}}$  is  obtained
from \eqref{eq:resyuk1} by the generalized $\CP$ transformation
\begin{equation}
  [\mathcal{M}(N \to  L^{\tilde{c}}\Phi^{\tilde{c}})]^k_{\ \alpha}\ =\
  \big([\mathcal{M}(N \to  L\Phi)]_{k}^{\phantom{k}\alpha}
  \big)^{\tilde{c}}
  \ =\
  [\mathcal{M}(L\Phi  \to  N)]^{k}_{\phantom{k}\alpha}\ =\
  [\mathbf{h}^{\tilde{c}}]^{k}_{\phantom{k}\alpha}\;.
\end{equation}
The                 $\widetilde{T}$-transformed                process
$L^{\tilde{c}}\Phi^{\tilde{c}} \to N$  can be  obtained from
\eqref{eq:resyuk2} via the generalized $\CP$ transformation
\begin{equation}
  [\mathcal{M}(L^{\tilde{c}}\Phi^{\tilde{c}}  \to N)]
  _k^{\phantom{k}\alpha}\ =\
  \big([\mathcal{M}(L\Phi  \to  N)]
  _{\phantom{k}\alpha}^{k}\big)^{\tilde{c}}\ =\
  [\mathcal{M}(N  \to  L\Phi)]_{k}^{\phantom{k}\alpha}\ =\
  \mathbf{h}^{\phantom{k}\alpha}_{k}\;.
\end{equation}

As in  Section~\ref{sec:4.1}, we  assume that the  charged-leptons are
massless, and hence, only one  of their helicity states are populated,
which  for concreteness,  we  choose  to be  $s=-\:  (+)$ for  leptons
(anti-leptons)    in   \eqref{eq:twodeltas}.     Summing    over   the
heavy-neutrino helicities, tracing over lepton and Higgs isospins, and
performing   the  momentum   integrals,  we   derive   the  tree-level
thermally-averaged rates for the processes $N\to L\Phi$ and $L\Phi \to
N$, as follows:
\begin{align}
  \label{eq:invdecrate}
  [\gamma(L\Phi  \to N)]
  _{k\phantom{l}\alpha}^{\phantom{k}l\phantom{\alpha}\beta}\ & =\
  \frac{1}{\mathcal{V}_4}
  \sum_{r,\,IJ}\int_{\mathbf{k},\,\mathbf{p},\,\mathbf{q}}
  e^{-\,(|\mathbf{p}|+|\mathbf{q}|)/T}\,[\Gamma^{0}_{r,-}
    (L\Phi  \to N;
    \mathbf{k},\mathbf{p},\mathbf{q})]
    _{k\phantom{l}\alpha}^{\phantom{k}l\phantom{\alpha}\beta}\;,
\end{align}
where  we have  divided  through by  the coordinate-space  four-volume
$\mathcal{V}_4  =  (2\pi)^4\delta^{(4)}(0)$  in  order to  remove  the
factor   arising  from  the   product  of   identical  energy-momentum
conserving delta functions from  the two transition matrix elements in
\eqref{eq:twodeltas}.   After  performing   the  summations  over  the
heavy-neutrino  helicities  and  the  traces  over  lepton  and  Higgs
isospin, we  find from \eqref{eq:invdecrate}  the heavy-neutrino decay
and inverse decay rates
\begin{align}
  \label{eq:NLPhi}
  [\gamma(N \to
  L\Phi)]_{k\phantom{l}\alpha}^{\phantom{k}l\phantom{\alpha}\beta}\ &
  =\ [\gamma(L^{\tilde{c}}\Phi^{\tilde{c}} \to
  N)]_{k\phantom{l}\alpha}^{\phantom{k}l\phantom{\alpha}\beta}\ 
  =\ \int_{NL\Phi} g_Lg_{\Phi}(2p_N\cdot
  p_L)\,\mathbf{h}^{\ \beta}_{k}\mathbf{h}^{l}_{\ \alpha}\;,\\ 
  \label{eq:LPhiN}
  [\gamma(L\Phi \: \to \: N)]
  _{k\phantom{l}\alpha}^{\phantom{k}l\phantom{\alpha}\beta}\ &
  =\ [\gamma(N \: \to \: L^{\tilde{c}}\Phi^{\tilde{c}})]
  _{k\phantom{l}\alpha}^{\phantom{k}l\phantom{\alpha}\beta}\
  =\ \int_{NL\Phi}g_Lg_{\Phi}
  (2p_N\cdot p_L)\,[\mathbf{h}^{\tilde{c}}]_{k}^{\phantom{k}\beta}
  [\mathbf{h}^{\tilde{c}}]^{l}_{\phantom{l}\alpha} \; ,
\end{align}
where we recall  that $g_L = g_\Phi = 2$  count the degenerate isospin
degrees of freedom of the  lepton and Higgs doublets.  In addition, we
have   used   the   notation    defined   in   \eqref{col}   for   the
thermally-weighted phase-space  integrals.  We note  that a relabeling
of indices  $k\: \leftrightarrow \: l$ and  $\alpha \: \leftrightarrow
\:  \beta$ has  been  performed in  \eqref{eq:LPhiN}  relative to  the
$\widetilde{T}$-transform of \eqref{eq:NLPhi}, that is
\begin{equation}
\label{eq:relab}
[\gamma(N \to
  L\Phi)]_{k\phantom{l}\alpha}^{\phantom{k}l\phantom{\alpha}\beta} \ = \
  \big([\gamma(L\Phi \: \to \: N)]
  _{l\phantom{k}\beta}^{\phantom{l}k\phantom{\beta}\alpha}\big)^{\widetilde{T}}\;.
\end{equation}
The  decay  rates in  \eqref{eq:NLPhi}  and \eqref{eq:LPhiN},  derived
using the  generalized optical theorem  are exactly the same  as those
given in \eqref{coll_nlp} and \eqref{coll_nlcpc}.

\subsection*{C.2\ Scatterings}

We  may write  the  homogeneous limit  of  the heavy-neutrino  Feynman
propagator  in a  general basis  by rotating  \eqref{C37} to  the mass
eigenbasis, i.e.
\begin{align}
  [i\Delta^{0}_{\mathrm{F},\,N}(k,k')]_{\alpha}^{\phantom{\alpha}\beta}\ &=\
  U^{\rho}_{\phantom{\rho}\alpha}
  [i\widehat{\Delta}_{\mathrm{F},\,N}^0(k,k')]_{\rho}
  \,\delta_{\rho}^{\phantom{\rho}\sigma}\,U_{\sigma}^{\phantom{\sigma}\beta}
\nonumber \\
 &  \to \ U^{\rho}_{\phantom{\rho}\alpha}
  [i\widehat{\Delta}_{\mathrm{F},\,N}^0(k)]_{\rho}
  (2\pi)^4\delta^{(4)}(k - k')U_{\rho}^{\phantom{\rho}\beta}\;.
\end{align}
Imposing kinetic equilibrium, the in-medium $L\Phi  \to
 L\Phi$ scattering rate \eqref{C33} can then be written
\begin{align}
  \Tdu{[\Gamma^{0}_{s_1}
     (L\Phi  \to  L\Phi; \mathbf{p}_1,\mathbf{q}_1)]}{k}{k'}{}{}
  &=\ \sum_{s_2}\frac{[n^{L}_{s_2}]_{l'}^{\phantom{l'}l}}
  {n^{L}_{\mathrm{eq}}}\int_{\mathbf{p}_2,\,\mathbf{q}_2}
      e^{-(|\mathbf{p}_2|+|\mathbf{q}_2|)/T}
      \nonumber \\&\qquad \times \: 
      \Tdu{[\Gamma^{0}_{s_1s_2}
      (L\Phi  \to  L\Phi; \mathbf{p}_1,\mathbf{p}_2,
      \mathbf{q}_1,\mathbf{q}_2)]}{l}{l'}{k}{k'}\; ,
\end{align}
using which we can define the thermally-averaged scattering rate as 
\begin{equation}
  \Tdu{[\gamma(L\Phi  \to  L\Phi)]}{l}{l'}{k}{k'}\ =\
  \frac{1}{\mathcal{V}_4}
  \sum_{I_1I_2J_1J_2}
  \int_{\mathbf{p}_{1,2},\mathbf{q}_{1,2}}
      e^{-(|\mathbf{p}_2|+|\mathbf{q}_2|)/T}
      \Tdu{[\Gamma^{0}_{--}
      (L\Phi \: \to \: L\Phi; \mathbf{p}_1,\mathbf{p}_2,
      \mathbf{q}_1,\mathbf{q}_2)]}{l}{l'}{k}{k'},
\end{equation}
in which  $I_{1,2}$, $J_{1,2}$ are  respectively the lepton  and Higgs
isospin indices,  and the flavour  indices have been  reordered.  
The rank-4 tensor rate is defined following \eqref{C34}:
\begin{align}
  \Tdu{[\Gamma^{0}_{s_1s_2}
    (L\Phi  \to  L\Phi; \mathbf{p}_1,\mathbf{p}_2,\mathbf{q}_1,
    \mathbf{q}_2)]}{l}{l'}{k}{k'}
  \ &=\ \Tud{[\mathcal{T}^{0}_{s_2s_1}
    (L\Phi  \to  L\Phi;
    \mathbf{p}_2,\mathbf{p}_1,\mathbf{q}_2,\mathbf{q}_1)]}{k'}{l}{}{}
    \nonumber\\&\qquad\times \:
    \Tdu{[\mathcal{T}^{0}_{s_1s_2}
    (L\Phi  \to  L\Phi;
    \mathbf{p}_1,\mathbf{p}_2,\mathbf{q}_1,\mathbf{q}_2)]}{k}{l'}{}{}\;.
\end{align}
Here, the transition  amplitudes are still defined in  a general basis
from \eqref{eq:scatT1} and \eqref{eq:scatT2}:
\begin{align}
  &\Tdu{[\mathcal{T}^{0}_{s_1s_2}
     (L\Phi  \to L\Phi;
     \mathbf{p}_1,\mathbf{p}_2,\mathbf{q}_1,\mathbf{q}_2)]}{k}{l'}{}{}
  \ =\ \braket{\mathbf{p}_1,s_1,L_k;\mathbf{q}_1,\Phi^{\dag}|\,
    \mathcal{T}^{1\,\dag}\,|
    \mathbf{p}_2,s_2,L^{l'};\mathbf{q}_2,\Phi^{\dag}}
  \nonumber\\ & \
  =\ -i\int_{k}
  [(2E_{L}(\mathbf{p}_1))^{-1/2}]^{i}_{\phantom{i}j}
  [(2E_{L}(\mathbf{p}_2))^{-1/2}]_{a}^{\phantom{a}b}
  (2E_{\Phi}(\mathbf{q}_1))^{-1/2}
  (2E_{\Phi}(\mathbf{q}_2))^{-1/2}
  \nonumber\\& \ \times 
  \DiFud{k  -  p_1  -  q_1}{j}{m}{\gamma}{\epsilon}
  \DiFdu{k  -  p_2  -  q_2}{b}{c}{\delta}{\sigma}
  \nonumber  \\& \ \times 
  [\bar{u}(\mathbf{p}_1,s_1)]^m_{\phantom{m}k}
  U^{\rho}_{\phantom{\rho}\epsilon}
  [i\widehat{\Delta}^{0}_{\mathrm{F},\,N}(k)]_{\rho}
  U_{\rho}^{\phantom{\rho}\sigma}
  [u(\mathbf{p}_2,s_2)]_{c}^{\phantom{c}l'}
  [\mathcal{M}^{0}(N  \to L\Phi)]_i^{\phantom{i}\gamma}
  [\mathcal{M}^{0}(L\Phi \to N)]^{a}_{\phantom{a}\delta} \; ,  
\\[3mm]
  &\Tud{[\mathcal{T}^{0}_{s_2s_1}
     (L\Phi  \to L\Phi;
     \mathbf{p}_2,\mathbf{p}_1,\mathbf{q}_2,\mathbf{q}_1)]}{k'}{l}{}{}
  \ =\ \braket{\mathbf{p}_2,s_2,L_{l};\mathbf{q}_2,\Phi^{\dag}|\,
    \mathcal{T}^{1}\,|
    \mathbf{p}_1,s_1,L^{k'};\mathbf{q}_1,\Phi^{\dag}}
  \nonumber\\& \
  =\ +i\int_{k'} 
  [(2E_{L}(\mathbf{p}_1))^{-1/2}]_{i'}^{\phantom{i'}j'}
  [(2E_{L}(\mathbf{p}_2))^{-1/2}]^{a'}_{\phantom{a'}b'}
  (2E_{\Phi}(\mathbf{q}_2))^{-1/2}
  (2E_{\Phi}(\mathbf{q}_1))^{-1/2}
  \nonumber\\& \ \times 
  \DiFdu{k'  -  p_1  -  q_1}{{j'}}{m'}{\gamma'}{\epsilon'}
  \DiFud{k'  -  p_2 -  q_2}{{b'}}{c'}{\delta'}{\sigma'}
  \nonumber \\&  \ \times 
  [\bar{u}(\mathbf{p}_2,s_2)]^{c'}_{\phantom{c'}l}
  U_{\rho'}^{\phantom{\rho'}\epsilon'}
  [i\widehat{\Delta}^{0}_{\mathrm{F},\,N}(k')]^{\rho'}
  U^{\rho'}_{\phantom{\rho'}\sigma'}
  [u(\mathbf{p}_1,s_1)]_{m'}^{\phantom{m'}k'}
  [\mathcal{M}^{0}(N  \to  L\Phi)]^{i'}_{\phantom{i'}\gamma'}
  [\mathcal{M}^{0}(L\Phi  \to N)]_{a'}^{\phantom{a'}\delta'} .
\end{align}
Rotating the matrix elements and four-momentum delta functions to
the heavy-neutrino mass eigenbasis, and working in the massless limit 
for the charged-leptons and Higgs, we obtain
\begin{align}
  &\Tdu{[\widehat{\mathcal{T}}^{0}_{s_1s_2}
     (L\Phi  \to  L\Phi;
     \mathbf{p}_1,\mathbf{p}_2,\mathbf{q}_1,\mathbf{q}_2)]}{k}{l'}{}{}
  \ =\
  -i(2|\mathbf{p}_1|)^{-1/2}
  (2|\mathbf{p}_2|)^{-1/2}
  (2|\mathbf{q}_1|)^{-1/2}
  (2|\mathbf{q}_2|)^{-1/2}
  \nonumber\\ & \qquad \qquad \times\:
  (2\pi)^4\delta^{(4)}(p_1  +  q_1  -  p_2  -  q_2) \: 
  \bar{u}(\mathbf{p}_1,s_1)
  [i\widehat{\Delta}^{0}_{\mathrm{F},\,N}(k)]_{\alpha}
  u(\mathbf{p}_2,s_2)
  \nonumber\\ & \qquad \qquad \times\:
  [\widehat{\mathcal{M}}^{0}(N  \to  L\Phi)]_k^{\phantom{k}\alpha}
  [\widehat{\mathcal{M}}^{0}(L\Phi  \to  N)]^{l'}_{\phantom{l'}\alpha}
 \;,\\
  &\Tud{[\widehat{\mathcal{T}}^{0}_{s_2s_1}
     (L\Phi  \to  L\Phi;
     \mathbf{p}_2,\mathbf{p}_1,\mathbf{q}_2,\mathbf{q}_1)]}{k'}{l}{}{}
  \ =\
  +i(2|\mathbf{p}_2|)^{-1/2}
  (2|\mathbf{p}_1|)^{-1/2}
  (2|\mathbf{q}_2|)^{-1/2}
  (2|\mathbf{q}_1|)^{-1/2}\nonumber\\&\qquad \times\:
  (2\pi)^4\delta^{(4)}(p_2  +  q_2  -  p_1  -  q_1) \:
  \bar{u}(\mathbf{p}_2,s_2)
  [i\widehat{\Delta}^{0}_{\mathrm{F},\,N}(k)]^{\beta}
  u(\mathbf{p}_1,s_1)
  \nonumber\\&\qquad \qquad \times\:
  [\widehat{\mathcal{M}}^{0}(N  \to  L\Phi)]^{k'}_{\phantom{k'}\beta}
  [\widehat{\mathcal{M}}^{0}(L\Phi  \to  N)]_{l}^{\phantom{l}\beta}\;,
\end{align}
where the four-momentum $k  =  p_1  +  q_1  =  p_2  + q_2$.
Finally, we  substitute for the  resummed Yukawas and  propagators and
extract the resonant part of the amplitude by writing
\begin{equation}
  [i\widehat{\Delta}_{\mathrm{F},\,N}(k)]_{\alpha}\
  \simeq\ {\mathrm{P_R}}(\slashed{k} \: + \: m_{\alpha}){\mathrm{P_L}}
  [i\widehat{\Delta}_{\mathrm{F},N}(k)]_{\mathrm{res},\,\alpha} \
  =\ {\mathrm{P_R}}\slashed{k}
  [i\widehat{\Delta}_{\mathrm{F},N}(k)]_{\mathrm{res},\,\alpha}\;.
\end{equation}
After  performing  the   helicity  summations  via  appropriate  Fierz
rearrangement, we obtain the following set of $\Delta L  = 0$ and $\Delta L
 = 2$ thermally-averaged scattering rates:
  \begin{align}
     \label{eq:scatrates1}
    \Tdu{[\widehat{\gamma}(L\Phi  \to  L\Phi)]}{k}{l}{m}{n}\  &= \
    \int_{L_1\Phi_1L_2\Phi_2}g_{L_1}g_{\Phi_1}g_{L_2}g_{\Phi_2}(2p_{L_2}\cdot p_{N})
    (2p_{N}\cdot p_{L_1})
    \nonumber \\ & \quad
    \times \: \big([i\widehat{\Delta}_{\mathrm{F},\,N}(p_N)]_{\alpha}
    [i\widehat{\Delta}_{\mathrm{F},\,N}(p_N)]^{\beta}\big)_{\mathrm{res}}
    \chrcst{k}{\beta}\chrct{l}{\alpha}
    \chr{m}{\alpha}\chrs{n}{\beta}\;,
    \\
    \Tdu{[\widehat{\gamma}(L^{\tilde{c}}\Phi^{\tilde{c}}  \to 
      L^{\tilde{c}}\Phi^{\tilde{c}})]}{k}{l}{m}{n}\ &
    = \ \int_{L_1\Phi_1L_2\Phi_2}g_{L_1}g_{\Phi_1}g_{L_2}g_{\Phi_2}
    (2p_{L_2}\cdot p_{N})
    (2p_{N}\cdot p_{L_1})
    \nonumber \\ & \quad
    \times \: \big([i\widehat{\Delta}_{\mathrm{F},\,N}(p_N)]_{\alpha}
    [i\widehat{\Delta}_{\mathrm{F},\,N}(p_N)]^{\beta}\big)_{\mathrm{res}}
    \chr{k}{\alpha}\chrs{l}{\beta}
    \chrcst{m}{\beta}\chrct{n}{\alpha}\;,
    \\
    \Tdu{[\widehat{\gamma}(L\Phi  \to  
      L^{\tilde{c}}\Phi^{\tilde{c}})]}{k}{l}{m}{n}\ &
    = \ \int_{L_1\Phi_1L_2\Phi_2}g_{L_1}g_{\Phi_1}g_{L_2}g_{\Phi_2}
    (2p_{L_2}\cdot p_{N})
    (2p_{N}\cdot p_{L_1})
    \nonumber \\ & \quad
    \times \: \big([i\widehat{\Delta}_{\mathrm{F},\,N}(p_N)]_{\alpha}
    [i\widehat{\Delta}_{\mathrm{F},\,N}(p_N)]^{\beta}\big)_{\mathrm{res}}
    \chrcst{k}{\beta}\chrct{l}{\alpha}
    \chrcst{m}{\beta}\chrct{n}{\alpha}\;,
    \\
    \Tdu{[\widehat{\gamma}(L^{\tilde{c}}\Phi^{\tilde{c}}  \to 
    L\Phi)]}{k}{l}{m}{n}\ &
    = \ \int_{L_1\Phi_1L_2\Phi_2}g_{L_1}g_{\Phi_1}g_{L_2}g_{\Phi_2}
    (2p_{L_2}\cdot p_{N})
    (2p_{N}\cdot p_{L_1})
    \nonumber \\ & \quad
    \times \: \big([i\widehat{\Delta}_{\mathrm{F},\,N}(p_N)]_{\alpha}
    [i\widehat{\Delta}_{\mathrm{F},\,N}(p_N)]^{\beta}\big)_{\mathrm{res}}
    \chr{k}{\alpha}\chrs{l}{\beta}
    \chr{m}{\alpha}\chrs{n}{\beta}\;,
     \label{eq:scatrates2}
\end{align}
in  which the  indices for  the  $\widetilde{T}$-transformed processes
have  been relabeled,  as in~\eqref{eq:relab}.  Tracing over lepton flavour  indices, the scattering rates relevant to
the heavy neutrino transport equations are, for instance,
\begin{align}
  \label{eq:Nscatrate}
  [\widehat{\gamma}(L\Phi\to L\Phi)]_{\alpha}^{\phantom{\alpha}\beta} \ &
  = \ \int_{L_1\Phi_1L_2\Phi_2}g_{L_1}g_{\Phi_1}g_{L_2}g_{\Phi_2}
    (2p_{L_2}\cdot p_{N})
   (2p_{N}\cdot p_{L_1})
    \nonumber \\ & \qquad
    \times \: \big([i\widehat{\Delta}_{\mathrm{F},\,N}(p_N)]_{\alpha}
    [i\widehat{\Delta}_{\mathrm{F},\,N}(p_N)]^{\beta}\big)_{\mathrm{res}}
  \chrcst{k}{\beta}\chrct{k}{\alpha}
  \chr{l}{\alpha}\chrs{l}{\beta} \; ,
\end{align}
in which  the heavy-neutrino flavour indices $\alpha$  and $\beta$ are
\emph{not} summed over. These rates  have been
used  in Section~\ref{sec:4.4}, along  with the  NWA for  the resonant
part of the Feynman  propagators, to derive the flavour-covariant rate
equations for scattering.

\section{Form Factors for LFV Decay Rates}
\label{app:loop}

For  completeness,  we  list  below  various  form  factors
appearing  in   the  expressions  \eqref{brmue},   \eqref{brmu3e}  and
\eqref{ratemue},    which    follow    from    the    results    given
in~\cite{Ilakovac:1994kj, Ilakovac:1995km}:
\begin{eqnarray} 
G_\gamma^{\mu e} &=& \sum_{\alpha} B_{e\alpha}
  B^*_{\mu \alpha}
  G_\gamma(x_{N_\alpha}) \; , \label{D1} \\
  F_\gamma^{\mu e} &=&
  \sum_{\alpha} B_{e\alpha} B^*_{\mu \alpha}
  F_\gamma(x_{N_\alpha}) \; ,
  \label{D2}\\
  F_Z^{\mu e} &=& \sum_{\alpha} B_{e\alpha}
  B^*_{\mu \alpha}
  \left[F_Z(x_{N_\alpha}) \; + \; 2G_Z(x_{N_\alpha},0)\right] \; ,
  \label{D3}\\
  F_{\rm Box}^{\mu euu} &=& \sum_\alpha B_{e\alpha}
  B^*_{\mu \alpha}
  \left[H_{\rm Box}(x_{N_\alpha},0) \; - \; H_{\rm Box}(0,0)\right] \; ,\\
  F_{\rm Box}^{\mu edd} &=& -\sum_\alpha B_{e\alpha}
  B^*_{\mu \alpha}
  \left[F_{\rm Box}(x_{N_\alpha},0) \; - \; F_{\rm Box}(0,0)\right] \; ,\\
  F_{\rm Box}^{\mu eee} &=& 2 \: \sum_\alpha B_{e\alpha}
  B^*_{\mu \alpha}
  \left[F_{\rm Box}(x_{N_\alpha},0) \; - \; F_{\rm Box}(0,0)\right] \; ,
  \label{D6}
\end{eqnarray}
where  $B_{l\alpha}$   denote  the  elements  of  the
light-heavy neutrino mixing matrix [cf.~\eqref{mixing}], and 
$x_{N_\alpha}\equiv (m_{N_\alpha}/M_W)^2$.  
In \eqref{D3} and
\eqref{D6}, we have ignored the ${\cal O}(\|\mat{\xi}\|^4)$ terms, 
which is a good approximation, since $\|\mat{\xi}\|^4\ll 1$ for all the benchmark points 
in our case (see Section~\ref{sec:6.1}).

The  loop  functions  appearing in  \eqref{D1}--\eqref{D6}  are
given by
\begin{eqnarray}
G_\gamma(x) &=& -\frac{x(2x^2+5x-1)}{4(1-x)^3}
  \; - \; \frac{3x^3}{2(1-x)^4}\ln x \; ,\\
  F_\gamma(x) &=& \frac{x(7x^2-x-12)}{12(1-x)^3}
  \; - \; \frac{x^2(x^2-10x+12)}{6(1-x)^4}\ln x \; , \\
  F_Z(x) &=& -\frac{5x}{2(1-x)}
  \; - \; \frac{5x^2}{2(1-x)^2}\ln x \; ,\\
  G_Z(x,0) &=& -\frac{x}{2(1-x)}\ln x \; ,\\
  H_{\rm Box}(x,0) &=& \frac{4}{1-x} \; + \; \frac{4x}{(1-x)^2}\ln x \; ,\\
  F_{\rm Box}(x,0) &=& \frac{1}{1-x} \; + \; \frac{x}{(1-x)^2}\ln x\; ,
\end{eqnarray}
with  the   limiting  values  $H_{\rm  Box}(0,0)  =   4$  and  $F_{\rm
Box}(0,0)=1$.

\begin{table}[t!]
  \begin{center}
    \begin{tabular}{c|c|c|c|c}\hline\hline
      Nucleus ($_Z^A X$) & $V^{(p)}$ & $V^{(n)}$ & $D$ &
      $\Gamma_{\rm capt}~(10^6~{\rm sec}^{-1})$\\ \hline
      $_{22}^{48}$Ti & 0.0396 & 0.0468 & 0.0864 & 2.59\\
      $_{79}^{197}$Au & 0.0974 & 0.146 & 0.189 & 13.07\\
      $_{82}^{208}$Pb & 0.0834 & 0.128 & 0.161 & 13.45\\
      \hline\hline
    \end{tabular}
  \end{center}
    \caption{The nuclear  form factors  and capture rates  for various
    nuclei of interest.   The numbers  were  taken from~\cite{Alonso:2012ji}  (see
    also~\cite{nucl1, nucl2}).}\label{tabn}
\end{table}

The  nuclear form factors  $D,V^{(p)}, V^{(n)}$  and the  capture rate
$\Gamma_{\rm  capt}$ appearing  in \eqref{ratemue}  are  summarized in
Table~\ref{tabn}  for  various  nuclei  relevant  to  $\mu\to e  $
conversion searches.


\end{document}